\pdfoutput=1
\documentclass[12pt,a4paper,twoside]{book}\input{twoside.def}

\begin{document}

\input{en.def}




\input{pre.def}



\begin{titlepage}
\begin{center}
{\LARGE Universidade de S\~ao Paulo}\\\bigskip
{\Large \it Instituto de F\'isica}
\end{center}

\vspace{0.5cm}
\begin{center}
 { \Large \textbf{\lang{Linear perturbations of black holes: stability, quasi-normal modes and tails.}{Perturba\c{c}\~oes lineares de buracos negros: estabilidade, modos quase-normais e caudas.}}}
\end{center}

\vspace{1.0cm}
\begin{center}
{\large \textbf{\lang{Alexander Zhidenko}{Olexandr Zhydenko}
} }
\end{center}

\vspace{1 cm}
\noindent
\begin{flushright}
Orientador : Prof. Dr. \'Elcio Abdalla \\
\end{flushright}
\vspace{1 cm}
\begin{flushright}
\parbox{3in}{
\begin{center}
Tese de doutorado apresentada no\\
Instituto de F\'isica para a obten\c{c}\~ao \\
do t\'itulo de Doutor em Ci\'encias.\\
\end{center}}
\end{flushright}
\vspace{0.5 cm}
\begin{flushleft}
\textbf{Comiss\~ao Examinadora}\\
Prof. Dr. \'Elcio Abdalla (Orientador, IFUSP) \\
Prof. Dr. Alberto Vasquez Saa (IMECC, Unicamp)\\
Prof. Dr. Laerte Sodr\'e Junior (IAGUSP)\\
Prof. Dr. Vilson Tonin Zanchin (UFABC)\\
Prof. Dr. Patr\'{\i}cio An\'{\i}bal Letelier Sotomayor (IMECC, Unicamp)
\end{flushleft}

\begin{center}
\vspace{1.5cm}

{S\~ao Paulo \\ 2009 }
\end{center}

\end{titlepage}

\newpage\thispagestyle{empty}
\begin{center}
\mbox{}\\
\mbox{}\\
\mbox{}\\
\mbox{}\\
\mbox{}\\
\mbox{}\\
\mbox{}\\
\mbox{}\\
\mbox{}\\
\mbox{}\\
\mbox{}\\
\mbox{}\\
\mbox{}\\
\mbox{}\\
\mbox{}\\
\mbox{}\\
\mbox{}\\
\mbox{}\\
{\bf FICHA CATALOGR\'AFICA\\
Preparada pelo Servi\c{c}o de Biblioteca e\\
Informa\c{c}\~ao\\
do Instituto de F\'{\i}sica da Universidade de S\~ao\\
Paulo}\\
\begin{tabular}{|l|}\hline\\
Zhydenko, Olexandr\\
~~~~~~Perturba\c{c}\~oes lineares de buracos negros: estabilidade,\\
~~~modos quase normais e caudas, S\~ao Paulo, 2009.\\
\\
~~~~~~Tese (Doutorado) -- Universidade de S\~ao Paulo.\\
~~~Institutos de F\'{\i}sica - Depto. de F\'{\i}sica Matem\'atica\\
\\
~~~~~~~Orientador: Prof. Dr. \'Elcio Abdalla\\
\\
~~~~~~\'Area de Concentra\c{c}\~ao: F\'{\i}sica\\
\\
~~~~~~Unitermos: 1. Relatividade (F\'{\i}sica); 2. F\'{\i}sica Te\'orica;\\
~~~3. Teoria de Campos e Ondas; 4. F\'{\i}sica computacional\\
\\
\\
USP/IF/SBI-046/2009\\
\hline
\end{tabular}
\end{center}

\newpage\setcounter{page}{1}

\langabstract{%
Black holes have their proper oscillations, which are called the quasi-normal modes. The proper oscillations of astrophysical black holes can be observed in the nearest future with the help of gravitational wave detectors. Quasi-normal modes are also very important in the context of testing of the stability of black objects, the anti-de Sitter/Conformal Field Theory (AdS/CFT) correspondence and in higher dimensional theories, such as the brane-world scenarios and string theory.

This dissertation reviews a number of works, which provide a thorough study of the quasi-normal spectrum of a wide class of black holes in four and higher dimensions for fields of various spin and gravitational perturbations. We have studied numerically the dependance of the quasi-normal modes on a number of factors, such as the presence of the cosmological constant, the Gauss-Bonnet parameter or the aether in the space-time, the dependance of the spectrum on parameters of the black hole and fields under consideration. By the analysis of the quasi-normal spectrum, we have studied the stability of higher dimensional Reissner-Nordstr\"om-de Sitter black holes, Kaluza-Klein black holes with squashed horizons, Gauss-Bonnet black holes and black strings. Special attention is paid to the evolution of massive fields in the background of various black holes. We have considered their quasi-normal ringing and the late-time tails.

In addition, we present two new numerical techniques: a generalisation of the Nollert improvement of the Frobenius method for higher dimensional problems and a qualitatively new method, which allows to calculate quasi-normal frequencies for black holes, which metrics are not known analytically. Also we considered a possibility of construction of the acoustic analogue of the Schwarzschild black hole.
}{%
Buracos negros t\^em as suas oscila\c{c}\~oes pr\'oprias, que s\~ao chamadas modos quase-normais. As oscila\c{c}\~oes pr\'oprias de buracos negros astrof\'isicos podem ser observadas no futuro mais pr\'oximo com a ajuda de detectores de ondas gravitacionais. Os modos quase-normais s\~ao tamb\'em muito importantes no contexto de teste da estabilidade de objetos negros, da correspond\^encia anti-de Sitter/Teoria Campos Conformes (AdS/CFT) e nas teorias em dimens\~oes mais altas, como os cen\'arios de mundo-brana e teoria das cordas.

Esta tese rev\^e um n\'umero de trabalhos, que fornecem um estudo completo do espectro quase-normal de grande classe de buracos negros em quatro e mais altas dimens\~oes para campos de v\'arios spins e perturba\c{c}\~oes gravitacionais. Foi estudada numericamente a depend\^encia dos modos quase-normais sobre um n\'umero de fatores, como a presen\c{c}a da constante cosmol\'ogica, o par\^ametro de Gauss-Bonnet ou o a\'eter no espa\c{c}o-tempo, a depend\^encia do espectro sobre os par\^ametros do buraco negro e os campos em considera\c{c}\~ao. Pela an\'alise do espectro quase-normal, foi estudada a estabilidade de buracos negros Reissner-Nordstr\"om-de Sitter em dimens\~oes mais altas, buracos negros de Kaluza-Klein  com horizontes esmagados, buracos negros de Gauss-Bonnet e cordas negras. Uma aten\c{c}\~ao especial foi prestada \`a evolu\c{c}\~ao de campos massivos no contexto de v\'arios buracos negros. Foram considerados os seus toques quase-normais e as caudas de tempo tardio.

Al\'em disso, foram apresentadas duas novas t\'ecnicas num\'ericas: uma generaliza\c{c}\~ao da melhora de Nollert para do m\'etodo de Frobenius para problemas em dimens\~oes mais altas e um m\'etodo qualitativamente novo, que permite calcular freq\"u\^encias quase-normais de buracos negros, cujas m\'etricas n\~ao s\~ao conhecidas analiticamente. Tamb\'em foi considerada uma possibilidade da constru\c{c}\~ao do an\'alogo ac\'ustico do buraco negro de Schwarzschild.
}%

\tableofcontents

\listoffigures

\langacknowledgement

\langpar{First, I would like to thank my supervisor Prof. Dr. \'Elcio Abdalla for all his help and support during my PhD study, for stimulating discussions and for the collaboration. I appreciate very much his invaluable help when reading and correcting Portuguese in the dissertation.}{Em primeiro lugar, eu gostaria de agradecer a meu orientador Prof. Dr. \'Elcio Abdalla por toda sua ajuda e apoio durante meu Doutorado, pelas discuss\~oes estimulantes e pela colabora\c{c}\~ao. Agrade\c{c}o muito pela sua inestim\'avel ajuda e por ter lido e corrigido o portugu\^es na tese.}

\langpar{My special thanks are to Dr. Roman Konoplya, whose contribution to all the presented results is impossible to exaggerate. He was (and stays) not only the best collaborator, but also a good friend, whose help and advise are very important to me. He also helped me very much with preparing of the text of the dissertation.}{Os meus agradecimentos especiais s\~ao para o Dr. Roman Konoplya, cuja contribui\c{c}\~ao para todos os resultados apresentados \'e imposs\'ivel exagerar. Ele foi (e \'e) n\~ao somente o melhor colaborador, mas tamb\'em um bom amigo, cujas ajuda e conselhos s\~ao muito importantes para mim. Ele tamb\'em me ajudou muito com a prepara\c{c}\~ao do texto da tese.}

\langpar{Next, I thank B\'arbara Idino Konoplya for all her help and, especially, for reading the first draft of the dissertation and correcting a lot of my mistakes in Portuguese.}{A seguir, eu agrade\c{c}o \`a B\'arbara Idino Konoplya por toda sua ajuda e, especialmente, por ler o primeiro esbo\c{c}o da tese e corrigir a maior parte dos meus erros em portugu\^es.}

\langpar{Also I would like to thank Prof. Dr. Carlos Molina for the collaboration and for helping me with adaptation to the Brazilian lifestyle.}{Eu tamb\'em gostaria de agradecer ao Prof. Dr. Carlos Molina pela colabora\c{c}\~ao e pela ajuda com a minha adapta\c{c}\~ao ao estilo de vida brasileiro.}

\langpar{I am grateful to my collaborators \'Elcio Abdalla, K.~H.~C. Castello-Branco, Hideki Ishihara, Panagiota Kanti, Masashi Kimura, Roman Konoplya, Carlos Molina, Keijo Murata, Jiro Soda for their contribution to the presented results.}{Agrade\c{c}o aos meus colaboradores \'Elcio Abdalla, K.~H.~C. Castello-Branco, Hideki Ishihara, Panagiota Kanti, Masashi Kimura, Roman~Konoplya, Carlos~Molina, Keijo Murata, Jiro Soda pelas suas contribui\c{c}\~oes aos resultados apresentados.}

\langpar{Many thanks to Jo\~ao da Silva Borges, Am\'elia Aparecida Ferrari Genova, Sybele Guedes de~Paulo Groff and Simone Toyoko Shinomiya for the help, to all the staff of the Department of the Mathematical Physics of IFUSP for their work and to the Funda\c{c}\~ao de Amparo \`a Pesquisa do Estado de S\~ao Paulo (FAPESP) for the financial support.}{Muito obrigado a Jo\~ao da Silva Borges, Am\'elia Aparecida Ferrari Genova, Sybele Guedes de~Paulo Groff e Simone Toyoko Shinomiya pela ajuda, a todos os funcion\'arios do Departamento de F\'isica Matem\'atica do IFUSP pelo trabalho deles e \`a Funda\c{c}\~ao de Amparo \`a Pesquisa do Estado de S\~ao Paulo (FAPESP) pelo apoio financeiro.}

\newpage 

\langchapter{Introduction}{Introdu\c{c}\~ao}
\langpar{The general relativity implies two qualitatively new phenomena, that cannot be described within Newton's gravity. First, there exist objects with masses so high, that the escape velocity from them exceeds the speed of light. These objects were called black holes. Second, the general relativity predicts gravitational waves, which appear due to finiteness of the gravitational interaction speed. These waves may be detected in the nearest future with the help of gravitational antennas. During last years there have been an advance in construction of the gravitational wave detectors, such as LIGO, GEO, VIRGO, TAMA and the planned space-based detector LISA, and the detection of gravitational waves is expected very soon \cite{Flanagan:1997td}.}{A relatividade geral implica dois fen\^omenos qualitativamente novos, que n\~ao podem ser descritos dentro da gravidade de Newton. Primeiro, existem objetos com massas t\~ao grandes, que a velocidade de escape deles excede a velocidade da luz. Esses objetos foram chamados buracos negros. Segundo, a relatividade geral prev\^e ondas gravitacionais, que aparecem por causa da limita\c{c}\~ao da velocidade de intera\c{c}\~oes gravitacionais. Essas ondas podem ser detectadas no futuro mais pr\'oximo com a ajuda de antenas gravitacionais. Durante os \'ultimos anos houve um avan\c{c}o na constru\c{c}\~ao dos detectores de ondas gravitacionais, como LIGO, GEO, VIRGO, TAMA e o detector planejado na base de espa\c{c}o LISA, e a detec\c{c}\~ao de ondas gravitacionais \'e esperada para logo \cite{Flanagan:1997td}.}

\langpar{One of the most promising sources of gravitational waves is collisions of black holes and/or of a star and a black hole. The result of these processes is a black hole with higher mass, which absorbs the gravitational waves. Therefore, the gravitational waves quickly decay and, at sufficiently late time, can be considered as small perturbations of the black hole metric. It means that one can study the linear perturbations, neglecting the higher order corrections. This approximation for the late-time behavior of the gravitational waves provides a good accuracy, which was checked by full non-linear simulations of collisions of two black holes \cite{PhysRevLett.71.2851}.}{Uma das fontes mais promissoras de ondas gravitacionais s\~ao choques entre buracos negros e/ou de uma estrela e um buraco negro. O resultado desses processos \'e um buraco negro com uma massa maior, que absorve as ondas gravitacionais. Por isso, as ondas gravitacionais rapidamente decaem e, no tempo suficientemente tarde, podem ser consideradas como pequenas perturba\c{c}\~oes da m\'etrica do buraco negro. Isto significa que pode-se estudar as perturba\c{c}\~oes lineares, negligenciando as corre\c{c}\~oes das ordens maiores. Essa aproxima\c{c}\~ao para o comportamento das ondas gravitacionais no tempo tardio fornece uma boa precis\~ao, que foi verificada por simula\c{c}\~oes completas n\~ao-lineares de colis\~oes de dois buracos negros \cite{PhysRevLett.71.2851}.}

\langpar{It turns out that the late-time behavior of the gravitational perturbations does not depend on the way they were induced. At the late-time stage of the evolution of gravitational perturbations we observe the damping oscillations, which give way to asymptotic tails at very late time. The damping oscillations are characterised by complex frequencies, which are called \emph{quasi-normal modes}. The real part of a complex frequency describes the actual oscillation frequency, while the imaginary part is the decay rate of the particular oscillation. The set of the quasi-normal frequencies form the spectrum. Being dependent only on the black hole parameters, the quasi-normal spectrum appears to be an important characteristic of a black hole, or as it is said its footprint. Therefore, detection of the quasi-normal modes allows us to determine the black hole parameters and compare them with those obtained by astronomical expectations \cite{Vishveshwara}.}{Acontece que o comportamento das perturba\c{c}\~oes gravitacionais no tempo tardio n\~ao depende do jeito que elas foram induzidas. Na etapa do tempo tardio da evolu\c{c}\~ao das perturba\c{c}\~oes gravitacionais observam-se oscila\c{c}\~oes decrescentes, que d\~ao lugar \`as caudas assint\'oticas no tempo muito tardio. As oscila\c{c}\~oes decrescentes s\~ao caracterizadas por freq\"u\^encias complexas, que s\~ao chamadas \emph{modos quase-normais}. A parte real de uma freq\"u\^encia complexa descreve a freq\"u\^encia real da oscila\c{c}\~ao, enquanto a parte imagin\'aria \'e a taxa de decaimento da oscila\c{c}\~ao particular. O conjunto das freq\"u\^encias quase-normais forma o espectro. Dependendo somente dos par\^ametros do buraco negro, o espectro quase-normal parece ser uma caracter\'istica importante de um buraco negro, ou como \'e dito a sua impress\~ao digital. Por isso, a detec\c{c}\~ao dos modos quase-normais permite determar os par\^ametros do buraco negro e compar\'a-los com os obtidos por expectativas astron\^omicas \cite{Vishveshwara}.}

\langpar{Despite they are well described by linear approximation, the quasi-normal modes, when detected, can be used to check the general relativity as a full non-linear theory. This allows to study some aspects of the gravitational theory that cannot be experimentally confirmed without considering the regime of strong gravity. A good example is a phenomenon of violation of the local Lorentz symmetry. If we assume that there exist the locally preferred state of rest at each point of the space-time, we can describe this state by a unit time-like vector field. This vector field is called ``aether'' \cite{Eling:2004dk}. The parameters of the theory of the aether can be bounded by post-Newton corrections found from astronomical observations. Nevertheless, these bounds do not answer the question if the aether exists or not. Thus, one of the possible ways to observe the effect of the aether is the determination of the corresponding shift in the quasi-normal spectrum of black holes.}{Apesar de serem bem descritos pela aproxima\c{c}\~ao linear, os modos quase-normais, quando detectados, podem ser usados para verificar a relatividade geral como uma teoria completamente n\~ao linear. Isto permite estudar alguns aspectos da teoria gravitacional que n\~ao podem ser confirmados experimentalmente sem considerar o regime de gravidade forte. Um bom exemplo \'e um fen\^omeno da viola\c{c}\~ao da simetria Lorentz local. Se assumir-se que existem o estado em repouso local preferido em cada ponto do espa\c{c}o-tempo, \'e poss\'ivel descrever este estado por um campo vetorial de unidade do tipo tempo. Este campo vetorial \'e chamado ``a\'eter'' \cite{Eling:2004dk}. Os par\^ametros da teoria do a\'eter podem ser limitados por corre\c{c}\~oes post-Newton achados a partir de observa\c{c}\~oes astron\^omicas. Todavia, esses limites n\~ao respondem \`a pergunta se o a\'eter existe ou n\~ao. Assim, um dos jeitos poss\'iveis de se obsevar o efeito do a\'eter \'e a determina\c{c}\~ao da desloca\c{c}\~ao correspondente no espectro quase-normal de buracos negros.}

\langpar{Another motivation for study of the black hole perturbations is checking of stability of the black holes. This is a very important property for higher dimensional theories, such as the brane-world scenarios and string theory \cite{Emparan:2008eg}. Since in higher than four dimensions there is no uniqueness theorem, stability may be the criteria which will select physical solutions among a variety of ``black objects'': black holes, black branes, black rings etc. It is easy to understand that the quasi-normal spectrum of a stable configuration contains only damping modes, while black objects, unstable under small perturbations, must contain at least one growing mode in their spectrum.}{Outra motiva\c{c}\~ao para estudar as perturba\c{c}\~oes de buracos negros \'e a verifica\c{c}\~ao da estabilidade deles. Isto \'e uma propriedade muito importante para teorias em dimens\~oes mais altas, como os cen\'arios de mundo-brana e a teoria das cordas \cite{Emparan:2008eg}. J\'a que em n\'umero de dimens\~oes maior do que quatro n\~ao h\'a o teorema de unicidade, a estabilidade pode ser o crit\'erio que selecionar\'a solu\c{c}\~oes f\'isicas entre v\'arios ``objetos negros'': buracos negros, branas negras, an\'eis negros etc. \'E f\'acil compreender que o espectro quase-normal de uma configura\c{c}\~ao est\'avel cont\'em somente modos decrescentes, enquanto os objetos negros, que s\~ao inst\'aveis com rela\c{c}\~ao \`as perturba\c{c}\~oes pequenas, devem conter pelo menos um modo crescente no seu espectro.}

\langpar{In addition, the quasi-normal spectrum can be interpreted in the context of the anti-de Sitter/Conformal Field Theory (AdS/CFT) correspondence \cite{Maldacena:1997re}. A black hole in the anti-de Sitter space-time corresponds to the thermal state in the dual Conformal Field Theory. The temperature of the thermal state coincides with the Hawking temperature of the black hole. Within the AdS/CFT duality the quasi-normal spectrum of the black hole in AdS corresponds to the poles of the retarded Green functions on the CFT side \cite{Horowitz:1999jd}. Due to the AdS/CFT correspondence we are able to calculate non-perturbative effects of the finite-temperature field theory at strong coupling by studying black holes in the anti-de Sitter space-times. The hydrodynamic parameters of the quark-gluon plasma were calculated within this approach, showing a good agreement with the results obtained in the experiments on the Relativistic Heavy Ion Collider (RHIC) \cite{Policastro:2002se}.}{Al\'em disso, o espectro quase-normal pode ser interpretado no contexto da correspond\^encia anti-de Sitter/Teoria Campos Conformes (AdS/CFT) \cite{Maldacena:1997re}. Um buraco negro no espa\c{c}o-tempo anti-de Sitter corresponde ao estado termal na Teoria dual de Campos Conformes. A temperatura do estado termal coincide com a temperatura de Hawking do buraco negro. Dentro da dualidade AdS/CFT o espectro quase-normal do buraco negro em AdS corresponde aos p\'olos das fun\c{c}\~oes retardadas de Green no lado de CFT \cite{Horowitz:1999jd}. Devido \`a correspond\^encia AdS/CFT \'e poss\'ivel calcular efeitos n\~ao-perturbativos da teoria de campos com a temperatura finita na regime de acoplamento forte estudando buracos negros nos espa\c{c}os-tempo anti-de Sitter. Os par\^ametros hidrodin\^amicos do plasma de quarks-gl\'uons foram calculados dentro desta aproxima\c{c}\~ao, mostrando um bom acordo com os resultados obtidos nos experimentos no Colisor Relativ\'istico de \'Ions Pesados (RHIC) \cite{Policastro:2002se}.}

\langpar{Within the present work we provide the detailed study of the linear perturbations of a wide class of black holes and find the quasi-normal spectrum for scalar, Dirac, Maxwell fields and the gravitational perturbations. We have studied quasi-normal spectra of these fields in four and higher dimensions, within the Einstein, Einstein-Aether theories, brane-world scenarios with and without the Gauss-Bonnet corrections. We have considered the influence of the cosmological constant on the quasi-normal frequencies and on the asymptotic behavior of high overtones. We have calculated quasi-normal modes of charged scalar and Dirac fields in the Kerr-Newman-de Sitter background.}{Dentro do trabalho presente fornece-se o estudo detalhado das perturba\c{c}\~oes lineares de uma classe grande de buracos negros e acha-se o espectro quase-normal dos campos escalar, Dirac, Maxwell e as perturba\c{c}\~oes gravitacionais. Foram Estudados os espectros quase-normais desses campos em quatro e maior n\'umero das dimens\~oes, dentro das teorias de Einstein, de Einstein-A\'eter, cen\'arios de mundo-brana com e sem as corre\c{c}\~oes de Gauss-Bonnet. Foi considerada a influ\^encia da constante cosmol\'ogica nas freq\"u\^encias quase-normais e no comportamento assint\'otico de sobretons altos. Foram calculados modos quase-normais de campos escalar carregado e de Dirac no contexto Kerr-Newman-de Sitter.}

\langpar{Also we have studied quasi-normal spectrum of the massive scalar field for Schwarzschild, Tangherlini and scalar hairy anti-de Sitter black holes and the massive vector field in the Schwarzschild background. We provided comprehensive discussions about properties of the quasi-normal spectrum and late-time tails of massive fields. We have considered appearing of the infinitely long-living oscillations (\emph{quasi-resonances}) for particular values of the masses of the black hole and of the field.}{Tamb\'em foi estudado o espectro quase-normal do campo escalar massivo para buracos negros de Schwarzschild, Tangherlini e anti-de Sitter com cabelo escalar e o campo vetorial massivo no contexto de Schwarzschild. Foram fornecidas discuss\~oes abrangentes sobre propriedades do espectro quase-normal e as caudas de tempo tardio de campos massivos. Foi considerado o aparecimento das oscila\c{c}\~oes da vida infinitamente longa (\emph{quase-resson\^ancias}) para determinados valores das massas do buraco negro e do campo.}

\langpar{We prove the instability of the higher dimensional Reissner-Nordstr\"om-de Sitter black holes for sufficiently large values of the black hole charge and of the cosmological constant. By the numerical analysis of the quasi-normal spectrum, we support the stability of the Kaluza-Klein black holes with squashed horizons, the instability region of the Gauss-Bonnet black holes and the long wavelength instability of black strings. We show the appearance of those instabilities in the time domain. Also we prove the stability of the massive scalar field in the background of the Kerr black hole.}{Foi comprovada a instabilidade dos buracos negros Reissner-Nordstr\"om-de Sitter para valores suficientemente grandes da carga do buraco negro e da constante cosmol\'ogica. Pela an\'alise num\'erica do espectro quase-normal, apoia-se a estabilidade dos buracos negros Kaluza-Klein com horizontes esmagados, a regi\~ao da instabilidade dos buracos negros Gauss-Bonnet e a instabilidade de cordas negras com rela\c{c}\~ao a ondas de comprimento longo. Mostra-se o aparecimento dessas instabilidades no dom\'inio de tempo. Tamb\'em foi comprovada a estabilidade do campo escalar massivo no contexto do buraco negro de Kerr.}

\langpar{We review the numerical methods used for our analysis: the time-domain simulation, the approximation by the P\"oschl-Teller potential, the WKB approach and the Frobenius method.}{Foram revisados os m\'etodos num\'ericos utilizados para a an\'alise: a simula\c{c}\~ao temporal, a aproxima\c{c}\~ao pelo potencial de P\"oschl-Teller, a aproxima\c{c}\~ao de WKB e o m\'etodo de Frobenius.}

\langpar{We propose a generalisation of the Nollert improvement of the Frobenius method for higher dimensional problems. This improvement provides better convergence of the numerical procedure, which is crucial for the case when the imaginary part of the quasi-normal frequency is much larger than its real part and for the calculation of the quasi-normal modes of massive fields.}{Foi proposta uma generaliza\c{c}\~ao da melhora de Nollert do m\'etodo de Frobenius para os problemas em dimens\~oes mais altas. Essa melhora fornece a melhor converg\^encia do procedimento num\'erico, que \'e crucial para o caso quando a parte imagin\'aria da freq\"u\^encia quase-normal \'e muito maior do que a sua parte real e para o c\'alculo dos modos quase-normais de campos massivos.}

\langpar{Also we describe a new numerical tool, which we developed for the calculation of the quasi-normal frequencies of a black hole, which metric is not known analytically, but can be found as a numerical solution of a set of differential equations. This technique supports the natural expectation that the dominant quasi-normal frequencies depend mainly on the region nearby the black hole and do not depend on the behavior of the metric at large distance. We have also checked this method by the time-domain integration.}{Tamb\'em foi descrito um novo instrumento num\'erico, que foi desenvolvido para o c\'alculo das freq\"u\^encias quase-normais de um buraco negro, cuja m\'etrica n\~ao \'e conhecida analiticamente, mas pode ser achada como uma solu\c{c}\~ao num\'erica de um conjunto de equa\c{c}\~oes diferenciais. Essa t\'ecnica apoia a expectativa natural que as freq\"u\^encias quase-normais dominantes dependem principalmente da regi\~ao pr\'oxima ao buraco negro e n\~ao dependem do comportamento da m\'etrica em grandes dist\^ancias. Esse m\'etodo foi verificado tamb\'em pela integra\c{c}\~ao temporal.}

\langpar{In addition, we discuss an interesting possibility of observation of the acoustic analogue of the Schwarzschild black hole in laboratories.}{Al\'em disso, discute-se uma possibilidade interessante da observa\c{c}\~ao do an\'alogo ac\'ustico do buraco negro Schwarzschild em laborat\'orios.}

\langpar{This work is aimed to clarify a number of questions about the behavior of perturbations of black holes. We provide a comprehensive analysis of the black hole perturbations in various theories in order to give the complete picture of quasi-normal modes, late-time tails and stability of black holes. The presented results were published in \cite{Zhidenko:2003wq,Konoplya:2004uk,CastelloBranco:2004nk,Konoplya:2004wg,Zhidenko:2005mv,Konoplya:2006br,Konoplya:2006gq,Konoplya:2006rv,Kanti:2006ua,Zhidenko:2006rs,Konoplya:2006ar,Konoplya:2007jv,Abdalla:2007dz,Konoplya:2007zx,Konoplya:2008ix,Ishihara:2008re,Zhidenko:2008fp,Konoplya:2008yy}\nocite{Konoplya:2008au,Zhidenko:2007sj}.}{Este trabalho visa esclarecer um n\'umero de perguntas sobre o comportamento de perturba\c{c}\~oes de buracos negros. Fornece-se uma an\'alise abrangente das perturba\c{c}\~oes de buracos negros em v\'arias teorias para dar o quadro completo de modos quase-normais, caudas de tempo tardio e estabilidade de buracos negros. Os resultados apresentados foram publicados em \cite{Zhidenko:2003wq,Konoplya:2004uk,CastelloBranco:2004nk,Konoplya:2004wg,Zhidenko:2005mv,Konoplya:2006br,Konoplya:2006gq,Konoplya:2006rv,Kanti:2006ua,Zhidenko:2006rs,Konoplya:2006ar,Konoplya:2007jv,Abdalla:2007dz,Konoplya:2007zx,Konoplya:2008ix,Ishihara:2008re,Zhidenko:2008fp,Konoplya:2008yy}\nocite{Konoplya:2008au,Zhidenko:2007sj}.}

\langchapter{Black hole perturbations}{Perturba\c{c}\~oes de um buraco negro}
\langsection{Equations of perturbations near a black hole}{Equa\c{c}\~ao de perturba\c{c}\~oes perto de um buraco negro}
\langpar{The dynamics of the general relativity in $D$ space-time dimensions is described by the Einstein-Hilbert action}{A din\^amica da relatividade geral em $D$ dimens\~oes de espa\c{c}o-tempo \'e descrita pela a\c{c}\~ao de Einstein-Hilbert}
\begin{equation}\label{E-Haction}
S=\int\sqrt{|g|}\left(\frac{1}{16\pi}(R-2\Lambda)+\mathcal{L}_M\right)d^Dx,
\end{equation}
\lang{where we use the geometrized unit system, so that the speed of light, $c$, and the gravitational constant, $\gamma$, are set equal to one. The metric signature is chosen as $(+---\ldots)$. $R$ is the Ricci scalar, $\mathcal{L}_M$ describes all matter fields $\phi_{(i)}$ appearing in the theory, and $\Lambda$ is the cosmological constant.}{onde foi usado o sistema de unidade geom\'etrico, para que a velocidade da luz, $c$, e a constante gravitacional, $\gamma$, fossem definidas iguais a um. A assinatura m\'etrica \'e escolhida como $(+---\ldots)$. $R$ \'e a curvatura escalar de Ricci, $\mathcal{L}_M$ descreve todos os campos de mat\'eria $\phi_{(i)}$ que aparecem na teoria e $\Lambda$ \'e a constante cosmol\'ogica.}

\langpar{The variation of the action (\ref{E-Haction}) allows to find the Einstein equations}{A varia\c{c}\~ao da a\c{c}\~ao (\ref{E-Haction}) permite encontrar as equa\c{c}\~oes de Einstein}
\begin{equation}\label{Einstein-equation}
R_{ab}-\frac{1}{2}R g_{ab}+\Lambda g_{ab}=8\pi T_{ab},
\end{equation}
\lang{and the field equations in the curved space-time}{e as equa\c{c}\~oes de campo no espa\c{c}o-tempo curvo}
\begin{equation}\label{field-equation}
\frac{\delta \mathcal{L}_M}{\delta \phi_{(i)}}=0.
\end{equation}
\lang{The energy-momentum tensor $T_{ab}$ in (\ref{Einstein-equation}) is defined as}{O tensor de energia-momento $T_{ab}$ em (\ref{Einstein-equation}) \'e definido como}
$$T_{ab}=-\frac{2}{\sqrt{-g}}\frac{\partial\sqrt{-g}\mathcal{L}_M}{\partial g^{ab}}.$$

\langpar{The general solution of the equations (\ref{Einstein-equation}, \ref{field-equation}) is some metric and fields in the background this metric}{A solu\c{c}\~ao geral das equa\c{c}\~oes (\ref{Einstein-equation}, \ref{field-equation}) \'e alguma m\'etrica e campos no contexto desta m\'etrica}
\begin{equation}\label{generalmetric}
ds^2 = g_{ab}(x)dx^adx^b, \qquad \phi_{(i)}=\phi_{(i)}(x).
\end{equation}

\langpar{We study a stationary black hole, for which $g_{ab}$ and $\phi_{(i)}$ do not depend on time in the appropriate coordinate system and gauge.}{Estudamos um buraco negro estacion\'ario, para o qual $g_{ab}$ e $\phi_{(i)}$ n\~ao dependem do tempo no sistema apropriado de coordenadas e calibre.}

%

\langpar{Let us consider perturbations of the metric and fields as a sum of the unperturbed background $g^0_{ab}$, $\phi^0_{(i)}$ and the actual perturbations $\delta g_{ab}$, $\delta\phi_{(i)}$}{Consideremos as perturba\c{c}\~oes da m\'etrica e dos campos como uma soma de um campo de fundo n\~ao-perturbado $g^0 _ {ab} $, $\phi^0 _ {i} $ e as perturba\c{c}\~oes reais $\delta g_{ab}$, $\delta\phi_{(i)}$}
\begin{eqnarray}\label{perturbing-equation}
\begin{array}{rcl}
g_{ab}&\rightarrow&g^0_{ab}+\delta g_{ab};\\
\phi_{(i)}&\rightarrow&\phi^0_{(i)}+\delta\phi_{(i)}.
\end{array}
\end{eqnarray}
\lang{The perturbations are assumed to be small, i. e. we neglect the contributions of order $\mathcal{O}(\delta g_{ab})^2$, ${\cal O}(\delta g_{ab}\delta\phi_{(i)})$, ${\cal O}(\delta\phi_{(i)})^2$ and higher.}{As perturba\c{c}\~oes s\~ao consideradas pequenas, isto \'e, descartam-se as contribui\c{c}\~oes da ordem $\mathcal {O} (\delta g _ {ab}) ^2$, $ {\cal O} (\delta g _ {ab} \delta\phi _ {i}) $ e $ {\cal O} (\delta\phi _ {i}) ^2$.}

\langpar{After substituting (\ref{perturbing-equation}) into (\ref{Einstein-equation}, \ref{field-equation}) and taking into account that $g^0_{ab}$ and $\phi^0_{(i)}$ satisfy the equations (\ref{Einstein-equation}, \ref{field-equation}) as well, we are able to find the set of linear equations for the perturbations $\delta g_{ab}$ and $\delta\phi_{(i)}$.}{Depois de se substituir (\ref{perturbing-equation}) em (\ref{Einstein-equation}, \ref{field-equation}) e levando-se em conta que $g^0_{ab} $ e $\phi^0_{i} $ tamb\'em satisfazem as equa\c{c}\~oes (\ref{Einstein-equation}, \ref{field-equation}), \'e poss\'ivel achar um conjunto de equa\c{c}\~oes lineares para as perturba\c{c}\~oes $\delta g _ {ab} $ e $\delta\phi _ {i} $.}

\langpar{In this chapter we consider the stationary spherically symmetric solution of the equation (\ref{Einstein-equation}) in 4 dimensions with $T_{ab}=0$, which describes a black hole in vacuum. This solution is given by the well-known metric}{Neste cap\'itulo consideramos a solu\c{c}\~ao estacion\'aria esfericamente sim\'etrica em 4 dimens\~oes com $T_{ab}=0$, que descreve um buraco negro no v\'acuo. Esta solu\c{c}\~ao \'e dada pela m\'etrica bem conhecida}
\begin{equation}\label{SdS-metric}
ds^2=f(r)dt^2-\frac{dr^2}{f(r)}-r^2(d\theta^2+\sin\theta^2 d\phi^2), \qquad f(r)=1-\frac{2M}{r}-\Lambda \frac{r^2}{3},
\end{equation}
\lang{where $M$ is the black hole mass and $\Lambda$ is the cosmological constant.}{onde $M$ \'e a massa do buraco negro e $\Lambda$ \'e a constante cosmol\'ogica.}

\langpar{For $\Lambda=0$ the metric (\ref{SdS-metric}) describes a Schwarzschild black hole. If the cosmological constant is positive ($\Lambda>0$), we obtain a Schwarzschild-de Sitter black hole, while for negative values of the cosmological constant ($\Lambda<0$) the black hole is called Schwarzschild-anti-de Sitter.}{Para $\Lambda=0$ a m\'etrica (\ref{SdS-metric}) descreve um buraco negro de Schwarzschild. Se a constante cosmol\'ogica for positiva ($\Lambda>0$), obtemos um buraco negro Schwarzschild-de Sitter, enquanto para valores negativos da constante cosmol\'ogica ($\Lambda<0$) o buraco negro chama-se Schwarzschild-anti-de Sitter.}

\langpar{Since $T_{ab}=0$, the field perturbations in such background are not coupled to the perturbations of the metric and, therefore, are equivalent to the test fields in the black hole background.}{Desde que $T _ {ab} =0$, as perturba\c{c}\~oes de campos em tal contexto n\~ao s\~ao ligadas \`as perturba\c{c}\~oes da m\'etrica e, por isso, s\~ao equivalentes para os campos de teste no contexto do buraco negro.}


\langpar{Let us describe the variable separation. As an example we consider the test scalar field which satisfies the Klein-Gordon equation in the curved space-time}{Descrevemos a separa\c{c}\~ao de vari\'aveis. Como um exemplo consideremos o campo escalar de teste que satisfaz \`a equa\c{c}\~ao de Klein-Gordon no espa\c{c}o-tempo curvo}
\begin{equation}\label{scalar-equation}
\frac{1}{\sqrt{-g}}\frac{\partial}{\partial x^a}g^{ab}\sqrt{-g}\frac{\partial\Phi(x)}{\partial x^b}=\frac{1}{f(r)}\frac{\partial^2\Phi}{\partial t^2}-\frac{\partial}{\partial r}f(r)\frac{\partial\Phi}{\partial r}-\frac{2f(r)}{r}\frac{\partial \Phi}{\partial r}-\frac{\Delta_{\theta,\phi}\Phi}{r^2}=0,
\end{equation}
\lang{where}{onde} $$\Delta_{\theta,\phi}=\frac{1}{\sin\theta}\frac{\partial}{\partial\theta}\sin\theta\frac{\partial}{\partial\theta}+\frac{1}{\sin^2\theta}\frac{\partial^2}{\partial\phi^2}$$ \lang{is the angular part of the Laplasian.}{\'e a parte angular do Laplasiano.}

\langpar{In order to separate the angular variables we choose the following ansatz:}{Para separar as vari\'aveis angulares foi escolido o seguinte ansatz:}
\begin{equation}\label{scalaranzatz}
\Phi(t,r,\theta,\phi)=\sum_{l=0}^\infty\sum_{m=-l}^l \frac{\Psi_{l,m}(t,r)}{r}Y_{l,m}(\theta,\phi),
\end{equation}
\lang{where $Y_{l,m}(\theta,\phi)$ are the spherical harmonics, which are eigenfunctions of the operator $\Delta_{\theta,\phi}$}{onde $Y_{l,m} (\theta,\phi)$ s\~ao os harm\^onicos esf\'ericos, que s\~ao fun\c{c}\~oes pr\'oprias do operador $\Delta_{\theta,\phi}$}
\begin{equation}
\Delta_{\theta,\phi}Y_{l,m}(\theta,\phi)=-l(l+1)Y_{l,m}(\theta,\phi).
\end{equation}
\lang{The integers $l\geq0$ and $|m|\leq l$ are called the multipole number and the azimuthal number respectively.}{Os n\'umeros inteiros $l\geq0$ e $|m|\leq l$ s\~ao chamados, respectivamente, o n\'umero de multipolo e o n\'umero de azimute.}

\langpar{After substitution (\ref{scalaranzatz}) into (\ref{scalar-equation}) we find the wave-like equation for the function $\Psi_{l,m}(t,r)$:}{Depois da substitui\c{c}\~ao (\ref{scalaranzatz}) em (\ref{scalar-equation}) encontramos a equa\c{c}\~ao de onda para a fun\c{c}\~ao $\Psi_{l,m}(t,r)$:}
\begin{equation}\label{wave-like}
\left(\frac{\partial^2}{\partial t^2}-\frac{\partial^2}{\partial r_\star^2}+V_l(r)\right)\Psi_{l,m}(t,r)=0,
\end{equation}
\lang{where the effective potential has the form}{onde o potencial efetivo tem a forma}
\begin{equation}\label{scalar-potential}
V_l(r)=f(r)\left(\frac{l(l+1)}{r^2}+\frac{f'(r)}{r}\right), \qquad l=0,1,2\ldots
\end{equation}
\lang{The variable $r_\star$ is called the \textit{tortoise coordinate}. It is defined up to an arbitrary constant as}{A vari\'avel $r_\star$ \'e a chamada \textit{coordenada tartaruga}. Ela \'e definida, a menos de uma constante arbitr\'aria, como}
\begin{equation}\label{tortoise-definition}
dr_\star=\frac{dr}{f(r)}.
\end{equation}

\langpar{If $\Lambda\geq0$ the tortoise coordinate maps all the region outside the black hole to the interval $(-\infty,\infty)$. In the anti-de Sitter space ($\Lambda<0$) the interval becomes $(-\infty, r_{\star\infty})$, where $r_{\star\infty}$ is a finite value that depends on the integration constant of (\ref{tortoise-definition}).}{Se $\Lambda\geq0$ a coordenada tartaruga faz o mapa de toda a regi\~ao fora do buraco negro no intervalo $(-\infty,\infty)$. No espa\c{c}o anti-de Sitter ($\Lambda<0$) o intervalo torna-se $(-\infty,r_{\star\infty})$, onde $r_{\star\infty}$ \'e um valor finito que depende da constante de integra\c{c}\~ao (\ref{tortoise-definition}).}

\langpar{We can also consider other test fields near the black hole. In the most cases, after separation of the angular variables, we are able to reduce the equations of motion for such fields (\ref{field-equation}) to the set of the wave-like equations of the form (\ref{wave-like}).}{Tamb\'em se podem considerar outros campos de teste perto do buraco negro. Na maior parte dos casos, depois da separa\c{c}\~ao das vari\'aveis angulares, somos capazes de reduzir as equa\c{c}\~oes do movimento para tais campos (\ref{field-equation}) para o conjunto das equa\c{c}\~oes de onda da forma (\ref{wave-like}).}

\langpar{One could find that the test massless Dirac field is described by the effective potentials \cite{Brill:1957fx}}{Pode-se achar que o campo de Dirac de teste sem massa \'e descrito pelos potenciais efetivos \cite{Brill:1957fx}}
\begin{equation}\label{Dirac-potential}
V_{D\pm}=f(r)\frac{\kappa_{\pm}^2}{r^2}\pm \frac{d}{dr_\star}\frac{\kappa_{\pm}\sqrt{f(r)}}{r}, \qquad \kappa_{\pm}=1,2,3\ldots
\end{equation}
\lang{and the electromagnetic field is described by the potential \cite{Mellor:1989ac}}{e o campo eletromagn\'etico \'e descrito pelo potencial \cite{Mellor:1989ac}}
\begin{equation}\label{EM-potential}
V_{EM}=f(r)\frac{l(l+1)}{r^2}, \qquad l=1,2,3\ldots
\end{equation}

\langpar{The equation (\ref{Einstein-equation}) allows us to find the equations of motion for the metric perturbations $\delta g_{ab}$ (\ref{perturbing-equation}). It is convenient to classify the tensor components $\delta g_{ab}$ with respect to the transformation law under rotations on the sphere around the black hole. They can be of \textit{scalar}, \textit{vector} and \textit{tensor} type, denoted by $s$, $v$ and $t$ respectively}{A equa\c{c}\~ao (\ref{Einstein-equation}) permite encontrar as equa\c{c}\~oes de movimento para as perturba\c{c}\~oes m\'etricas $\delta g_{ab}$ (\ref{perturbing-equation}). \'E conveniente classificar os componentes do tensor $\delta g_{ab}$ com rela\c{c}\~ao \`a lei de transforma\c{c}\~ao em rota\c{c}\~oes sobre a esfera em torno do buraco negro. Eles podem ser dos tipos \textit{escalar}, \textit{vetorial} e \textit{tensorial}, denotados por $s$, $v$ e $t$ respectivamente}
\begin{equation}\label{gravitational-classification}
\delta g_{ab} = \left(\begin{array}{cccc}
s & s & v & v \\
s & s & v & v \\
v & v & t & t \\
v & v & t & t \\
\end{array}\right).
\end{equation}

\langpar{In order to simplify the equations for the metric perturbations we can use the invariance under infinitesimal coordinate transformations. These transformations act as gauge transformations upon the metric perturbations}{Para simplificar as equa\c{c}\~oes de perturba\c{c}\~oes da m\'etrica utilizamos a invari\^ancia com rela\c{c}\~ao \`as transforma\c{c}\~oes infinitesimais das coordenadas. Essas transforma\c{c}\~oes atuam como transforma\c{c}\~oes de calibre nas perturba\c{c}\~oes m\'etricas}
$$x^a\rightarrow x^a+\delta x^a\qquad\Longrightarrow\qquad \delta g_{ab}\rightarrow \delta g_{ab}+\delta x_{a;b}+\delta x_{b;a}.$$
\lang{The gauge freedom allows to simplify perturbation equations \cite{PhysRev.108.1063}. After the Regge-Wheeler gauge fixing, the perturbations are described by the potentials \cite{Mellor:1989ac}}{A liberdade de calibre permite simplificar as equa\c{c}\~oes de perturba\c{c}\~ao \cite{PhysRev.108.1063}. Depois da fixa\c{c}\~ao do calibre de Regge-Wheeler, as perturba\c{c}\~oes s\~ao descritas pelos potenciais \cite{Mellor:1989ac}}
\begin{eqnarray}\label{polar-potential}
V_s(r)&=&\frac{2f(r)}{r^3}\frac{9M^3 + 3c^2Mr^2 + c^2(1 + c)r^3 + 3M^2(3cr - \Lambda r^3)}{(3M+cr)^2}, \\
\label{axial-potential}
V_v(r)&=&f(r)\left(\frac{l(l+1)}{r^2}-\frac{6M}{r^3}\right),
\\\nonumber &&c=\frac{l(l+1)}{2}-1, \qquad l=2,3,4\ldots
\end{eqnarray}
\lang{for the scalar and vector types respectively. The perturbations of the tensor type in 4 dimensions are eliminated by this gauge fixing. However, the perturbations of the higher dimensional black hole metric can be of all the three types \cite{Kodama:2003jz}.}{para os tipos escalar e vetorial respectivamente. As perturba\c{c}\~oes do tipo tensorial, em 4 dimens\~oes, s\~ao eliminadas por esta fixa\c{c}\~ao do calibre. Contudo, as perturba\c{c}\~oes da m\'etrica de buracos negros em dimens\~oes mais altas podem ser de todos os tr\^es tipos \cite{Kodama:2003jz}.}

\langpar{According to the Chandrasekhar \cite{Chandrasekhar} classification, the perturbations of vector type are called \emph{axial} because they impart the differential rotation to the black hole. Perturbations of scalar type are called \emph{polar}. They are related with infinitesimal deformations of the event horizon. Any perturbations of a spherically symmetric black hole in 4 dimensions can be divided to its polar and axial part in any gauge \cite{Nollert:1999ji}.}{Segundo a classifica\c{c}\~ao de Chandrasekhar \cite{Chandrasekhar}, as perturba\c{c}\~oes do tipo vetorial s\~ao chamadas \emph{axiais} porque elas transmitem a rota\c{c}\~ao diferencial ao buraco negro. As perturba\c{c}\~oes do tipo escalar s\~ao chamadas \emph{polares}. Eles e\~ao relacionados com deforma\c{c}\~oes infinit\'esimas do horizonte de eventos. Quaisquer perturba\c{c}\~oes de buraco negro esfericamente sim\'etrico em 4 dimens\~oes pode ser dividida em sua parte polar e axial em qualquer calibre \cite{Nollert:1999ji}.}

\langpar{One should note that the multipole numbers $l=0,1$ for gravitational perturbations and $l=0$ for the electromagnetic field perturbations are not dynamical. For instance, $l=0$ gravitational perturbations are spherically symmetric and, therefore, obey the Birkhoff theorem. They correspond to infinitesimal change of the black hole mass. Similarly, $l=0$ perturbations of the electromagnetic field and $l=1$ gravitational perturbations correspond, respectively, to the infinitesimal shift of the charge and the position of the black hole \cite{PhysRev.108.1063}.}{Observemos que os n\'umeros de multipolo $l=0,1$ para perturba\c{c}\~oes gravitacionais e $l=0$ para perturba\c{c}\~oes de campo eletromagn\'etico n\~ao s\~ao din\^amicas. Por exemplo, $l=0$ perturba\c{c}\~oes gravitacionais s\~ao esfericamente sim\'etricas e, por isso, obedecem o teorema de Birkhoff. Elas correspondem a uma modifica\c{c}\~ao infinitesimal da massa do buraco negro, levando a outra solu\c{c}\~ao est\'atica de buraco negro. Semelhantemente, perturba\c{c}\~oes $l=0$ de campo eletromagn\'etico e perturba\c{c}\~oes gravitacionais $l=1$ correspondem, respectivamente, a uma mudan\c{c}a infinitesimal da carga e da posi\c{c}\~ao do buraco negro \cite{PhysRev.108.1063}.}

\langsectionlabel{Time evolution of perturbations}{Evolu\c{c}\~ao temporal das perturba\c{c}\~oes}{sec:time-domain}
\langpar{In order to integrate the equation (\ref{wave-like}) numerically we use the technique developed by Gundlach, Price and Pullin \cite{Gundlach:1993tp}. We rewrite the wave-like equation (\ref{wave-like}) in terms of the so called light-cone coordinates $du = dt - dr_\star$ and $dv = dt + dr_\star$}{Para integrar a equa\c{c}\~ao (\ref{wave-like}) numericamente usamos a t\'ecnica desenvolvida por Gundlach, Price e Pullin \cite{Gundlach:1993tp}. Reescrevemos a equa\c{c}\~ao de onda (\ref{wave-like}) nos termos das chamadas coordenadas de cone de luz, $du = dt - dr_\star$ e $dv = dt + dr_\star$}
\begin{equation}\label{light-cone}
\left(4\frac{\partial^2}{\partial u\partial v}+V(u,v)\right)\Psi(u,v)=0.
\end{equation}
\lang{Let us consider the operator of time evolution in these coordinates}{Consideremos o operador da evolu\c{c}\~ao de tempo nessas coordenadas}
\begin{eqnarray}\nonumber
\exp\left(h\frac{\partial}{\partial t}\right)&=&\exp\left(h\frac{\partial}{\partial u}+h\frac{\partial}{\partial v}\right)=\exp\left(h\frac{\partial}{\partial u}\right)+\exp\left(h\frac{\partial}{\partial v}\right) - 1 + \\\nonumber&& + \frac{h^2}{2}\left(\exp\left(h\frac{\partial}{\partial u}\right)+\exp\left(h\frac{\partial}{\partial v}\right)\right)\frac{\partial^2}{\partial u\partial v} + \mathcal{O}(h^4).
\end{eqnarray}
\lang{Acting by this operator on $\Psi$, and taking into account (\ref{light-cone}), we find}{Atuando este operador em $\Psi$ e
levando em conta (\ref{light-cone}), encontramos}
\begin{equation}\label{integration-scheme}
\Psi(N)=\Psi(W)+\Psi(E)-\Psi(S)-\frac{h^2}{8}V(S)\left(\Psi(W)+\Psi(E)\right) + \mathcal{O}(h^4),
\end{equation}
\lang{where we introduced letters to mark the points as $S=(u,v)$, $W=(u+h,v)$, $E=(u,v+h)$, $N=(u+h,v+h)$.}{onde introduzimos as letras para marcar os pontos como $S=(u,v)$, $W=(u+h,v)$, $E=(u,v+h)$, $N=(u+h,v+h)$.}

\begin{figure}
\begin{center}
\setlength{\unitlength}{1.4cm}
\begin{picture}(10,10)
  \thicklines
  \put(0,0){\vector(0,1){9}}
  \put(0,0){\vector(1,0){9}}
  \put(0,9.25){$u$}
  \put(9.25,0){$v$}
  \thicklines
  \linethickness{0.08mm}
  \multiput(0,0)(1,0){8}{\line(0,1){7}}
  \multiput(0,0)(0,1){8}{\line(1,0){7}}
  \put(3.6,1.6){$S$}
  \put(3.56,3.1){$W$}
  \put(5.1,1.6){$E$}
  \put(5.1,3.1){$N$}
  \put(4,2){\circle*{0.25}}
  \put(4,3){\circle*{0.25}}
  \put(5,2){\circle*{0.25}}
  \put(5,3){\circle*{0.25}}
\end{picture}
\end{center}
\langfigurecaption{The integration grid. Each cell of the grid represents an integration step. The thick points illustrate the choice of ($S$, $W$, $E$, $N$) for the particular step of the integration. The initial data are specified on the left and bottom sides of the rhombus.}{A grade de integra\c{c}\~ao. Cada c\'elula da grade representa um passo de integra\c{c}\~ao. Os pontos grossos ilustram a escolha de ($S$, $W$, $E$, $N$) para o passo particular da integra\c{c}\~ao. Os dados iniciais est\~ao especificados ao lado esquerdo e ao fundo do losango.}\label{integration-square}
\end{figure}

\langpar{The equation (\ref{integration-scheme}) allows us to calculate the values of $\Psi$ inside the rhombus, which is built on the two null-surfaces $u=u_0$ and $v=v_0$ (see fig. \ref{integration-square}), starting from the initial data specified on them.}{A equa\c{c}\~ao (\ref{integration-scheme}) permite calcular os valores de $\Psi$ dentro do losango, constru\'ido nas duas superf\'icies nulas $u=u_0$ e $v=v_0$ (ver o fig. \ref{integration-square}), come\c{c}ando pelos dados iniciais especificados nelas.}

\begin{figure}
\includeonegraph{D=4_l=2_grav_vector_r=11}
\langfigurecaption{The three stages of the evolution of the Schwarzschild black hole gravitational perturbations ($l=2$ vector type, $\Lambda=0$, $R=22M$). Time is measured in units of the horizon radius.}{As tr\^es etapas da evolu\c{c}\~ao das perturba\c{c}\~oes gravitacionais do buraco negro de Schwarzschild ($l=2$ tipo vetorial, $\Lambda=0$, $R=22M$). O tempo \'e medido em unidades do raio do horizonte.}\label{time-domain-profile}
\end{figure}

\langpar{Let us study, as a qualitative example, the time-domain profile of the vector type gravitational perturbations of the Schwarzschild black hole $\Psi(t,r=R)$ (fig. \ref{time-domain-profile}). We can divide the evolution of the perturbations into three stages. The first stage depends on the initial conditions and on the point $R$. At late time ($t\sim R$) we see exponential damping of the amplitude of the perturbations, which is followed by the so-called tails at asymptotically late time ($t\gg M$).}{Vamos estudar, como um exemplo qualitativo, o perfil temporal das perturba\c{c}\~oes gravitacionais do tipo vetorial do buraco negro de Schwarzschild $\Psi(t,r=R)$ (o fig. \ref{time-domain-profile}). Podemos dividir a evolu\c{c}\~ao das perturba\c{c}\~oes em tr\^es etapas. A primeira etapa depende das condi\c{c}\~oes iniciais e do ponto $R$. No tempo tardio ($t\sim R$) vemos o amortecimento exponencial da amplitude das perturba\c{c}\~oes, que \'e seguida pelas assim chamadas caudas no tempo tardio assintoticamente ($t\gg M$).}

\langpar{One can observe, that, after the initial outburst at the first stage, the behavior of the amplitude of the perturbations does not depend on the initial conditions. Being independent on the source of the perturbations, it depends only on the parameters of the field and the black hole. Therefore, the late-time damping law appears to be an important characteristic of the black hole.}{Pode-se observar, que, depois da irrup\c{c}\~ao inicial na primeira etapa, o comportamento da amplitude das perturba\c{c}\~oes n\~ao depende das condi\c{c}\~oes iniciais. Sendo independente da fonte das perturba\c{c}\~oes, ele depende s\'o dos par\^ametros do campo e do buraco negro. Por isso, a lei de amortecimento no tempo tardio parece ser uma caracter\'istica importante do buraco negro.}

\langpar{The exponential damping of the perturbations is called \textit{quasi-normal ringing}. It can be split to the superposition of exponentially damping oscillations, that can be represented as a set of complex frequencies, which are called \textit{quasi-normal modes}. The real part of a quasi-normal frequency describes the actual frequency of the oscillation, while the imaginary part is its damping rate. In the next section we will study the basic properties of the quasi-normal spectrum of black holes.}{O amortecimento exponencial das perturba\c{c}\~oes \'e chamado \textit{toque quase-normal}. Ele pode ser dividido em uma superposi\c{c}\~ao de oscila\c{c}\~oes exponencialmente decrescentes, que podem ser representadas como um conjunto de freq\"u\^encias complexas, chamadas \textit{modos quase-normais}. A parte real de uma freq\"u\^encia quase-normal descreve a freq\"u\^encia real da oscila\c{c}\~ao, enquanto a parte imagin\'aria \'e a taxa de decaimento dela. Na se\c{c}\~ao seguinte estudaremos as propriedades b\'asicas do espectro quase-normal de buracos negros.}

\langpar{In the case under consideration, the late-time tails decay according to the inverse power law. The qualitatively similar behavior is observed for the decay of all massless fields in asymptotically flat backgrounds. Yet, the late-time tails appear to be very sensitive to the asymptotical behavior of the potential: in the asymptotically anti-de Sitter space-times the quasi-normal ringing governs the decay of perturbations at all times \cite{Horowitz:1999jd}. The black hole perturbations in the de Sitter space-time have exponential tails, if the cosmological constant is large, and both power and exponential tails, if the cosmological constant is small \cite{Brady:1996za,Brady:1999wd}.}{No caso em considera\c{c}\~ao, as caudas de tempo tardio decaem segundo a lei de pot\^encia inversa. O comportamento qualitativamente semelhante foi observado para o decaimento de todos campos sem massa em contextos assintoticamente planos. Mas as caudas de tempo tardio parecem ser muito sens\'iveis ao comportamento assint\'otico do potencial: em espa\c{c}os-tempo assintoticamente anti-de Sitter o toque quase-normal governa o decaimento de perturba\c{c}\~oes todo tempo \cite{Horowitz:1999jd}. As perturba\c{c}\~oes de buraco negro no espa\c{c}o-tempo de Sitter t\^em caudas exponenciais, se a constante cosmol\'ogica for grande, e ambas as caudas, potenciais e exponenciais, se a constante cosmol\'ogica for pequena \cite {Brady:1996za, Brady:1999wd}.}

\langsection{Properties of the quasi-normal spectrum}{Propriedades do espectro quase-normal}
\langsubsection{Boundary conditions}{Condi\c{c}\~oes de contorno}
\langpar{In order to study quasi-normal spectrum of a black hole it is convenient to make Fourier transform for the function}{Para estudar o espectro quase-normal de um buraco negro \'e conveniente fazer a transformada de Fourier da fun\c{c}\~ao}
\begin{equation}\label{eigenfunction-expansion}
\Psi(t,r)=\sum_{n=0}^\infty\exp(-\imo\omega_nt)Q_{\omega_n}(r).
\end{equation}

\lang{Let us note that generally the functions $\{\exp(-\imo\omega_n)\}$, where $\{\omega_n\}$ is the quasi-normal spectrum, do not form basis in the vector space of the solutions of the equation (\ref{wave-like}). Therefore, the signal cannot be expanded in terms of these functions for all time. Indeed, as we see on fig. \ref{time-domain-profile}, the exponential decay is replaced by slower power-law decay at asymptotically late time. Yet, the expansion (\ref{eigenfunction-expansion}) is appropriate to describe the quasi-normal ringing epoch.}{Podemos constatar que geralmente as fun\c{c}\~oes $\{\exp(-\imo\omega_n)\}$, onde $\{\omega_n\}$ \'e espectro quase-normal, n\~ao formam a base no espa\c{c}o vetorial das solu\c{c}\~oes da equa\c{c}\~ao (\ref{wave-like}). Por isso, o sinal n\~ao pode ser expandido em termos dessas fun\c{c}\~oes para tempo todo. De fato, como vemos na fig. \ref{time-domain-profile}, o decaimento exponencial no tempo tardio assintoticamente \'e trocada pela lei de decaimento tipo pot\^encia, que \'e mais lenta. Al\'em disso, a expans\~ao (\ref{eigenfunction-expansion}) \'e apropriada para descrever a \'epoca de toque quase-normal.}

\langpar{Henceforward, we will omit the index $n$ of $\omega$, implying that $\omega$ is any frequency from the quasi-normal spectrum.}{Daqui em diante, omitiremos o \'indice $n$ de $\omega$. Suporemos que $\omega$ seja qualquer freq\"u\^encia do espectro quase-normal.}

\langpar{The function $Q_\omega$ satisfies the linear equation}{A fun\c{c}\~ao $Q_\omega$ satisfaz \`a equa\c{c}\~ao linear}
\begin{equation}\label{eigenfunction-equation}
\left(\frac{d^2}{dr_\star^2}+\omega^2-V(r)\right)Q_\omega(r)=0.
\end{equation}

\langpar{Thus, the searching of the quasi-normal modes is reduced to the problem of finding eigenfrequencies of the equation (\ref{eigenfunction-equation}). Since we are interested in what happens in the $R$-region outside the black hole, we have to impose the boundary conditions at the event horizon and the cosmological horizon for $\Lambda>0$ (or the spatial infinity for $\Lambda\leq0$). In order do this we must study the structure of the singularities of the equation (\ref{eigenfunction-equation}) at those points.}{Assim, a procura dos modos quase-normais \'e reduzida ao problema de achar as freq\"u\^encias pr\'oprias da equa\c{c}\~ao (\ref{eigenfunction-equation}). Como estamos interessados no que acontece na regi\~ao $R$ fora do buraco negro, temos que impor as condi\c{c}\~oes de contorno no horizonte de eventos e no horizonte cosmol\'ogico para $\Lambda>0$ (ou no infinito espacial para $\Lambda\leq0$). Para fazer isto precisamos estudar as singularidades da equa\c{c}\~ao (\ref{eigenfunction-equation}) naqueles pontos.}

\langpar{Let us study the general properties of the singular points of the equation (\ref{eigenfunction-equation}).
First we note that at any horizon $r\rightarrow h$}{Estudaremos as propriedades gerais dos pontos singulares da equa\c{c}\~ao (\ref{eigenfunction-equation}). Primeiro observamos que em qualquer horizonte}
\begin{equation}\label{horizon-potential}
V\propto f(r)=o\left(\frac{1}{r_\star}\right).
\end{equation}

\langpar{To prove (\ref{horizon-potential}) we consider two types of the horizon:}{Para comprovar (\ref{horizon-potential}) foram considerados dois tipos de horizonte:}

1. $f(r\rightarrow h) = (r-h)f'(h) + o(r-h)$\\
\lang{then}{ent\~ao}
$$r_\star = \frac{1}{f'(h)}\ln\left|\frac{r}{h}-1\right|\left(1+o(r-h)\right).$$
\lang{Thus, we find that the dominant contribution is}{Assim, achamos que a contribui\c{c}\~ao dominante \'e}
$$f(r\rightarrow h) \sim h f'(h)\exp\left(f'(h)r_\star\right)\sim h f'(h)\exp\left(\ln\left|\frac{r}{h}-1\right|\right)=o\left(\frac{1}{r_\star}\right)$$
\lang{This term decays exponentially with respect to $r_\star$ because}{Este termo decai exponencialmente com $r_\star$ porque}
$$\lim_{r\rightarrow h}\ln\left|\frac{r}{h}-1\right|=-\infty.$$

2. $f(r\rightarrow h) = A(r-h)^\alpha+o(r-h)^\alpha$, $\alpha>1$
\lang{then}{ent\~ao}
$$r_\star = -\frac{1}{A(\alpha-1)(r-h)^{\alpha-1}}\left(1+o(r-h)\right)$$
\lang{and}{e}
$$f(r\rightarrow h) = \frac{r-h}{(\alpha-1)r_\star}+o(r-h)^\alpha=o\left(\frac{1}{r_\star}\right).$$

\langpar{If $\Lambda=0$, the condition (\ref{horizon-potential}) at spatial infinity is also satisfied for the potentials (\ref{scalar-potential}, \ref{Dirac-potential}, \ref{EM-potential}, \ref{polar-potential}, \ref{axial-potential}). Indeed, $r\sim r_\star$ and}{Se $\Lambda=0$, a condi\c{c}\~ao (\ref{horizon-potential}) no infinito espacial tamb\'em \'e satisfeita para os potenciais (\ref{scalar-potential}, \ref{Dirac-potential}, \ref{EM-potential}, \ref{polar-potential}, \ref{axial-potential}). De fato, $r\sim r_\star$ e}
$$V(r)=\mathcal{O}\left(\frac{1}{r^2}\right)=\mathcal{O}\left(\frac{1}{r_\star^2}\right)=o\left(\frac{1}{r_\star}\right).$$

\langpar{Since the potential satisfies (\ref{horizon-potential}), the behavior of the eigenfrequency at the boundaries can be}{Como o potencial satisfaz (\ref{horizon-potential}), o comportamento da freq\"u\^encia pr\'opria nos limites pode ser}
\begin{equation}
Q_\omega \propto \exp(\pm\imo\omega r_\star).
\end{equation}
\lang{The appropriate boundary conditions for the problem under consideration are purely ingoing wave at the event horizon and outgoing wave at the cosmological horizon (spatial infinity)}{As condi\c{c}\~oes de contorno apropriadas do problema em considera\c{c}\~ao s\~ao puramente onda, que entra no horizonte de eventos, e onda, que sai do horizonte cosmol\'ogico (infinito espacial)}
\begin{equation}\label{QNMbc}
\begin{array}{rcl}
r_\star\rightarrow-\infty: &\quad& Q_\omega\propto \exp(-\imo\omega r_\star),\\
r_\star\rightarrow+\infty: &\quad& Q_\omega\propto \exp(+\imo\omega r_\star).
\end{array}
\end{equation}

\lang{This choice of the boundary conditions imply that, after the black hole is perturbed, there is no signal, which comes from the black hole or from any remote source \cite{Chandrasekhar}. It means that the source of the perturbations acts only before the first stage of the evolution. After this, we consider only the black hole response upon the perturbations.}{Esta escolha das condi\c{c}\~oes de contorno implicam que, depois que o buraco negro foi perturbado, n\~ao h\'a nenhum sinal, que vem do buraco negro ou de qualquer fonte remota \cite{Chandrasekhar}. Esto significa que a fonte das perturba\c{c}\~oes atua s\'o antes da primeira etapa da evolu\c{c}\~ao. Depois disso, consideremos s\'o a resposta do buraco negro, provocada pelas perturba\c{c}\~oes.}

\langpar{For the 4-dimensional Schwarzschild-anti-de Sitter black hole ($\Lambda<0$), one can find that the linear independent solutions at spatial infinity ($r\rightarrow\infty$) are}{Para um buraco negro 4 dimensional de Schwarzschild-anti-de Sitter ($\Lambda<0$), pode-se achar que as solu\c{c}\~oes linearmente independentes no infinito espacial ($r\rightarrow\infty$) s\~ao}
$$Q_1\propto r,\qquad Q_2\propto r^{-2}.$$
\lang{Since we do not suppose to have an infinite amplitude of the perturbation at the spatial infinity, we must choose $Q_2$ as the appropriate boundary condition there, i. e. we impose the Dirichlet condition at the AdS boundary \cite{Horowitz:1999jd}}{Como n\~ao supomos ter uma amplitude infinita de perturba\c{c}\~ao no infinito espacial, devemos selecionar $Q_2$ como a condi\c{c}\~ao de contorno apropriada l\'a, isto \'e, impomos a condi\c{c}\~ao de Dirichlet no limite de AdS \cite{Horowitz:1999jd}}
\begin{equation}\label{AdSbc}
Q_\omega(r=\infty)=0.
\end{equation}
\lang{The Dirichlet boundary conditions are usually imposed at the spatial infinity in asymptotically anti-de Sitter backgrounds (see e. g. \cite{Kovtun:2005ev}).}{As condi\c{c}\~oes de contorno de Dirichlet s\~ao normalmente impostas no infinito espacial em contextos assintoticamente anti-de Sitter (ver por exemplo \cite{Kovtun:2005ev}).}

\langsubsectionlabel{Quasi-normal spectrum and black hole stability}{O espectro quase-normal e a estabilidade de buracos negros}{sec:stability}
\langpar{Black holes do not exist in nature, that is why, the important property of any black hole solution is its stability against perturbations. Unfortunately, it is possible to prove stability analytically only of some relatively simple solutions of the Einstein equations. That is why the numerical test of stability is important for the black hole study. The instability implies the existing of growing modes in the quasi-normal spectrum. It is clear, that the linear approximation is enough for this test. Indeed, if the black hole is stable, any perturbations will decay until the linear approximation is valid. If the quasi-normal spectrum has a growing mode, the amplitude of perturbations will grow until we are compelled to consider the non-linear back reaction of the perturbations upon the metric. In this case, we state, at least, that in order to obtain the stable solution we must take into account this non-linear correction, which could be also non-stationary.}{Buracos negros inst\'aveis n\~ao existem na natureza, ent\~ao a propriedade importante de qualquer solu\c{c}\~ao de buraco negro \'e a estabilidade dela contra perturba\c{c}\~oes. Infelizmente, \'e poss\'ivel de comprovar analiticamente a estabilidade s\'o de algumas solu\c{c}\~oes das equa\c{c}\~oes de Einstein, que s\~ao relativamente simples. Por isso o teste num\'erico da estabilidade \'e importante para o estudo de buracos negros. A instabilidade implica a exist\^encia de modos crescentes no espectro quase-normal. \'E claro, que a aproxima\c{c}\~ao linear \'e suficiente para esse teste. De fato, se o buraco negro for est\'avel, quaisquer perturba\c{c}\~oes decair\~ao at\'e que a aproxima\c{c}\~ao linear seja v\'alida. Se o espectro quase-normal tiver um modo crescente, a amplitude de perturba\c{c}\~oes crescer\'a at\'e que sejamos compelidos para considerar a rea\c{c}\~ao de volta n\~ao linear das perturba\c{c}\~oes na m\'etrica. Neste caso, afirmamos, pelo menos, que para obter a solu\c{c}\~ao est\'avel precisamos levar em considera\c{c}\~ao esta corre\c{c}\~ao n\~ao linear, que pode ser tamb\'em n\~ao-estacion\'aria.}

\langpar{It is important to note that the black hole is unstable if there is only one growing mode in its spectrum. Therefore, in order to prove the black hole stability we must show that the quasi-normal spectrum does not contain any growing mode, i. e. frequency with positive imaginary part. It makes the numerical proof of the black hole stability extremely complicated.}{\'E importante constatar que o buraco negro \'e inst\'avel mesmo que tenha s\'o um modo crescente no seu espectro. Por isso, para comprovar a estabilidade do buraco negro, devemos mostrar que o espectro quase-normal n\~ao cont\'em nenhum modo crescente, isto \'e, a freq\"u\^encia com a parte imagin\'aria positiva. Isto faz a prova num\'erica da estabilidade de um buraco negro extremamente complicada.}

\langpar{Yet, in some cases we are able to prove the stability analytically. Let us multiply the equation (\ref{eigenfunction-equation}) by the complex conjugated $Q_\omega^\star$ and integrate the first term by parts}{Mas em alguns casos somos capazes de comprovar a estabilidade analiticamente. Vamos multiplicar a equa\c{c}\~ao (\ref{eigenfunction-equation}) pelo complexo conjugado $Q_\omega^\star$ e integrar o primeiro termo pelas partes}
$$Q_\omega^\star(r_\star)\frac{dQ_\omega(r_\star)}{dr_\star}\Biggr|_{-\infty}^\infty + \intop_{-\infty}^\infty \left(\omega^2|Q_\omega(r_\star)|^2 -V|Q_\omega(r_\star)|^2 - \left|\frac{dQ_\omega(r_\star)}{dr_\star}\right|^2\right)dr_\star = 0.$$

\langpar{Taking into account the boundary conditions (\ref{QNMbc}) we find}{Levando em considera\c{c}\~ao as condi\c{c}\~oes de contorno (\ref {QNMbc}) achamos}
\begin{equation}\label{ppot-ineq}
\imo\omega A + \omega^2 B =  \intop_{-\infty}^\infty \left(V(r)|Q_\omega(r_\star)|^2 + \left|\frac{dQ_\omega(r_\star)}{dr_\star}\right|^2\right)dr_\star,
\end{equation}
\lang{where}{onde} $\displaystyle A=|Q_\omega(r_\star=\infty)|^2+|Q_\omega(r_\star=-\infty)|^2>0$, $\displaystyle B=\intop_{-\infty}^\infty|Q_\omega(r_\star)|^2 dr_\star >0$.

\langpar{The imaginary part of (\ref{ppot-ineq}) reads}{A parte imagin\'aria de (\ref{ppot-ineq}) se l\^e}
\begin{equation}\label{basic-ppot}
\re(\omega) A + 2\re(\omega)\im(\omega) B = 0.
\end{equation}
\lang{We can see that $\im(\omega)<0$ except for the case $\re(\omega)=0$. Thus, we conclude that \textit{the growing modes do not oscillate} (see sec. \ref{sec:Gauss-Bonnet}).}{Pode-se ver que $\im(\omega)<0$ excetuando o caso $\re(\omega)=0$. Assim, conclui-se que \textit {os modos crescentes n\~ao oscilam} (ver a se\c{c}\~ao. \ref{sec:Gauss-Bonnet}).}

\langpar{If the potential is positive in the $R$-region outside the event horizon and $\re(\omega)=0$, the real part of (\ref{ppot-ineq}) implies that the imaginary part of the frequency remains negative}{Se o potencial for positivo na regi\~ao $R$ fora do horizonte de eventos e $\re(\omega)=0$, a parte real de (\ref{ppot-ineq}) implica que a parte imagin\'aria da freq\"u\^encia permanece negativa}
$$-\im(\omega)A-\im(\omega)^2B>0\quad\Rightarrow\quad \im(\omega)<0.$$

\langpar{It is clear, that the righthand side of (\ref{ppot-ineq}) can be positive even though $V(r)$ is not positive everywhere. By introducing the new derivative $\displaystyle D=\frac{d}{dr_\star}+S(r_\star)$ we can rewrite the integral as}{\'E claro, que o lado direito de (\ref{ppot-ineq}) pode ser positivo embora $V(r)$ n\~ao seja positivo em cada ponto. Introduzindo a nova derivada $\displaystyle D=\frac{d}{dr_\star}+S(r_\star)$ podemos reescrever a integral como}
$$\intop_{-\infty}^\infty \left(V(r)|Q_\omega(r_\star)|^2 + \left|\frac{dQ_\omega(r_\star)}{dr_\star}\right|^2\right)dr_\star= \intop_{-\infty}^\infty \left(\tilde{V}|Q_\omega(r_\star)|^2 + \left|DQ_\omega\right|^2\right)dr_\star - \Biggr|_{r_\star=-\infty}^{r_\star=\infty}S(r_\star)|Q_\omega(r_\star)|^2,$$
\lang{where $\displaystyle \tilde{V}=V+\frac{dS}{dr_\star}-S^2$. Thus, we conclude that if we find the function $S(r_\star)$ such as $S(r_\star=\infty)\leq0$, $S(r_\star=-\infty)\geq0$ and $\tilde{V}\geq0$, the righthand side of (\ref{ppot-ineq}) stays positive and, therefore, $\im(\omega)<0$.}{onde $\displaystyle \tilde{V}=V+\frac{dS}{dr_\star}-S^2$. Assim, conclu\'imos que se acharmos a fun\c{c}\~ao $S(r_\star)$ tal que $S(r_\star=\infty)\leq0$, $S(r_\star=-\infty)\geq0$ e $\tilde{V}\geq0$, o lado direito de (\ref{ppot-ineq}) fica positivo e, por isso, $\im(\omega)<0$.}

\langpar{This technique is called \emph{$S$-deformation} and allows us to prove stability for some cases, when the potential is not positive definite \cite{Kodama:2003ck}. It is important to note that if we find an appropriate function $S$, we prove the black hole stability. Otherwise we do not know if the black hole is stable or not.}{Esta t\'ecnica \'e chamada \emph{deforma\c{c}\~ao $S$} e permite comprovar a estabilidade para alguns casos, quando o potencial n\~ao \'e definido positivo \cite{Kodama:2003ck}. \'E importante notar que se acharmos uma fun\c{c}\~ao $S$ apropriada, comprovamos a estabilidade do buraco negro. No caso contr\'ario n\~ao saber\'iamos se o buraco negro \'e est\'avel ou n\~ao.}

\langpar{The same approach can be used in a similar way for anti-de Sitter black holes. We must just put $r_{\star\infty}$ as the upper bound in the integrals and use (\ref{AdSbc}). Then we obtain $A=|Q_\omega(r_\star=-\infty)|^2>0$ and the requirement for $S$ at the spatial infinity reads $S(r_\star=r_{\star\infty})<\infty$.}{A mesma abordagem pode ser usada no modo semelhante para os buracos negros anti-de Sitter. Devemos p\^or somente $r_{\star\infty}$ como o limite superior das integrais e utilizar (\ref{AdSbc}). Ent\~ao obtemos $A=|Q_\omega(r_\star=-\infty)|^2>0$ e a exig\^encia para $S$ no infinito espacial se l\^e $S(r_\star=r_{\star\infty})<\infty$.}

\langpar{Because we must guess the appropriate function $S$, the technique of $S$-deformation allows to proof stability only for relatively simple potentials. For more complicated cases we are able to prove the black hole stability only numerically.}{Como devemos adivinhar a fun\c{c}\~ao $S$ apropriada, a t\'ecnica da deforma\c{c}\~ao $S$ permite a prova de estabilidade s\'o para potenciais relativamente simples. Para casos mais complicados somos capazes provar a estabilidade do buraco negro s\'o numericamente.}

\langsubsection{Isospectrality}{Isoespectralidade}
\langpar{Let us consider two equations (\ref{eigenfunction-equation}) with the effective potentials, taken as $V^+$ and $V^-$, with}{Vamos considerar duas equa\c{c}\~oes (\ref{eigenfunction-equation}) com os potenciais efetivos, tomados como $V^+$ e $V^-$, com}
\begin{equation}
V^\pm=W^2(r_\star)\pm\frac{dW(r_\star)}{dr_\star}+\beta,
\end{equation}
\lang{where $W(r_\star)$ is some finite function and $\beta$ is a constant. Then, if $Q^+_{\omega}$ is an eigenfunction of (\ref{eigenfunction-equation}) for the potential $V^+$, the eigenfunction for the potential $V^-$ is given (up to an arbitrary factor) by}{onde $W(r_\star)$ \'e alguma fun\c{c}\~ao finita e $\beta$ \'e uma constante. Ent\~ao, se $Q^+_{\omega}$ \'e uma fun\c{c}\~ao pr\'opria de (\ref{eigenfunction-equation}) para o potencial $V^+$, a fun\c{c}\~ao pr\'opria do potencial $V^-$ \'e dada (at\'e um fator arbitr\'ario) por}
\begin{equation}
Q^-_{\omega}=\left(W-\frac{d}{dr_\star}\right)Q^+_{\omega},
\end{equation}
\lang{corresponding to the same eigenvalue $\omega$. Thus we conclude that the quasi-normal spectrum is the same for the potentials $V^+$ and $V^-$.}{correspondente ao mesmo valor pr\'oprio de $\omega$. Assim conclu\'imos que o espectro quase-normal \'e o mesmo para os potenciais $V^+$ and $V^-$.}

\langpar{Let us consider the examples of isospectrality for 4-dimensional Schwarzschild((-anti)-de Sitter) black holes. One can check \cite{Chandrasekhar} that the potentials (\ref{polar-potential}) and (\ref{axial-potential}) can be obtained by taking}{Consideremos os exemplos de isospectrality para uns buracos negros de Schwarzschild((-anti)-de Sitter). Pode-se verificar \cite{Chandrasekhar} que os potenciais (\ref{polar-potential}) e (\ref{axial-potential}) podem ser obtidos tomando}
$$W=\frac{2M}{r^2}-\frac{3+2c}{3r}+\frac{3c^2+2c^3-9\Lambda M^2}{3c(3M+cr)}-\frac{1}{3M}\left(c^2+c-\frac{3\Lambda M^2}{c}\right), \qquad \beta = -\frac{c^2(c+1)^2}{9M^2}.$$
\lang{The potentials for the Dirac field (\ref{Dirac-potential}) are, obviously, isospectral too.}{Os potenciais para o campo de Dirac (\ref{Dirac-potential}) s\~ao, obviamente, tamb\'em isoespectrais.}

\langpar{It is clear that if one of the isospectral potentials is positive outside the event horizon, as one can notice for $V_s(r)$ ($c\geq2$), both do not lead to growing modes, implying stability against the perturbations. The other possible case of stable potentials is $\beta\geq0$, as it happens for (\ref{Dirac-potential}). In order to prove the stability we just use the $S$-deformation technique with $S=\pm W$. In fact, all the potentials (\ref{scalar-potential}, \ref{Dirac-potential}, \ref{EM-potential}, \ref{polar-potential}, \ref{axial-potential}) are positive for $-\infty<r_\star<\infty$, thereby, the stability of the Schwarzschild black hole is evident.}{\'E claro que se um dos potenciais isoespectrais for positivo fora do horizonte de eventos, como podemos notar para $V_s(r)$ ($c\geq2$), ambos n\~ao levam a modos crescentes, implicando estabilidade contra a perturba\c{c}\~ao. Outro caso poss\'ivel de potenciais est\'aveis \'e $\beta\geq0$, como acontece para (\ref{Dirac-potential}). Para comprovar a estabilidade use-se somente a t\'ecnica da deforma\c{c}\~ao $S$ com $S=\pm W$. De fato, todos os potenciais (\ref{scalar-potential}, \ref{Dirac-potential}, \ref{EM-potential}, \ref{polar-potential}, \ref{axial-potential}) s\~ao positivos para $-\infty<r_\star<\infty$, por meio disso, a estabilidade do buraco negro de Schwarzschild \'e evidente.}

\langchapter{Numerical methods of calculation of the quasi-normal modes}{M\'etodos num\'ericos para o c\'alculo de modos quase-normais}
\langsectionlabel{Fitting time-domain data}{Ajustando dados do perfil temporal}{sec:time-domain-fit}
\langpar{The most direct approach to finding quasi-normal modes is the numerical integration of the equation (\ref{wave-like}) as described in the section \ref{sec:time-domain}. The result of the time-domain integration is a time profile data $\{\Psi(t=0),\Psi(t=h),\Psi(t=2h)\ldots\}$, which can be used to calculate the quasi-normal modes.}{A abordagem mais direta \`a procura de modos quase-normais \'e a integra\c{c}\~ao num\'erica da equa\c{c}\~ao (\ref{wave-like}) como descrito na se\c{c}\~ao \ref{sec:time-domain}. Os resultados da integra\c{c}\~ao temporal s\~ao dados de perfil temporal $\{\Psi(t=0),\Psi(t=h),\Psi(t=2h)\ldots\}$ que podem ser utilizados para calcular os modos quase-normais.}

\langpar{Let us describe the simplest Prony method of fitting the profile data by superposition of damping exponents (see e. g.  \cite{Berti:2007dg} and references therein)}{O m\'etodo simples de Prony \'e descrito para ajustar os dados do perfil pela superposi\c{c}\~ao de expoentes decrescentes (ver por exemplo \cite{Berti:2007dg} e suas refer\^encias)}
\begin{equation}\label{damping-exponents}
\Psi(t)\simeq\sum_{i=1}^pC_ie^{-\imo\omega_i t}.
\end{equation}

\langpar{We suppose that the quasi-normal ringing epoch starts at $t=0$ and ends at $t=Nh$, where integer $N\geq2p-1$. Then the formula (\ref{damping-exponents}) is valid for each value from the profile data}{Sup\~oe-se que a \'epoca de toque quase-normal come\c{c}a em $t=0$ e termina em $t=Nh$, onde $N\geq2p-1$ \'e de n\'umero inteiro. Ent\~ao a f\'ormula (\ref{damping-exponents}) \'e v\'alida para cada valor dos dados do perfil}
\begin{equation}
x_n\equiv\Psi(nh)=\sum_{j=1}^pC_je^{-\imo\omega_j nh}=\sum_{j=1}^pC_jz_j^n.
\end{equation}

\langpar{The Prony method allows to find $z_i$ in terms of known $x_n$ and, since $h$ is also known, to calculate the quasi-normal frequencies $\omega_i$. In order to do this, we define a polynomial function $A(z)$ as}{O m\'etodo de Prony permite achar $z_i$ nos termos dos conhecidos $x_n$ e, como $h$ tamb\'em \'e conhecido, tamb\'em permite calcular as freq\"u\^encias quase-normais $\omega_i$. Para fazer isto, \'e definida uma fun\c{c}\~ao polinomial $A(z)$ como}
\begin{equation}
A(z)=\prod_{j=1}^p(z-z_j)=\sum_{m=0}^{p}\alpha_m z^{p-m}, \qquad \alpha_0=1.
\end{equation}
\lang{Let us consider the sum}{Consideremos a soma}
$$\sum_{m=0}^p\alpha_mx_{n-m}=\sum_{m=0}^p\alpha_m\sum_{j=1}^pC_jz_j^{n-m}=\sum_{j=1}^pC_jz_j^{n-p}\sum_{m=0}^p\alpha_mz_j^{p-m}=\sum_{j=1}^pC_jz_j^{n-p}A(z_j)=0.$$

\langpar{Taking into account that $\alpha_0=1$, we find}{Levando em conta que $\alpha_0=1$ encontra-se}
\begin{equation}\label{Prony-equation}
\sum_{m=1}^p\alpha_mx_{n-m}=-x_{n}.
\end{equation}
\lang{Substituting $n=p..N$ into (\ref{Prony-equation}) we obtain $N-p+1\geq p$ linear equations for $p$ unknown coefficients $\alpha_m$.}{Substituindo $n=p..N$ em (\ref{Prony-equation}), obt\^em-se $N-p+1\geq p$ equa\c{c}\~oes lineares para $p$ coeficientes desconhecidos $\alpha_m$.}

\langpar{Let us rewrite these equations in the matrix form}{Reescrevendo essas equa\c{c}\~oes na forma de matriz}
$$\left(\begin{array}{llll}
    x_{p-1} & x_{p-2} & \ldots & x_0 \\
    x_p & x_{p-1} & \ldots & x_1 \\
    \vdots & \vdots & \ddots & \vdots \\
    x_{N-1} & x_{N-2} & \ldots & x_{N-p} \\
  \end{array}\right)
\left(\begin{array}{c}
    \alpha_1 \\
    \alpha_2 \\
    \vdots \\
    \alpha_p\\
\end{array}\right)=-
\left(\begin{array}{c}
    x_p \\
    x_{p+1} \\
    \vdots \\
    x_N\\
\end{array}\right).
$$
\lang{Such matrix equation}{Tal equa\c{c}\~ao de matriz}
$$X\alpha=-x$$
\lang{can be solved in the least-squares sense}{pode ser resolvida no sentido de m\'inimos-quadrados}
\begin{equation}
\alpha=-(X^+X)^{-1}X^+x,
\end{equation}
\lang{where $X^+$ denotes the Hermitian transposition of the matrix $X$.}{onde $X^+$ denota a transposi\c{c}\~ao Hermitiana da matriz $X$.}

\langpar{After the coefficients $\alpha_m$ of the polynomial function $A(z)$ are found, we can calculate numerically the roots $z_j$ of the polynom and the quasi-normal frequencies}{Depois que os coeficientes $\alpha_m$ da fun\c{c}\~ao polinomial $A(z)$ s\~ao achados, pode-se calcular numericamente as ra\'izes $z_j$ do polin\^omio e as freq\"u\^encias quase-normais}
$$\omega_j=\frac{\imo}{h}\ln(z_j).$$

\langpar{Because the quasi-normal stage is not a precisely defined time interval, in practice, it is difficult to determine when the quasi-normal ringing begins. In fact, when we observe explicitly damped oscillations, we usually see only the fundamental mode, while higher overtones, which damp quickly, are already exponentially suppressed. Being a small corrections to the signal, such higher damped oscillations are indistinguishable from numerical errors within the described approach. Thus, the higher overtones are difficult to find. Usually, the Prony method allows to calculate at most six roots of the polynom $A(z)$, including the complex conjugated ones, which correspond to the symmetry $\omega\leftrightarrow (-\omega^*)^*$. This symmetry exists just because the wave-like equation (\ref{wave-like}) is real. Therefore, in fact, we are able to calculate only two or, sometimes, three dominant frequencies.}{Como o est\'agio quase-normal n\~ao \'e um intervalo temporal precisamente definido, na pr\'atica, \'e dif\'icil determinar quando o toque quase-normal come\c{c}a. De fato, quando observamos oscila\c{c}\~oes explicitamente decrescentes, normalmente \'e visto somente o modo fundamental, enquanto os sobretons mais altos, que se decaem rapidamente, j\'a est\~ao exponencialmente suprimidos. Como s\~ao pequenas corre\c{c}\~oes ao sinal, tais oscila\c{c}\~oes decrescentes mais altas s\~ao indistingu\'iveis de erros num\'ericos nessa aproxima\c{c}\~ao. Ent\~ao, \'e dificil achar os sobretons mais altos. Normalmente, o m\'etodo de Prony permite calcular no m\'aximo seis ra\'izes do polin\^omio $A(z)$, inclusive os complexos conjugados, que correspondem \`a simetria $\omega\leftrightarrow (-\omega^*)^*$. Esta simetria existe somente porque a equa\c{c}\~ao de onda (\ref{wave-like}) \'e real. Por isso, de fato, \'e possivel calcular somente duas ou, \`as vezes, tr\^es freq\"u\^encias dominantes.}

\langpar{In order to determine the beginning of the quasi-normal ringing epoch more precisely, we can use the following technique \cite{Abdalla:2006fv}. Let us find the dominant quasi-normal mode $\omega_1$ and the corresponding coefficient $C_1$ at some late time interval. Then we can subtract this oscillation from the numerical data, and obtain the profile data without the contribution of the dominant mode. After this, one can see ringing for the first overtone. If the lifetime of the quasi-normal ringing is long enough, we are able to find, step by step, higher overtones, making sure, that the numerical error of the initial data is less than the signal after removing the contributions of the lower-damping modes.}{Para determinar o come\c{c}o da \'epoca de toque quase-normal mais precisamente, pode-se utilizar a seguinte t\'ecnica \cite{Abdalla:2006fv}: encontra-se o modo $\omega_1$ quase-normal dominante e o coeficiente correspondente $C_1$ em algum intervalo de tempo tardio. Ent\~ao pode-se subtrair esta oscila\c{c}\~ao dos dados num\'ericos e obter os dados do perfil sem a contribui\c{c}\~ao do modo dominante. Depois disto, pode-se ver o toque do primeiro sobretom. Se a vida do toque quase-normal for longa o suficiente, \'e possivel de achar, passo a passo, sobretons mais altos, assegurando-se de que o erro num\'erico dos dados iniciais \'e menor do que o sinal depois de retirar as contribui\c{c}\~oes dos modos mais lento-decrescentes.}

\langsectionlabel{Approximation by the P\"oschl-Teller potential}{Aproxima\c{c}\~ao pelo potencial de P\"oschl-Teller}{sec:Poschl-Teller}
\langpar{The easiest method of calculation of the quasi-normal modes in frequency domain is approximation of the effective potential by the P\"oschl-Teller potential. This method was suggested by Bahram Mashhoon \cite{Phys.Lett.A100.5.231,PhysRevLett.52.1361}.}{O m\'etodo mais f\'acil para o c\'alculo dos modos quase-normais no dom\'inio de freq\"u\^encia \'e a aproxima\c{c}\~ao do potencial efetivo pelo potencial de P\"oschl-Teller. Este m\'etodo foi sugerido por Bahram Mashhoon \cite{Phys.Lett.A100.5.231,PhysRevLett.52.1361}.}

\langpar{Suppose that the potential in the equation (\ref{eigenfunction-equation}) is invariant under the following transformation}{Supondo-se que o potencial na equa\c{c}\~ao (\ref{eigenfunction-equation}) seja invariante sob a transforma\c{c}\~ao}
$$V(r^*,\alpha)=V(-\imo r^*,\alpha'),$$
\lang{where $\alpha$ is some parameter and $\alpha'$ depends on $\alpha$.}{onde $\alpha$ \'e um par\^ametro e $\alpha'$ depende de $\alpha$.}

\langpar{Let us consider the solution of the equation with the invese potential}{Consideremos a solu\c{c}\~ao da equa\c{c}\~ao com o potencial invertido}
\begin{equation}
\frac{d^2 Q_\Omega}{dr_*^2}+ (-\Omega^2 + V) Q_\Omega= 0,
\end{equation}
\lang{with the boundary conditions that are characteristic of bound states if $\re(\Omega)>0$}{com as condi\c{c}\~oes de contorno que s\~ao caracter\'isticas dos estados limitados, se $\re(\Omega)>0$}
\begin{equation}
Q_\Omega \propto e^{ \mp\Omega r_{*}},  \qquad  r_{*} \rightarrow \pm \infty.
\end{equation}

\langpar{It is easy to see, that this solution is related with the solution of (\ref{eigenfunction-equation}) with the quasi-normal boundary conditions (\ref{QNMbc}) in a simple way}{\'E f\'acil de ver que esta solu\c{c}\~ao est\'a relacionada \`a solu\c{c}\~ao de (\ref{eigenfunction-equation}) com condi\c{c}\~oes de contorno quase-normais (\ref{QNMbc}) de um modo simples,}
\begin{equation}
Q_\omega(r_*,\alpha) = Q_\Omega(-\imo r_*, \alpha'), \qquad \omega(\alpha)=\Omega(\alpha').
\end{equation}

\langpar{Thus, the quasi-normal mode problem is reduced now to the bound states problem for an inverse potential $V\rightarrow-V$, which is smooth potential gap, approaching zero at the infinite boundaries. This gap can be approximated by the P\"oschl-Teller potential}{Assim, o problema dos modos quase-normais \'e reduzido agora ao problema de estados limitados para um potencial invertido $V\rightarrow-V$, que \'e um espa\c{c}o suave, aproximando-se de zero nos contornos infinitos. Este espa\c{c}o pode ser aproximado pelo potencial de P\"oschl-Teller}
\begin{equation}\label{Pochl-Teller-potential}
V_{PT} = \frac{V_0}{\cosh^{2}\alpha(r_*-r_*^0)},
\end{equation}
\lang{where $V_{0}$ is the height of the effective potential and $-2V_{0}\alpha^2$ is the curvature of the potential at its maximum. The bound states of the P\"oschl-Teller potential are known in analytical form \cite{PoschlTeller}}{onde $V_{0}$ \'e a altura do potencial efetivo e $-2V_{0}\alpha^2$ \'e a curvatura do potencial no seu m\'aximo. Os estados limitados do potencial de P\"oschl-Teller s\~ao conhecidos na forma anal\'itica \cite{PoschlTeller}}
\begin{equation}
\Omega = \alpha\left(-\left(n+\frac{1}{2}\right) + \left(\frac{1}{4} + \frac{V_{0}}{\alpha^2} \right)^{1/2} \right),\qquad n=0, 1, 2, \dots
\end{equation}

\langpar{The potential (\ref{Pochl-Teller-potential}) is obviously invariant under the transformation}{O potencial (\ref{Pochl-Teller-potential}) \'e obviamente invariante sob a transforma\c{c}\~ao}
$$V_{PT}(r_*,\alpha)=V_{PT}(-\imo r_*,\imo\alpha).$$
\lang{Therefore, the quasi-normal modes for the inverted P\"oschl-Teller potential are}{Por isso, os modos quase-normais para o potencial inverso de P\"oschl-Teller s\~ao}
\begin{equation}
\omega(\alpha) = \Omega(\imo\alpha)=\pm \left(V_{0} - \frac{\alpha^2}{4} \right)^{1/2} - i \alpha \left(n + \frac{1}{2} \right),\qquad n=0, 1, 2, \dots
\end{equation}

\langpar{Technically one has to fit a given effective potential to an inverted P\"oschl-Teller potential. In the chapter \ref{sec:non-analytical} one shall see that in many cases the behavior of the effective potential only near the black hole is essential for determining the dominating quasi-normal modes. So that the fit of the height of the effective potential $V_0$ and of its curvature $-2V_{0}\alpha^2$ is indeed sufficient to estimate quasi-normal frequencies.}{Tecnicamente precisa-se ajustar um potencial efetivo dado a um potencial inverso de P\"oschl-Teller. No cap\'itulo \ref{sec:non-analytical} ser\'a visto que em muitos casos, para se determinar os modos quase-normais dominantes, \'e essencial o comportamento do potencial efetivo somente perto do buraco negro. Por isso, o ajuste da altura do potencial efetivo $V_0$ e da sua curvatura $-2V_{0}\alpha^2$ s\~ao de fato suficientes para estimar as freq\"u\^encias quase-normais.}

\langpar{This method gives quite accurate results for the regime of high multipole numbers $l$. In particular, for gravitational perturbations of the $D=4$ Schwarzschild black holes, the fundamental quasi-normal modes obtained by the Mashhoon method gives relative error of not more than $2\%$ for the lowest multi-pole $l=2$, and of about fractions of one percent for higher multipoles.}{Esse m\'etodo d\'a resultados precisos o suficiente para o regime de n\'umeros multipolares $l$ altos. Particularmente, para perturba\c{c}\~oes gravitacionais dos buracos negros de  Schwarzschild $D=4$, os modos quase-normais fundamentais, obtidos pelo m\'etodo de Mashhoon, d\~ao um erro relativo de n\~ao mais de $2\%$ para o multipolo mais baixo $l=2$, e aproximadamente de fra\c{c}\~oes de um por cento para os multipolos mais altos.}

\langpar{There are cases when the effective potential of a black hole is \emph{exactly} the P\"oschl-Teller potential. These are Schwarzschild-de Sitter \cite{Cardoso:2003sw} and Reissner-Nordstr\"om-de Sitter black holes with extremal value of the $\Lambda$-term in $D\geq 4$ space-time dimensions \cite{Molina:2003ff}.}{H\'a casos quando o potencial efetivo de um buraco negro \'e \emph{exatamente} o potencial de P\"oschl-Teller. Esses s\~ao os buracos negros Schwarzschild-de Sitter \cite{Cardoso:2003sw} e Reissner-Nordstr\"om-de Sitter com um valor extremo do termo $\Lambda$ em $D\geq 4$ dimens\~oes de espa\c{c}o-tempo \cite{Molina:2003ff}.}

\langsectionlabel{WKB method}{O m\'etodo de WKB}{sec:WKB}
\langpar{In order to evaluate quasi-normal modes for more complicated effective potentials it is convenient to use the WKB (Wentzel, Kramers, Brillouin) method, which provides good accuracy. The WKB technique was applied to finding of the quasi-normal modes of black holes for the first time by Schutz and Will \cite{Astrophys.J.Lett.291.L33.1985}.}{Para avaliar modos quase-normais de potenciais efetivos mais complicados \'e conveniente utilizar o m\'etodo de WKB (Wentzel, Kramers, Brillouin), que fornece uma boa precis\~ao. A t\'ecnica de WKB foi aplicada para achar os modos quase-normais de buracos negros pela primeira vez por Schutz e Will \cite{Astrophys.J.Lett.291.L33.1985}.}

\langpar{In order to simplify our notations, let us re-write the wave-like equation (\ref{eigenfunction-equation}) in the following form}{Para simplificar as nota\c{c}\~oes deste trabalho, reescreve-se a equa\c{c}\~ao de onda (\ref{eigenfunction-equation}) na forma}
\begin{equation}\label{WKBwave}
\frac{d^2 \Psi}{d x^2} + Q(x) \Psi(x) = 0,
\end{equation}
\lang{i.e. we identify $x \equiv r_{*}$, $Q \equiv \omega^2 - V$, and $\Psi\equiv Q_\omega$.}{isto \'e, identifica-se $x \equiv r_{*}$, $Q \equiv \omega^2 - V$, e $\Psi\equiv Q_\omega$.}

\langpar{Let us introduce the WKB parameter $\epsilon$ in order to track orders of the WKB expansion. The asymptotic WKB approximation at both infinities has the following general form}{Introduzindo o par\^ametro WKB $\epsilon$ para monitorar as ordens da expans\~ao WKB. A aproxima\c{c}\~ao assint\'otica de WKB em ambos infinitos tem a forma geral}
\begin{equation}\label{WKBexpansion}
\Psi(x) \propto \exp\left(\sum_{n=0}^{\infty} \frac{S_{n}(x) \epsilon^n}{\epsilon} \right)\,.
\end{equation}
\lang{Substituting the expansion (\ref{WKBexpansion}) into the wave equation (\ref{WKBwave}), and equating the same powers of $\epsilon$, we find}{Substituindo a expans\~ao (\ref{WKBexpansion}) na equa\c{c}\~ao de onda (\ref{WKBwave}), e egualando as mesmas pot\^encias de $\epsilon$, acha-se}
\begin{equation}\label{s0}
S_{0}(x) = \pm i \intop^{x}_{x_0} Q(\eta)^{1/2} d \eta\,,
\end{equation}
\begin{equation}\label{s1}
S_{1}(x) = -\frac{1}{4} \ln Q(x)\,.
\end{equation}
\lang{The two choices of the sign in (\ref{s0}) correspond to either incoming or outgoing waves at either of the infinities $x=\pm\infty$.}{As duas escolhas de sinal em (\ref{s0}) correspondem \`as ondas que entram ou que saem em quaisquer dos infinitos $x=\pm\infty$.}

\begin{figure}
\includeonegraph{WKBMethod}
\langfigurecaption{The three regions separated by the two turning points $Q(x)=0$.}{As tr\^es regi\~oes separadas pelos dois pontos de virada $Q(x)=0$.}\label{WKBmethod}
\end{figure}

\langpar{Thus, at $x\rightarrow+\infty$ (region I), $Q(x)\rightarrow\omega^2$ in the dominant order, so that $S_{0}\rightarrow+\imo\omega x$ for a wave outgoing at the infinity and $S_{0}\rightarrow -\imo\omega x$ for a wave in-coming from infinity. In a similar fashion, at the event horizon $x \rightarrow - \infty$ (region III), $S_{0} \rightarrow +\imo\omega x$ is for wave in-coming from the event horizon, while $S_{0} \rightarrow -\imo \omega x$ is for a wave out-going to the event horizon. We shall designate these four solutions as $\Psi^{I}_{+}$, $\Psi^{I}_{-}$, $\Psi^{III}_{+}$ and $\Psi^{III}_{-}$ respectively for plus and minus signs in $S_{0}$ in I and III regions (see fig. \ref{WKBmethod}). Thus}{Assim, em $x\rightarrow+\infty$ (regi\~ao I), $Q(x)\rightarrow\omega^2$ na ordem dominante, para que $S_{0}\rightarrow+\imo\omega x$ para uma onda que sai no infinito e $S_{0}\rightarrow -\imo\omega x$ para uma onda que entra do infinito. De maneira semelhante, no horizonte de eventos $x\rightarrow-\infty$ (regi\~ao III), $S_{0}\rightarrow\imo\omega x$ para uma onda que entra do horizonte de eventos, enquanto $S_{0}\rightarrow-\imo\omega x$ para uma onda que sai ao horizonte de eventos. Essas quatro solu\c{c}\~oes ser\~ao designadas como $\Psi^{I}_{+}$, $\Psi^{I}_{-}$, $\Psi^{III}_{+}$ e $\Psi^{III}_{-}$ respectivamente para sinais positivos e negativos em $S_{0}$ nas regi\~oes I e III (ver o fig. \ref{WKBmethod}). Assim,}
\begin{equation}
\Psi^{I}_{+} \sim e^{+ \imo\omega x}, \quad x \rightarrow + \infty, \quad  \Psi^{I}_{-} \sim e^{- \imo\omega x}, \quad x \rightarrow + \infty
\end{equation}
\begin{equation}
\Psi^{III}_{+} \sim e^{+ \imo\omega x}, \quad x \rightarrow - \infty, \quad  \Psi^{III}_{-} \sim e^{- \imo\omega x}, \quad x \rightarrow - \infty
\end{equation}
\lang{The general solutions in the regions I and III are}{As solu\c{c}\~oes gerais nas regi\~oes I e III s\~ao}
\begin{eqnarray}\label{solution_I}
\Psi &=& Z_{in}^{I} \Psi^{I}_{-} + Z_{out}^{I} \Psi^{I}_{+},
\\\label{solution_III}
\Psi &=& Z_{in}^{III} \Psi^{III}_{+} + Z_{out}^{III} \Psi^{III}_{-},
\end{eqnarray}
\lang{The amplitudes at $x\rightarrow+\infty$ are connected with the amplitudes at  $x\rightarrow-\infty$ through the linear matrix relation}{As amplitudes em $x\rightarrow\infty$ s\~ao conectadas com as amplitudes em $x\rightarrow-\infty$ pela rela\c{c}\~ao linear de matriz}
\begin{equation}\label{Mmatrix}
\left(\begin{array}{c}
Z_{out}^{III}\\
Z_{in}^{III}
\end{array}\right)= \left(\begin{array}{cc}S_{11} & S_{12}\\ S_{21} & S_{22}\end{array}\right) \left(\begin{array}{c}Z_{out}^{I}\\  Z_{in}^{I}\end{array}\right).
\end{equation}
\lang{Now we need to match both WKB solutions of the form (\ref{WKBexpansion}) in the regions I and III with a solution in region II, through the two turning points $Q(x)=0$.}{Agora \'e preciso combinar ambas as solu\c{c}\~oes WKB da forma (\ref{WKBexpansion}) nas regi\~oes I e III com uma solu\c{c}\~ao na regi\~ao II, pelos dois pontos de virada $Q(x)=0$.}

\langpar{If the turning points are closely spaced, i.e. if $-Q(x)_{max} \ll Q(\pm \infty)$, then the solution in the region II can be well approximated by the Taylor series}{Se os pontos de virada forem pr\'oximos, isto \'e se $-Q(x)_{max} \ll Q(\pm \infty)$, ent\~ao a solu\c{c}\~ao na regi\~ao II pode ser bem aproximada pela s\'erie de Taylor}
\begin{equation}\label{Taylor}
Q(x) = Q_{0} + \frac{1}{2} Q_0^{\prime \prime}(x- x_{0})^{2} + {\cal O}\left((x- x_{0})^3\right),
\end{equation}
\lang{where $x_{0}$ is the point of maximum of the function $Q(x)$, $Q_{0} = Q(x_{0})$, and $Q_{0}^{\prime \prime}$ is the second derivative with respect to $x$ taken at the point $x = x_{0}$. Region II corresponds to}{onde $x_{0}$ \'e o ponto de m\'aximo da fun\c{c}\~ao $Q(x)$, $Q_{0} = Q(x_{0})$, e $Q_{0}^{\prime \prime}$ \'e a derivada segunda com rela\c{c}\~ao a $x$ no ponto $x = x_{0}$. A regi\~ao II corresponde a}
\begin{equation}\label{regionII}
| x - x_{0} |  < \sqrt{\frac{-2 Q_{0}}{Q_{0}^{\prime \prime}}} \approx \epsilon^{1/2}.
\end{equation}
\lang{The latter relation gives also the region of validity of the WKB approximation: $\epsilon$ must be small.}{A \'ultima rela\c{c}\~ao d\'a tamb\'em a regi\~ao de validade da aproxima\c{c}\~ao de WKB: $\epsilon$ deve ser pequeno.}

\langpar{Let us introduce new functions}{Introduzimos as novas fun\c{c}\~oes}
\begin{eqnarray}\label{newfunctions1}
t &=& (2 Q_{0}^{\prime \prime})^{1/4} e^{\imo\pi/4} (x - x_{0}),
\\\label{newfunctions2}
\nu + \frac{1}{2} &=& - \imo Q_{0}/(2 Q_{0}^{\prime \prime})^{1/2}.
\end{eqnarray}
\lang{Then the wave equation (\ref{WKBwave}) takes the form}{Ent\~ao a equa\c{c}\~ao de onda (\ref{WKBwave}) assume a forma}
\begin{equation}\label{parabolic}
\frac{d^{2}\Psi}{d t^{2}} + \left(\nu + \frac{1}{2} - \frac{1}{4} t^2 \right)\Psi = 0.
\end{equation}
\lang{The general solution of this equation can be expressed in terms of parabolic cylinder functions $D_{\nu}(t)$,}{A solu\c{c}\~ao geral desta equa\c{c}\~ao pode ser expressa em termos das fun\c{c}\~oes de cilindro parab\'olico $D_{\nu}(t)$,}
\begin{equation}\label{solution_parabolic}
\Psi = A D_{\nu}(t) + B D_{- \nu -1}(\imo t).
\end{equation}
\lang{Large $|t|$ asymptotics of this solution are}{As assint\'oticas de grande $|t|$ dessa solu\c{c}\~ao s\~ao}
$$ \Psi \approx B e^{- 3 i \pi (\nu + 1)/4} (4 k)^{-(\nu + 1)/4} (x-x_{0})^{-(\nu +1)} e^{i k^{1/2} (x-x_0)^{2}/2} + $$
\begin{equation}\label{large_t_sol1}
(A + B (2 \pi)^{1/2 e^{- i \nu \pi/2}}/\Gamma(\nu + 1)) e^{i \pi \nu/4} (4 k)^{\nu/4} (x-x_{0})^{\nu} e^{-i k^{1/2} (x-x_0)^{2}/2}, \quad x \gg x_2,
\end{equation}
$$ \Psi \approx A e^{- 3 i \pi \nu /4} (4 k)^{\nu/4} (x-x_{0})^{\nu} e^{-i k^{1/2} (x-x_0)^{2}/2} + $$
\begin{equation}\label{large_t_sol2}
(B - i A(2 \pi)^{1/2} e^{- i \nu \pi/2}/\Gamma(-\nu)) e^{i \pi (\nu + 1) /4} (4 k)^{-(\nu + 1)/4} (x-x_{0})^{-(\nu + 1)} e^{i k^{1/2} (x-x_0)^{2}/2}, \quad x \ll x_1,
\end{equation}
\lang{where $\displaystyle k = \frac{1}{2} Q_{0}^{\prime \prime}$.}{onde $\displaystyle k = \frac{1}{2} Q_{0}^{\prime \prime}$.}

\langpar{Equating the corresponding coefficients in (\ref{large_t_sol1}), (\ref{large_t_sol2})  and eliminating $A$ and $B$, we obtain the elements of the $S$ matrix,}{Equiparando os coeficientes correspondentes em (\ref{large_t_sol1}), (\ref{large_t_sol2}) e eliminando $A$ e $B$, obt\^em-se os elementos da matriz $S$,}
\begin{equation}\label{Mmatrixres}
\left(\begin{array}{c}
Z_{out}^{III}\\
Z_{in}^{III}
\end{array}\right)= \left(\begin{array}{cc} e^{i \pi \nu} & i R^2  e^{i \pi \nu} (2 \pi)^{1/2}/\Gamma(\nu +1) \\
R^{-2} (2 \pi)^{1/2}/\Gamma(-\nu) & - e^{i \pi \nu}  \end{array}\right) \left(\begin{array}{c}Z_{out}^{I}\\  Z_{in}^{I}\end{array}\right),
\end{equation}
\lang{where}{onde}
\begin{equation}
R = (\nu +1)^{(\nu + 1/2)/2} e^{-(\nu + 1/2)/2}.
\end{equation}
\lang{When expanding in higher WKB orders, the $S$ matrix has the same general form with just other $R$, still depending only on $\nu$.
Let us note that for a black hole there are no waves ``reflected by the horizon'', so that $Z_{in}^{III} = 0$, and
due to quasi-normal mode boundary conditions, there are no waves coming from infinity, i.e. $Z_{in}^{I} = 0$.
Both these conditions are satisfied by (\ref{Mmatrixres}), only if}{Quando se faz a expans\~ao em ordens de WKB mais altas, a matriz $S$ tem a mesma forma geral, somente com outro $R$, ainda dependendo somente de $\nu$. Nota-se que para um buraco negro n\~ao h\'a qualquer onda ``refletida pelo horizonte'', ent\~ao $Z_{in}^{III} = 0$, e por causa das condi\c{c}\~oes de contorno quase-normais, n\~ao h\'a ondas que v\^em do infinito, isto \'e $Z_{in}^{I} = 0$. Ambas condi\c{c}\~oes s\~ao satisfeitas por (\ref{Mmatrixres}) somente quando}
\begin{equation}
\Gamma(-\nu) = \infty,
\end{equation}
\lang{and, consequently, $\nu$ must be integer. Then, from the relation (\ref{newfunctions2}) we find}{e, conseq\"uentemente, $\nu$ deve ser um n\'umero inteiro. Ent\~ao, da rela\c{c}\~ao (\ref{newfunctions2}) acha-se}
\begin{equation}\label{QNM_WKB1}
n + \frac{1}{2} = - i Q_{0}/(2 Q_{0}^{\prime \prime})^{1/2}, \quad n = 0, 1, 2,\ldots\,.
\end{equation}
\lang{The latter relation gives us the complex quasi-normal modes labeled by an overtone number $n$ with the accuracy of the first WKB order \cite{Astrophys.J.Lett.291.L33.1985}. Later this approach was extended to the third WKB order beyond the eikonal approximation by Iyer and Will \cite{PhysRevD.35.3621} and to the sixth order by Konoplya \cite{Konoplya:2003ii,Konoplya:2003dd}. In order to make the higher order WKB extension it is sufficient to take higher orders in $\epsilon$ WKB series (\ref{WKBexpansion}) and to take appropriate number of consequent terms in the Taylor expansion (\ref{Taylor}). Since the $S$-matrix (\ref{Mmatrix}) depends only on $\nu$, its elements $S_{i j}$  can be found simply by solving the interior problem in region II at higher orders in $\epsilon$ \cite{PhysRevD.35.3621}, and without explicit matching of the interior solution with WKB solutions in regions I and III in each order.}{A \'ultima rela\c{c}\~ao d\'a os modos quase-normais complexos rotulados por um n\'umero de sobretom $n$ com a precis\~ao da primeira ordem de WKB \cite{Astrophys.J.Lett.291.L33.1985}. Mais tarde, esta aproxima\c{c}\~ao foi extendida \`a terceira ordem de WKB al\'em da aproxima\c{c}\~ao eikonal por Iyer e Will \cite {PhysRevD.35.3621} e \`a sexta ordem por Konoplya \cite{Konoplya:2003ii,Konoplya:2003dd}. Para fazer a extens\~ao de WKB da ordem superior \'e suficiente expandir a s\'erie de WKB at\'e ordens mais altas de $\epsilon$ (\ref{WKBexpansion}) e achar o n\'umero apropriado de termos consecutivos na expans\~ao de Taylor (\ref{Taylor}). Como a matriz $S$ (\ref{Mmatrix}) depende somente de $\nu$, pode-se achar os seus elementos $S_{i j}$ simplesmente resolvendo o problema interior na regi\~ao II em ordens mais altas de $\epsilon$ \cite{PhysRevD.35.3621}, e sem uma combina\c{c}\~ao expl\'icita da solu\c{c}\~ao interior com solu\c{c}\~oes WKB nas regi\~oes I e III para cada ordem.}

\langpar{Going over from $Q$ to the effective potential $V$, the sixth order WKB formula has the form}{Passando de $Q$ ao potencial efetivo $V$,  a f\'ormula de WKB da sexta ordem tem a forma}
\begin{equation}\label{QNM_WKB6}
\frac{i (\omega^2 - V_{0})}{\sqrt{- 2 V_{0}^{\prime \prime}}} - \Lambda_2 - \Lambda_3 - \Lambda_4 - \Lambda_5 - \Lambda_6 = n + \frac{1}{2}, \quad n = 0, 1, 2,..
\end{equation}
\lang{where the correction terms $\Lambda_i$ depend on the value of the effective potential and its derivatives (up to the $2 i$-th order) in the maximum.}{onde os termos de corre\c{c}\~ao $\Lambda_i$ dependem do valor do potencial efetivo e de suas derivadas (at\'e a ordem $2i$) no m\'aximo dele.}

\langpar{It was shown in \cite{Konoplya:2004ip} that WKB formula, extended to the sixth order, gives the relative error which is about two orders less than that of the 3rd WKB order. Yet one should remember that strictly speaking the WKB series converge only asymptotically and the consequent decreasing of the relative error in each WKB order is not guaranteed. Therefore it is reasonable to develop a modified WKB technique in the so-called optimal order \cite{Froeman:1992gp}. The latter gives better results for moderately higher overtones $n$ and especially when $n>l$. Yet, in many cases when $n \leq l$ the 6th order WKB formula gives better results than the optimal order treatment.}{Foi mostrado em \cite{Konoplya:2004ip} que a f\'ormula de WKB, extendida \`a sexta ordem, d\'a o erro relativo, que \'e aproximadamente duas ordens menor do que o erro da $3^a$ ordem de WKB. Ainda deve-se lembrar que, na realidade, a s\'erie de WKB converge s\'o assintoticamente e a redu\c{c}\~ao conseq\"uente do erro relativo em cada ordem de WKB n\~ao \'e garantida. Por isso, \'e razo\'avel desenvolver uma t\'ecnica WKB modificada na chamada ordem \'otima \cite{Froeman:1992gp}. Esta \'ultima, d\'a melhores resultados de sobretons $n$ moderadamente mais altos e, sobretudo, quando $n>l$. Ainda, em muitos casos quando $n\leq l$, a f\'ormula de WKB da $6^a$ ordem d\'a melhores resultados do que a abordagem da \'otima ordem.}

\langpar{In some cases, the WKB approach needs modifications: for instance when considering a massive scalar field in a black hole background, the effective potential has a local minimum far from a black hole. This local minimum induces two changes in the WKB procedure.
First, there are three turning points which separate all space into four regions, so that three matchings are required. Second, an influent subdominant term in the asymptotic WKB expansion at spatial infinity (\ref{solution_III}) appears (see sec. \ref{Frobenius-massive}).}{Em alguns casos, a aproxima\c{c}\~ao de WKB precisa de modifica\c{c}\~oes: por exemplo, quando considerando um campo escalar massivo no contexto de um buraco negro, o potencial efetivo tem um m\'inimo local longe de um buraco negro. Este m\'inimo local induz duas modifica\c{c}\~oes na abordagem de WKB. Primeiro, h\'a tr\^es pontos de virada que separam todo o espa\c{c}o em quatro regi\~oes, de modo que tr\^es combina\c{c}\~oes sejam necess\'arias. Segundo, um termo subdominante influente na expans\~ao assint\'otica WKB no infinito espacial (\ref{solution_III}) aparece (ver a se\c{c}\~ao \ref{Frobenius-massive}).}

\langsectionlabel{Frobenius method}{M\'etodo de Frobenius}{sec:Frobenius}

\langsubsection{Frobenius series}{S\'erie de Frobenius}
\langpar{The most accurate method of searching of eigenvalues of the equation (\ref{eigenfunction-equation}) is Frobenius method, which allows to find quasi-normal modes with arbitrary precision and does not require special form of the effective potential. This method allows to calculate also higher overtones of black holes. It was done for the first time by E. W. Leaver for Schwarzschild and Kerr black holes \cite{Leaver:1985ax}.}{O m\'etodo mais preciso para buscar valores pr\'oprios da equa\c{c}\~ao (\ref{eigenfunction-equation}) \'e o m\'etodo de Frobenius, que permite achar modos quase-normais com precis\~ao arbitr\'aria e n\~ao exige a forma especial do potencial efetivo. Este m\'etodo permite calcular tamb\'em sobretons mais altos de buracos negros. Isto foi feito pela primeira vez por E. W. Leaver para buracos negros de Schwarzschild e Kerr \cite{Leaver:1985ax}.}

\langpar{Let us consider the second order differential equation of more general form than (\ref{wave-like})}{Consideremos a equa\c{c}\~ao diferencial da segunda ordem de forma mais geral do que (\ref{wave-like})}
\begin{equation}\label{radial}
\left(\frac{d^2}{dr^2}+p(r)\frac{d}{dr}+q(r)\right)R(r)=0,
\end{equation}
\lang{where the functions $p(r)$ and $q(r)$ depend on the eigenfrequency $\omega$.}{onde as fun\c{c}\~oes $p(r)$ e $q(r)$ dependem da freq\"u\^encia pr\'opria $\omega$.}

\langpar{Let us start from the analysis of character of singular points of this equation. There are two points, which are always singular: the event horizon $r=r_+$ and the cosmological horizon (or the spatial infinity) $r=r_\infty$. Usually, there are also other singular points, that depend on $p(r)$ and $q(r)$. By definition, the quasi-normal modes are eigenvalues of $\omega$ with the boundary conditions that correspond to the outgoing wave at the cosmological horizon (spatial infinity) and the ingoing wave at the black hole event horizon. So, we are able to define the function $R(r)$ as a multiplication of some factor and the Frobenius series. The factor is divergent at these singular points. It is chosen in order to the series be convergent in the region $r_+\leq r\leq r_\infty$. If $p(r)$ and $q(r)$ are rational functions of $r$, we can construct such series in terms of the rational functions,}{Comecemos pela an\'alise do car\'ater de pontos singulares desta equa\c{c}\~ao. H\'a dois pontos, que s\~ao sempre singulares: o horizonte de eventos $r=r_+$ e o horizonte cosmol\'ogico (ou o infinito espacial) $r=r_\infty$. Normalmente, h\'a tamb\'em outros pontos singulares, que dependem de $p(r) $ e $q(r)$. Por defini\c{c}\~ao, os modos quase-normais s\~ao valores pr\'oprios de $\omega$ com as condi\c{c}\~oes de contorno que correspondem \`a onda que sai do horizonte cosmol\'ogico (infinito espacial) e a onda que entra no horizonte de eventos de buraco negro. Assim, \'e possivel definir a fun\c{c}\~ao $R(r)$ como uma multiplica\c{c}\~ao de algum fator e a s\'erie de Frobenius. O fator \'e divergente nesses pontos singulares. Ele \'e escolhido para a s\'erie ser convergente na regi\~ao $r_+\leq r\leq r_\infty$. Se $p(r)$ e $q(r)$ forem fun\c{c}\~oes racionais de $r$, pode-se construir tal s\'erie nos termos das fun\c{c}\~oes racionais,}
\begin{equation}\label{Frobenius}
R(r)=\left\{
  \begin{array}{ll}
    \displaystyle\left(\frac{r-r_\infty}{r-r_0}\right)^{\imo\Omega}\left(\frac{r-r_+}{r-r_0}\right)^{-\imo\beta} \sum_{k=0}^\infty b_k\left(\frac{r-r_+}{r-r_0}\frac{r_\infty-r_0}{r_\infty-r_+}\right)^k, & r_\infty<\infty, \\
    \displaystyle e^{\imo\Omega r}(r-r_0)^{\sigma}\left(\frac{r-r_+}{r-r_0}\right)^{-\imo\beta} \sum_{k=0}^\infty b_k\left(\frac{r-r_+}{r-r_0}\right)^k,  & r_\infty=\infty.
  \end{array}
\right.
\end{equation}
\lang{The values $\Omega$, $\sigma$ and $\beta$ are defined in order to satisfy (\ref{radial}) in the singular points $r=r_+$ and $r=r_\infty$. The quasi-normal boundary conditions fix $Re(\Omega)$ and $Re(\beta)$, which must be chosen of the same sign as $Re(\omega)$.}{Os valores $\Omega$, $\sigma$ e $\beta$ s\~ao definidos para satisfazer (\ref{radial}) nos pontos singulares $r=r_+$ e $r=r_\infty$. As condi\c{c}\~oes de contorno quase-normais fixam $Re(\Omega)$ e $Re(\beta)$, que devem ser escolhidos do mesmo sinal que $Re(\omega)$.}

\langpar{Let us consider the series}{Consideremos a s\'erie}
\begin{equation}\label{Frobenius_series}
u(z)=\sum_{k=0}^\infty b_k z^k.
\end{equation}
\lang{If all the singular points of the equation (\ref{radial}) satisfy $|z|>1$, the series (\ref{Frobenius_series}) are convergent at $z=1$ ($r=r_\infty$), if and only if the value of $\omega$ is the eigenfrequency of the equation (\ref{radial}). If there is at least one singular point inside the unit circle, one has to continue the series (\ref{Frobenius_series}) through some midpoints (see sec. \ref{sec:midpoints}) in order to test the convergence at the cosmological horizon or at the spatial infinity.}{Se todos os pontos singulares da equa\c{c}\~ao (\ref{radial}) satisfizerem $|z|>1$, a s\'erie (\ref {Frobenius_series}) \'e convergente em $z=1$ ($r=r_\infty$), se e somente se o valor de $\omega$ for a freq\"u\^encia pr\'opria da equa\c{c}\~ao (\ref{radial}). Se houver pelo menos um ponto singular dentro do c\'irculo de unidade, \'e preciso continuar a s\'erie (\ref{Frobenius_series}) atrav\'es de alguns pontos centrais (ver a se\c{c}\~ao \ref{sec:midpoints}) para testar a converg\^encia no horizonte cosmol\'ogico ou no infinito espacial.}

\langpar{Note, that the definition of $z$ contains an arbitrary parameter $r_0<r_+$. In most cases, it can be chosen in order to move all the singularities outside the unit circle.}{Observe, que a defini\c{c}\~ao de $z$ cont\'em um par\^ametro arbitr\'ario $r_0<r_+$. Na maior parte dos casos, este par\^ametro pode ser escolhido para mover todas as singularidades para fora do c\'irculo de unidade.}

\langsubsection{Method of a continued fraction}{O m\'etodo de uma fra\c{c}\~ao cont\'inua}
\langpar{Substituting (\ref{Frobenius}) into (\ref{radial}), one can obtain an $N$-term recurrence relation for the coefficients $b_i$}{Substituindo (\ref{Frobenius}) em (\ref{radial}), pode-se obter uma rela\c{c}\~ao de recorr\^encia de $N$ termos para os coeficientes $b_i$}
\begin{equation}\label{rrelation}
\sum_{j=0}^{min(N-1,i)} c_{j,i}^{(N)}(\omega)\,b_{i-j}=0,\quad
\,\,i>0\,,
\end{equation}
\lang{where the coefficients $c_{j,i}^{(N)}(\omega)$ ($0\leq j\leq min(N-1,i)$) depend on $\omega$.}{onde os coeficientes $c_{j,i}^{(N)}(\omega)$ ($0\leq j\leq min(N-1,i)$) dependem de $\omega$.}

\langpar{We now decrease the number of terms in the recurrence relation}{Agora \'e reduzido o n\'umero de termos na rela\c{c}\~ao de recorr\^encia}
\begin{equation}\label{srcRE}
\sum_{j=0}^{min(k,i)}c_{j,i}^{(k+1)}(\omega)\,b_{i-j}=0
\end{equation}
\lang{by one, i.\ e.\ we find $c_{j,i}^{(k)}(\omega)$, which satisfy
the equation}{por um, isto \'e, achamos $c_{j,i}^{(k)}(\omega)$ que satisfazem a equa\c{c}\~ao}
\begin{equation}\label{finRE}
\sum_{j=0}^{min(k-1,i)}c_{j,i}^{(k)}(\omega)\,b_{i-j}=0\,.
\end{equation}
\lang{For $i\geq k$, we can rewrite the above expression as}{Para $i\geq k$, pode-se reescrever a express\~ao acima como}
\begin{equation}\label{subsRE}
\frac{c_{k,i}^{(k+1)}(\omega)}{c_{k-1,i-1}^{(k)}(\omega)}
\sum_{j=1}^{k}c_{j-1,i-1}^{(k)}(\omega)\,b_{i-j}=0.
\end{equation}
\lang{Subtracting (\ref{subsRE}) from (\ref{srcRE}) we find the relation (\ref{finRE}) explicitly. Thus we obtain,}{Subtraindo (\ref{subsRE}) de (\ref{srcRE}) acha-se a rela\c{c}\~ao  (\ref{finRE}) explicitamente. Assim obt\'em-se,}
\begin{eqnarray}
&&c_{j,i}^{(k)}(\omega) = c_{j,i}^{(k+1)}(\omega),\qquad
\,\,j=0,\,\,\mbox{\ou}\,\,i<k,\nonumber \\[2mm]
&&c_{j,i}^{(k)}(\omega) = c_{j,i}^{(k+1)}(\omega)-\frac{c_{k,i}^{(k+1)}(\omega)\,
c_{j-1,i-1}^{(k)}(\omega)}{c_{k-1,i-1}^{(k)}(\omega)}\,.
\nonumber
\end{eqnarray}
\lang{This procedure is called \emph{Gaussian eliminations}, and allows us to determine the coefficients in the three-term recurrence relation numerically for a given $\omega$ up to any finite $i$}{Este procedimento \'e chamado \emph{elimina\c{c}\~oes de Gauss}, e permite determinar os coeficientes na rela\c{c}\~ao de recorr\^encia de tr\^es termos, numericamente, para um $\omega$ dado at\'e qualquer $i$ finito}
\begin{subequations}
\begin{eqnarray}\label{three-terms}
&&c_{0,i}^{(3)}\,b_i+c_{1,i}^{(3)}\,b_{i-1}+c_{2,i}^{(3)}\,b_{i-2}=0, \quad
\,i>1\\\label{two-terms}
&&c_{0,1}^{(3)}\,b_1+c_{1,1}^{(3)}\,b_0=0.
\end{eqnarray}\label{full-three-term}
\end{subequations}
\lang{The complexity of the procedure is \emph{linear} with respect to $i$ and $N$.}{A complexidade do procedimento \'e \emph{linear} com rela\c{c}\~ao a $i$ e $N$.}

\langpar{If the the Frobenius series are convergent, we are able to find $b_1/b_0$ from (\ref{two-terms}) and substitute it into (\ref{three-terms})}{Se a s\'erie de Frobenius for convergente, pode-se achar $b_1/b_0$ de (\ref{two-terms}) e substitui-lo em (\ref{three-terms})}
\begin{equation}
\frac{b_1}{b_0}=-\frac{c_{1,1}^{(3)}}{c_{0,1}^{(3)}} = -\frac{c_{2,2}^{(3)}}{c_{1,2}^{(3)}-}
\,\frac{c_{0,2}^{(3)}c_{2,3}^{(3)}}{c_{1,3}^{(3)}-}\,\frac{c_{0,3}^{(3)}c_{2,4}^{(3)}}{c_{1,4}^{(3)}-}\ldots\,.
\end{equation}
\lang{Finally we find}{Finalmente acha-se}
\begin{equation}
0=c_{1,1}^{(3)}-\frac{c_{0,1}^{(3)}c_{2,2}^{(3)}}{c_{1,2}^{(3)}-}
\,\frac{c_{0,2}^{(3)}c_{2,3}^{(3)}}{c_{1,3}^{(3)}-}\ldots\,,
\end{equation}
\lang{what can be inverted $n$ times}{o que pode ser invertido $n$ vezes}
\begin{equation}
c_{1,n+1}^{(3)}-\frac{c_{2,n}^{(3)}c_{0,n-1}^{(3)}}{c_{1,n-1}^{(3)}-}
\,\frac{c_{2,n-1}^{(3)}c_{0,n-2}^{(3)}}{c_{1,n-2}^{(3)}-}\ldots\,
\frac{c_{2,2}^{(3)}c_{0,1}^{(3)}}{c_{1,1}^{(3)}}=\frac{c_{0,n+1}^{(3)}c_{2,n+2}^{(3)}}{c_{1,n+2}^{(3)}-}
\frac{c_{0,n+2}^{(3)}c_{2,n+3}^{(3)}}{c_{1,n+3}^{(3)}-}\ldots\,.\label{invcf}
\end{equation}

\langpar{The equation (\ref{invcf}) with the \emph{infinite continued fraction} on the right-hand side can be solved numerically by minimising the absolute value of the difference between the left- and right-hand sides. The equation has an infinite number of roots (corresponding to the quasi-normal spectrum), but for each $n$, the most stable root is different. In general, we have to use the $n$ times inverted equation to find the $n$-th quasi-normal mode. The requirement that the continued fraction be itself convergent allows us to restrict its depth by some large value, always ensuring that an increase in this value does not change the final results within the desired precision.}{A equa\c{c}\~ao (\ref{invcf}) com a \emph{fra\c{c}\~ao cont\'inua infinita} do lado direito pode ser resolvida numericamente minimizando o valor absoluto da diferen\c{c}a entre os lados esquerdo e direito. A equa\c{c}\~ao tem um n\'umero infinito de ra\'izes, que correspondem ao espectro quase-normal, mas para cada $n$ a raiz mais est\'avel \'e diferente. Em geral, tem-se que usar a equa\c{c}\~ao $n$ vezes invertida para achar o $n^o$ modo quase-normal. A exig\^encia de que a fra\c{c}\~ao cont\'inua seja convergente, permite restringir a sua profundidade por algum valor grande, sempre assegurando-se de que um aumento neste valor n\~ao modifica os resultados finais dentro da precis\~ao desejada.}

\langsubsection{Nollert improvement}{Melhora de Nollert}
\langpar{It turns out, that the convergence of the infinite continued fraction becomes worse, if the imaginary part of $\omega$ increases with respect to the real part. It means that in order to calculate the higher overtones correctly, we must increase the depth of the continued fraction, what dramatically increases the time of calculation. The convergence is bad also if $r_0$ in (\ref{Frobenius}) is not a singular point. Such fixing of $r_0$ is necessary to move all the singular points outside the unit circle for higher-dimensional Schwarzschild black holes.}{Acontece que a converg\^encia da fra\c{c}\~ao cont\'inua infinita fica pior, se a parte imagin\'aria de $\omega$ aumentar com rela\c{c}\~ao \`a parte real. Isto significa que para calcular os sobretons mais altos corretamente, deve-se aumentar a profundidade da fra\c{c}\~ao cont\'inua, o que aumenta dramaticamente o tempo do c\'alculo. A converg\^encia \'e ruim tamb\'em se $r_0$ em (\ref{Frobenius}) n\~ao for um ponto singular. Tal fixa\c{c}\~ao de $r_0$ \'e necess\'aria para mover todos os pontos singulares para fora do c\'irculo de unidade para buracos negros de Schwarzschild em dimens\~oes mais altas.}

\langpar{The problem of slow convergence was circumvented in \cite{PhysRevD.47.5253} for the three-term recurrence relation and generalised for higher $N$ in \cite{Zhidenko:2006rs}. Let us consider}{O problema da converg\^encia lenta foi contornado em \cite{PhysRevD.47.5253} para a rela\c{c}\~ao de recorr\^encia de tr\^es termos e generalizada para $N$'s mais altos em \cite{Zhidenko:2006rs}. Consideremos}
\begin{equation}\label{NollertR}
-\frac{b_n}{b_{n-1}}=R_n=\frac{c_{2,n+1}^{(3)}}{c_{1,n+1}^{(3)}-}
\frac{c_{0,n+1}^{(3)}c_{2,n+2}^{(3)}}{c_{1,n+2}^{(3)}-}\ldots\,,
\end{equation}
\lang{that for large $n$ can be expanded as}{que para $n$ grande pode ser expandida como}
\begin{equation}\label{NollertExp}
R_n(\omega)=C_0(\omega)+\frac{C_1(\omega)}{\sqrt{n}}+\frac{C_2(\omega)}{n}+\ldots\,.
\end{equation}

\langpar{In order to find the coefficients $C_j$ of (\ref{NollertExp}), we divide the equation (\ref{rrelation}) by $b_{i-N+1}$ and use the definition $R_n=-b_n/b_{n-1}$. We find the equation with respect to $R_n$}{Para achar os coeficientes $C_j$ de (\ref{NollertExp}), divide-se a equa\c{c}\~ao (\ref{rrelation}) por $b_{i-N+1}$ e utiliza-se a defini\c{c}\~ao $R_n=-b_n/b_{n-1}$. Encontra-se a equa\c{c}\~ao}
\begin{equation}\label{NollertHD}
\sum_{j=0}^{N-1}(-1)^j\,c_{j,i}^{(N)}(\omega)\,\prod_{k=0}^{N-2-j}R_{i-k}=0.
\end{equation}
\lang{For large $n$, $c_{j,n}^{(N)}(\omega)\propto n^2$. Thus, substituting the expansion (\ref{NollertExp}) into (\ref{NollertHD}), we find}{Para $n$ grande, $c_{j,n}^{(N)}(\omega)\propto n^2$. Assim, substituindo a expans\~ao (\ref{NollertExp}) em (\ref{NollertHD}), acha-se}
\begin{equation}\label{C0eq}
\lim_{n\rightarrow\infty}\frac{1}{n^2}\sum_{j=0}^{N-1}(-1)^j\,c_{j,n}^{(N)}(\omega)\,C_0^{N-1-j}(\omega)=0\,.
\end{equation}

\langpar{In general, the equation (\ref{C0eq}) has $N-1$ roots (in fact there are multiple roots). One of the roots (also multiple) is \textit{always} $C_0=-1$, implying the unit radius of convergence of the series (\ref{Frobenius}). Other roots appear due to the existing of additional singular points of the equation (\ref{radial}). Thus we choose $C_0=-1$.}{Em geral, a equa\c{c}\~ao (\ref{C0eq}) tem $N-1$ ra\'izes (de fato h\'a ra\'izes m\'ultiplas). Uma das ra\'izes (tamb\'em m\'ultipla) \'e \textit{sempre} $C_0=-1$, implicando que o raio de converg\^encia da s\'erie (\ref{Frobenius}) \'e unit\'ario. Outras ra\'izes aparecem por causa da exist\^encia de pontos singulares adicionais da equa\c{c}\~ao (\ref{radial}). Assim seleciona-se $C_0 =-1$.}

\langpar{After fixing $C_0=-1$ one can find an equation with respect to $C_1^2$. In order to fix the sign of $C_1$ we can use the convergence of the series (\ref{Frobenius}) at $z=1$. Therefore,}{Depois de fixar $C_0=-1$ pode-se achar uma equa\c{c}\~ao com rela\c{c}\~ao a $C_1^2$. Para fixar o sinal de $C_1$ pode-se utilizar a converg\^encia da s\'erie (\ref{Frobenius}) em $z=1$. Por isso,}
$$\lim_{n\rightarrow\infty}b_n=0,\ \qquad\hbox{i. e.}\ \nexists N:~\forall n>N~|b_n|>|b_{n-1}|.$$
\lang{Since for large $n$ we have}{Como para $n$ grande tem-se}
$$\frac{b_n}{b_{n-1}}\sim -R_n\sim -C_0-\frac{C_1}{\sqrt{n}}=1-\frac{C_1}{\sqrt{n}}\,,$$
\lang{we find out that the real part of $C_1$ \emph{cannot be negative}.}{descobre-se que a parte real de $C_1$ \emph{n\~ao pode ser negativa}.}

\langpar{After the sign of $C_1$ is fixed, the other coefficients in (\ref{NollertExp}) can be found step by step from (\ref{NollertHD}) without encountering indeterminations.}{Depois que o sinal de $C_1$ for fixado, os outros coeficientes em (\ref{NollertExp}) podem ser achados passo a passo a partir de (\ref{NollertHD}) sem encontrar indetermina\c{c}\~oes.}

\langpar{Since we can calculate the coefficients $C_j$, the expansion (\ref{NollertExp}) could be used as an initial approximation for the ``remaining'' infinite continued fraction. In order to ensure the convergence of (\ref{NollertExp}) for a given value of $\omega$, one has to start from the found approximation deeply enough inside the continued fraction (\ref{invcf}). The expansion gives a good approximation for $R_n$. Therefore, the required depth is less than it would be, if we started from some arbitrary value.}{Como pode-se calcular os coeficientes $C_j$, a expans\~ao (\ref{NollertExp}) p\^ode ser utilizada como uma aproxima\c{c}\~ao inicial do ``restante'' da fra\c{c}\~ao cont\'inua infinita. Para assegurar a converg\^encia de (\ref{NollertExp}) para um valor dado de $\omega$, tem-se que come\c{c}ar da aproxima\c{c}\~ao encontrada dentro da fra\c{c}\~ao cont\'inua (\ref {invcf}) profundamente o suficiente. A expans\~ao d\'a uma boa aproxima\c{c}\~ao para $R_n$. Por isso, a profundidade necess\'aria \'e menor do que ela seria, se fosse come\c{c}ada de algum valor arbitr\'ario.}

\langsubsectionlabel{Continuation of the Frobenius series through midpoints}{Continua\c{c}\~ao da s\'erie de Frobenius por pontos centrais}{sec:midpoints}
\langpar{Let us consider the case when we are unable to fix the parameter $r_0$ in (\ref{Frobenius}) in such a way that all the singularities, except $r=r_+$ and $r=r_\infty$, move outside the unit circle. In this case there is at least one singularity, for which $|z|<1$. This singularity implies smaller radius of convergence for the series (\ref{Frobenius_series}). In order to test that the function $u(z)$ is convergent at $z=1$, we must continue the series analytically, by constructing iteratively the expansions of $u(z)$ at some midpoints \cite{Rostworowski:2006bp}.}{Consideremos o caso quando n\~ao \'e poss\'ivel fixar o par\^ametro $r_0$ em (\ref{Frobenius}) de tal jeito que todas as singularidades, exceto $r=r_+$ e $r=r_\infty$, se movem para fora do c\'irculo de unidade. Neste caso existe pelo menos uma singularidade, para qual $|z|<1$. Esta singularidade implica menor raio de converg\^encia para a s\'erie (\ref{Frobenius_series}). Para testar que a fun\c{c}\~ao $u(z)$ \'e convergente em $z=1$, deve-se continuar a s\'erie analiticamente, construindo iterativamente as expans\~oes de $u(z)$ em alguns pontos centrais \cite{Rostworowski:2006bp}.}

\langpar{Namely, we equate the series expansion at two points}{Equaciona-se a expans\~ao de s\'erie em dois pontos}
\begin{equation}\label{Frobenius_series-midpoint}
u(z)=\sum_{n=0}^\infty b_n z^n = \sum_{n=0}^\infty \tilde{b}_n(z-z_0)^n,
\end{equation}
\lang{where $z=z_0$ is a midpoint inside the radius of convergence of (\ref{Frobenius_series}).}{onde $z=z_0$ \'e um ponto central dentro do raio de converg\^encia de (\ref{Frobenius_series}).}

\langpar{The coefficients $\tilde{b}_n$ also satisfy the $N$-term recurrence relation, which could be reduced to the three-term one}{Os coeficientes $\tilde{b}_n$ tamb\'em satisfazem a rela\c{c}\~ao de recorr\^encia de $N$ termos, que poderia ser reduzida \`a rela\c{c}\~ao de tr\^es termos}
\begin{equation}\label{three-terms-midpoint}
\tilde{c}_{0,i}^{(3)}\,\tilde{b}_i+\tilde{c}_{1,i}^{(3)}\,\tilde{b}_{i-1}+\tilde{c}_{2,i}^{(3)}\,\tilde{b}_{i-2}=0, \quad
i>1.
\end{equation}
\lang{In order to find $\tilde{b}_1/\tilde{b}_2$, we must use the condition at the event horizon by taking into account (\ref{Frobenius_series-midpoint}),}{Para achar $\tilde{b}_1/\tilde{b}_2$, deve-se utilizar a condi\c{c}\~ao no horizonte de eventos levando em conta}
\begin{equation}
\tilde{b}_0=\sum_{n=0}^\infty b_n z_0^n, \qquad \tilde{b}_1=\sum_{n=1}^\infty n b_n z_0^{n-1}.
\end{equation}
\lang{From (\ref{two-terms}) and (\ref{three-terms}) we find the coefficients $b_n$ and substitute them into (\ref{three-terms-midpoint}). If $z=1$ is the closest singular point to $z=z_0$, we obtain the equation with respect to $\omega$ as}{De (\ref{two-terms}) e (\ref{three-terms}) acham-se os coeficientes $b_n$ substituindo-os em (\ref{three-terms-midpoint}). Se $z=1$ for o ponto singular mais pr\'oximo a $z=z_0$, obt\'em-se a equa\c{c}\~ao com rela\c{c}\~ao a $\omega$ como}
\begin{equation}\label{continued-fraction-midpoint}
\frac{\tilde{b}_1}{\tilde{b}_0} = -\frac{\tilde{c}_{2,2}^{(3)}}{\tilde{c}_{1,2}^{(3)}-}
\,\frac{\tilde{c}_{0,2}^{(3)}\tilde{c}_{2,3}^{(3)}}{\tilde{c}_{1,3}^{(3)}-}\,\frac{\tilde{c}_{0,3}^{(3)}\tilde{c}_{2,4}^{(3)}}{\tilde{c}_{1,4}^{(3)}-}\ldots\,.
\end{equation}
\lang{Otherwise one has to repeat the procedure, by constructing the series (\ref{Frobenius_series-midpoint}) for the next midpoints $z_1,\,z_2,\,z_3,\,\ldots$, until the cosmological horizon (or spatial infinity) appears to be inside the radius of convergence.}{Sen\~ao, tem-se que repetir o procedimento, construindo a s\'erie (\ref{Frobenius_series-midpoint}) para os pontos centrais seguintes $z_1,\,z_2,\,z_3,\,\ldots$, at\'e que o horizonte cosmol\'ogico (ou infinito espacial) pare\c{c}a estar dentro do raio de converg\^encia.}

\langpar{One should note, if the convergence of the continued fraction on the right-hand side of (\ref{continued-fraction-midpoint}) is slow, one can use the Nollert improvement. Since the radius of convergence of the Frobenius series is now less than one ($R<1$), we must choose $C_0=-R^{-1}$ in (\ref{NollertExp}).}{\'E preciso observar que, se a converg\^encia da fra\c{c}\~ao cont\'inua do lado direito de (\ref{continued-fraction-midpoint}) for lenta, pode-se utilizar a melhora de Nollert. Como o raio da converg\^encia da s\'erie de Frobenius \'e agora menor que um ($R<1$), deve-se selecionar $C_0=-R^{-1}$ em (\ref{NollertExp}).}

\langsubsectionlabel{Generalisation of the Frobenius series}{Generaliza\c{c}\~ao da s\'erie de Frobenius}{sec:Frobenius-angular}
\langpar{The series expansion (\ref{Frobenius_series}) is not necessarily to be done in terms of power of a rational function of $r$. For some purposes, the more convenient choice is expansion in terms of an other full set of functions in the appropriate Hilbert space. Here we consider, as an example, a Kerr-Newman-de Sitter black hole, that is described by the line element}{A expans\~ao da s\'erie (\ref{Frobenius_series}) n\~ao deve necessariamente ser feita nos termos da pot\^encia de uma fun\c{c}\~ao racional de $r$. Para alguns objetivos, a escolha mais conveniente \'e a expans\~ao nos termos de um outro conjunto completo de fun\c{c}\~oes no espa\c{c}o de Hilbert apropriado. Aqui considera-se, como um exemplo, um buraco negro Kerr-Newman-de Sitter, que \'e descrito pelo elemento de linha}
\begin{equation}\label{Kerr-Newman-deSitter-metric}
ds^2 = -\rho^2
\left(\frac{dr^2}{\Delta_r}+\frac{d\theta^2}{\Delta_\theta}\right)
-\frac{\Delta_\theta \sin^2\theta}{(1+\alpha)^2 \rho^2}
[adt-(r^2+a^2)d\varphi]^2  +\frac{\Delta_r}{(1+\alpha)^2 \rho^2}(dt-a\sin^2\theta d\varphi)^2,
\end{equation}
\lang{where}{onde}
\begin{equation}
\begin{array}{cc}
\multicolumn{2}{c}{{\displaystyle
\Delta_r=(r^2+a^2)\left(1-\frac{\alpha}{a^2}r^2\right)-2Mr+Q^2,}} \\
\Delta_\theta=1+\alpha\cos^2\theta, &
{\displaystyle \alpha=\frac{\Lambda a^2}{3}}, \\
\end{array}
\end{equation}
\lang{$M$ is the black hole mass, $Q$ is the charge, $a$ is the rotation parameter, and $\Lambda$ is the cosmological constant.}{$M$ \'e a massa do buraco negro, $Q$ \'e a carga, $a$ \'e o par\^ametro de rota\c{c}\~ao, e $\Lambda$ \'e a constante cosmol\'ogica.}

\langpar{After separation of the variables, the angular part of the massless (charged) field equation of motion can be reduced to \cite{Suzuki:1998vy}}{Depois da separa\c{c}\~ao das vari\'aveis, a parte angular da equa\c{c}\~ao do movimento para o campo (carregado) sem massa reduze-se a}
\begin{eqnarray}
\left(\frac{d}{dx}(1 + \alpha x^2)(1 - x^2)\frac{d}{dx}+\lambda-s(1-\alpha)+\frac{(1+\alpha)^2}{\alpha}\xi^2-2\alpha x^2 + \right.\nonumber\\ \label{angular-Kerr}\frac{1+\alpha}{1+\alpha x^2}\left(2s(\alpha m - (1+\alpha)\xi)x-\frac{(1+\alpha)^2}{\alpha}\xi^2-2m(1+\alpha)\xi+s^2\right)-\\\nonumber
\left.-\frac{(1+\alpha)^2m^2}{(1+\alpha x^2)(1-x^2)}-\frac{(1+\alpha)(s^2+2smx)}{1-x^2}\right)S(x)=0,
\end{eqnarray}
\lang{where $\lambda$ is the separation constant $\xi=a\omega$, $x=\cos(\theta)$, $s$ is the field spin and $m$ is the projection of the angular momentum of the field onto the axis of the black hole rotation, $0\leq s\leq 2$ and $m$ are (half)integers.}{onde $\lambda$ \'e a constante de separa\c{c}\~ao $\xi=a\omega$, $x=\cos(\theta)$, $s$ \'e o spin do campo e $m$ \'e a proje\c{c}\~ao do momento angular do campo para o eixo de rota\c{c}\~ao do buraco negro, $0\leq s\leq2$ e $m$ s\~ao n\'umeros (semi)inteiros.}

\langpar{The appropriate series for the function $S$ are \cite{Suzuki:1998vy}}{A s\'erie apropriada para a fun\c{c}\~ao $S$ \'e \cite{Suzuki:1998vy}}
\begin{equation}\label{angular-Kerr-Frobenius}
S(z)=z^{A_1}(z-1)^{A_2}(z-z_s)^{A_3}(z-z_\infty)\sum_{n=0}^\infty b_n u_n(z),
\end{equation}
\lang{where}{onde}
$$z=\frac{\sqrt{\alpha}-\imo}{2}\frac{x+1}{x\sqrt{\alpha}-\imo}, \qquad z_s = -\frac{\imo(1+\imo\sqrt{\alpha})^2}{4\sqrt{\alpha}}, \qquad z_\infty = -\frac{\imo(1+\imo\sqrt{\alpha})}{2\sqrt{\alpha}},$$
$$A_1=\frac{|m-s|}{2}, \qquad A_2=\frac{|m+s|}{2}, \qquad A_3 = \pm\frac{\imo}{2}\left(\frac{1+\alpha}{\sqrt{\alpha}}\xi-\sqrt{\alpha}\xi -\imo s\right),$$
\lang{and the expansion is done in terms of the Jacobi polynomials}{e a expans\~ao \'e feita nosem termos dos polin\^omios de Jacobi}
$$u_n(z)=F(-n,n+\bar{\omega};\gamma;z)=(-1)^n\frac{\Gamma(2n+\bar{\omega})n!}{\Gamma(n+\gamma)}P_n^{(\bar{\omega}-\gamma,\gamma-1)}(2z-1),$$
\lang{where $\bar{\omega} = 2A_1+2A_2+1$ and $\gamma = 2A_1+1$.}{onde $\bar{\omega} = 2A_1+2A_2+1$ e $\gamma = 2A_1+1$.}

\langpar{The coefficients $b_n$ in (\ref{angular-Kerr-Frobenius}) satisfy the three-term recurrence relation (\ref{three-terms}, \ref{two-terms}) with}{Os coeficientes $b_n$ em (\ref{angular-Kerr-Frobenius}) satisfazem a rela\c{c}\~ao de recorr\^encia de tr\^es termos (\ref{three-terms}, \ref{two-terms}) com}
\begin{eqnarray}
c^{(3)}_{0,n} &=&
\pm \frac{i}{\sqrt{\alpha}}\xi \
\frac{n(n+A_1+A_2 \mp s )(n+2A_2)}
{2(2n+2A_1+2A_2+1)(n+A_1+A_2)}, \\
c^{(3)}_{1,n} &=&
\frac{i}{\sqrt{\alpha}} \left\{
\pm \xi \frac{J_n}
{2(n+A_1+A_2)(n+A_1+A_2-1)} \right.\nonumber \\
&& \makebox[10mm]{} + \frac{(n-1)(n+2A_1+2A_2)}{4} \\
&& \makebox[10mm]{} \left. -\frac{1}{4}\left[
\lambda-2A_1 A_2 -A_1 -A_2 +2\big(m+s \mp (2A_1 +1)\big)\xi
-\frac{m^2-s^2}{2}-s \right] \right\}, \nonumber \\
c^{(3)}_{2,n} &=&
\mp\frac{i}{\sqrt{\alpha}}\xi \
\frac{(n-1+A_1+A_2\pm s)(n-1+2A_1)(n-1+2A_1+2A_2)}
{2(2n+2A_1+2A_2-3)(n-1+A_1+A_2)},
\end{eqnarray}
\lang{where}{onde}
$$J_n = (n-1)(n+2A_1+2A_2)(A_1-A_2) +(A_1+A_2 \pm s+1)\left((n-1)(n+2A_1+2A_2)+(2A_1+1)(A_1+A_2)\right).$$

\langpar{Since the series (\ref{angular-Kerr-Frobenius}) must be convergent at $z=1$, we can solve numerically the equation (\ref{invcf}), and find, thereby, the separation constant as a function of frequency.}{Como a s\'erie (\ref{angular-Kerr-Frobenius}) deve ser convergente em $z=1$, \'e poss\'ivel resolver numericamente a equa\c{c}\~ao (\ref{invcf}), e achar, por meio desta, a constante de separa\c{c}\~ao como uma fun\c{c}\~ao da freq\"u\^encia.}

\langpar{In the Reissner-Nordstr\"om-de Sitter limit ($a\rightarrow0$) the equation (\ref{invcf}) is reduced to $c^{(3)}_{1,n}=0$. In this case the value of $\lambda$ does not depend on $\omega$}{No limite de Reissner-Nordstr\"om-de Sitter ($a\rightarrow0$) a equa\c{c}\~ao (\ref{invcf}) reduze-se a $c^{(3)}_{1,n}=0$. Neste caso o valor de $\lambda$ n\~ao depende de $\omega$}
$$\lambda = (l-s+1)(l+s), \qquad l = n+A_1+A_2\geq max(|m|,|s|).$$

\langsubsectionlabel{Frobenius series for the radial part of the charged field equation in the Kerr-Newman-de Sitter background}{S\'erie de Frobenius para a parte radial da equa\c{c}\~ao de campo carregado no contexto de Kerr-Newman-de Sitter}{sec:Frobenius-radial}
\langpar{The radial part of the massless (charged) field equation of motion is of the form (\ref{radial}) and reads \cite{Suzuki:1998vy}}{A parte radial da equa\c{c}\~ao de movimento de campo (carregado) sem massa \'e da forma (\ref{radial}) e l\^e-se \cite{Suzuki:1998vy}}
\begin{eqnarray}
&&\Bigg\{ \
\Delta_r^{-s}\frac{d}{dr}\Delta_r^{s+1}\frac{d}{dr}
+\frac{1}{\Delta_r}\left(K^2
- is K \frac{d\Delta_r}
{dr} \right)
\nonumber\\
&&+4\imo s(1+\alpha)\omega r -\frac{2\alpha}{a^2}(s+1)(2s+1) r^2
+2s(1-\alpha)-2 i s q Q -\lambda \ \Bigg\} R(r) = 0,
\label{eqn:Rr}
\end{eqnarray}
\lang{where $K = [\omega(r^2+a^2)- am](1+\alpha)- q Q r$, $q$ is the field charge.}{onde $K = [\omega(r^2+a^2)- am](1+\alpha)- q Q r$, $q$ \'e a carga do campo.}

\langpar{The appropriate Frobenius series are found to be \cite{Konoplya:2007zx}}{A s\'erie apropriada de Frobenius foi achada na forma \cite{Konoplya:2007zx}}
\begin{equation}\label{Frobenius-KNdD}
R(r)=\left(\frac{r-r_+}{r-r_-}\right)^{-s-2\imo
K(r_+)/\Delta_r'(r_+)}e^{\imo B(r)}r^{-2s-1}u\left(\frac{r-r_+}{r-r_-}\frac{r_\infty-r_-}{r_\infty-r_+}\right)
\end{equation}
\lang{Note, that, in order to obtain the recurrence relation for both types of the boundaries (asymptotically flat and de Sitter), we introduce the exponent $\displaystyle e^{\imo B(r)}$ such that $\displaystyle\frac{dB(r)}{dr}=\frac{K}{\Delta_r}$. This exponent describes an outgoing wave at the horizons and spatial infinity. Thus we have to compensate the outgoing wave at the event horizon. That is why the factor $2$ appears in the power of the first multiplier in (\ref{Frobenius-KNdD}). The parameter $r_0$ is fixed to be the inner horizon $r_-$, in order to move all the singularities outside the unit circle and, at the same time, to provide the best convergence of the infinite continued fraction (\ref{invcf}).}{H\'a que se observar, que para se obter a rela\c{c}\~ao de recorr\^encia para ambos os tipos de limites (assintoticamente plano e de Sitter), introduz-se o expoente $\displaystyle e^{\imo B(r)}$ tal que $\displaystyle\frac{dB(r)}{dr}=\frac{K}{\Delta_r}$. Este expoente descreve uma onda que sai nos horizontes e no infinito espacial. Assim, \'e preciso compensar a onda que sai no horizonte de eventos. Por isso, o fator $2$ aparece na pot\^encia do primeiro multiplicador em (\ref{Frobenius-KNdD}). O par\^ametro $r_0$ foi fixado para ser o horizonte interior $r_-$, para mover todas as singularidades para fora do c\'irculo de unidade e, ao mesmo tempo, fornecer a melhor converg\^encia da fra\c{c}\~ao cont\'inua infinita (\ref{invcf}).}

\langpar{Since $\lambda$ can be calculated numerically as a function of $\omega$ (see section \ref{sec:Frobenius-angular}), we are able to solve the equation (\ref{invcf}) with respect to $\omega$.}{Como $\lambda$ pode ser calculado numericamente como uma fun\c{c}\~ao de $\omega$ (ver a se\c{c}\~ao \ref{sec:Frobenius-angular}), \'e poss\'ivel resolver a equa\c{c}\~ao (\ref {invcf}) com rela\c{c}\~ao a $\omega$.}

\langpar{In the Reissner-Nordstr\"om-de Sitter limit ($a=0$) for neutral field ($q=0$) we obtain $K=\omega r^2$ and}{No limite de Reissner-Nordstr\"om-de Sitter ($a=0$) para campo neutro ($q=0$) obt\^em-se $K=\omega r^2$ e}
\begin{equation}\label{Frobenius-RNdD}
R(r)=\left(\frac{r-r_+}{r-r_-}\right)^{-s-\imo
\omega/\kappa}e^{\imo \omega r_\star}r^{-2s-1}u\left(\frac{r-r_+}{r-r_-}\frac{r_\infty-r_-}{r_\infty-r_+}\right).
\end{equation}
\lang{The tortoise coordinate is defined by $\displaystyle dr_\star=\frac{r^2 dr}{\Delta_r}$, and $\displaystyle\kappa=\frac{\Delta_r'(r_+)}{2r_+^2}$ is the surface gravity on the event horizon.}{A coordenada tartaruga \'e definida por $\displaystyle dr_\star=\frac{r^2 dr}{\Delta_r}$, e $\displaystyle\kappa=\frac{\Delta_r'(r_+)}{2r_+^2}$ \'e a gravidade da superf\'icie no horizonte de eventos.}

\langsubsectionlabel{Frobenius series for the massive scalar field equation in the higher-dimensional Reissner-Nordstr\"om-de Sitter background}{S\'erie de Frobenius para a equa\c{c}\~ao do campo escalar massivo no contexto Reissner-Nordstr\"om-de Sitter de dimens\~oes mais altas}{Frobenius-massive}
\langpar{A $D$-dimensional Reissner-Nordstr\"om-de Sitter black hole is described by the metric}{Um buraco negro Reissner-Nordstr\"om-de Sitter $D$-dimensional descreve-se pela m\'etrica}
\begin{equation}
ds^2=f(r)dt^2-\frac{dr^2}{f(r)}-r^2d\Omega_{D-2},
\end{equation}
\lang{where $d\Omega_{D-2}$ is the line element of a $(D-2)$-dimensional sphere,}{onde $d\Omega_{D-2}$ \'e o elemento de linha de uma esfera $(D-2)$-dimensional,}
\begin{equation}
f(r)=1-\frac{2M}{r^{D-3}}+\frac{Q^2}{r^{2D-6}}-\frac{2\Lambda r^2}{(D-1)(D-2)}.
\end{equation}

\langpar{After separation of the angular and time variables, the radial part of the massive scalar field equation of motion $(\Box+\mu^2)\Psi=0$ is reduced to the wave-like equation}{Depois da separa\c{c}\~ao das vari\'aveis angulares e vari\'avel de tempo, a parte radial da equa\c{c}\~ao de movimento para o campo escalar massivo  $(\Box+\mu^2)\Psi=0$ \'e reduzida \`a equa\c{c}\~ao de onda}
\begin{equation}
\label{potential-tensortype}
\left(\frac{d^2}{dr_\star^2}+\omega^2-f(r)\left(\mu^2+\frac{l(l+D-3)}{r^2}+\frac{f'(r)(D-2)}{2r}+\frac{f(r)(D-2)(D-4)}{4r^2}\right)\right)r^{\frac{D-2}{2}}R(r) = 0,
\end{equation}
\lang{where integer $l=0,1,2\ldots$ parameterises the angular separation constant.}{onde o n\'umero inteiro $l=0,1,2\ldots$ parametra a constante angular de separa\c{c}\~ao.}

\langpar{The Frobenius series for this case are}{A s\'erie de Frobenius neste caso \'e}
\begin{equation}\label{Frobenius-msRNdD}
R(r) = \left(\frac{r-r_+}{r-r_0}\right)^{-\frac{\imo\omega}{\kappa}}e^{\imo A(r)}r^{-\frac{D-2}{2}}u\left(\frac{r-r_+}{r-r_0}\frac{r_\infty-r_0}{r_\infty-r_+}\right),
\end{equation}
\lang{where $\displaystyle\kappa=\frac{1}{2}f'(r_+)$, $\displaystyle e^{\imo A(r)}$ describes the outgoing wave for the spatial infinity and the horizons $\displaystyle\frac{dA(r)}{dr}=\frac{\sqrt{\omega^2-\mu^2f(r)}}{f(r)}$. The sign in the exponent is fixed by the quasi-normal boundary condition: the real part of $A(r\rightarrow r_\infty)$ must be of the same sign as the real part of the eigenfrequency $\omega$. This choice of the sign makes the wave outgoing at the cosmological horizon (or spatial infinity).}{onde $\displaystyle\kappa=\frac{1}{2}f'(r_+)$, $\displaystyle e^{\imo A(r)}$ descreve a onda que sai para o infinito espacial e horizontes $\displaystyle\frac{dA(r)}{dr}=\frac{\sqrt{\omega^2-\mu^2f(r)}}{f(r)}$. O sinal no expoente foi fixado pela condi\c{c}\~ao de contorno quase-normal: a parte real de $A(r\rightarrow r_\infty)$ deve ser do mesmo sinal que a parte real da freq\"u\^encia pr\'opria $\omega$. Esta escolha do sinal faz a onda sair para o horizonte cosmol\'ogico (ou infinito espacial).}

\langpar{For massless field ($\mu=0$) this exponent is $e^{\imo\omega r_\star}$ \cite{Konoplya:2007jv} and the Frobenius series read}{Para um campo sem massa ($\mu=0$) esse expoente \'e $e^{\imo\omega r_\star}$ \cite{Konoplya:2007jv} e a s\'erie de Frobenius l\^e-se}
\begin{equation}\label{Frobenius-sRNdD}
R(r) = \left(\frac{r-r_+}{r-r_0}\right)^{-\frac{\imo\omega}{\kappa}}e^{\imo \omega r_\star}r^{-\frac{D-2}{2}}u\left(\frac{r-r_+}{r-r_0}\frac{r_\infty-r_0}{r_\infty-r_+}\right).
\end{equation}
\lang{Since for $D=4$ we can choose $r_0=r_-$, we come to (\ref{Frobenius-RNdD}) ($s=0$).}{Como para $D=4$ pode-se escolher $r_0=r_-$, vem-se a (\ref{Frobenius-RNdD}) ($s=0$).}

\langpar{At the cosmological horizon we can observe the same asymptotical behavior of the exponent}{No horizonte cosmol\'ogico \'e poss\'ivel observar o mesmo comportamento assint\'otico do expoente}
$$e^{\imo A(r)}\sim e^{\imo\omega r_\star}, \qquad r\rightarrow r_\infty<\infty.$$
\lang{If $\Lambda=0$, $D\geq5$, $f(r)=1+o(r^{-1})$ we can write the Frobenius series as \cite{Zhidenko:2006rs}}{Se $\Lambda=0$, $D\geq5$, $f(r)=1+o(r^{-1})$, pode-se escrever a s\'erie de Frobenius como \cite{Zhidenko:2006rs}}
\begin{equation}\label{Frobenius-msDRN}
R(r) = \left(\frac{r-r_+}{r-r_0}\right)^{-\frac{\imo\omega}{2\kappa}}e^{\imo r\sqrt{\omega^2-\mu^2}}r^{-\frac{D-2}{2}}u\left(\frac{r-r_+}{r-r_0}\right).
\end{equation}
\lang{For $D=4$ the term of order $\sim r^{-1}$ in $f(r)$ leads to the non-trivial contribution \cite{Ohashi:2004wr}}{Para $D=4$ o termo de ordem $\sim r^{-1}$ em $f(r)$ leva \`a contribui\c{c}\~ao n\~ao trivial \cite{Ohashi:2004wr}}
\begin{equation}\label{Frobenius-msRN}
R(r) = \left(\frac{r-r_+}{r-r_0}\right)^{-\frac{\imo\omega}{2\kappa}}e^{\imo r\sqrt{\omega^2-\mu^2}}(r-r_0)^{\imo r_+\sqrt{\omega^2-\mu^2}+\imo r+2\mu^2/2\sqrt{\omega^2-\mu^2}}r^{-\frac{D-2}{2}}u\left(\frac{r-r_+}{r-r_0}\right).
\end{equation}

\langpar{The same approach could be applied for the equations of motion for the Maxwell field and the gravitational perturbations, because their radial part can be reduced to the form (\ref{radial}) \cite{Kodama:2003kk,Ishibashi:2003ap}. Practically, the continued fraction coefficients appear to be so complicated for such cases, that we are unable to compute quasi-normal modes with this method during reasonable time.}{A mesma abordagem pode ser aplicada para as equa\c{c}\~oes de movimento para o campo de Maxwell e as perturba\c{c}\~oes gravitacionais, porque as suas partes radiais podem ser reduzidas \`a forma (\ref{radial}) \cite{Kodama:2003kk,Ishibashi:2003ap}. Na pr\'atica, os coeficientes de fra\c{c}\~ao cont\'inua parecem ser t\~ao complicados para tais casos, que n\~ao \'e poss\'ivel computar modos quase-normais com este m\'etodo durante um tempo aceit\'avel.}

\langsectionlabel{Horowitz-Hubeny method}{O m\'etodo de Horowitz-Hubeny}{sec:Horowitz-Hubeny}
\langpar{In order to find quasi-normal modes in the asymptotically anti-de Sitter space-times, usually we need to impose the Dirichlet boundary conditions at the spatial infinity. Thus, we can find the appropriate expansion for the function $R(r)$ in (\ref{radial}) without consideration of the singularity point at the infinity. This method was proposed by Horowitz and Hubeny \cite{Horowitz:1999jd}. Namely, we define}{Para achar modos quase-normais nos espa\c{c}os-tempo assintoticamente anti-de Sitter, normalmente \'e preciso impor as condi\c{c}\~oes de contorno de Dirichlet no infinito espacial. Assim, pode-se achar a expans\~ao apropriada para a fun\c{c}\~ao $R(r)$ em (\ref{radial}) sem a considera\c{c}\~ao do ponto de singularidade no infinito. Este m\'etodo foi proposto por Horowitz e Hubeny \cite{Horowitz:1999jd}. Definimos}
\begin{equation}\label{HH-singularity}
R = z^{-\imo\omega/(2\kappa)}\psi(z), \qquad z=\frac{r-r_+}{r-r_-},
\end{equation}
\lang{where $\kappa$ is the surface gravity at the event horizon.}{onde $\kappa$ \'e a gravidade da superf\'icie no horizonte de eventos.}

\langpar{If we substitute (\ref{HH-singularity}) into  (\ref{radial}), we find that $\psi(z)$ satisfies the equation}{Se substituirmos  (\ref{HH-singularity}) em (\ref{radial}), descobrimos que $\psi(z)$ satisfaz a equa\c{c}\~ao}
\begin{equation}\label{HH-radial}
s(z)\psi''(z)+\frac{t(z)}{z}\psi'(z)+\frac{u(z)}{z^2}\psi(z)=0,
\end{equation}
\lang{where}{onde}
$$s(z)=\sum_{n=0}^{N_s}s_nz^n, \quad t(z)=\sum_{n=0}^{N_t}t_nz^n, \quad u(z)=\sum_{n=1}^{N_u}u_nz^n,$$
\lang{are polynomial functions of $z$.}{s\~ao fun\c{c}\~oes polinomiais de $z$.}

\langpar{Since the function $\psi(z)$ is regular at the event horizon $z=0$, we can expand it as}{Como a fun\c{c}\~ao $\psi(z)$ \'e regular no horizonte de eventos $z=0$, pode-se expandi-la como}
\begin{equation}\label{HH-Frobenius}
\psi(z)=\sum_{n=0}^\infty a_n z^n.
\end{equation}

\langpar{The Dirichlet boundary condition at the spatial infinity $\psi(1)=0$ reads}{A condi\c{c}\~ao de contorno de Dirichlet no infinito espacial $\psi(1)=0$ l\^e-se}
\begin{equation}\label{HH-equation}
\sum_{n=0}^\infty a_n=0.
\end{equation}
\lang{Substituting (\ref{HH-Frobenius}) into (\ref{HH-radial}), we can find the recurrence relation for the coefficients $a_n$}{Substituindo (\ref{HH-Frobenius}) em (\ref{HH-radial}), pode-se encontrar a rela\c{c}\~ao de recorr\^encia para as coeficientes $a_n$}
\begin{equation}\label{HH-recurrence}
a_n=-\sum_{k=0}^{n-1}a_k\frac{k(k-1)s_{n-k}+kt_{n-k}+u_{n-k}}{n(n-1)s_0+nt_0}.
\end{equation}

\langpar{The equation (\ref{HH-recurrence}) allows to calculate all $a_n$ starting from an arbitrary $a_0$, which fixes the scale of $\psi(z)$. Substituting $a_n$ into (\ref{HH-equation}), we find the equation with respect to the eigenvalue $\omega$. Practically, since the series (\ref{HH-equation}) are convergent, we sum over some finite number of terms and solve (\ref{HH-equation}) with respect to $\omega$. In order to ensure that this truncation does not cause incorrect result we have to increase the number of terms until the value of $\omega$ does not change within the required precision.}{A equa\c{c}\~ao (\ref{HH-recurrence}) permite calcular todos $a_n$ come\c{c}ando de um $a_0$ arbitr\'ario, que fixa a escala de $\psi(z)$. Substituindo $a_n$ em (\ref{HH-equation}), encontra-se a equa\c{c}\~ao com rela\c{c}\~ao \`a freq\"u\^encia pr\'opria $\omega$. Na pr\'atica, como a s\'erie (\ref{HH-equation}) \'e convergente, soma-se um n\'umero finito de termos e resolve-se (\ref{HH-equation}) com rela\c{c}\~ao a $\omega$. Para assegurar que este truncamento n\~ao causa um resultado incorreto deve-se aumentar o n\'umero de termos at\'e que o valor de $\omega$ n\~ao se modifique em rela\c{c}\~ao \`a precis\~ao de que se necessita.}

\langpar{Note, that, for the sum (\ref{HH-equation}) to be convergent, $r_-$ has to be a singular point of the equation (\ref{radial}), and all the other singularities, except $r=r_+$ and $r=r_\infty$, must lay outside the unit circle $|z|>1$. If both of these conditions are impossible to satisfy, we must use the continued fraction method with the appropriate fixing of the behavior of $R$ at spatial infinity.}{H\'a que se observar, que, para a soma (\ref{HH-equation}) ser convergente, $r_-$ tem que ser um ponto singular da equa\c{c}\~ao (\ref{radial}), e todas as outras singularidades, exceto $r=r_+$ e $r=r_\infty$, devem p\^or-se fora do c\'irculo de unidade $|z|>1$. Se ambas essas condi\c{c}\~oes forem imposs\'iveis de se satisfazer, deve-se usar o m\'etodo de fra\c{c}\~ao cont\'inua com a fixa\c{c}\~ao apropriada do comportamento de $R$ no infinito espacial.}

\langchapter{Perturbations of four-dimensional black holes}{Perturba\c{c}\~oes de buracos negros quadridimensionais}
\langsectionlabel{Quasi-normal modes of Schwarzschild and Schwarzschild-de Sitter black holes}{Modos quase-normais de buracos negros de Schwarzschild e de Schwarzschild-de Sitter}{sec:SdS}
\begin{figure}
\includetwograph{D=4_l=2_Re}{D=4_l=2_Im}
\langfigurecaption{Real and imaginary parts of the fundamental quasi-normal frequency of the Schwarzschild-de Sitter black hole for scalar ($l=2$), electromagnetic ($l=2$), gravitational ($l=2$), and Dirac ($\kappa=2$) perturbations plotted with solid blue, dashed red, dot-dashed green and dotted black lines respectively.}{As partes real e imagin\'aria da freq\"u\^encia quase-normal fundamental do buraco negro Schwarzschild-de Sitter para perturba\c{c}\~oes escalares ($l=2$), eletromagn\'eticas ($l=2$), gravitacionais ($l=2$), e Dirac ($\kappa=2$) s\~ao plotadas com linhas azuis s\'olidas, vermelhas tracejadas, verdes pontilhadas-tracejadas e pretas pontilhadas, respectivamente.}\label{dSQNMs}
\end{figure}
\langpar{Perturbations of four-dimensional black holes were studied extensively in the context of possibility to observe quasi-normal ringing with the help of gravitational wave detectors (see reviews \cite{Nollert:1999ji,Kokkotas:1999bd}). Since in our Universe the cosmological constant appears to be positive, its correction to the quasi-normal spectrum must be taken into account. Gravitational perturbations of spherically symmetric black holes in the de Sitter background were studied for the first time by the numerical integration (see sec. \ref{sec:time-domain}) in \cite{Mellor:1989ac}.}{Perturba\c{c}\~oes de buracos negros quadridimensionais foram estudadas extensivamente no contexto da possibilidade de observar o toque quase-normal com a ajuda de detectores de ondas gravitacionais (ver revis\~oes \cite{Nollert:1999ji,Kokkotas:1999bd}). Como no nosso Universo a constante cosmol\'ogica parece ser positiva, a sua corre\c{c}\~ao ao espectro quase-normal deve ser levada em conta. Perturba\c{c}\~oes gravitacionais de buracos negros esfericamente sim\'etricos no contexto de Sitter
foram estudadas pela primeira vez por integra\c{c}\~ao num\'erica (ver a se\c{c}\~ao \ref{sec:time-domain}) in \cite{Mellor:1989ac}.}

\langpar{In \cite{Zhidenko:2003wq} we study for the first time quasi-normal modes of massless Dirac and electromagnetic field in the Schwarzschild-de Sitter background. Also we study massless scalar and gravitational perturbations of Schwarzschild-de Sitter black holes, using the 6-th order WKB approach (see sec. \ref{sec:WKB}), which provides more accurate results. Namely, we substitute the effective potentials for the scalar, electromagnetic, Dirac and gravitational perturbations (\ref{scalar-potential}, \ref{Dirac-potential}, \ref{EM-potential}, \ref{polar-potential}, \ref{axial-potential}) into the WKB formula (\ref{QNM_WKB6}) and find, that the presence of the cosmological constant leads to decrease of the real oscillation frequency and to a slower decay (see fig. \ref{dSQNMs}).}{Em \cite{Zhidenko:2003wq} estudam-se pela primeira vez modos quase-normais de campo Dirac sem massa e campo eletromagn\'etico no contexto de Schwarzschild-de Sitter. Tamb\'em estudam-se perturba\c{c}\~oes escalares sem massa e gravitacionais de buracos negros Schwarzschild-de Sitter, usando a $6^a$ ordem da aproxima\c{c}\~ao de WKB (ver a se\c{c}\~ao \ref{sec:WKB}), que fornece resultados mais precisos. Substituindo os potenciais efetivos para as perturba\c{c}\~oes escalares, eletromagn\'eticas, Dirac e gravitacionais (\ref{scalar-potential}, \ref{Dirac-potential}, \ref{EM-potential}, \ref{polar-potential}, \ref{axial-potential}) na f\'ormula WKB (\ref{QNM_WKB6}), foi achado que a presen\c{c}a da constante cosmol\'ogica leva a diminuir a freq\"u\^encia real de oscila\c{c}\~ao e ao decaimento mais lento (ver a fig. \ref{dSQNMs}).}

\langpar{For large $l$ the following analytical expressions were found \cite{Zhidenko:2003wq}}{Para grande $l$ as express\~oes anal\'iticas foram achadas \cite{Zhidenko:2003wq}}
\begin{subequations}
\begin{eqnarray}
\omega &=& \frac{\sqrt{1 - 9M^2\Lambda}}{3\sqrt 3 M}\left(l + \frac 1 2 - \left(n+\frac{1}{2}\right)\imo\right) + {\cal O}\left(\frac{1}{l}\right),
\\\omega &=& \frac{\sqrt{1 - 9M^2\Lambda}}{3\sqrt 3 M}\left(\kappa_{\pm} - \left(n+\frac{1}{2}\right)\imo\right) + {\cal O}\left(\frac{1}{\kappa_{\pm}}\right).
\end{eqnarray}
\end{subequations}
\lang{As the black hole mass approaches its extremal value, the effective potentials (\ref{scalar-potential}, \ref{Dirac-potential}, \ref{EM-potential}, \ref{polar-potential}, \ref{axial-potential}) look like the P\"oshl-Teller potential (see sec. \ref{sec:Poschl-Teller}). Thus, using the P\"oshl-Teller approach, it was found in \cite{Cardoso:2003sw} that in the near-extremal limit of the Schwarzschild-de Sitter black hole the quasi-normal frequencies are proportional to the surface gravity at the event horizon $\kappa=\frac{1}{2}f'(r_+)$}{Como a massa do buraco negro se aproxima de seu valor extremo, os potenciais efetivos (\ref{scalar-potential}, \ref{Dirac-potential}, \ref{EM-potential}, \ref{polar-potential}, \ref{axial-potential}) se parecem com o potencial de P\"oshl-Teller (ver a se\c{c}\~ao \ref{sec:Poschl-Teller}). Assim, usando a abordagem de P\"oshl-Teller, foi encontrado em \cite{Cardoso:2003sw} que, no limite pr\'oximo ao extremo do buraco negro Schwarzschild-de Sitter, as freq\"u\^encias quase-normais s\~ao proporcionais \`a gravidade da superf\'icie no horizonte de eventos $\kappa=\frac{1}{2}f'(r_+)$}
\begin{subequations}
\begin{eqnarray}
\omega&=&\kappa\left(\sqrt{(l+1-s)(l+s)-\frac{1}{4}}-\imo\left(n+\frac{1}{2}\right)\right) + o(\kappa),
\\
\omega&=&\kappa\left(\sqrt{\kappa_\pm^2-\frac{1}{4}}-\imo\left(n+\frac{1}{2}\right)\right) + o(\kappa).
\end{eqnarray}
\end{subequations}

\langpar{Despite the dominant contribution of the imaginary part does not depend on the field spin $s$, the metric perturbations decay slower among the considered perturbations for all values of $\Lambda$.}{Apesar da contribui\c{c}\~ao dominante da parte imagin\'aria n\~ao depender do spin $s$ do campo, as perturba\c{c}\~oes m\'etricas decaem mais lentamente entre as perturba\c{c}\~oes consideradas para todos os valores de $\Lambda$.}

\langsectionlabel{High overtones in the quasi-normal spectrum}{Sobretons altos no espectro quase-normal}{sec:highovertones}
\langpar{Though, practically, we can observe only low-damping oscillations, the frequencies with large imaginary part attracted some attention in the context of the black hole thermodynamics. Such frequencies are called high overtones. Since their imaginary part is very large, the WKB approach is not appropriate and one should use the more accurate Frobenius method (see sec. \ref{sec:Frobenius}).}{Embora, na pr\'atica, possam-se observar somente as oscila\c{c}\~oes que decaem lentamente, as freq\"u\^encias com grande parte imagin\'aria atra\'iram alguma aten\c{c}\~ao no contexto da termodin\^amica de buracos negros. Tais freq\"u\^encias s\~ao chamadas sobretons altos. Como a sua parte imagin\'aria \'e muito grande, a aproxima\c{c}\~ao de WKB n\~ao \'e apropriada e deve-se usar o m\'etodo de Frobenius, que \'e mais preciso (ver a se\c{c}\~ao \ref{sec:Frobenius}).}

\langpar{It has been suggested in \cite{PhysRevLett.81.4293} that the asymptotic value of the real part of the quasi-normal frequency (i. e. when the imaginary part approaches infinity) coincides with the so-called Barbero-Immirzi parameter \cite{Barbero:1994ap,Immirzi:1996dr}. This parameter must be fixed in order to predict the Bekenstein-Hawking formula for entropy within the framework of Loop Quantum Gravity. For Schwarzschild black holes it was found numerically \cite{PhysRevD.47.5253} and analytically \cite{Motl:2002hd} that the asymptotical behavior of the quasi-normal frequency is given by}{Foi sugerido em \cite{PhysRevLett.81.4293} que o valor assint\'otico da parte real da freq\"u\^encia quase-normal (isto \'e, quando a parte imagin\'aria aproxima-se do infinito) coincide com o chamado par\^ametro de Barbero-Immirzi \cite{Barbero:1994ap,Immirzi:1996dr}. Este par\^ametro deve ser fixado para prever a f\'ormula de Bekenstein-Hawking para a entropia dentro da estrutura de Gravidade qu\^antica em loop. Para buracos negros de Schwarzschild foi achado numericamente \cite{PhysRevD.47.5253} e analiticamente \cite{Motl:2002hd} que o comportamento assint\'otico da freq\"u\^encia quase-normal \'e dado por}
$$\omega_n r_+=\frac{\ln(3)}{4\pi}-\frac{\imo(n+1/2)}{2}+{\cal O}(n^{-1/2})$$
\lang{for the gravitational perturbations and for the test scalar field, while for the electromagnetic perturbations the real part asymptotically approaches zero}{para as perturba\c{c}\~oes gravitacionais e para o campo escalar de teste, enquanto que, para as perturba\c{c}\~oes eletromagn\'eticas, a parte real aproxima-se assintoticamente de zero}
$$\omega_n r_+=-\frac{\imo n}{2}+{\cal O}(n^{-1/2}).$$

\begin{figure}
\includeonegraph{D=4_k=1_high_Dirac}
\langfigurecaption{Real part of the highly damping quasi-normal frequencies as a function of imaginary part ($\kappa_\pm=1$ perturbations of the massless Dirac field in the Schwarzschild background). The frequencies are measured in inverse units of the black hole mass.}{A parte real das freq\"u\^encias quase-normais que decaem rapidamente como uma fun\c{c}\~ao da parte imagin\'aria ($\kappa_\pm=1$ perturba\c{c}\~oes do campo de Dirac sem massa no contexto de Schwarzschild). As freq\"u\^encias foram medidas em unidades inversas da massa do buraco negro.}\label{SHD}
\end{figure}

\langpar{We have shown numerically in \cite{CastelloBranco:2004nk} that for perturbations of the massless Dirac field, the asymptotical value of the real part of the highly damping quasi-normal frequencies is also zero (see fig. \ref{SHD}). The correct spacing of the imaginary part was found later in \cite{Jing:2005dt}, with an alternative form of the effective potential for the Dirac perturbations.}{Foi mostrado numericamente em \cite{CastelloBranco:2004nk} que para perturba\c{c}\~oes do campo de Dirac sem massa, o valor assint\'otico da parte real das freq\"u\^encias quase-normais que decaem rapidamente \'e tamb\'em zero (ver o fig. \ref{SHD}). O espa\c{c}amento correto da parte imagin\'aria foi achado depois em \cite{Jing:2005dt}, utilizando uma forma alternativa do potencial efetivo para as perturba\c{c}\~oes de Dirac.}
\lang{The asymptotical formula for the high overtones of the massless Dirac field in the Schwarzschild background reads}{A f\'ormula assint\'otica para os sobretons altos do campo de Dirac sem massa no contexto de Schwarzschild l\^e-se}
$$\omega_n r_+\sim-\frac{\imo n}{2}.$$


\begin{figure}
\includeonegraph{l=1_Lambda=0_02_high_em}
\langfigurecaption{Real part of highly damping quasi-normal frequencies as a function of imaginary part for Schwarzschild-de Sitter black hole ($l=1$ electromagnetic perturbations, $\Lambda M^2 = 0.02$). The frequencies are measured in inverse units of the black hole mass.}{A parte real das freq\"u\^encias quase-normais que decaem rapidamente como uma fun\c{c}\~ao da parte imagin\'aria para o buraco negro Schwarzschild-de Sitter ($l=1$ perturba\c{c}\~oes eletromagn\'eticas, $\Lambda M^2 = 0.02$). As freq\"u\^encias foram medidas em unidades inversas da massa do buraco negro.}\label{SdSHEM}
\end{figure}

\langpar{Using the Frobenius method (see sec. \ref{sec:Frobenius}), for Schwarzschild-de Sitter black holes we have found for the first time in \cite{Konoplya:2004uk} that again the real part of the quasi-normal frequencies for electromagnetic perturbations asymptotically approaches zero, satisfying}{Usando o m\'etodo de Frobenius (ver a se\c{c}\~ao \ref{sec:Frobenius}), para buracos negros Schwarzschild-de Sitter foi achado pela primeira vez em \cite{Konoplya:2004uk} que novamente a parte real das freq\"u\^encias quase-normais para perturba\c{c}\~oes eletromagn\'eticas aproxima-se assintoticamente de zero, satisfazendo}
$$\omega_n\sim\imo\kappa n \quad\bigcup\quad \omega_n\sim\imo\kappa_\infty n,$$
\lang{where $\kappa=\frac{1}{2}f'(r_+)$ and $\kappa_\infty=\frac{1}{2}f'(r_\infty)$ are the surface gravities at the black hole horizon and the cosmological horizon respectively (see fig. \ref{SdSHEM}). This result was later confirmed analytically in \cite{Cardoso:2004up}.}{onde $\kappa=\frac{1}{2}f'(r_+)$ e $\kappa_\infty=\frac{1}{2}f'(r_\infty)$ s\~ao as gravidades da superf\'icie no horizonte do buraco negro e no horizonte cosmol\'ogico, respectivamente (ver o fig. \ref{SdSHEM}). Esse resultado foi depois confirmado analiticamente em \cite{Cardoso:2004up}.}

\begin{figure}
\includeonegraph{l=2_Lambda=0_02_high_grav}
\langfigurecaption{Real part of highly damping quasi-normal frequencies as a function of imaginary part ($l=2$ metric perturbations of the Schwarzschild-de Sitter black hole, $\Lambda M^2 = 0.02$). The frequencies are measured in inverse units of the black hole mass.}{A parte real das freq\"u\^encias quase-normais que decaem rapidamente como uma fun\c{c}\~ao da parte imagin\'aria ($l=2$ perturba\c{c}\~oes m\'etricas do buraco negro Schwarzschild-de Sitter, $\Lambda M^2 = 0.02$). As freq\"u\^encias foram medidas em unidades inversas da massa do buraco negro.}\label{SdSHgrav}
\end{figure}

\langpar{The real part of the quasi-normal modes for metric perturbations does not approach a constant. Frequencies with high imaginary part satisfies the non-algebraic equation}{A parte real dos modos quase-normais para perturba\c{c}\~oes m\'etricas n\~ao se aproxima de uma constante. As freq\"u\^encias com a parte imagin\'aria grande satisfazem a equa\c{c}\~ao n\~ao-alg\'ebrica}
$$\cosh\left(\frac{\pi\omega}{\kappa}-\frac{\pi\omega}{\kappa_\infty}\right)+3\cosh\left(\frac{\pi\omega}{\kappa}+\frac{\pi\omega}{\kappa_\infty}\right)\sim0,$$
\lang{which implies oscillation of the real part as a function of the imaginary part (see fig. \ref{SdSHgrav}).}{que implica oscila\c{c}\~ao da parte real como uma fun\c{c}\~ao da parte imagin\'aria (ver o fig. \ref{SdSHgrav}).}

\langsectionlabel{Decay of charged scalar and Dirac fields in the Kerr-Newman-de Sitter background}{O decaimento dos campos carregados escalar e Dirac no contexto de Kerr-Newman-de Sitter}{sec:KNdS4D}
\langpar{For charged black holes, the scalar field electrodynamics can describe the interaction of a charged field with the electromagnetic background of the black hole. When the influence of the spin of the field is neglected, we can consider the charged scalar field. The late-time tails of the charged scalar hair were studied for the first time by Hod \cite{Hod:1997mt,Hod:1997fy} in the context of the gravitational collapse. Quasi-normal modes of the charged scalar field for various black holes were studied for the first time in \cite{Konoplya:2002ky}. Quasi-normal modes of the charged massive scalar field for Reissner-Nordstr\"om black holes were considered in \cite{Konoplya:2002wt}. Decay of the charged Dirac field in the Reissner-Nordstr\"om background was studied in \cite{Zhou:2003ts}. The decay law for the late-time tails of the charged massive Dirac field also was found for Reissner Nordstr\"om \cite{Jing:2004zb} and Kerr-Newman \cite{He:2006jv} black holes. Yet, the calculations of the quasi-normal modes in the above papers were limited by the third order WKB accuracy. The WKB accuracy does not allow to study the quasi-normal modes of the near extremal charged black holes. Namely, the decay rate of charged fields in the near extremal Reissner-Nordstr\"om background is the same within the WKB accuracy.}{Para buracos negros carregados, a eletrodin\^amica de campo escalar pode descrever a intera\c{c}\~ao de um campo carregado com um campo eletromagn\'etico de fundo do buraco negro. Quando a influ\^encia do spin do campo \'e ignorada, pode-se considerar o campo escalar carregado. As caudas de tempo tardio dos cabelos escalares carregados foram estudadas pela primeira vez por Hod \cite{Hod:1997mt,Hod:1997fy} no contexto de colapso gravitacional. Os modos quase-normais de campo escalar carregado para v\'arios buracos negros foram estudados pela primeira vez em \cite{Konoplya:2002ky}. Os modos quase-normais de campo escalar massivo carregado para buracos negros de Reissner-Nordstr\"om foram considerados em \cite{Konoplya:2002wt}. O decaimento de campo de Dirac carregado no contexto de Reissner-Nordstr\"om foi estudado em \cite{Zhou:2003ts}. A lei de decaimento das caudas de tempo tardio de campo de Dirac massivo carregado tamb\'em foi achada para buracos negros de Reissner Nordstr\"om \cite{Jing:2004zb} e Kerr-Newman \cite{He:2006jv}. Mas os c\'alculos dos modos quase-normais nos trabalhos acima mencionados foram limitados pela precis\~ao da terceira ordem de WKB. A precis\~ao de WKB n\~ao permite estudar os modos quase-normais dos buracos negros carregados pr\'oximo ao extremo. A taxa de decaimento de campos carregados no contexto Reissner-Nordstr\"om pr\'oximo ao extremo \'e a mesma dentro da precis\~ao de WKB.}

\begin{figure}
\includetwograph{l=0_chargedRN_scalar_Re}{l=0_chargedRN_scalar_Im}
\langfigurecaption{Real part and imaginary part of the fundamental quasi-normal frequency of charged ($q$) scalar field ($l=0$) for the charged ($Q$) black hole.}{A parte real e a parte imagin\'aria da freq\"u\^encia quase-normal fundamental do campo escalar ($l=0$) carregado ($q$) para  o buraco negro carregado ($Q$).}\label{RNchargedscalar}
\end{figure}

\langpar{The numerical study of quasi-normal frequencies of the charged fields (both scalar and Dirac) in the Kerr-Newman-de Sitter background was done in \cite{Konoplya:2007zx} within a much more accurate Frobenius method (see sec. \ref{sec:Frobenius-angular}, \ref{sec:Frobenius-radial}). We have shown that for not very large value of $Q$, the charged field decays quicker than the neutral one. For the near extremal value of $Q$, the charged field decays slower than the neutral one (see fig. \ref{RNchargedscalar}).}{O estudo num\'erico de freq\"u\^encias quase-normais dos campos carregados (tanto escalar como Dirac) no contexto de Kerr-Newman-de Sitter foi feito em \cite{Konoplya:2007zx} utilizando um m\'etodo de Frobenius muito mais preciso (ver as se\c{c}\~oes \ref{sec:Frobenius-angular}, \ref{sec:Frobenius-radial}). Foi mostrado que para um valor de $Q$ n\~ao muito grande, o campo carregado decai mais rapidamente do que o campo neutro. Para o valor de $Q$ pr\'oximo ao extremo, o campo carregado decai mais devagar do que o campo neutro (ver a fig. \ref{RNchargedscalar}).}

\langpar{Let us summarise briefly what happens if a black hole has charge and rotation. We measure all the quantities in units of the size of the event horizon.}{Resumimos brevemente o que acontece, se um buraco negro tiver carga e rota\c{c}\~ao. Mede-se todas as quantidades em unidades do tamanho do horizonte de eventos.}
\langlong{\begin{itemize}
\item As we know from previous sections, the presence of the cosmological constant decreases the absolute values of the real and imaginary parts of $\omega$.
\item We already know, that as the multipole number $l$ increases, the real part of $\omega$ grows, while the imaginary part of $\omega$ approaches a constant.
\item Metric perturbations decay slower than any field perturbations of the same $l$.
\item The dependance of $\omega$ on charge of the field $q$ and of the black hole $Q$ is shown on figure \ref{RNchargedscalar}
\begin{itemize}
\item If $qQ>0$, the real part of $\omega$ monotonically grows as the field charge $q$ increases. It grows also as the black hole charge $Q$ increases, attains some maximum value at a large (close to extremal) value of $Q$, and then decreases.
\item If $qQ<0$, the real part of $\omega$ decreases when either the black hole charge or the field charge increases.
\item The imaginary part of $\omega$ has more difficult behavior, in most cases decreasing its absolute value as $qQ$ grows.
\end{itemize}
\item The influence of the rotation parameter depends significantly on the projection of the field momentum on the axis of the black hole rotation $m$. We will consider this dependance later for the massive scalar field in the Kerr background (see sec. \ref{sec:massive-Kerr}). Qualitatively the same behavior was observed for Kerr black holes, projected on the brane (see sec. \ref{sec:Kerr-projected}).
\end{itemize}}{\begin{itemize}
\item Como visto em se\c{c}\~oes pr\'evias, a presen\c{c}a da constante cosmol\'ogica diminui os valores absolutos das partes real e imagin\'aria de $\omega$.
\item J\'a \'e sabido, tamb\'em, que quando o n\'umero multipolar $l$ aumenta, a parte real de $\omega$ cresce, enquanto a parte imagin\'aria de $\omega$ se aproxima de uma constante.
\item Perturba\c{c}\~oes m\'etricas decaem mais devagar do que qualquer perturba\c{c}\~ao de campo de mesmo $l$.
\item A depend\^encia de $\omega$ na carga do campo $q$ e do buraco negro $Q$ \'e mostrada na figura \ref{RNchargedscalar}
\begin{itemize}
\item Se $qQ>0$, a parte real de $\omega$ cresce monotonicamente j\'a que a carga de campo $q$ aumenta. Ela cresce tamb\'em, j\'a que a carga do buraco negro $Q$ aumenta, alcan\c{c}a um valor m\'aximo em um alto (perto do extremo) valor de $Q$, e depois diminui.
\item Se $qQ<0$, a parte real de $\omega$ diminui quando a carga do buraco negro ou do campo cresce.
\item A parte imagin\'aria de $\omega$ tem um comportamento mais dif\'icil, na maior parte dos casos, reduzindo o seu valor absoluto com o crescimento de $qQ$.
\end{itemize}
\item A influ\^encia do par\^ametro de rota\c{c}\~ao depende significativamente da proje\c{c}\~ao do momento de campo no eixo de rota\c{c}\~ao do buraco negro $m$. Depois esta depend\^encia ser\'a considerada para o campo escalar massivo no contexto de Kerr (ver a se\c{c}\~ao \ref{sec:massive-Kerr}). Qualitativamente o mesmo comportamento foi observado para buracos negros de Kerr, projetados na brana (ver a se\c{c}\~ao  \ref{sec:Kerr-projected}).
\end{itemize}}

\langchapter{Perturbations of higher-dimensional black holes}{Perturba\c{c}\~oes de buracos negros em dimens\~oes mais altas}
\langsectionlabel{Stability and quasi-normal modes of Reissner-Nordstr\"om-de Sitter black holes}{Estabilidade e modos quase-normais de buracos negros Reissner-Nordstr\"om-de Sitter}{sec:HDRNdS}
\langpar{Last years higher-dimensional black holes have attracted considerable interest in the context of the string theory and higher-dimensional brane-world models \cite{ArkaniHamed:1998rs,ArkaniHamed:1998nn,Antoniadis:1998ig,Randall:1999ee,Randall:1999vf,Abdalla:2006qj}. Some of such models allow compactification radius of extra dimensions to be of macroscopic size \cite{Cremades:2002dh,Kokorelis:2002qi}. That is why the extra dimensions are important if we study small black holes. Such black holes could be produced at the next-generation particle colliders, probably at energies of order $\sim1TeV$ \cite{Argyres:1998qn,Banks:1999gd,Giddings:2001bu,Dimopoulos:2001hw}.}{Nos \'ultimos anos, os buracos negros em dimens\~oes mais altas atra\'iram um interesse consider\'avel no contexto da teoria das cordas e em modelos de mundo-brana \cite{ArkaniHamed:1998rs,ArkaniHamed:1998nn,Antoniadis:1998ig,Randall:1999ee,Randall:1999vf,Abdalla:2006qj}. Alguns desses modelos permitem que um raio de compactifica\c{c}\~ao das dimens\~oes extras seja de um tamanho macrosc\'opico \cite{Cremades:2002dh,Kokorelis:2002qi}. Por isso as dimens\~oes extras s\~ao importantes para estudar buracos negros pequenos. Tais buracos negros podem ser produzidos nos colisores de part\'iculas de nova gera\c{c}\~ao, provavelmente em energias da ordem $\sim1TeV$ \cite{Argyres:1998qn,Banks:1999gd,Giddings:2001bu,Dimopoulos:2001hw}.}

\langpar{Being an important characteristic of a black hole, its quasi-normal spectrum can be used in future experiments to find number of extra dimensions and other parameters of the theory. That is why we study quasi-normal spectrum of black holes within different scenarios with various possible numbers of extra dimensions.}{Sendo uma caracter\'istica importante de buraco negro, o seu espectro quase-normal pode ser usado em futuros experimentos para achar o n\'umero de dimens\~oes extra e outros par\^ametros da teoria. Por isso estuda-se o espectro quase-normal de buracos negros dentro de cen\'arios diferentes com v\'arios n\'umeros poss\'iveis de dimens\~oes extras.}

\langpar{When a black hole is much smaller than the size of extra dimensions, it can be described by $D$-dimensional Einstein theory. The $D$-dimensional generalisation of the Schwarzschild metric was done by Tangherlini \cite{Tangherlini:1963bw}.}{Quando um buraco negro \'e muito menor do que o tamanho das dimens\~oes extras, ele pode ser descrito pela teoria de Einstein $D$-dimensional. A generaliza\c{c}\~ao $D$-dimensional da m\'etrica de Schwarzschild foi feita por Tangherlini \cite{Tangherlini:1963bw}.}

\langpar{Let us consider the $D$-dimensional Reissner-Nordstr\"om-de Sitter black hole, which is described by the line element}{Consideremos o buraco negro Reissner-Nordstr\"om-de Sitter $D$-dimensional, que \'e descrito pelo elemento de linha}
\begin{equation}\label{HDBHmetric}
ds^2=f(r)dt^2-\frac{dr^2}{f(r)}-r^2d\Omega_d^2,
\end{equation}
\lang{where $d\Omega_d$ is the line element of $d$-sphere, $d=D-2$, $f(r)=1-X+Z-Y$. The dimensionless quantities $X$, $Y$, $Z$ are defined as}{onde $d\Omega_d$ \'e o elemento de linha de $d$-esfera, $d=D-2$, $f(r)=1-X+Z-Y$. As quantidades adimensionais $X$, $Y$, $Z$ s\~ao definidas como}
$$X=\frac{dM{\cal A}_d}{4\pi r^{d-1}},\qquad Y=\frac{2\Lambda r^2}{d(d+1)}, \qquad Z=\frac{Q^2}{r^{2d-2}},$$
\lang{where $\displaystyle{\cal A}_d=\frac{2\pi^{(d+1)/2}}{\Gamma\left((d+1)/2\right)}$ is the area of a unit $d$-sphere, $M$ is the black hole mass, $Q$ is the black hole charge and $\Lambda$ is the cosmological constant.}{onde $\displaystyle{\cal A}_d=\frac{2\pi^{(d+1)/2}}{\Gamma\left((d+1)/2\right)}$ \'e a \'area de uma $d$-esfera unit\'aria, $M$ \'e a massa do buraco negro, $Q$ \'e a carga do buraco negro e $\Lambda$ \'e a constante cosmol\'ogica.}

\langpar{Perturbations of the Einstein-Maxwell equations can be reduced to the wave-like form (\ref{wave-like}). The corresponding effective potential depends on their transformation law under rotations on the $d$-sphere (\ref{gravitational-classification}). Thus, there are three types of perturbations described by three effective potentials: tensor ($V_T$), vector ($V_V$) and scalar ($V_S$) type. The explicit form of the potentials was derived in \cite{Kodama:2003kk}}{As perturba\c{c}\~oes das equa\c{c}\~oes de Einstein-Maxwell podem ser reduzidas \`a forma de onda (\ref{wave-like}). O potencial efetivo correspondente depende da lei de transforma\c{c}\~ao deles, sob rota\c{c}\~oes na $d$-esfera (\ref{gravitational-classification}). Assim, h\'a tr\^es tipos de perturba\c{c}\~oes, descritas por tr\^es potenciais efetivos: tensorial ($V_T$), vetorial ($V_V$) e escalar ($V_S$). A forma expl\'icita dos potenciais foi derivada em \cite{Kodama:2003kk}}
\begin{eqnarray}
V_T(r)&=&\frac{f(r)}{r^2}\left(\lambda+d+\frac{f'(r)rd}{2}+\frac{f(r)d(d-2)}{4}\right),\\
V_V(r)&=&\frac{f(r)}{r^2}\left(\lambda+d\pm\sqrt{X^2\frac{(d^2-1)^2}{4}+2Z\lambda d(d-1)}\times\right.\nonumber\\&&\left.\times\frac{d(d-2)(1-Y)+Zd(5d-2)-X(d^2+2)}{4}\right),\\
V_S(r)&=&f(r)\frac{U_\pm}{64r^2H_\pm^2}\label{potential-scalartype}
\end{eqnarray}
\lang{where}{onde}
\begin{eqnarray}\nonumber
H_+&=&1-\frac{d(d+1)}{2}\delta X,\\\nonumber
H_-&=&\lambda+\frac{d(d+1)}{2}(1+\lambda\delta)X,\\
U_+&=& \left[-4 d^3 (d+2) (d+1)^2 \delta^2 X^2-48 d^2 (d+1) (d-2) \delta X-16 (d-2) (d-4)\right] Y
\notag\\
&&   -\delta^3 d^3 (3 d-2) (d+1)^4 (1+\lambda \delta) X^4 +4 \delta^2 d^2 (d+1)^2
   \left\{(d+1)(3d-2) \lambda \delta+4 d^2+d-2\right\} X^3
\notag\\
&&   +4 \delta (d+1)\left\{
   (d-2) (d-4) (d+1) (\lambda+d^2) \delta-7 d^3+7 d^2-14 d+8
   \right\}X^2
\notag\\
&&  + \left\{16 (d+1) \left(-4 \lambda+3 d^2(d-2) \right) \delta
     -16 (3 d-2) (d-2) \right\}X +64 \lambda+16 d(d+2) ,\notag\\
U_-&=& \left[-4 d^3 (d+2) (d+1)^2 (1+\lambda \delta)^2 X^2
      +48 d^2 (d+1) (d-2) \lambda (1+\lambda \delta) X  -16 (d-2) (d-4) \lambda^2\right] Y
\notag\\
&&
     -d^3 (3 d-2) (d+1)^4 \delta (1+\lambda \delta)^3 X^4 -4 d^2 (d+1)^2 (1+\lambda \delta)^2
     \left\{(d+1)(3 d-2) \lambda \delta-d^2\right\} X^3
\notag\\
&&  +4 (d+1) (1+\lambda \delta)\left\{ \lambda (d-2) (d-4) (d+1) (\lambda+d^2 ) \delta
 +4 d (2 d^2-3 d+4) \lambda\right. \notag\\ &&\left.+ d^2 (d-2) (d-4) (d+1) \right\}X^2+64 \lambda^3+16 d(d+2)\lambda^2
\notag\\
&&  -16\lambda \left\{ (d+1) \lambda \left(-4 \lambda+3 d^2(d-2) \right) \delta +3 d (d-4) \lambda+3 d^2 (d+1) (d-2) \right\}X \,.\notag
\end{eqnarray}
\lang{Here we defined $\lambda=(l+d)(l-1)$ through the multipole number $l=2,3,4\ldots\,$. The value $\delta$ is a dimensionless constant}{Aqui foi definido $\lambda=(l+d)(l-1)$ pelo n\'umero de multipolo $l=2,3,4\ldots\,$. O valor $\delta$ \'e uma constante adimensional}
$$2\lambda\delta=\sqrt{1+\frac{16\lambda Z}{(d+1)^2X^2}}-1.$$

\langpar{Note, that the effective potential for tensor-type perturbations coincides with the potential for a test scalar field (\ref{potential-tensortype}) for $l\geq2$. Vector-type and scalar-type metric perturbations are coupled to perturbations of the electromagnetic field. For neutral ($Q=0$) black holes the type ``+'' potentials are reduced to the effective potential for the test Maxwell field and for the type ``-'' $V_V$ and $V_S$ describe pure vector-type and scalar type metric perturbations respectively.}{Observe, que o potencial efetivo para perturba\c{c}\~oes do tipo tensorial coincide com o potencial para um campo escalar de teste (\ref{potential-tensortype}) para $l\geq2$. Perturba\c{c}\~oes m\'etricas dos tipos vetorial e escalar s\~ao ligadas a perturba\c{c}\~oes do campo eletromagn\'etico. Para buracos negros neutros ($Q=0$) os tipos ``+'' potentciais s\~ao reduzidos ao potencial efetivo para o campo de Maxwell de teste e os tipos ``-'' $V_V$ e $V_S$ descrevem as perturba\c{c}\~oes m\'etricas puras do tipo vetorial e escalar, respectivamente.}

\begin{figure}
\includetwograph{potentials_HD}{profiles_HD}
\langfigurecaption{Effective potentials and time-domain profiles for scalar-type gravitational perturbations, $D = 5$ (blue)\ldots$D = 11$ (red) ($l = 2$, $Q = 0$, $\Lambda = 0$). For higher $D$ both the peak and the negative gap of the potential increase. Profile for higher $D$ decays quicker. All quantities are measured in units of the event horizon $r_+$.}{Potenciais efetivos e perfis temporais para perturba\c{c}\~oes gravitacionais do tipo escalar, $D = 5$ (azul)\ldots$D = 11$ (vermelho) ($l = 2$, $Q = 0$, $\Lambda = 0$). Para $D$ mais alto tanto o pico como o espa\c{c}o negativo do potencial aumentam. Perfil para $D$ mais alto decai mais rapidamente. Todas as quantidades s\~ao medidas em unidades do horizonte de eventos $r_+$.}\label{HD.scalargrav}
\end{figure}

\begin{figure}
\includeonegraph{profiles_D=11_rho}
\langfigurecaption{Time-domain profiles for scalar-type gravitational perturbations
of $11$-dimensional Schwarzschild-de Sitter black hole ($Q = 0$, $D = 11$, $\rho=r_+/r_\infty$) for $\rho = 0.3$ (blue), $\rho = 0.5$
(green), $\rho = 0.7$ (yellow), $\rho = 0.8$ (orange), $\rho = 0.9$ (red). Profile for higher $\rho$ decays slower. All quantities are measured in units of the event horizon $r_+$.}{Perfis temporais para perturba\c{c}\~oes gravitacionais do tipo escalar de buraco negro Schwarzschild-de Sitter $11$-dimensional ($Q = 0$, $D = 11$, $\rho=r_+/r_\infty$) para $\rho = 0.3$ (azul), $\rho = 0.5$
(verde), $\rho = 0.7$ (amarelo), $\rho = 0.8$ (laranja), $\rho = 0.9$ (vermelho). Perfil para $\rho$ mais alto decai mais lentamente. Todas as quantidades s\~ao medidas em unidades do horizonte de eventos $r_+$.}\label{D=11.rho.scalargrav}
\end{figure}

\langpar{It was proven analytically that the black hole is stable against tensor-type and vector-type gravitational perturbations \cite{Kodama:2003kk}. The effective potential of the scalar type (\ref{potential-scalartype}) is not positive definite, having a negative gap for small $l$. Since the effective potential has an extremely complicated form, the appropriate ansatz for the $S$-deformation technique (see sec. \ref{sec:stability}) was not found. That is why, the black hole stability against scalar-type perturbations was not proven analytically.}{Foi provado analiticamente que o buraco negro \'e est\'avel contra perturba\c{c}\~oes gravitacionais dos tipos tensorial e vetorial \cite{Kodama:2003kk}. O potencial efetivo do tipo escalar (\ref{potential-scalartype}) n\~ao \'e definido positivo, tendo um espa\c{c}o negativo para $l$ pequeno. Como o potencial efetivo tem uma forma extremamente complicada, o ansatz apropriado para a t\'ecnica de deforma\c{c}\~ao $S$ (ver a se\c{c}\~ao \ref{sec:stability}) n\~ao foi encontrado. Por isso, a estabilidade de buracos negros contra perturba\c{c}\~oes do tipo escalar n\~ao foi provada analiticamente.}

\langpar{The stability of higher dimensional Schwarzschild-de Sitter black holes was proven numerically in \cite{Konoplya:2007jv}. We have shown in time domain that the gravitational perturbations of scalar type decay for arbitrary black hole mass and $\Lambda$-term (see figs. \ref{HD.scalargrav},\ref{D=11.rho.scalargrav}).}{A estabilidade de buracos negros Schwarzschild-de Sitter em dimens\~oes mais altas foi provada numericamente em \cite{Konoplya:2007jv}. Foi mostrado no dom\'inio temporal, que as perturba\c{c}\~oes gravitacionais do tipo escalar decaem para a massa do buraco negro e o termo $\Lambda$ arbitrarios (ver as figs. \ref{HD.scalargrav},\ref{D=11.rho.scalargrav}).}



\langpar{Also we studied the quasi-normal modes for all types of gravitational perturbations of $D$-dimensional spherically symmetric black holes. Since for the observational purposes only the dominant quasi-normal frequency is essential we were limited by the fundamental mode ($l=2$) only. However, using the WKB approximation, we are able to find the large $l$ formula for any given black hole parameters.}{Tamb\'em foram estudados os modos quase-normais para todos os tipos de perturba\c{c}\~oes gravitacionais para buracos negros $D$-dimensionais esfericamente sim\'etricos. Como para os objetivos de observa\c{c}\~ao somente a freq\"u\^encia quase-normal dominante \'e essencial, os calculos foram restritos somente ao modo fundamental ($l=2$). Contudo, usando a aproxima\c{c}\~ao WKB, \'e poss\'ivel achar a f\'ormula para $l$ grande para quaisquer par\^ametros de buraco negro dados.}

\langpar{The dependance of the quasi-normal modes on the $\Lambda$-term is qualitatively the same as for the four-dimensional black holes: both real and imaginary parts of the quasi-normal frequency decrease their absolute values as the cosmological constant grows (see fig. \ref{D=11.rho.scalargrav}). We also observe that the real and imaginary parts of the quasi-normal frequency are non-monotonic functions of the black hole charge, in most cases decreasing their absolute values as $Q$ grows.}{A depend\^encia de modos quase-normais no termo $\Lambda$ \'e qualitativamente a mesma da dos buracos negros quadridimensionais: tanto a parte real como a parte imagin\'aria da freq\"u\^encia quase-normal reduzem os seus valores absolutos enquanto a constante cosmol\'ogica cresce (ver a fig. \ref{D=11.rho.scalargrav}). Tamb\'em foi observado que as partes real e imagin\'aria da freq\"u\^encia quase-normal s\~ao fun\c{c}\~oes n\~ao-monot\^onicas da carga do buraco negro, na maior parte dos casos reduzindo os seus valores absolutos enquanto $Q$ cresce.}

\begin{figure}
\includeonegraph{RNdS_profiles_D=11_rho=0_8}
\langfigurecaption{Time-domain profile for gravitational perturbations of scalar type ``-'' of the Reissner-Nordstr\"om-de Sitter black hole ($D=11$, $\rho=r_+/r_\infty=0.8$, $l=2$) for various values of the black hole charge $q=Q/Q_{ext}$:
q=0.4 (brown) q=0.5 (blue) q=0.6 (green) q=0.7 (orange) q=0.8 (red) q=0.9 (magenta). The smaller $q$, the slower growth of the profile is.}{Perfis temporais para perturba\c{c}\~oes gravitacionais do tipo ``-'' escalar do buraco negro Reissner-Nordstr\"om-de Sitter ($D=11$, $\rho=r_+/r_\infty=0.8$, $l=2$) para v\'arios valores da carga do buraco negro $q=Q/Q_{ext}$: q=0.4 (marrom) q=0.5 (azul) q=0.6 (verde) q=0.7 (laranja) q=0.8 (vermelho) q=0.9 (magenta). Para menor $q$ o crescimento do perfil \'e mais lento.}\label{RNdS.profiles.D=11.rho=0.8}
\end{figure}

\langpar{In \cite{Konoplya:2008au} we have shown for the first time that Reissner-Nordstr\"om-de Sitter black holes are gravitationally unstable for large values of the electric charge and cosmological constant in $D \geq 7$ space-time dimensions. On the figure \ref{RNdS.profiles.D=11.rho=0.8} we see the time-domain profiles for the linear gravitational perturbations of scalar type ``-'' (\ref{potential-scalartype}) of the near extremal Reissner-Nordstr\"om-de Sitter black hole in $D=11$ dimensions. For sufficiently small values of the black hole charge we observe usual picture of the quasi-normal ringing. Then, as the black hole charge increases, a purely imaginary (non-oscillating) mode becomes dominant, decreasing its decay rate until the threshold point of instability is reached. After crossing the instability point we observe the growing non-oscillating mode (see sec. \ref{sec:stability}). Its growth rate increases as the black hole charge grows. Therefore, we conclude that exactly at the threshold point of instability there is some static solution ($\omega=0$) of the perturbation equation. The static solution at the threshold point of instability was observed also in time-domain for the black strings (see sec. \ref{sec:Gregory-Laflamme}).}{Em \cite{Konoplya:2008au} foi mostrado pela primeira vez que buracos negros Reissner-Nordstr\"om-de Sitter s\~ao gravitacionalmente inst\'aveis para valores grandes da carga el\'etrica e da constante cosmol\'ogica em $D\geq7$ dimens\~oes de espa\c{c}o-tempo. Na figura \ref{RNdS.profiles.D=11.rho=0.8} pode-se ver os perfis temporais para as perturba\c{c}\~oes gravitacionais lineares do tipo escalar ``-'' (\ref{potential-scalartype}) do buraco negro Reissner-Nordstr\"om-de Sitter pr\'oximo ao extremo em $D=11$ dimens\~oes. Para os valores suficientemente pequenos da carga do buraco negro pode-se observar o quadro habitual do toque quase-normal. Ent\~ao, como a carga do buraco negro cresce, um modo puramente imagin\'ario (sem oscila\c{c}\~ao) torna-se dominante, reduzindo a sua taxa de decaimento at\'e conseguir o ponto de limiar da instabilidade. Depois de cruzar o ponto da instabilidade observa-se o modo crescente que n\~ao oscila (ver a se\c{c}\~ao \ref{sec:stability}). A sua taxa de crescimento aumenta como a carga do buraco negro cresce. Por isso, pode-se concluir, que exatamente no ponto de limiar da instabilidade h\'a alguma solu\c{c}\~ao est\'atica ($\omega=0$) da equa\c{c}\~ao de perturba\c{c}\~ao. A solu\c{c}\~ao est\'atica no ponto de limiar da instabilidade foi observada tamb\'em no dom\'inio de tempo para as cordas negras (ver a se\c{c}\~ao \ref{sec:Gregory-Laflamme}).}

\begin{figure}
\includeonegraph{RNdS_instability}
\langfigurecaption{The parametric region of instability in the right upper corner of the square ($\rho=r_+/r_\infty$, $q=Q/Q_{ext}$) for $D=7$ (top, black), $D=8$ (blue), $D=9$ (green), $D=10$ (red), $D=11$ (bottom, magenta).}{A regi\~ao param\'etrica de instabilidade na esquina topo direita do quadrado ($\rho=r_+/r_\infty$, $q=Q/Q_{ext}$) para $D=7$ (topo, preto), $D=8$ (azul), $D=9$ (verde), $D=10$ (vermelho), $D=11$ (abaixo, magenta).}\label{parametrs_instability}
\end{figure}

\langpar{The parametric region of instability is shown on the figure \ref{parametrs_instability}. The larger number of space-time dimensions $D$ is, the bigger region of instability we observe. Though the region of instability increases with $D$, the charged black holes in the asymptotically flat space-time are stable at least for $D\leq11$. The instability occurs if \emph{both} the black hole charge and the cosmological constant are large enough.}{A regi\~ao param\'etrica da instabilidade \'e mostrada na figura \ref{parametrs_instability}. Quanto maior \'e o n\'umero de dimens\~oes $D$, maior \'e a regi\~ao da instabilidade observada. Embora a regi\~ao da instabilidade aumente com $D$, os buracos negros carregados no espa\c{c}o-tempo assintoticamente plano s\~ao est\'aveis pelo menos para $D\leq11$. A instabilidade ocorre se \emph{tanto} a carga do buraco negro \emph{como} a constante cosmol\'ogica forem suficientemente grandes.}

\langsectionlabel{(In)stability of $D$-dimensional black holes in the Gauss-Bonnet theory}{(In)estabilidade de buracos negros $D$-dimensionais na teoria de Gauss-Bonnet}{sec:Gauss-Bonnet}
\langpar{Higher dimensional quantum gravity implies corrections to classical general relativity. The dominant order correction to the Lagrangian is called the Gauss-Bonnet term \cite{Boulware:1985wk}. This term is squared in curvature and vanishes for $D=4$. The effective action is given by}{A gravidade qu\^antica em dimens\~oes mais altas implica corre\c{c}\~oes \`a relatividade geral cl\'assica. A corre\c{c}\~ao da ordem dominante ao Lagrangiano \'e chamada de termo de Gauss-Bonnet \cite{Boulware:1985wk}. Este termo \'e quadr\'atico em curvatura e desaparece para $D=4$. A a\c{c}\~ao efetiva \'e dada por}
\begin{equation}
S = \frac{1}{16 \pi G_{D}} \int{d^{D} x \sqrt{-g} \left(R + \alpha(R_{abcd} R^{abcd}- 4 R_{cd}R^{cd} + R^2)\right)},
\end{equation}
\lang{where $\alpha$ is a positive coupling constant, $[\alpha]=L^2$.}{onde $\alpha$ \'e uma constante positiva de acoplamento, $[\alpha]=L^2$.}

\langpar{The spherically symmetric black hole solution, which satisfies the corresponding equations of motion, is described by the line element (\ref{HDBHmetric}) with}{A solu\c{c}\~ao de buraco negro esfericamente sim\'etrico, que satisfaz as correspondentes equa\c{c}\~oes do movimento, \'e descrita pelo elemento de linha (\ref{HDBHmetric}) com}
\begin{equation}\label{GBmetricfunction}
f(r)=1+\frac{r^2}{\alpha(D-3)(D-4)}\left(1-q(r)\right), \qquad
q(r)=\sqrt{1+\frac{\alpha(D-2)(D-3)(D-4)M{\cal A}_{D-2}}{2\pi r^{D-1}}},
\end{equation}
\lang{which is reduced to the Tangherlini metric \cite{Tangherlini:1963bw} in the limit of $\alpha\rightarrow0$.}{que se reduz \`a m\'etrica de Tangherlini \cite{Tangherlini:1963bw} no limite de $\alpha\rightarrow0$.}

\langpar{Quasi-normal modes and late-time tails of the test scalar field for black holes in the Gauss-Bonnet theory were studied in \cite{Konoplya:2004xx,Abdalla:2005hu}.}{Os modos quase-normais e as caudas de tempo tardio do campo escalar de teste para buracos negros na teoria de Gauss-Bonnet foram estudados em \cite{Konoplya:2004xx,Abdalla:2005hu}.}

\langpar{The quantum corrections imply that the effective potential for tensor-type gravitational perturbations ($V_T$) does not coincide with the potential for the test scalar field \cite{Dotti:2005sq}. The effective potentials for vector-type ($V_V$) and scalar-type ($V_S$) gravitational perturbations were found in \cite{Gleiser:2005ra}. They are given by the formulae}{As corre\c{c}\~oes qu\^anticas implicam que o potencial efetivo para as perturba\c{c}\~oes gravitacionais do tipo tensorial ($V_T$) n\~ao coincide com o potencial para o campo escalar de teste \cite{Dotti:2005sq}. Os potenciais efetivos para perturba\c{c}\~oes gravitacionais do tipo vetorial ($V_V$) e do tipo escalar ($V_S$) foram achados em \cite{Gleiser:2005ra}. Eles s\~ao dados pelas f\'ormulas}
\begin{eqnarray}
V_T(r)&=&f(r)\frac{(D-2)(c+1)}{r^2}\left(3-\frac{B(r)}{A(r)}\right)+\frac{1}{\sqrt{r^{D-2}A(r)q(r)}}\frac{d^2}{dr_\star^2}\sqrt{r^{D-2}A(r)q(r)},\\
V_V(r)&=&f(r)\frac{(D-2)c}{r^2}A(r)+\sqrt{r^{D-2}A(r)q(r)}\frac{d^2}{dr_\star^2}\frac{1}{\sqrt{r^{D-2}A(r)q(r)}},\\
V_S(r)&=&\frac{f(r)U(r)}{64 r^2(D-3)^2A(r)^2q(r)^8(4c q(r) + (D-1)R (q(r)^2-1))^2},\\\nonumber
\end{eqnarray}
\lang{where we used the following dimensionless quantities}{onde foram usadas as quantidades adimensionais}
\begin{eqnarray}\nonumber
A(r)&=&\frac{1}{q(r)^2}\left(\frac{1}{2}+\frac{1}{D-3}\right)+\left(\frac{1}{2}-\frac{1}{D-3}\right),\\\nonumber
B(r)&=&A(r)^2\left(1+\frac{1}{D-4}\right)+\left(1-\frac{1}{D-4}\right),\\\nonumber
R&=&\frac{r^2}{\alpha(D-3)(D-4)},\\\nonumber
c&=&\frac{l(l + D - 3)}{D-2}-1, \qquad l=2,3,4\ldots\,,
\end{eqnarray}
\begin{eqnarray}\nonumber
U(r)&=&5 (D - 1)^6 R^2 (1 + R) - 3 (D - 1)^5 R ((D - 1) R^2 + 24 c (1 + R)) q(r) + \\\nonumber&& +
 2 (D - 1)^4 (24 c (D - 1) R^2 +
    168 c^2 (1 + R) - (D - 1) R^2 (-3 + 5 R + 7 D (1 + R))) q(r)^2 + \\\nonumber&& +
 2 (D - 1)^4 R (-184 c^2 + (D - 1) (13 + D) R^2 +
    c (-84 + 44 R + 84 D (1 + R))) q(r)^3 + \\\nonumber&& +
    (D - 1)^3 (384 c^3 - 48 c (2 + D (3 D - 5)) R^2 +
    192 c^2 (-11 + D + (-15 + D) R) + \\\nonumber&& + (D - 1) R^2 (-3 (7 + 55 R) +
       D (26 + 106 R + 7 D (1 + R)))) q(r)^4 + \\\nonumber&& +
       (D - 1)^3 R (-64 c^2 (D - 38) + (D - 1) (71 + D (7 D - 90)) R^2 + \\\nonumber&& +
    16 c (303 + 255 R + 13 D^2 (1 + R) - 2 D (73 + 81 R))) q(r)^5 + \\\nonumber&& +
 4 (D - 1)^2 (96 c^3 (-7 + D) -
    8 c (D - 1) (145 - 74 D + 6 D^2) R^2 - \\\nonumber&& -
    8 c^2 (9 - 175 R + D (-58 - 34 R + 11 D (1 + R))) + (D - 1) R^2 (-5 (79 + 23 R) + \\\nonumber&& +
       D (5 (57 + 41 R) + D (-81 - 89 R + 7 D (1 + R))))) q(r)^6 - \\\nonumber&& -
 4 (D - 1)^2 R (8 c^2 (43 + (72 - 13 D) D) + (D - 1) (-63 + D (99 + D (-49 + 5 D))) R^2 + \\\nonumber&& +
    4 c (321 + 465 R + D (121 - 39 R + D (-123 - 107 R + 17 D (1 + R))))) q(r)^7 + \\\nonumber&& +
       (D - 1) (128 c^3 (-9 + D) (D - 5) + 32 c (D - 1) (246 + D (9 + D (-55 + 8 D))) R^2 + \\\nonumber&& +
    64 c^2 (D - 5) (D^2 - 3 + (49 + (D - 4) D) R) - \\\nonumber&& -
    (D - 1) R^2 (1173 + 565 R + D (-4 (997 + 349 R) + D (6 (393 + 217 R)+
\\\nonumber&& + D (-548 - 452 R + 45 D (1 + R)))))) q(r)^8 + \\\nonumber&& +
    (D - 1) R (-64 c^2 (D - 5) (36 + D (-13 + 3 D)) +
\\\nonumber&& + (D - 1) (635 + D (-1204 + 3 D (294 + D (-92 + 9 D)))) R^2 - \\\nonumber&& -
    8 c (D - 5) (63 + 31 R + D (127 + 191 R + D (-47 + D + (-79 + D) R)))) q(r)^9 + \\\nonumber&& +
 2 (D - 5) (64 c^3 (D - 5) (D - 3) + 8 c (D - 1) (-27 + D (141 + (-43 + D) D)) R^2 + \\\nonumber&& +
    8 c^2 (D - 5) (-3 + 77 R + D (D - 2 + (D - 18) R)) + (D - 1)^2 R^2 (-33 (R - 7) + \\\nonumber&& +
       D (59 + 43 R + D (-59 - 35 R + 9 D (1 + R))))) q(r)^{10} - \\\nonumber&& -
 2 (D - 5) R (24 c^2 (-11 + D) (D - 5) (D - 3) + (D - 1)^2 (-65 + D (81 + D (7 D - 39))) R^2 + \\\nonumber&& +
    12 c (-7 + D) (D - 5) (D - 3) (D - 1) (1 + R)) q(r)^{11} + \\\nonumber&& +
    (D - 5)^2 (-1 + D) R^2 (16 c (26 + (D - 9) D) + \\\nonumber&& + (D - 1) (77 - 3 R + D (-18 + D + (D - 2) R))) q(r)^{12} +
    \\\nonumber&& + (D - 5)^2 (D - 3)^2 (D - 1)^2 R^3 q(r)^{13}.
\end{eqnarray}

\begin{figure}
\includeonegraph{profile_D=6_l=8,12,16_a=1_4_gauss-bonnet_tensor}
\langfigurecaption{The picture of instability of tensor-type of gravitational perturbations of Gauss-Bonnet black holes, developing at large multipole numbers: $D=6$, $l= 8$ (red), $l=12$ (green), $l=16$ (blue), $\alpha=1.3$. All quantities are measured in units of the event horizon $r_+$.}{O quadro de instabilidade de perturba\c{c}\~oes gravitacionais de buracos negros Gauss-Bonnet do tipo tensorial, que se desenvolvem para grandes n\'umeros de multipolo: $D=6$, $l=8$ (vermelho), $l=12$ (verde), $l=16$ (azul), $\alpha=1.3$. Todas as quantidades s\~ao medidas em unidades do horizonte de eventos $r_+$.}\label{D=6.l=8,12,16.a=1.4.gauss-bonnet.tensor}
\end{figure}
\begin{figure}
\includeonegraph{D=6_gauss-bonnet_tensor_threshold}
\langfigurecaption{The threshold $\alpha$ as a function of the inverse multipole number $l$ for tensor type of gravitational perturbations of Gauss-Bonnet black holes $D=6$. The points $l=16,20,32,40,50,64$ were fit by the line $\alpha = 2.627l^{-1}+1.005$. The theoretical result is $\alpha_t\approx1.006$ (the value of $\alpha$ is measured in units of the event horizon $r_+$).}{O limiar $\alpha$ como uma fun\c{c}\~ao do n\'umero de multipolo $l$ inverso para perturba\c{c}\~oes de buracos negros Gauss-Bonnet ($D=6$) do tipo tensorial. Os pontos $l=16,20,32,40,50,64$ foram ajustados pela linha $\alpha = 2.627l^{-1}+1.005$. O resultado te\'orico \'e $\alpha_t\approx1.006$ (o valor de $\alpha$ \'e medido em unidades do horizonte de eventos $r_+$).}\label{D=6.gauss-bonnet.tensor.threshold}
\end{figure}

\langpar{The quasi-normal modes of gravitational perturbations of Einstein-Gauss-Bonnet black holes were found in time-domain (see sec. \ref{sec:time-domain-fit}) for the first time in \cite{Konoplya:2008ix}. Consequently, we confirm the instability of these black holes in five- and six-dimensional space-times, proven analytically in \cite{Beroiz:2007gp}.}{Os modos quase-normais de perturba\c{c}\~oes gravitacionais de buracos negros Einstein-Gauss-Bonnet foram achados no dom\'inio temporal (ver a se\c{c}\~ao \ref{sec:time-domain-fit}) pela primeira vez em \cite{Konoplya:2008ix}. Conseq\"uentemente, foi confirmada a instabilidade desses buracos negros em espa\c{c}os-tempo de cinco e seis dimens\~oes, provada analiticamente em \cite{Beroiz:2007gp}.}

\langpar{Namely, in five dimensional case scalar-type perturbations are unstable for $\alpha>0.207r_+^2$ while in six dimensions tensor-type perturbations are unstable for $\alpha>1.006r_+^2$. We can see on the figure \ref{D=6.l=8,12,16.a=1.4.gauss-bonnet.tensor} that the larger $l$, at the earlier times instability growth occurs, and the stronger the growth rate is. In the region near the threshold value of $\alpha$, one can observe the growth only for large enough $l$, while perturbations of lower multipole number are not growing. For each $l$ there is some maximal value of $\alpha$ for which the perturbations are not growing. In order to find the threshold value of $\alpha$ numerically we extrapolate this value for $l\rightarrow\infty$ (see fig. \ref{D=6.gauss-bonnet.tensor.threshold}). Einstein-Gauss-Bonnet black holes in $D\geq7$ space-times are stable.}{Perturba\c{c}\~oes do tipo escalar s\~ao inst\'aveis em cinco dimens\~oes para $\alpha>0.207r_+^2$, enquanto que perturba\c{c}\~oes do tipo tensorial s\~ao inst\'aveis em seis dimens\~oes para $\alpha>1.006r_+^2$. Pode-se ver na figura \ref{D=6.l=8,12,16.a=1.4.gauss-bonnet.tensor} que quanto maior $l$, mais cedo o crescimento de instabilidade ocorre, e mais forte ser\'a a taxa de crescimento. Na regi\~ao perto do valor de limiar de $\alpha$, pode-se observar o crescimento somente para $l$ grande o suficiente, enquanto as perturba\c{c}\~oes do n\'umero multipolar menor n\~ao est\~ao crescendo. Para cada $l$ h\'a algum valor m\'aximo de $\alpha$ para o qual as perturba\c{c}\~oes n\~ao est\~ao crescendo. Para achar o valor de limiar de $\alpha$ numericamente extrapola-se esse valor para $l\rightarrow\infty$ (ver a fig. \ref{D=6.gauss-bonnet.tensor.threshold}). Os buracos negros Einstein-Gauss-Bonnet em espa\c{c}o-tempos $D\geq7$ s\~ao est\'aveis.}

\begin{figure}
\includeonegraph{profile_D=10_l=2_a=0_01_qauss-bonnet_scalar}
\langfigurecaption{The picture of time-domain evolution for scalar-type gravitational perturbations of Gauss-Bonnet black holes $D=10$, $l=2$, $\alpha=0.01$. One can see that two modes are dominating at the different stages. All quantities are measured in units of the event horizon $r_+$.}{O quadro de evolu\c{c}\~ao no dom\'inio de tempo para perturba\c{c}\~oes gravitacionais de buracos negros de Gauss-Bonnet do tipo escalar  $D=10$, $l=2$, $\alpha=0.01$. Pode-se ver que dois modos est\~ao dominando nas diferentes etapas. Todas as quantidades s\~ao medidas em unidades do horizonte de eventos $r_+$.}\label{D=10.l=2.a=0.01.qauss-bonnet.scalar}
\end{figure}

\langpar{In units of the event horizon radius, the imaginary part of fundamental quasi-normal modes decreases, when $\alpha$ increases, for all numbers of $D$ and all types of perturbations. Unlike the imaginary part, the real oscillation frequency does not behave uniformly: it decreases as $\alpha$ grows for most cases of tensor and vector modes. The behavior of the scalar mode is different: there are \emph{two} competing for the domination modes at different stages of the quasi-normal ringing. This superposition of modes, also with competing excitation coefficients, makes dependence of the fundamental scalar type quasi-normal modes on $\alpha$ and $D$ non-monotonic. On the figure \ref{D=10.l=2.a=0.01.qauss-bonnet.scalar} one can see that at the first stage the actual frequency of the dominant mode is much larger than at the second stage, while their damping rate stays almost the same.}{Em unidades do raio do horizonte de eventos, a parte imagin\'aria de modos quase-normais fundamentais diminui, quando $\alpha$ aumenta, para todos os n\'umeros de $D$ e todos os tipos de perturba\c{c}\~oes. Diferentemente da parte imagin\'aria, a freq\"u\^encia real de oscila\c{c}\~ao n\~ao se comporta uniformemente: ela diminui enquanto $\alpha$ cresce para a maior parte de casos de modos tensorial e vetorial. O comportamento do modo escalar \'e diferente: h\'a \emph{dois} modos, que competem pela domin\^ancia, em etapas diferentes do toque quase-normal. Esta superposi\c{c}\~ao de modos, tamb\'em com os coeficientes de excita\c{c}\~ao competidores, faz a depend\^encia do modo quasi-normal fundamental do tipo escalar n\~ao monot\^onico em $\alpha$ e $D$. Na figura \ref{D=10.l=2.a=0.01.qauss-bonnet.scalar} pode-se ver que na primeira etapa a freq\"u\^encia real do modo dominante \'e muito maior do que na segunda etapa, enquanto a sua taxa de decaimento fica quase a mesma.}

\langsectionlabel{Quasi-normal modes of brane-localised Standard Model fields}{Modos quase-normais de campos do Modelo Padr\~ao localizados na brana}{sec:BH-projected}
\langpar{The other possible scenario is that the Standard Model particles (scalars, fermions and gauge bosons) are restricted to live on the $3+1$-brane, which is embedded in the higher-dimensional bulk, while the gravitons can propagate also in the bulk. Therefore, if we study propagation of the fields near a $D$-dimensional black hole, we must consider the induced-on-the-brane gravitational background. If the size of extra dimensions is large, comparing to the size of the black hole, the induced background is given by the projection of the $D$-dimensional black hole metric onto the brane by fixing the values of the additional angular coordinates that describe the $(D-4)$ extra spacelike dimensions \cite{Kanti:2002nr,Kanti:2002ge}.}{Outro cen\'ario poss\'ivel \'e que as part\'iculas do Modelo Padr\~ao (escalares, fermions e bosons de calibre) s\~ao restringidas para viver na $3+1$-brana, que \'e encaixada no volume de dimens\~oes mais altas, enquanto os gr\'avitons tamb\'em podem propagar no volume. Por isso, quando estuda-se a propaga\c{c}\~ao dos campos perto de um buraco negro $D$-dimensional, deve-se considerar o contexto gravitacional induzido na brana. Se o tamanho das dimens\~oes extras for grande, comparando-se com o tamanho do buraco negro, o contexto induzido \'e dado pela proje\c{c}\~ao da m\'etrica do buraco negro $D$-dimensional para a brana, fixando os valores das coordenadas angulares adicionais que descrevem os $(D-4)$ dimens\~oes espaciais extra \cite{Kanti:2002nr,Kanti:2002ge}.}

\langsubsectionlabel{Kerr black holes}{Buracos negros de Kerr}{sec:Kerr-projected}
\langpar{The quasi-normal ringing of brane-localised fields propagating in Schwarzschild, Reissner-Nordstr\"om and Schwarzschild-(anti) de Sitter induced gravitational backgrounds was studied in \cite{Kanti:2005xa}. Yet, during the high-energy collisions of elementary particles resulting in the creation of black holes, it is unnatural to expect that only head-on collisions, leading to spherically symmetric black holes, would take place. Collisions with a non-vanishing impact parameter are most likely to occur, and, in addition, it is for these collisions that the black-hole production cross-section is maximised \cite{Kanti:2004nr}. Therefore, microscopic rotating black holes should be the
most generic situation, and the effect of the angular momentum of the black hole on the quasi-normal spectra of brane-localised fields is essential and cannot be neglected. This effect has been studied in \cite{Kanti:2006ua}.}{O toque quase-normal de campos localizados na brana que propagam no contexto gravitacional, induzido por buracos negros de Schwarzschild, Reissner-Nordstr\"om e Schwarzschild-(anti) de Sitter, foi estudado em \cite{Kanti:2005xa}. Mas, durante as colis\~oes de alta energia de part\'iculas elementares que resultam na cria\c{c}\~ao de buracos negros, n\~ao \'e natural esperar que somente as colis\~oes frontais se realizariam, levando a buracos negros esfericamente sim\'etricos. As colis\~oes com um par\^ametro de impacto n\~ao-nulo, s\~ao as que ocorrem com maior probabilidade, e, al\'em do mais, para essas colis\~oes a sec\c{c}\~ao transversal de produ\c{c}\~ao do buraco negro \'e maximizada \cite{Kanti:2004nr}. Por isso, os buracos negros microsc\'opicos com rota\c{c}\~ao deveriam ser a situa\c{c}\~ao mais gen\'erica, e o efeito do momento angular do buraco negro nos espectros quase-normais de campos localizados na brana \'e essencial e n\~ao pode ser desprezado. Este efeito foi estudado em \cite{Kanti:2006ua}.}

\langpar{The line-element, describing a higher-dimensional rotating neutral black hole, is given by the Myers-Perry solution \cite{Myers:1986un}. After the projection of the Myers-Perry metric onto the brane, the brane background assumes the form \cite{Kanti:2004nr}}{O elemento de linha, descrevendo um buraco negro neutro com rota\c{c}\~ao em dimens\~oes mais altas, \'e dado pela solu\c{c}\~ao de Myers-Perry \cite{Myers:1986un}. Depois da proje\c{c}\~ao da m\'etrica de Myers-Perry para a brana, o contexto na brana assume a forma \cite{Kanti:2004nr}}
\begin{equation}\label{brane-localised}
ds^2=\left(1-\frac{\mu}{\Sigma\,r^{D-5}}\right)dt^2+\frac{2 a\mu\sin^2\theta}
{\Sigma\,r^{D-5}}\,dt\,d\varphi-\frac{\Sigma}{\Delta}dr^2
-\Sigma\,d\theta^2-\left(r^2+a^2+\frac{a^2\mu\sin^2\theta}{\Sigma\,r^{D-5}}\right)
\sin^2\theta\,d\varphi^2,
\end{equation}
\lang{where}{onde}
\begin{equation}
\Delta=r^2+a^2-\frac{\mu}{r^{D-5}}, \quad\Sigma=r^2+a^2\cos^2\theta\,.
\label{master}
\end{equation}
\lang{The parameters $\mu$ and $a$ are related to the mass $M$ and the angular momentum $J$, respectively, of the black hole through the definitions \cite{Myers:1986un}}{Os par\^ametros $\mu$ e $a$ s\~ao relacionados \`a massa $M$ e ao momento angular $J$, respectivamente, do buraco negro, pelas defini\c{c}\~oes \cite{Myers:1986un}}
\begin{equation}
M=\frac{(D-2)\,\pi^{(D-1)/2}}{\kappa^2_D\,\Gamma[(D-1)/2]}\,\mu\,, \qquad J=\frac{2Ma}{D-2}\,,
\end{equation}
\lang{where $\kappa^2_D=8\pi G=8 \pi/M_*^{D-2}$ is the $D$-dimensional Newton's constant. The radius of the event horizon $r_+$ parameterises the black hole mass}{onde $\kappa^2_D=8\pi G=8 \pi/M_*^{D-2}$ \'e a constante de Newton $D$-dimensional. O raio do horizonte de eventos $r_+$ parametriza a massa do buraco negro}
$$\mu=r_+^{D-5}(r_+^2+a^2).$$

\langpar{We should note here that the higher-dimensional black hole is assumed to have only one non-vanishing component of angular momentum, about an axis in the brane. This is due to the simplifying assumption that the particles that created the black hole were restricted to live on an infinitely-thin brane, therefore, during collision they had a non-vanishing impact parameter only on a 2-dimensional plane along our brane.}{Observemos aqui, que o buraco negro em dimens\~oes mais altas \'e considerado tendo somente um componente do momento angular n\~ao-nulo sobre um eixo na brana. Isto \'e por causa da suposi\c{c}\~ao simplificada de que as part\'iculas que criaram o buraco negro foram restringidas para viver em uma brana infinitamente fina, por isso, durante a colis\~ao, elas tiveram um par\^ametro de impacto n\~ao-nulo somente em um plano bidimensional ao longo desta brana.}

\langpar{The equations of motion of the Standard Model fields in the background (\ref{brane-localised}) can be reduced to the following to coupled equations for the angular}{As equa\c{c}\~oes de movimento dos campos do Modelo Padr\~ao no contexto (\ref{brane-localised}) podem ser reduzidas \`as equa\c{c}\~oes ligadas \`as partes angular}
\begin{eqnarray}\label{brane-localised-angular}
&&\frac{1}{\sin\theta}\,\frac{d \,}{d \theta}\,\biggl(\sin\theta\, \frac{d S_{s,\ell}^m}{d \theta}\biggr) + \\\nonumber&&+ \biggl[-\frac{2 m s \cot\theta} {\sin\theta} - \frac{m^2}{\sin^2\theta} - a^2 \omega^2 \sin^2\theta - 2 a \omega s \cos\theta -s - s^2 \cot^2\theta + 2a\omega m + \lambda \biggr]\,S(\theta)=0\,,
\end{eqnarray}
\lang{and radial parts}{e radial}
\begin{equation}\label{brane-localised-radial}
\Delta^{s}\,\frac{d \,}{dr}\,\biggl(\Delta^{1-s}\,\frac{d P_s}{dr}\,\biggr) + \biggl(\frac{K^2-isK \Delta'}{\Delta} + 4i s\,\omega\,r -  \lambda \biggr)\,R(r)=0\,,
\end{equation}
\lang{where $\lambda$ is the separation constant, $m$ is the azimuthal number, $s = 0, 1/2, 1$ for scalar, fermion and gauge boson fields respectively and $K$ is defined as}{onde $\lambda$ \'e a constante de separa\c{c}\~ao , $m$ \'e o n\'umero azimutal, $s=0,1/2,1$ para campos escalar, fermion e boson de calibre, respectivamente, e $K$ \'e definido como}
$$K=(r^2+a^2)\,\omega - a m.$$

\langpar{In order to solve numerically the equations (\ref{brane-localised-angular}) and (\ref{brane-localised-radial}) we use the Frobenius method. Since (\ref{angular-Kerr}) coincides with (\ref{brane-localised-angular}) in the limit of $\alpha\rightarrow0$, the equation for the angular part can be solved as described in the section \ref{sec:Frobenius-angular}. The analysis of the singular points of the equation (\ref{brane-localised-radial}) allows to find the Frobenius series for the function $R(r)$ \cite{Kanti:2006ua}}{Para resolver numericamente as equa\c{c}\~oes (\ref{brane-localised-angular}) e (\ref{brane-localised-radial}) usa-se o m\'etodo de Frobenius. Como (\ref{angular-Kerr}) coincide com (\ref{brane-localised-angular}) no limite de $\alpha\rightarrow0$, a equa\c{c}\~ao para a parte angular pode ser resolvida como descrito na se\c{c}\~ao \ref{sec:Frobenius-angular}. A an\'alise dos pontos singulares da equa\c{c}\~ao (\ref{brane-localised-radial}) permite achar a s\'erie de Frobenius para a fun\c{c}\~ao $R(r)$ \cite{Kanti:2006ua}}
\begin{equation}
R(r)=\frac{e^{\imo\omega r}}{r-r_0}\left(\frac{r-r_+}{r-r_0}\right)^{-i\beta}
\,\sum_{i=0}^\infty b_i\left(\frac{r-r_+}{r-r_0}\right)^i,
\end{equation}
\lang{where $\beta$ is fixed (see sec. \ref{sec:Frobenius-radial}) as}{onde $\beta$ foi fixado (ver a se\c{c}\~ao \ref{sec:Frobenius-radial}) como}
$$\beta=\frac{K(r_+)}{\Delta'(r_+)}=\frac{\omega r_+ (r_+^2+a^2)-m a r_+}{(D-5)(r_+^2+a^2)+2r_+^2}.$$

\begin{figure}
\includeonegraph{QNM_D=6_fundamental_projected}
\langfigurecaption{Fundamental quasi-normal modes for the 6-dimensional black hole projected on the $4$-brane.}{Modos quase-normais fundamentais do buraco negro de 6 dimens\~oes projetado na $4$-brana.}\label{QNM.D=6.fundamental.projected}
\end{figure}

\langpar{From the figure \ref{QNM.D=6.fundamental.projected}, one easily observes that, as $a$ increases, the absolute values of both the real and imaginary parts of the fundamental quasi-normal frequency decrease. This makes the damping time longer and the field oscillations on the brane longer-lived. For large $a$ one can observe that the lifetime of a fermion signal is longer for the positive value of the azimuthal number $m$. The same behavior was observed also for the scalar field: the field oscillation lifetime grows as $m$ increases.}{Da figura \ref{QNM.D=6.fundamental.projected}, facilmente observa-se que, como $a$ aumenta, os valores absolutos tanto da parte real como da parte imagin\'aria da freq\"u\^encia quase-normal fundamental diminuem. Isto faz o tempo de decaimento ser mais longo e as oscila\c{c}\~oes de campo na brana serem de longa vida. Para grande $a$ pode-se observar que o tempo de vida de um sinal de fermion \'e mais longo para o valor positivo do n\'umero azimutal $m$. O mesmo comportamento foi observado tamb\'em para o campo escalar: o tempo de vida da oscila\c{c}\~ao do campo cresce enquanto $m$ aumenta.}

\begin{figure}
\includeonegraph{QMM_D=6,s=1,l=3,m=0_projected}
\langfigurecaption{A few higher overtones of the electromagnetic field for the 6-dimensional Kerr black hole projected on the $4$-brane ($l=3$, $m=0$) for the range $(0,r_+)$ of the angular momentum parameter $a$ with a step of $r_+/8$.}{Alguns sobretons mais altos do campo eletromagn\'etico do buraco negro de Kerr de 6 dimens\~oes projetado na $4$-brana ($l=3$, $m=0$) para o intervalo $(0,r_+)$ do par\^ametro do momento angular $a$ com um passo de $r_+/8$.}\label{QNM.D=6.projected}
\end{figure}

\langpar{On the figure \ref{QNM.D=6.projected}, we display a few of the higher overtones for a brane-localised gauge field (with $l=3$, $m=0$). For all higher overtones, it was found that an increase in $a$ leads again to the decrease of the absolute values of both the real and imaginary parts.}{Na figura \ref{QNM.D=6.projected}, foram expostos alguns dos sobretons mais altos de um campo de calibre localizado na brana (com $l=3$, $m=0$). Para todos os sobretons mais altos, foi achado que um aumento em $a$ conduz novamente \`a redu\c{c}\~ao dos valores absolutos tanto das partes reais como das partes imagin\'arias.}

\begin{figure}
\includeonegraph{QNM_a=1_fundamental_projected}
\langfigurecaption{Fundamental quasi-normal modes for the higher-dimensional Kerr black hole ($a=r_+$) projected on the $4$-brane.}{Modos quase-normais para buracos negros de Kerr em dimens\~oes mais altas ($a=r_+$) projetado na $4$-brana.}\label{QNM.a=1.fundamental.projected}
\end{figure}

\langpar{For fixed $a$ we find that, as $D$ increases, the imaginary part of the quasi-normal frequency decreases while the real part changes insignificantly  (see fig. \ref{QNM.a=1.fundamental.projected}). This behavior of the real part is different from the case of $D$-dimensional black holes, which are not projected on the brane (see sec. \ref{sec:HDRNdS}). For those black holes the real part of the quasi-normal frequency also grows as $D$ increases. Therefore, we conclude that higher dimensional black holes are better oscillators than the black holes, projected on the $4$-brane.}{Para $a$ fixo acha-se que, como $D$ aumenta, a parte imagin\'aria da freq\"u\^encia quase-normal diminui enquanto a parte real modifica-se insignificantemente (ver a fig. \ref{QNM.a=1.fundamental.projected}). Este comportamento da parte real \'e diferente do caso de buracos negros $D$-dimensionais, que n\~ao s\~ao projetados na brana (ver a se\c{c}\~ao \ref{sec:HDRNdS}). Para aqueles buracos negros a parte real da freq\"u\^encia quase-normal tamb\'em cresce enquanto $D$ aumenta. Por isso, conclui-se que os buracos negros em dimens\~oes mais altas s\~ao melhores osciladores do que os buracos negros projetados na $4$-brana.}

\langsubsection{Gauss-Bonnet black holes}{Buracos negros de Gauss-Bonnet}
\langpar{Within large extra dimensions scenarios of TeV-scale gravity, the classical space-time, induced by mini black holes, has large curvature along the transverse collision plane. Thus quantum gravity effects, and in particular higher curvature corrections to the Einstein gravity, cannot be ignored \cite{Rychkov:2004sf}. Therefore, the quantum corrections, provided by the Gauss-Bonnet theory (see sec. \ref{sec:Gauss-Bonnet}), must be taken into account, when mini black holes are considered.}{Dentro de cen\'arios com grandes dimens\~oes extra da gravidade de escala TeV, o espa\c{c}o-tempo cl\'assico, induzido por mini buracos negros, tem uma grande curvatura ao longo do plano de colis\~ao transversal. Assim, os efeitos de gravidade qu\^antica e, em particular, corre\c{c}\~oes mais altas da curvatura \`a gravidade de Einstein, n\~ao podem ser ignoradas \cite{Rychkov:2004sf}. Por isso, as corre\c{c}\~oes qu\^anticas, fornecidas pela teoria de Gauss-Bonnet (ver a se\c{c}\~ao \ref{sec:Gauss-Bonnet}), devem ser levadas em conta, quando mini buracos negros s\~ao considerados.}

\langpar{In this context, the propagation of brane-localised Standard Model fields in the background, induced by the Gauss-Bonnet black hole, has been studied numerically for the first time in \cite{Zhidenko:2008fp}.}{Neste contexto, a propaga\c{c}\~ao de campos do Modelo Padr\~ao localizado na brana no contexto induzido pelo buraco negro de Gauss-Bonnet foi estudada numericamente pela primeira vez em \cite{Zhidenko:2008fp}.}

\langpar{The metric induced on the brane is given by the following line element}{A m\'etrica induzida na brana \'e dada pelo elemento de linha}
\begin{equation}
ds^2=f(r)dt^2-f(r)^{-1}dr^2-r^2(d\theta^2+\sin\theta^2 d\phi^2),
\end{equation}
\lang{with the function $f(r)$ defined in (\ref{GBmetricfunction}).}{com a fun\c{c}\~ao $f(r)$ definida em (\ref{GBmetricfunction}).}

\langpar{The effective potentials for the Standard Model fields can be derived in the same way as in the Schwarzschild-de Sitter case. Their explicit form can be obtained by substituting the function $f(r)$ from (\ref{GBmetricfunction}) into the effective potentials for the test scalar (\ref{scalar-potential}), massless Dirac (\ref{Dirac-potential}), and Maxwell (\ref{EM-potential}) fields.}{Os potenciais efetivos para os campos do Modelo Padr\~ao podem ser derivados do mesmo jeito que no caso de Schwarzschild-de Sitter. A sua forma expl\'icita pode ser obtida substituindo a fun\c{c}\~ao $f(r)$ de (\ref{GBmetricfunction}) nos potenciais efetivos para o campo escalar de teste (\ref{scalar-potential}), o campo de Dirac sem massa (\ref{Dirac-potential}), e o campo de Maxwell (\ref{EM-potential}).}

\begin{figure}
\includeonegraph{Quality_GBprojected_D=9_s=0_l=1}
\langfigurecaption{The quality factor of the scalar field localised on the $4$-brane as a function of the Gauss-Bonnet parameter $\alpha$ ($D = 9$, $l = 1$).}{O fator de qualidade do campo escalar localizado na $4$-brana como uma fun\c{c}\~ao do par\^ametro de Gauss-Bonnet $\alpha$ ($D = 9$, $l = 1$).}\label{Quality.GBprojected.D=9.s=0.l=1}
\end{figure}

\langpar{As well as for the Schwarzschild black holes, the oscillations of brane-localised Standard Model fields decay faster for higher $D$. On the other hand, the Gauss-Bonnet term causes the perturbations decay slower. The real part of the quasi-normal frequencies has a more complicated behavior: for $D=5$ it decreases as $\alpha$ grows, but for higher-dimensional cases it starts growing first and then decreases after some value of $\alpha$ is reached. However, the quality factor $Q=\frac{\re(\omega)}{2\im(\omega)}$ increases as $\alpha$ grows for all fields and all values of $D$ (see fig. \ref{Quality.GBprojected.D=9.s=0.l=1}).}{Bem como para os buracos negros de Schwarzschild, as oscila\c{c}\~oes de campos do Modelo Padr\~ao localizados na brana decaem mais rapidamente para valores de $D$ mais altos. Por outro lado, o termo de Gauss-Bonnet faz com que as perturba\c{c}\~oes decaiam mais lentamente. A parte real das freq\"u\^encias quase-normais tem um comportamento mais complicado: para $D=5$ ela diminui enquanto $\alpha$ cresce, mas em dimens\~oes mais altas ela come\c{c}a a crescer primeiramente e ent\~ao diminui depois que um valor de $\alpha$ \'e alcan\c{c}ado. Contudo, o fator de qualidade $Q=\frac{\re(\omega)}{2\im(\omega)}$ aumenta j\'a que $\alpha$ cresce para todos os campos e todos os valores de $D$ (ver a fig. \ref{Quality.GBprojected.D=9.s=0.l=1}).}

\begin{figure}
\includeonegraph{D=7_GBprojected_a=5_0_scalar_profiles}
\langfigurecaption{The time-domain profiles of the brane-localised massless scalar field for the Gauss-Bonnet black hole ($D=7$, $\alpha=5$) for $l=0$ (blue), $l=1$ (green), $l=2$ (orange), $l=4$ (red). The bigger $l$ corresponds to the longer life of quasi-normal ringing and the quicker tail decay.}{Os perfis temporais do campo escalar sem massa localizado na brana para o buraco negro de Gauss-Bonnet ($D=7$, $\alpha=5$) para $l=0$ (azul), $l=1$ (verde), $l=2$ (laranja), $l=4$ (vermelho). O maior $l$ corresponde \`a vida mais longa do toque quase-normal e ao decaimento mais r\'apido da cauda.}\label{D=7.GBprojected.a=5.0.scalar.profiles}
\end{figure}

\langpar{As we can see in time-domain (see fig. \ref{D=7.GBprojected.a=5.0.scalar.profiles}), the late-time tails decay according to the inverse power law, which is found to be \cite{Zhidenko:2008fp}}{Como pode-se ver no dom\'inio de tempo (ver a fig. \ref{D=7.GBprojected.a=5.0.scalar.profiles}), as caudas de tempo tardio decaem segundo a lei de pot\^encia inversa, que foi achada na forma \cite{Zhidenko:2008fp}}
\begin{equation}\label{GBpowerlaw}
\Psi\propto t^{-(2l+D-1)}.
\end{equation}
\lang{This law depends only on the multipole number $l$ and the number of extra dimensions $D$.}{Esta lei depende somente do n\'umero de multipolo $l$ e do n\'umero de dimens\~oes extra $D$.}

\langpar{Using the WKB formula, we are able to find the Gauss-Bonnet corrections to the large multipole limit}{Utilizando a f\'ormula de WKB, \'e poss\'ivel achar as corre\c{c}\~oes de Gauss-Bonnet ao limite de multipolo grande}
\begin{gather}
\nonumber\omega=\Omega_R(1+\alpha A_1+\ldots)\left(l+\frac{1}{2}\right)-\imo\Omega_I\left(1+\alpha B_1+\ldots\right)\left(n+\frac{1}{2}\right)+ {\cal O}\left(\frac{1}{l}\right),\qquad n=0,1,2\ldots\\
\nonumber\omega=\Omega_R(1+\alpha A_1+\ldots)\kappa_{\pm}-\imo\Omega_I\left(1+\alpha B_1+\ldots\right)\left(n+\frac{1}{2}\right)+ {\cal O}\left(\frac{1}{\kappa_{\pm}}\right),\qquad n=0,1,2\ldots
\end{gather}
\begin{equation}\label{GB-largemultipole}
\Omega_R=\frac{1}{R_0}\sqrt{\frac{D-3}{D-1}},\quad\Omega_I=\frac{1}{R_0}\frac{D-3}{\sqrt{D-1}},
\end{equation}
\lang{where $R_0$ is the point, where the effective potential reaches its maximum for large $l$}{onde $R_0$ \'e o ponto, onde o potencial efetivo chega o seu m\'aximo para $l$ grande}
$$\left(\frac{R_0}{r_+}\right)^{D-3}=\frac{D-1}{2}+{\cal O}\left(\frac{1}{l}\right).$$
\lang{We see that for large multipoles the quality factor also decreases as $D$ grows}{V\^e-se que para multipolos grandes o fator de qualidade tamb\'em diminui enquanto $D$ cresce}
$$Q\sim\frac{\Omega_R}{\Omega_I}=\frac{1}{\sqrt{D-3}}.$$

\langpar{The corrections of the first order of $\alpha$ are given by}{As corre\c{c}\~oes da primeira ordem de $\alpha$ s\~ao dadas por}
\begin{eqnarray}\nonumber
A_1&=&-\frac{1}{r_+^2}\frac{D-4}{D-1}\left(\frac{D-1}{2}-\frac{r_+^2}{R_0^2}\right)<0,\\\nonumber
B_1&=&-\frac{1}{r_+^2}\frac{D-4}{D-1}\left(\frac{D-1}{2}+(D-2)\frac{R_0^2}{r_+^2}\right)<0,
\end{eqnarray}
\lang{implying quicker decreasing of the absolute value of the imaginary part than decreasing of the real part. It means that the quality factor grows as $\alpha$ increases. Thus we conclude, that the large multipole limit resembles the main properties of the fundamental quasi-normal modes for small multipole number.}{implicando redu\c{c}\~ao mais r\'apida do valor absoluto da parte imagin\'aria do que redu\c{c}\~ao da parte real. Isto significa que o fator de qualidade cresce enquanto $\alpha$ aumenta. Assim conclui-se, que o limite de multipolo grande se assemelha com as propriedades principais dos modos quase-normais fundamentais para n\'umero de multipolo pequeno.}

\langsectionlabel{Perturbations of squashed Kaluza-Klein black holes}{Perturba\c{c}\~oes de buracos negros esmagados de Kaluza-Klein}{sec:squashedKK}
\langpar{If the compactification radius of the extra dimensions is comparable with the size of the black holes we must take into account the size of extra dimensions. Such higher dimensional model of black holes has the asymptotic structure of the Kaluza-Klein type. The simplest example of five-dimensional black objects with the Kaluza-Klein geometry is the black string, the direct product of four-dimensional black hole and a circle. These objects look different from four-dimensional black holes only at sufficiently high energies, when Kaluza-Klein modes are excited. Therefore within these space-times we need high energy regime to see the extra dimensions. At the same time, there exist exact solutions of Kaluza-Klein black holes with squashed horizons, that look like five-dimensional squashed black holes near the event horizon, and like a Kaluza-Klein space-time at spatial infinity. Owing to the non-trivial bundle structure, the size of the extra dimension might be observed even at low energies by detecting e. g. their Hawking radiation \cite{Ishihara:2007ni}.}{Se o raio de compactifica\c{c}\~ao das dimens\~oes extras for compar\'avel com o tamanho dos buracos negros deve-se levar em conta o tamanho de dimens\~oes extras. Tal modelo de buracos negros em dimens\~oes mais altas tem a estrutura assint\'otica do tipo Kaluza-Klein. O exemplo mais simples de objetos negros de cinco dimens\~oes com a geometria Kaluza-Klein \'e a corda negra, que \'e o produto direto de um buraco negro em quatro dimens\~oes e um c\'irculo. Esses objetos parecem diferentes de buracos negros de quatro dimens\~oes somente em energias suficientemente altas, quando os modos de Kaluza-Klein s\~ao excitados. Por isso, dentro desses espa\c{c}os-tempo precisa-se de um regime de alta energia para se poder observar as dimens\~oes extra. Ao mesmo tempo, existem solu\c{c}\~oes exatas de buracos negros de Kaluza-Klein com horizontes esmagados, que se parecem com buracos negros esmagados de cinco dimens\~oes perto do horizonte de eventos, e com um espa\c{c}o-tempo Kaluza-Klein no infinito espacial. Devido \`a estrutura de fibrado n\~ao trivial, o tamanho da dimens\~ao extra poderia ser observado at\'e nas energias baixas, por exemplo, pela detec\c{c}\~ao de sua radia\c{c}\~ao de Hawking \cite{Ishihara:2007ni}.}

\langpar{To the best of my knowledge, quasi-normal frequencies of such black holes have been studied for the first time in \cite{Ishihara:2008re}.}{Ao meu conhecimento, as freq\"u\^encias quase-normais de tais buracos negros foram estudadas pela primeira vez em \cite{Ishihara:2008re}.}

\langsubsection{Quasi-normal modes of the scalar field for rotating squashed Kaluza-Klein black holes}{Os modos quase-normais de campo escalar para buracos negros esmagados de Kaluza-Klein com rota\c{c}\~ao}
\langpar{The five-dimensional rotating squashed Kaluza-Klein black hole with two equal angular momenta is described by}{O buraco negro esmagado de Kaluza-Klein de cinco dimens\~oes com rota\c{c}\~ao com dois momentos angulares iguais \'e descrito por}
\begin{eqnarray}
ds^2=dt^2-\frac{\Sigma_0}{\Delta_0}k(r)^2dr^2-\frac{r^2+a^2}{4}
[k(r)(d\sigma^2_1+d\sigma^2_2)+d\sigma^2_3]-\frac{M}{r^2+a^2}(dt-\frac{a}{2}d\sigma_3)^2,
\label{metric.KKR}
\end{eqnarray}
\lang{with}{com}
\begin{eqnarray}
d\sigma_1 = -\sin{\psi} d\theta+\cos{\psi} \sin{\theta}d\phi \ , \quad
d\sigma_2 = \cos{\psi} d\theta+\sin{\psi} \sin{\theta}d\phi \ , \quad
d\sigma_3 = d\psi+\cos{\theta}d\phi \ ,
\end{eqnarray}
\lang{where $0<\theta<\pi$, $0<\phi<2\pi$ and $0<\psi<4\pi$. The parameters are given by}{onde $0<\theta<\pi$, $0<\phi<2\pi$ e $0<\psi<4\pi$. Os par\^ametros s\~ao dados por}
\begin{eqnarray}
\Sigma_0&=&r^2(r^2+a^2),\nonumber\\
\Delta_0&=&(r^2+a^2)^2-Mr^2,\\
k(r)&=&\frac{(r^2_{\infty}-r^2_+)(r^2_{\infty}-r^2_-)}{(r^2_{\infty}-r^2)^2}.\nonumber
\end{eqnarray}
\lang{Here $M$ and $a$ correspond to mass and angular momenta, respectively. Values $r=r_+$ and $r=r_-$ are outer and inner horizons of the black hole. They relate to $M$ and $a$ by $a^4=(r_+r_-)^2, M-2a^2=r^2_++r^2_-$. The parameter $r_{\infty}$ corresponds to the spatial infinity. In the parameter space $0 < r_-\leq r_+ < r_{\infty}$, $r$ is restricted within the range $0<r<r_{\infty}$. The shape of black hole horizon is deformed by the parameter $k(r_+)$.}{Aqui $M$ e $a$ correspondem \`a massa e momentos angulares, respectivamente. Valores $r_+$ e $r_-$ s\~ao os horizontes exterior e interior do buraco negro. Eles relacionam-se a $M$ e $a$ por $a^4=(r_+r_-)^2, M-2a^2=r^2_++r^2_-$. O par\^ametro $r_{\infty}$ corresponde ao infinito espacial. No espa\c{c}o de par\^ametros $0 < r_-\leq r_+ < r_{\infty}$, $r$ \'e restringido pela variedade $0<r<r_{\infty}$. A forma do horizonte do buraco negro \'e deformada pelo par\^ametro $k(r_+)$.}

\langpar{The wave equation for the massless scalar field $\Phi(t,r,\theta,\phi,\psi)$ in the background (\ref{metric.KKR}) is given by}{A equa\c{c}\~ao de onda para o campo escalar sem massa $\Phi(t,r,\theta,\phi,\psi)$ no contexto (\ref{metric.KKR}) \'e dado por}
\begin{eqnarray}
\frac{1}{\sqrt{-g}}\partial_{\mu}\sqrt{-g}g^{\mu\nu}\partial_{\nu}
\Phi(t,r,\theta,\phi,\psi)=0.\label{WE}
\end{eqnarray}
\lang{Taking the ansatz}{Selecionando o ansatz}
$$\Phi(t,r,\theta,\phi,\psi)=e^{-i\omega t}R(\rho)e^{i m\phi+i\lambda \psi}S(\theta),$$
\lang{where $S(\theta)$ is the so-called spheroidal harmonics, the radial and time variables can be decoupled from angular ones, so that the final wave-like equation reads}{onde $S(\theta)$ \'e a chamada harm\^onica esferoidal, as vari\'aveis radial e temporal podem ser separadas de vari\'aveis angulares, de modo a equa\c{c}\~ao final de onda leia-se}

\begin{eqnarray}
\frac{d}{d\rho}\bigg[\Delta\frac{d R(\rho)}{d\rho}\bigg]
+\bigg[\frac{\tilde{H}^2}{\Delta}+\Lambda-l(l+1)+\lambda^2\bigg]R(\rho)=0,\label{radial.KKR}
\end{eqnarray}
\lang{where $l$ is the non-negative integer multipole number, $|m|<l$ and $|2\lambda|<2l$ are integers,}{onde $l$ \'e o n\'umero inteiro de multipolo n\~ao negativo, $|m|<l$ e $|2\lambda|<2l$ s\~ao n\'umeros inteiros,}
\begin{eqnarray}
&&\tilde{H}^2=\frac{Mr^2_{\infty}(\rho+\rho_0)^4}{H^4(r^2_{\infty}+a^2)^2}
\bigg[\omega-\frac{\lambda a
H^2(r^2_{\infty}+a^2)}{\rho_0r^3_{\infty}}\bigg]^2,
\\
&&\Lambda=\frac{4\rho^2_0r^6_{\infty}(\rho+\rho_0)^2}{H^2(r^2_{\infty}+a^2)^4}\omega^2-
\frac{4\lambda^2(\rho+\rho_0)^2}{r^2_{\infty}+a^2},
\\
&&H^2=\frac{\rho+\rho_0}{\rho+\frac{a^2}{r^2_{\infty}+a^2}\rho_0}.
\end{eqnarray}

\langpar{The radial coordinate $\rho$ is given by}{A coordenada radial $\rho$ \'e dada por}
\begin{eqnarray}
\rho=\rho_0\frac{r^2}{r^2_{\infty}-r^2},
\end{eqnarray}
\lang{with}{com}
\begin{eqnarray}
\rho^2_0&=&\frac{k_0}{4}(r^2_{\infty}+a^2),\nonumber \\
k_0&=&k(r=0)=\frac{(r^2_{\infty}+a^2)^2-Mr^2_{\infty}}{r^4_{\infty}}.
\end{eqnarray}

\langpar{Note that the three parameters $\rho_0$ and $\rho_\pm=\rho_0r_\pm^2/(r^2_{\infty}-r_\pm^2)$ can define the metric (\ref{metric.KKR}) if $r_\infty<\infty$. In some papers they are used to parameterise the black hole, instead of the parameters $r_\infty$, $r_\pm$.}{Observe que os tr\^es par\^ametros $\rho_0$ e $\rho_\pm=\rho_0r_\pm^2/(r^2_{\infty}-r_\pm^2)$ podem definir a m\'etrica (\ref{metric.KKR}) se $r_\infty <\infty$. Em alguns trabalhos eles s\~ao usados para parametrizar o buraco negro a em vez dos par\^ametros $r_\infty$, $r_\pm$.}

\langpar{The quasi-normal modes have been found with Frobenius method (see sec. \ref{sec:Frobenius}), using the following expansion \cite{Ishihara:2008re}}{Os modos quase-normais foram achados com o m\'etodo de Frobenius (ver a se\c{c}\~ao \ref{sec:Frobenius}), utilizando a expans\~ao \cite{Ishihara:2008re}}
\begin{equation}
R = \left(\frac{r^2-r_+^2}{r^2-r_-^2}\right)^{-\imo\beta}e^{\imo\rho\Omega}\rho^{\imo\nu-1}\sum_{n=0}^\infty a_n\left(\frac{r^2-r_+^2}{r^2-r_-^2}\right)^n,
\end{equation}
\lang{where $\beta$, $\nu$ and $\Omega$ are chosen in order to eliminate the singularities at $r=r_+$ and $r=r_\infty$. The sign of $\alpha$ and $\Omega$ is chosen in order to remain them in the same complex quadrant as $\omega$.}{onde $\beta$, $\nu$ e $\Omega$ s\~ao escolhidos para eliminar as singularidades em $r=r_+$ e $r=r_\infty$. O sinal de $\alpha$ e $\Omega$ \'e escolhido para que permane\c{c}am no mesmo quadrante complexo que $\omega$.}

\begin{figure}
\includetwograph{KKR_l=0_scalar_Re}{KKR_l=0_scalar_Im}
\langfigurecaption{Real and imaginary part of the fundamental quasi-normal frequency of the test scalar field ($l=0$) for rotating squashed Kaluza-Klein black holes.}{Partes real e imagin\'aria da freq\"u\^encia quase-normal fundamental do campo escalar de teste ($l=0$) para buracos negros esmagados de Kaluza-Klein com rota\c{c}\~ao.}\label{QNM.KKR.l=0.scalar}
\end{figure}

\langpar{The fundamental quasi-normal modes are presented on the figure \ref{QNM.KKR.l=0.scalar}. One can see that the real oscillation frequency exerts some irregular growth (with local minimums) when $r_\infty$ is increasing until some moderately large values of $r_\infty$. At larger $r_\infty$ the growth of $\re(\omega)$ changes into monotonic decay. The imaginary part of $\omega$ that determines the damping rate also has some initial irregular growth when $r_\infty$ increases, but at larger $r_\infty$ the two scenarios are possible: either monotonic decay (for large values of $r_-$) or monotonic growth (for small and moderate $r_-$). Thus, for a given mass and angular momentum of the black hole, one can learn the size of extra dimension $r_\infty$ from values of quasi-normal modes of the emitted radiation.}{Os modos quase-normais fundamentais s\~ao apresentados na figura \ref{QNM.KKR.l=0.scalar}. Pode-se ver que a freq\"u\^encia real de oscila\c{c}\~ao exerce um crescimento irregular (com m\'inimos locais) quando $r_\infty$ est\'a aumentando at\'e valores de $r_\infty$ moderadamente grandes. Para $r_\infty$ maior, o crescimento de $\re(\omega)$ modifica-se em decrescimento monot\^onico. A parte imagin\'aria de $\omega$ que determina a taxa de decaimento tamb\'em tem um crescimento irregular inicial quando $r_\infty$ aumenta, mas para $r_\infty$ maior, os dois cen\'arios s\~ao poss\'iveis: ou decrescimento monot\^onico (para valores de $r_-$ grandes) ou crescimento monot\^onico (para $r_-$ pequeno e moderado). Assim, para uma massa e momento angular do buraco negro dados, pode-se obter o tamanho da dimens\~ao extra $r_\infty$ de valores de modos quase-normais da radia\c{c}\~ao emitida.}

\langsubsection{Gravitational quasi-normal modes for non-rotating squashed Kaluza-Klein black holes}{Modos quase-normais gravitacionais para buracos negros esmagados de Kaluza-Klein sem rota\c{c}\~ao}
\langpar{In the previous subsection, we have considered the scalar field in the background of squashed Kaluza-Klein black holes. However, the tensor perturbations are more interesting from the point of view of the stability and possibility to observe gravitational waves from black holes.}{Na subse\c{c}\~ao pr\'evia, o campo escalar foi considerado no contexto de buracos negros esmagados de Kaluza-Klein. Contudo, as perturba\c{c}\~oes tensoriais s\~ao mais interessantes do ponto de vista da estabilidade e da possibilidade de observar ondas gravitacionais de buracos negros.}

\langpar{The metric of the uncharged non-rotating squashed Kaluza-Klein black hole is a particular case of the metric (\ref{metric.KKR})}{A m\'etrica do buraco negro esmagado de Kaluza-Klein n\~ao carregado sem rota\c{c}\~ao \'e um caso particular da m\'etrica (\ref{metric.KKR})}
\begin{eqnarray}
ds^2 =F(\rho)d\tau^2 - \frac{G(\rho)^2}{F(\rho)}d\rho^2 - 4 \rho^2 G(\rho)^2 d\sigma^+ d\sigma^- -\frac{r_{\infty}^2}{4 G(\rho)^2}(d\sigma^3)^2,
\end{eqnarray}
\lang{where we have defined $\tau = 2\rho_0 t /r_\infty$ and}{onde define-se $\tau = 2\rho_0 t /r_\infty$ e}
$$
F(\rho) = 1-\frac{\rho_+}{\rho} \ , \quad
G(\rho)^2 = 1+\frac{\rho_0}{\rho} \ , \quad
r_{\infty}^2 = 4 \rho_0(\rho_+ + \rho_0) \ .
$$
\lang{Here, we have used a basis}{Aqui, foi usada a base}
$$
 d\sigma^{\pm} = \frac{1}{2}(d\sigma^1 \mp i d\sigma^2)\ .
$$

\langpar{Since the space-time has the symmetry $SU(2)\times U(1)$, the metric perturbations can be classified by eigenvalues $J,M$ for $SU(2)$ and $K$ for $U(1)$. Here we consider only zero modes $J=M=0$. Even in this case, since $d\sigma^\pm$ carry eigenvalues $K=\pm 1$, each component could have different eigenvalue $K$. It is important to recognise that the components with different $K$ are decoupled. That is why we have the master equations for each $K$. To obtain master equations, we choose the gauge condition as}{Como o espa\c{c}o-tempo tem a simetria $SU(2)\times U(1)$, as perturba\c{c}\~oes m\'etricas podem ser classificadas por valores pr\'oprios $J, M$ para $SU(2)$ e $K$ para $U(1)$. Aqui foram considerados somente modos nulos $J=M=0$. Mesmo neste caso, como $d\sigma^\pm$ leva a valores pr\'oprios $K=\pm 1$, cada componente pode ter valor pr\'oprio $K$ diferente. \'E importante reconhecer que os componentes com $K$ diferente s\~ao separados. Por isso t\^em-se as equa\c{c}\~oes mestras para cada $K$. Para obter equa\c{c}\~oes mestras, escolheu-se a condi\c{c}\~ao de calibre como}
\begin{eqnarray}
h_{3+} = h_{3-} = h_{+-} = h_{tt} = h_{t3}  =0\,.
\end{eqnarray}
%
\lang{As is shown in \cite{Kimura:2007cr}, the perturbation equations for the $|K|=2$ mode can be reduced to the wave equation for $h_{++}$ with the effective potential in the form}{Como foi mostrado em \cite{Kimura:2007cr}, as equa\c{c}\~oes de perturba\c{c}\~ao para o modo $|K|=2$ podem ser reduzidas \`a equa\c{c}\~ao de onda para $h_{++}$ com o potencial efetivo na forma}
\begin{eqnarray}
V_2 &=& \frac{
         -(\rho_+-\rho)}{16\rho^3 \rho_0 (\rho_++\rho_0)(\rho+\rho_0)^3}
\Big[64\rho^5 + 256\rho^4\rho_0-32\rho^3 \rho_0(\rho_+-11\rho_0)
                            +9\rho_+ \rho_0^3(\rho_+ +\rho_0)
\nonumber\\
  &&  \qquad           +8\rho^2\rho_0(2\rho_+^2 -5\rho_+\rho_0+25\rho_0^2)
                            +\rho \rho_0^2 (20\rho_+^2-9\rho_+\rho_0+35\rho_0^2)
                       \Big] \ .
\end{eqnarray}
%
%
\lang{Similarly, the perturbation equations  for the $|K|=1$ mode can be reduced to the wave equation for $h_{\rho +}$ with the effective potential}{Semelhantemente, as equa\c{c}\~oes de perturba\c{c}\~ao para o modo $|K|=1$ podem ser reduzidas \`a equa\c{c}\~ao de onda para $h_{\rho +}$ com o potencial efetivo}
\begin{eqnarray}
V_1 &=& \left(1-\frac{\rho_+}{\rho}\right)
\bigg[\frac{1}{\rho_+\rho_0}
+\frac{7(-\rho_++\rho)}{16(\rho+\rho_0)^3}
+\frac{17 \rho_+-19\rho}{8 \rho(\rho+\rho_0)^2}
+\frac{9(-3\rho_++7\rho)}{16\rho^2(\rho+\rho_0)}
+\frac{\rho_+-\rho}{\rho_+^2\rho +\rho_+\rho \rho_0}
\nonumber\\ && \qquad
-\frac{8(\rho_+-\rho)^2\rho}{(\rho_+-2\rho)(\rho_+\rho_0-\rho(\rho+2\rho_0))^2}
-\frac{2(\rho_+ +2\rho)}{(\rho_+-2\rho)(\rho_+\rho_0-\rho(\rho+2\rho_0))} \bigg] \ .
\end{eqnarray}
\lang{Finally, for the $K = 0$ mode, we obtain the wave equation for $h_{33}$
with the effective potential}{Finalmente, para o modo $K = 0$, obt\'em-se a equa\c{c}\~ao de onda para $h_{33}$
com o potencial efetivo}
\begin{eqnarray}
V_0 &=& \frac{-(\rho_+-\rho)}{16\rho^3 (\rho+\rho_0)^3(4\rho+3\rho_0)^2}
\Big[256\rho_+\rho^4 + 64\rho^3 (17\rho_++2\rho)\rho_0 +48\rho^2(32\rho_+ +11\rho)\rho_0^2
\nonumber\\ &&  \qquad
+60\rho(13\rho_++12\rho)\rho_0^3+9(9\rho_+ +35\rho)\rho_0^4
\Big] \ .
\end{eqnarray}

\begin{figure}
\includetwograph{KK_0grav_Re}{KK_0grav_Im}
\langfigurecaption{Real and imaginary parts of the fundamental quasi-normal frequency for metric perturbations ($M=J=0$) $|K|=0$ (blue), $|K|=1$ (red), $|K|=2$ (yellow) of squashed Kaluza-Klein black holes. Higher values of $K$ correspond to higher oscillation frequency and slower damping.}{Partes real e imagin\'aria da freq\"u\^encia quase-normal fundamental para perturba\c{c}\~oes m\'etricas ($M=J=0$) $|K|=0$ (azuis), $|K|=1$ (vermelhas), $|K|=2$ (amarelas) de buracos negros esmagados de Kaluza-Klein. Os valores mais altos de $K$ correspondem \`a freq\"u\^encia de oscila\c{c}\~ao mais alta e decaimento mais lento.}\label{QNM.KK.0grav}
\end{figure}

\langpar{The fundamental quasi-normal modes calculated with 6th order WKB method (see sec. \ref{sec:WKB}) are presented on the figure \ref{QNM.KK.0grav}. The real part of $\omega$ decreases for higher values of $\rho_0$. The absolute value of imaginary part has maximum for $K\neq0$. For not large values of $\rho_0$ the quasi-normal modes of such Kaluza-Klein black hole perturbations are longer lived. Their behavior for small values of $\rho_0$ in time-domain looks like massive field tails (see chapter \ref{sec:massive-fields}), which occur at much earlier time.}{Os modos quase-normais fundamentais calculados com o m\'etodo de WKB da $6^a$ ordem (ver a se\c{c}\~ao \ref{sec:WKB}) s\~ao apresentados na figura \ref{QNM.KK.0grav}. A parte real de $\omega$ diminui para valores de $\rho_0$ mais altos. O valor absoluto da parte imagin\'aria tem o m\'aximo para $K\neq0$. Para valores de $\rho_0$ n\~ao grandes os modos quase-normais de tais perturba\c{c}\~oes de buracos negros de Kaluza-Klein t\^em uma vida mais longa. O seu comportamento para valores de $\rho_0$ pequenos no dom\'inio de tempo parece-se com caudas de campo massivo (ver o cap\'itulo \ref{sec:massive-fields}), que ocorrem muito mais cedo.}

\langpar{It is important to note, that if we know the quasi-normal frequency we can find the size of the extra dimension in the considered black hole model, so that quasi-normal modes give a kind of opportunity to "look into'' an extra dimension at low energies. In detail, when a dominant quasi-normal mode is measured, one can compare it with the numerically found one and find out which is the value of $\rho_0$ and the radius of the event horizon that corresponds to the observed quasinormal mode. In this way we can determine the parameters of the black hole and the size of the extra dimension, assuming that there exist no other Kaluza-Klein black holes with similar features. Also, we do not observe any growing mode, what supports stability of such black holes.}{\'E importante observar que se a freq\"u\^encia quase-normal for conhecida, pode-se achar o tamanho da dimens\~ao extra no modelo de buraco negro considerado, de modo que os modos quase-normais levam a uma oportunidade de se ``investigar'' uma dimens\~ao extra em energias baixas. Detalhadamente, quando um modo quase-normal dominante \'e medido, pode-se compar\'a-lo com a freq\"u\^encia achada numericamente e descobrir qual \'e o valor de $\rho_0$ e o raio do horizonte de eventos que corresponde ao modo quase-normal observado. Deste modo, pode-se determinar os par\^ametros do buraco negro e o tamanho da dimens\~ao extra, assumindo que n\~ao existe qualquer outro buraco negro de Kaluza-Klein com caracter\'isticas semelhantes. Tamb\'em n\~ao se observa qualquer modo crescente, o que sustenta a estabilidade de tais buracos negros.}

\langchapterlabel{Massive fields around black holes}{Campos massivos em torno de buracos negros}{sec:massive-fields}
\langsectionlabel{Evolution of massive fields. Quasi-resonances}{Evolu\c{c}\~ao de campos massivos. Quase-resson\^ancias}{sec:massive-properties}
\langpar{In the previous chapters we have considered massless fields propagating in the background of various black holes. Let us go further and study evolution of massive fields.}{Nos cap\'itulos pr\'evios foram considerados campos sem massa que propagam no contexto de v\'arios buracos negros. Indo al\'em disso, agora ser\'a apresentada a evolu\c{c}\~ao de campos massivos.}

\langpar{One should note, that a scalar field with the mass term can be interpreted as a self-interacting massless scalar field within regime of small perturbations \cite{Hod:1998ra}. A massless scalar field, when considered in models with extra dimensions of Randall-Sundrum type \cite{Randall:1999ee,Randall:1999vf}, gains a large effective mass due to the Kaluza-Klein momentum. Also, the effective mass is acquired by a massless scalar field in the vicinity of the magnetised black holes \cite{Konoplya:2007yy}.}{Observa-se que um campo escalar com o termo de massa pode ser interpretado como um campo escalar sem massa que auto-interage dentro do regime de perturba\c{c}\~oes pequenas \cite{Hod:1998ra}. Um campo escalar sem massa, quando considerado em modelos com dimens\~oes extra do tipo de Randall-Sundrum \cite{Randall:1999ee,Randall:1999vf}, ganha uma massa efetiva grande por causa do momento de Kaluza-Klein. Tamb\'em, a massa efetiva \'e adquirida por um campo escalar sem massa nas proximidades de buracos negros magnetizados \cite{Konoplya:2007yy}.}

\langpar{Let us start from the consideration of a test massive scalar field in the $D$-dimensional Schwarzschild background, given by the metric (\ref{HDBHmetric}) with}{Comecemos pela considera\c{c}\~ao de um campo escalar massivo de teste no contexto Schwarzschild $D$-dimensional, dado pelo m\'etrico (\ref{HDBHmetric}) com}
$$f(r)=1-\left(\frac{r_+}{r}\right)^{D-3}.$$

\langpar{The Klein-Gordon equation,}{A equa\c{c}\~ao de Klein-Gordon}
\begin{equation}\label{massive-scalar-equation}
\frac{1}{\sqrt{-g}}\frac{\partial}{\partial x^a}g^{ab}\sqrt{-g}\frac{\partial\Phi(x)}{\partial x^b}=-\mu^2\Phi(x),
\end{equation}
\lang{which governs the massive scalar field evolution in the curved background, can be reduced to the wave-like equation (\ref{wave-like}) with the effective potential}{que governa a evolu\c{c}\~ao do campo escalar massivo no contexto curvo, pode ser reduzida \`a equa\c{c}\~ao de onda (\ref{wave-like}) com o potencial efetivo}
\begin{equation}\label{massive-scalar-potential}
V(r)=f(r)\left(\mu^2+\frac{l(l+D-3)}{r^2}+\frac{(D-4)(D-2)}{4r^2}f(r)+\frac{D-2}{2r}f'(r)\right),
\end{equation}
\lang{where $\mu$ is the field mass and $l=0,1,2\ldots$ is the multipole number.}{onde $\mu$ \'e a massa do campo e $l=0,1,2\ldots$ \'e o n\'umero de multipolo.}

\langpar{One can see, that the potential (\ref{massive-scalar-potential}) does not vanish at the spatial infinity. It changes the behavior of the eigenfunction $Q_\omega$ (\ref{wave-like}), so that}{Pode-se ver que o potencial (\ref{massive-scalar-potential}) n\~ao desaparece no infinito espacial. Ele modifica o comportamento das fun\c{c}\~oes pr\'oprias $Q_\omega$ (\ref{wave-like}), de modo que}
\begin{equation}
r_\star\rightarrow+\infty: \quad Q_\omega\propto \exp(\imo\chi r_\star)
\end{equation}
\lang{where $\chi=\sqrt{\omega^2-\mu^2}$ (see sec. \ref{Frobenius-massive}).}{onde $\chi=\sqrt{\omega^2-\mu^2}$ (ver a se\c{c}\~ao \ref{Frobenius-massive}).}

\langpar{Repeating the calculations of the section \ref{sec:stability}, instead of (\ref{basic-ppot}) we find \cite{Konoplya:2004wg}}{Repetindo os c\'alculos da se\c{c}\~ao \ref{sec:stability}, ao inv\'es de (\ref{basic-ppot}) encontra-se \cite{Konoplya:2004wg}}
\begin{equation}
\re(\chi) |Q_\omega(r_\star=\infty)|^2+\re(\omega) |Q_\omega(r_\star=-\infty)|^2 + 2\re(\omega)\im(\omega) \intop_{-\infty}^\infty|Q_\omega(r_\star)|^2 dr_\star = 0.
\end{equation}
\lang{Since $\re(\chi)$ has the same sign as $\re(\omega)$, the unstable modes do not oscillate (see sec. \ref{sec:Gregory-Laflamme}).}{Como $\re(\chi)$ tem o mesmo sinal que $\re(\omega)$, os modos inst\'aveis n\~ao oscilam (ver a se\c{c}\~ao \ref{sec:Gregory-Laflamme}).}

\langpar{For some values of the black hole mass $M$ and the scalar field mass $\mu$ purely real frequencies were observed. Such oscillations have infinitely long lifetime and were called quasi-resonances \cite{Ohashi:2004wr}. Indeed, for massive fields quasi-normal frequencies are not required to be complex. Let us suppose that for some parameters we have $\im(\omega)=0$, so that $\re(\chi)=0$, implying}{Para alguns valores da massa do buraco negro $M$ e da massa do campo escalar $\mu$ as freq\"u\^encias puramente reais foram observadas. Tais oscila\c{c}\~oes t\^em vida infinitamente longa e foram chamadas quase-resson\^ancias \cite{Ohashi:2004wr}. De fato, para campos massivos as freq\"u\^encias quase-normais n\~ao precisam ser complexas. Suponhamos que para alguns par\^ametros tem-se $\im(\omega)=0$, de modo que $\re (\chi) =0$. Isto implica em}
\begin{equation}\label{quasi-resonance-condition}
Q_\omega(r_\star=-\infty)=0.
\end{equation}
\lang{It means, that there is no wave at the event horizon and, because $\re(\chi)=0$, no energy transmission to the spatial infinity. That is why such oscillations do not decay. The situation is similar to the standing waves on a fixed string. The requirement for $\chi$ to be imaginary, bounds the purely real frequencies by the field mass, what was pointed out in \cite{Konoplya:2004wg},}{Isto significa que n\~ao h\'a qualquer onda no horizonte de eventos e, porque $\re(\chi)=0$, nenhuma transmiss\~ao de energia ao infinito espacial. Por isso tais oscila\c{c}\~oes n\~ao decaem. A situa\c{c}\~ao \'e semelhante \`as ondas permanentes em uma corda fixa. A exig\^encia para $\chi$ ser imagin\'ario, limita as freq\"u\^encias puramente reais pela massa do campo, o que foi indicado em \cite{Konoplya:2004wg},}
\begin{equation}\label{quasiresonance-constraint}
\omega_{QRM}<\mu.
\end{equation}

\langpar{The phenomenon of quasi-resonance exists since the effective potential is not zero at the spatial infinity. If the potential vanishes at $r_\star\rightarrow\pm\infty$ the condition $\im(\omega)=0$ does not satisfy (\ref{basic-ppot}). Thus, for instance, there is no quasi-resonances in the spectrum of the massive scalar field in the Schwarzschild-de Sitter background.}{O fen\^omeno da quase-resson\^ancia existe j\'a que o potencial efetivo n\~ao \'e zero no infinito espacial. Se o potencial desaparecer em $r_\star\rightarrow\pm\infty$ a condi\c{c}\~ao $\im(\omega)=0$ n\~ao satisfaz (\ref{basic-ppot}). Assim, por exemplo, n\~ao h\'a nenhuma quase-resson\^ancia no espectro do campo escalar massivo no contexto de Schwarzschild-de Sitter.}

\langsectionlabel{Quasi-normal spectrum of the massive scalar field around Schwarzschild black holes}{Espectro quase-normal do campo escalar massivo em torno de buracos negros Schwarzschild}{sec:quasi-resonance}
\langpar{The quasi-normal modes of the massive scalar field around 4-dimensional Schwarzschild, Kerr \cite{Simone:1991wn} and Reissner-Nordstr\"om black holes \cite{Konoplya:2002wt} and of the massive Dirac field in the 4-dimensional Schwarzschild background \cite{Cho:2003qe} were calculated within the 3rd order WKB approach. Yet, the WKB formula, used for the calculations, is valid only for small field mass. For large mass of the field the WKB approach needs modifications (see sec. \ref{sec:WKB}), that were not taken into account in the earlier research \cite{Simone:1991wn,Konoplya:2002wt,Cho:2003qe}.}{Os modos quase-normais do campo escalar massivo em torno de buracos negros 4-dimensionais de Schwarzschild, Kerr \cite{Simone:1991wn} e Reissner-Nordstr\"om  \cite{Konoplya:2002wt} e do campo massivo de Dirac no contexto de Schwarzschild 4-dimensional \cite{Cho:2003qe} foram calculados dentro da $3^a$ ordem da aproxima\c{c}\~ao de WKB. Mas a f\'ormula de WKB, usada para os c\'alculos, \'e v\'alida somente para um campo com massa pequena. Para uma massa grande a aproxima\c{c}\~ao de WKB precisa de modifica\c{c}\~oes (ver a se\c{c}\~ao \ref{sec:WKB}), que n\~ao foram levadas em conta anteriormente \cite{Simone:1991wn,Konoplya:2002wt,Cho:2003qe}.}

\langpar{The quasi-normal spectrum of the massive scalar field for Reissner-Nordstr\"om black holes were studied for the first time by the more accurate Frobenius method by Ohashi and Sakagami \cite{Ohashi:2004wr}. They have found the quasi-resonances and supposed that they appear in a limiting situation and quasi-normal modes can disappear when the field mass exceeds a certain value. We found \cite{Konoplya:2004wg} that after reaching a quasi-resonance only one quasi-normal frequency disappears. The higher overtones remain in the spectrum for any finite field mass.}{O espectro quase-normal do campo escalar massivo para buracos negros de Reissner-Nordstr\"om foi estudado pela primeira vez pelo m\'etodo de Frobenius, que \'e mais preciso, por Ohashi e Sakagami \cite{Ohashi:2004wr}. Eles encontraram as quase-resson\^ancias e supuseram que elas aparecem em uma situa\c{c}\~ao restritiva e os modos quase-normais podem desaparecer quando a massa do campo excede um certo valor. Achamos \cite{Konoplya:2004wg} que depois de chegar uma quase-resson\^ancia somente uma freq\"u\^encia quase-normal desaparece. Os sobretons mais altos permanecem no espectro para qualquer massa do campo finita.}

\langpar{The quasi-normal spectrum of massive scalar fields in $D$-dimensional Schwarzschild background was studied for the first time in \cite{Zhidenko:2006rs} within Frobenius method.}{O espectro quase-normal de campos escalares massivos no contexto de Schwarzschild $D$-dimensional foi estudado pela primeira vez em \cite {Zhidenko:2006rs} utilizando o m\'etodo de Frobenius.}

\begin{figure}
\includeonegraph{QNM_D=6_l=0_massive-scalar}
\langfigurecaption{Three higher quasi-normal modes ($l=0$) of the Schwarzschild black hole ($D=6$) for the massive scalar field of various $\mu$. The frequency for $\mu=0$ has the largest imaginary part. The points were plotted with the step of $\Delta\mu r_+ = 1/2$. Solid lines mark the same overtone number.}{Tr\^es modos quase-normais mais altos ($l=0$) do buraco negro de Schwarzschild ($D=6$) para o campo escalar massivo de v\'arios $\mu$. A freq\"u\^encia para $\mu=0$ tem a maior parte imagin\'aria. Os pontos foram tra\c{c}ados com o passo de $\Delta\mu r_+ = 1/2$. As linhas s\'olidas marcam o mesmo n\'umero de sobretom.}\label{QNM.D=6.l=0.massive-scalar}
\end{figure}

\langpar{On the figure \ref{QNM.D=6.l=0.massive-scalar} one can see, that increasing of the field mass gives rise to decreasing of the imaginary part of the quasi-normal mode until reaching the vanishing damping rate. When some threshold values of $\mu r_+$ are exceeded, the particular quasi-normal modes ``disappear''. The larger field mass is, the more first overtones share this destiny. The disappearing of a finite number of modes implies that for large field mass some higher overtone becomes a fundamental frequency, having the longest lifetime among the rest of the modes. Nevertheless, the quasi-normal spectrum remains infinite for any finite field mass.}{Na figura \ref{QNM.D=6.l=0.massive-scalar} pode-se ver, que o aumento da massa do campo d\'a a origem \`a redu\c{c}\~ao da parte imagin\'aria do modo quase-normal at\'e alcan\c{c}ar a taxa de decaimento nula. Quando alguns valores de limiar de $\mu r_+$ s\~ao excedidos, determinados modos quase-normais ``desaparecem''. Quanto maior \'e a massa do campo, mais os primeiros sobretons compartilham este destino. O desaparecimento de um n\'umero finito de modos implica que para uma massa grande algum sobretom mais alto transforma-se em uma freq\"u\^encia fundamental, tendo a mais longa vida entre o resto dos modos. Todavia, o espectro quase-normal permanece infinito para qualquer massa finita de campo.}

\begin{figure}
\includeonegraph{QNM_l=0_massive-scalar}
\langfigurecaption{Fundamental quasi-normal frequencies for the $D$-dimensional Schwarzschild black hole for the massive scalar field ($l=0$) of various $\mu$. The frequency for $\mu=0$ has the largest imaginary part. The points were plotted with the step of $\Delta \mu r_+ = 1/10$. Solid lines mark the same number of $D$.}{Freq\"u\^encias quase-normais fundamentais do buraco negro de Schwarzschild $D$-dimensional para o campo escalar massivo ($l=0$) de v\'arios $\mu$. A freq\"u\^encia para $\mu=0$ tem a maior parte imagin\'aria. Os pontos foram tra\c{c}ados com o passo de $\Delta \mu r_+ = 1/10$. As linhas s\'olidas marcam o mesmo n\'umero de $D$.}\label{QNM.l=0.massive-scalar}
\end{figure}

\langpar{The behavior of the fundamental mode depends qualitatively on the number $D$. For $D=4,5$, the fundamental mode behaves like higher overtones: decreases its imaginary part, reaches quasi-resonance for some value of $\mu$, and disappears. For $D\geq6$ we observe qualitatively different picture (see fig. \ref{QNM.l=0.massive-scalar}). As $\mu$ grows the imaginary part of the fundamental mode tends to zero asymptotically while the real part approaches $\mu$ (see fig. \ref{QNM.D=6.massive-scalar}).}{O comportamento do modo fundamental depende qualitativamente do n\'umero $D$. Para $D=4,5$, o modo fundamental comporta-se como sobretons mais altos: reduz a sua parte imagin\'aria, chega \`a quase-resson\^ancia para algum valor de $\mu$, e desaparece. Para $D\geq6$ observa-se um quadro qualitativamente diferente (ver a fig. \ref{QNM.l=0.massive-scalar}). Como $\mu$ cresce, a parte imagin\'aria do modo fundamental tende a zero assintoticamente enquanto a parte real aproxima-se de $\mu$ (ver a fig. \ref{QNM.D=6.massive-scalar}).}

\lang{This property of the fundamental frequency behavior for large $D$ leads to another remarkable fact. Since for high field mass the imaginary part of the fundamental mode tends to zero only asymptotically (see fig. \ref{QNM.l=0.massive-scalar}), and the imaginary part of higher overtones reaches zero for some finite value of $\mu r_+$ (see fig. \ref{QNM.D=6.massive-scalar}), there are some values of $\mu r_+$ for which the imaginary parts of two overtones are the same. After one of these values is reached the overtones can be distinguished only by their real part and the quasi-normal ringing has two dominant frequencies in its spectrum. Thus one could observe the superposition of that two frequencies at late times of the quasi-normal ringing.}{Essa propriedade do comportamento da freq\"u\^encia fundamental para $D$ grande leva a outro fato not\'avel. Como para a massa do campo alta a parte imagin\'aria do modo fundamental tende a zero s\'o assintoticamente (ver a fig. \ref{QNM.l=0.massive-scalar}), e a parte imagin\'aria de sobretons mais altos alcan\c{c}a zero para algum valor finito de $\mu r_+$ (ver a fig. \ref{QNM.D=6.massive-scalar}), h\'a alguns valores de $\mu r_+$ para os quais as partes imagin\'arias de dois sobretons s\~ao as mesmas. Depois que um desses valores \'e alcan\c{c}ado, os sobretons podem ser distinguidos somente pela sua parte real e o toque quase-normal tem duas freq\"u\^encias dominantes no seu espectro. Assim pode-se observar a superposi\c{c}\~ao daquelas duas freq\"u\^encias em tempos tardios do toque quase-normal.}

\begin{figure}
\includeonegraph{QNM_D=6_massive-scalar}
\langfigurecaption{Fundamental quasi-normal frequencies of the Schwarzschild black hole ($D=6$) for the massive scalar field as function of the field mass $\mu$ for $l=0,1,2$ (represented as dots, rhombuses and triangles respectively). The dashed line corresponds to $\re(\omega)=\mu$.}{Freq\"u\^encias quase-normais fundamentais do buraco negro de Schwarzschild ($D=6$) para o campo escalar massivo como fun\c{c}\~ao da massa do campo $\mu$ para $l=0,1,2$ (representado como pontos, losangos e tri\^angulos respectivamente). A linha tracejada corresponde a $\re(\omega)=\mu$.}\label{QNM.D=6.massive-scalar}
\end{figure}

\langpar{The dependance on the field mass is qualitatively the same for all multipole numbers (see fig. \ref{QNM.D=6.massive-scalar}). Since the real part of the quasi-normal frequency of higher $l$ is larger, the quasi-resonances are reached for larger $\mu$ in order to satisfy the constraint (\ref{quasiresonance-constraint}).}{A depend\^encia na massa \'e qualitativamente a mesma para todos os n\'umeros de multipolo (ver a fig. \ref{QNM.D=6.massive-scalar}). Como a parte real da freq\"u\^encia quase-normal de $l$ mais alto \'e maior, as quase-resson\^ancias s\~ao alcan\c{c}adas para maior $\mu$ para satisfazer o v\'inculo (\ref{quasiresonance-constraint}).}

\begin{figure}
\includeonegraph{QNM_D=4_l=0_massive-scalar_high}
\langfigurecaption{High overtones of the Schwarzschild black hole for the massive scalar field ($D=4$, $l=0$) for $\mu r_+=0.6$ (dots) and $\mu r_+=6.0$ (boxes).}{Sobretons altos do buraco negro de Schwarzschild para o campo escalar massivo ($D=4$, $l=0$) para $\mu r_+=0.6$ (pontos) e $\mu r_+=6.0$ (caixas).}\label{QNM.D=4.l=0.massive-scalar.high}
\end{figure}

\langpar{It was shown both analytically and numerically \cite{Konoplya:2004wg,Zhidenko:2006rs} that asymptotically high overtones do not depend on the field mass and satisfy the same formula as for the massless case}{Foi mostrado tanto analiticamente como numericamente \cite{Konoplya:2004wg,Zhidenko:2006rs} que os sobretons assintoticamente altos n\~ao dependem da massa do campo e satisfazem a mesma f\'ormula como no caso sem massa}
$$\omega_n=\frac{f'(r_+)}{4\pi}\left(\pm\ln(3)-2\pi\imo \left(n+\frac{1}{2}\right)\right)=T_{Hawking}\left(\pm\ln(3)-2\pi\imo \left(n+\frac{1}{2}\right)\right).$$

\langpar{It is important to note, that despite in most cases quasi-normal modes of the fields of different spin show qualitatively the same behavior, when the field is massive, the field spin becomes crucial. The quasi-normal modes of the massive vector field for 4-dimensional black holes were studied in \cite{Konoplya:2005hr}. It was found that the fundamental mode shows correlation with the field mass totally different from all the remaining higher overtones. For massive scalar and massive vector fields the behavior of the fundamental frequency is qualitatively similar: the real part grows with $\mu$, while the absolute value of the imaginary part decreases until reaching zero (quasi-resonance) and then disappears. The higher overtones have their real part decreasing to tiny values, and, the absolute value of the imaginary part is growing with $\mu$, leading to existence of almost pure imaginary modes which just damp without oscillations.}{\'E importante observar que apesar de na maior parte de casos os modos quase-normais dos campos de spin diferente mostrarem qualitativamente o mesmo comportamento, quando o campo \'e massivo o spin do campo torna-se crucial. Os modos quase-normais do campo vetorial massivo para buracos negros 4-dimensionais foram estudados em \cite{Konoplya:2005hr}. Foi achado que o modo fundamental mostra correla\c{c}\~ao com a massa do campo totalmente diferente de todos os sobretons mais altos remanescentes. Para campos escalar massivo e vetorial massivo o comportamento da freq\"u\^encia fundamental \'e qualitativamente semelhante: a parte real cresce com $\mu$, enquanto o valor absoluto da parte imagin\'aria diminui at\'e chegar a zero (quase-resson\^ancia) e depois desaparece. A parte real dos sobretons mais altos, diminui a valores muito pequenos, e o valor absoluto da parte imagin\'aria est\'a crescendo com $\mu$, levando \`a exist\^encia de modos imagin\'arios quase puros que somente decaem sem oscila\c{c}\~oes.}

\langsectionlabel{Stability and quasi-normal modes of the massive scalar field around Kerr black holes}{
Estabilidade e modos quase-normais do campo escalar massivo em torno de buracos negros de Kerr}{sec:massive-Kerr}

\langpar{Analysis of the quasi-normal spectrum allows also to prove stability of massive fields in the Kerr background numerically \cite{Konoplya:2006br}. The stability was proved analytically only for high field mass \cite{Beyer:2000fz}:}{A an\'alise do espectro quase-normal permite tamb\'em comprovar numericamente a estabilidade de campos massivos no contexto de Kerr \cite{Konoplya:2006br}. A estabilidade foi comprovada analiticamente apenas para valores altos da massa do campo \cite{Beyer:2000fz}:}
\begin{equation}\label{stability-bound}
\mu \geq \frac{| m | a}{2 M r_{+}} \sqrt{1 + \frac{2 M}{r_{+}} + \frac{a^{2}}{r_{+}^{2}}}.
\end{equation}
\lang{Strictly speaking, because the azimuthal number appears in the righthand side, for any finite field mass one could find such kind of perturbations, that is not proven to be stable. In practice, the numerical proof is reduced to checking that there is no growing mode in the quasi-normal spectrum for small field mass and high enough value of $m$ \cite{Konoplya:2006br}.}{Em princ\'ipio, como o n\'umero azimutal aparece do lado direito, para qualquer massa de campo finita pode-se encontrar tais tipos de perturba\c{c}\~oes, para as quais, a estabilidade n\~ao foi provada. Na pr\'atica, a prova num\'erica \'e reduzida \`a verifica\c{c}\~ao de que n\~ao h\'a qualquer modo crescente no espectro quase-normal para massa de campo pequena e valor de $m$ alto o suficiente \cite{Konoplya:2006br}.}

\langpar{The Kerr metric is described by (\ref{Kerr-Newman-deSitter-metric}), if we put $Q=0$, $\Lambda=0$. The Klein-Gordon equation for the massive scalar field (\ref{massive-scalar-equation}) in this background, after substituting the ansatz}{A m\'etrica de Kerr \'e descrita por  (\ref{Kerr-Newman-deSitter-metric}), colocando-se $Q=0$, $\Lambda=0$. A equa\c{c}\~ao de Klein-Gordon do campo escalar massivo (\ref{massive-scalar-equation}) neste contexto, depois de substituir o ansatz}
$$\Phi(t,r,\theta,\phi) = e^{-\imo\omega t}e^{\imo m\phi} R(r) S(\theta),$$
\lang{allows to separate variables. The radial part satisfies}{permite separar vari\'aveis. A parte radial satisfaz}
\begin{equation}\label{radial-massive-kerr}
\frac{d}{dr}\left(\Delta_r\frac{dR(r)}{dr}\right)+\left(\frac{K^2}{\Delta_r}-\lambda-\mu^2r^2\right)R(r)=0,
\end{equation}
\lang{where $\lambda$ is the separation constant,}{onde $\lambda$ e a constante de separa\c{c}\~ao,}
\begin{eqnarray}
\Delta_r=r^2+a^2-2Mr, \qquad K= \omega(r^{2} + a^2) - a m.
\end{eqnarray}

\langpar{The angular part}{A parte angular}
\begin{equation}
\frac{1}{\sin\theta}\frac{d}{d\theta}\left(\sin\theta\frac{dS(\theta)}{d\theta}\right)+
\Biggr(-\frac{m^2}{\sin^2\theta}\label{angular-massive-kerr}-a^2\omega^2\sin^2\theta
-a^2\mu^2\cos^2\theta+2am\omega+\lambda\Biggr)S(\theta)=0
\end{equation}
\lang{is equivalent to (\ref{brane-localised-angular}) up to the redefinition of variables and can be solved as described in the section \ref{sec:Frobenius-angular}.}{\'e equivalente a (\ref{brane-localised-angular}) at\'e a redefini\c{c}\~ao de vari\'aveis e pode ser resolvida como descrito na se\c{c}\~ao \ref{sec:Frobenius-angular}.}

\langpar{The appropriate Frobenius series for (\ref{radial-massive-kerr}) are given by}{A s\'erie de Frobenius apropriada para (\ref{radial-massive-kerr}) \'e dada por}
\begin{equation}\label{massive-Kerr-Frobenius-series}
R(r)=\exp(\imo\chi r)\left(\frac{r}{r_+}-\frac{a^2}{r_+^2}\right)^{\imo\sigma-1}\left(\frac{rr_+-r_+^2}{rr_+-a^2}\right)^{-\imo\alpha}\sum_{k=0}^\infty a_k\left(\frac{rr_+-r_+^2}{rr_+-a^2}\right)^k,
\end{equation}
\lang{where $\chi=\sqrt{\omega^2-\mu^2}$.}{onde $\chi=\sqrt{\omega^2-\mu^2}$.}

\langpar{The coefficients $a_k$ of (\ref{massive-Kerr-Frobenius-series}) satisfy the three-term recurrence relation (\ref{full-three-term}), which allows to find the spectrum numerically and prove, that it does not contain unstable modes \cite{Konoplya:2006br}.}{Os coeficientes $a_k$ de (\ref{massive-Kerr-Frobenius-series}) satisfazem a rela\c{c}\~ao de recorr\^encia de tr\^es termos (\ref{full-three-term}), que permite achar o espectro numericamente e comprovar que ele n\~ao cont\'em modos inst\'aveis \cite{Konoplya:2006br}.}

\begin{figure}
\includeonegraph{QNM_massiveKerr_Re}\\\includeonegraph{QNM_massiveKerr_Im}
\langfigurecaption{The fundamental quasi-normal frequency of Kerr black holes for the massive scalar field as a function of azimuthal number $m$ for $l=6$, $a=0.15r_+$, $\mu r_+=0.2$.}{A freq\"u\^encia quase-normal fundamental de buracos negros de Kerr para o campo escalar massivo como uma fun\c{c}\~ao do n\'umero azimutal $m$ para $l=6$, $a=0.15r_+$, $\mu r_+=0.2$.}\label{QNM.massiveKerr}
\end{figure}

\langpar{We can see that despite the lifetime of the quasi-normal oscillation increases monotonously as the azimuthal number $m$ grows, its imaginary part remains bounded (see fig. \ref{QNM.massiveKerr}) and, thereby, does not show any tendency to instability. Therefore, we conclude that the formula (\ref{stability-bound}) implies stability for the large field mass.}{Pode-se ver que apesar da vida da oscila\c{c}\~ao quase-normal aumentar monotonicamente com o n\'umero azimutal $m$, a sua parte imagin\'aria permanece limitada (ver a fig. \ref{QNM.massiveKerr}) e, por meio disso, n\~ao mostra qualquer tend\^encia \`a instabilidade. Por isso, conclui-se que a f\'ormula (\ref{stability-bound}) implica em estabilidade para massas grandes.}

\langpar{The dependance of the quasi-normal frequency on the field mass $\mu$ is qualitatively the same as for the Schwarzschild case: larger $\mu$ leads to considerable decreasing of the damping rate and finally to the quasi-resonances (see sec. \ref{sec:quasi-resonance}).}{A depend\^encia da freq\"u\^encia quase-normal na massa $\mu$ do campo \'e qualitativamente a mesma para o caso de Schwarzschild: maior $\mu$ leva \`a redu\c{c}\~ao consider\'avel da taxa de decaimento e, finalmente, \`as quase-resson\^ancias (ver a se\c{c}\~ao \ref{sec:quasi-resonance}).}

\langpar{The dependance on the rotation parameter $a$ is more complicated. For $m=0$ the imaginary part of quasi-normal frequencies decreases its absolute value as $a$ grows, while the real part increases, reaches maximum and then decreases.}{A depend\^encia do par\^ametro de rota\c{c}\~ao $a$ \'e mais complicada. Para $m=0$ a parte imagin\'aria de freq\"u\^encias quase-normais reduz o seu valor absoluto j\'a que $a$ cresce enquanto a parte real aumenta, alcan\c{c}a o m\'aximo e depois diminui.}

\langsectionlabel{Quasi-normal modes of black strings and the Gregory-Laflamme instability}{Modos quase-normais de cordas negras e a instabilidade de Gregory-Laflamme}{sec:Gregory-Laflamme}

\langpar{Unlike four dimensional Einstein gravity, which allows existence of black holes, higher dimensional theories, such as the brane-world scenarios and string theory, allow existence of a number of ``black'' objects: higher dimensional black holes, black strings and branes, black rings and saturns and others. In higher than four dimensions we lack the uniqueness theorem, so that stability may be the criteria which will select physical solutions among this variety of solutions.}{Diferentemente da gravidade de Einstein de quatro dimens\~oes, que permite a exist\^encia de buracos negros, as teorias em dimens\~oes mais altas, como os cen\'arios de mundo na brana e a teoria das cordas, permitem a exist\^encia de um n\'umero de objetos ``negros'': buracos negros em dimens\~oes mais altas, cordas e branas negras, an\'eis e saturnos negros entre outros. Em mais do que quatro dimens\~oes, falta o teorema de unicidade, ent\~ao a estabilidade pode ser o crit\'erio que selecionar\'a solu\c{c}\~oes f\'isicas entre esta variedade de solu\c{c}\~oes.}

\langpar{According to the brane-world scenarios, if the matter localised on the brane undergoes gravitational collapse, a black hole with the horizon extended to the transverse extra direction will form. This object looks like a black hole on the brane, but is, in fact, a black string in the full $D$-dimensional theory. Such black strings suffer from the so-called Gregory-Laflamme instability \cite{Gregory:1993vy,Gregory:1994bj}, which is the long-wavelength gravitational instability of the scalar type of the metric perturbations. The threshold values of the wave vector $k$ at which the instability appears were found in \cite{Hovdebo:2006jy}.}{Segundo os cen\'arios de mundo na brana, se a mat\'eria localizada na brana sofrer um colapso gravitacional, um buraco negro com o horizonte extendido \`a dire\c{c}\~ao transversal extra se formar\'a. Este objeto parece um buraco negro na brana, mas \'e, de fato, uma corda negra na teoria completa $D$-dimensional. Tais cordas negras sofrem da chamada instabilidade de Gregory-Laflamme \cite{Gregory:1993vy,Gregory:1994bj}, que \'e a instabilidade gravitacional das perturba\c{c}\~oes m\'etricas do tipo escalar de ondas de comprimento longo. Os valores de limiar do vetor da onda $k$ no qual a instabilidade aparece foram achados em \cite{Hovdebo:2006jy}.}

\langpar{The evolution of the spherically symmetric linear perturbations of $D$-dimensional black strings in time and frequency domains was studied in \cite{Konoplya:2008yy}. For the first time the quasi-normal modes and time-domain profiles were studied in the stable sector. Also the appearance of the Gregory-Laflamme instability was shown in the time domain.}{A evolu\c{c}\~ao das perturba\c{c}\~oes lineares esfericamente sim\'etricas de cordas negras $D$-dimensionais em dom\'inio de tempo e de freq\"u\^encia foi estudada em \cite{Konoplya:2008yy}. Pela primeira vez os modos quase-normais e os perfis temporais foram estudados no setor est\'avel. Tamb\'em a apar\^encia da instabilidade de Gregory-Laflamme foi mostrada no dom\'inio de tempo.}

\langpar{Let us note, that the equations for gravitational perturbations contain the Kaluza-Klein momentum. The contribution of this momentum in the corresponding effective potential looks like the mass term.}{Observemos, que as equa\c{c}\~oes para perturba\c{c}\~oes gravitacionais cont\'em o momento de Kaluza-Klein. A contribui\c{c}\~ao deste momento no correspondente potencial efetivo se parece com o termo de massa.}

\langpar{For the static black string in $D\geq5$ space-time dimensions, the background metric can be written as}{Para a corda negra est\'atica em $D\geq5$ dimens\~oes de espa\c{c}o-tempo, o contexto m\'etrico pode ser descrito como}
\begin{equation}
ds^2=g_{\mu\nu}dx^\mu dx^\nu=f(r)dt^2-\frac{dr^2}{f(r)}-r^2d\Omega^2_{D-3}-dz^2,
\end{equation}
\lang{where}{onde}
$$f(r)=1-\left(\frac{r_+}{r}\right)^n,\qquad n=D-4.$$
\lang{and $d\Omega^2_{D-3}$ is the line element on a unit $(D-3)$-sphere.}{e $d\Omega^2_{D-3}$ \'e o elemento de linha em uma $(D-3)$-esfera unit\'aria.}

\langpar{The $z$-direction is periodically identified by the relation $z=z+2\pi R$. Let us study the $(D-3)$-spherically symmetric perturbations, which we can write in the following form}{A dire\c{c}\~ao $z$ \'e periodicamente identificada pela rela\c{c}\~ao $z=z+2\pi R$. Ser\~ao estudadas as perturba\c{c}\~oes $(D-3)$-esfericamente sim\'etricas, que podem-se escritas da forma}
$$\delta g_{\mu\nu}=e^{\imo kz}a_{\mu\nu}(t,r), \qquad k=\frac{m}{R}, \quad m\in\mathbb{Z}.$$
\lang{The perturbed vacuum Einstein equations have the form}{As equa\c{c}\~oes de Einstein perturbadas no v\'acuo t\^em a forma}
\begin{equation}
\delta R_{\mu \nu} = 0\,.
\end{equation}
\lang{The perturbations can be reduced to the form, where the only non-vanishing components of $a_{\mu\nu}$ are}{As perturba\c{c}\~oes podem ser reduzidas \`a forma, em que os \'unicos componentes n\~ao-nulos de $a_{\mu\nu}$, s\~ao}
$$a_{tt}=h_t, \quad a_{rr}=h_r, \quad a_{zz}=h_z, \quad a_{tr}=\dot{h}_v, \quad a_{zr}=-\imo k h_v.$$

\langpar{The linearised Einstein equations give a set of coupled equations determining the four radial profiles above.
However, we may eliminate $h_v$, $h_r$ and $h_t$ from these equations in order to produce a single second order equation for $h_z$:}{As equa\c{c}\~oes de Einstein linearizadas levam a um conjunto de equa\c{c}\~oes ligadas que determinam os quatro perfis radiais acima.
Contudo, pode-se eliminar $h_v$, $h_r$ e $h_t$ dessas equa\c{c}\~oes para produzir uma \'unica equa\c{c}\~ao de segunda ordem para $h_z$:}
\begin{equation}\label{HovdeboMyers}
\ddot{h}_z=f(r)^2h_z''+p(r)h_z'+q(r)h_z,
\end{equation}
\lang{where}{onde}
\begin{eqnarray}\nonumber
p(r)&=&\frac{f(r)^2}{r}\left(1+\frac{n}{f(r)}-\frac{4(2+n)k^2r^2}{2k^2r^2+n(n+1)(r_+/r)^n}\right),\\\nonumber
q(r)&=&-k^2f(r)\frac{2k^2r^2-n(n+3)(r_+/r)^n}{2k^2r^2+n(n+1)(r_+/r)^n}.
\end{eqnarray}

\langpar{Defining}{Definindo}
$$h_z(t,r)=\frac{r^{-(n-1)/2}}{2k^2r^2+n(n+1)(r_+/r)^n}\Psi(t,r),$$
\lang{we can reduce the equation (\ref{HovdeboMyers}) to the wave-like equation}{pode-se reduzir a equa\c{c}\~ao (\ref{HovdeboMyers}) \`a equa\c{c}\~ao de onda}
\begin{equation}\label{wavelike}
\left(\frac{\partial^2}{\partial t^2}-\frac{\partial^2}{\partial r_\star^2} + V(r)\right)\Psi=0,
\end{equation}
\lang{where $\displaystyle dr_\star=\frac{dr}{f(r)}$ is the tortoise coordinate. Here, the effective potential $V(r)$ is given by}{onde $\displaystyle dr_\star=\frac{dr}{f(r)}$ \'e a coordenada tartaruga. Aqui, o potencial efetivo $V(r)$ \'e dado por}
$$V(r)=\frac{f(r)}{4 r^2}\frac{U(r)}{\left(2k^2r^2+n(n+1)(r_+/r)^n\right)^2} \ , $$
\lang{where}{onde}
\begin{eqnarray}\nonumber
U(r)&=&16 k^6 r^6+ 4 k^4 r^4 (n+5) (3f(r)-2 n + 3n f(r))-\\\nonumber&&-4k^2r^2n(n+1)\left(n(n+5)+f(r)(2n^2+7n+9)\right)\left(\frac{r_+}{r}\right)^n-
\\\nonumber&&-n^2(n+1)^3\left(f(r)-2n+nf(r)\right)\left(\frac{r_+}{r}\right)^{2n}.
\end{eqnarray}

\begin{figure}
\includeonegraph{blackstring_k=2_5_tails}
\langfigurecaption{Time-domain profiles of black string perturbations for $kr_+ = 2.5$: $n = 2$ (red, top), $n = 3$ (orange), $n = 4$ (green),
$n = 5$ (blue, bottom). Late-time decay of perturbations for $n \geq 3$ is $\Psi\propto t^{-(n+6)/2}\sin(kt)$.}{Perfis temporais das perturba\c{c}\~oes de cordas negras para $kr_+ = 2.5$: $n = 2$ (vermelho, topo), $n = 3$ (laranja), $n = 4$ (verde),
$n = 5$ (azul, abaixo). O decaimento de tempo tardio para $n \geq 3$ \'e $\Psi\propto t^{-(n+6)/2}\sin(kt)$.}\label{blackstring.k=2.5.tails}
\end{figure}

\langpar{One can see that $k$ plays the role of the effective mass. At asymptotically late time we observe power-law damped tails, which have the oscillation frequency equal to $k$ (see fig. \ref{blackstring.k=2.5.tails}), resembling asymptotical behavior of massive fields near Schwarzschild black holes (see sec. \ref{sec:late-time-tails}). The behavior of the first overtone is qualitatively similar to that of the fundamental mode for massive fields of higher-dimensional Schwarzschild black holes (see sec. \ref{sec:quasi-resonance}).}{Pode-se ver que $k$ desempenha o papel da massa efetiva. No tempo assintoticamente tardio observam-se caudas que decaem segundo a lei de pot\^encia e t\^em a freq\"u\^encia de oscila\c{c}\~ao igual a $k$ (ver a fig. \ref{blackstring.k=2.5.tails}), parecendo-se com o comportamento assint\'otico de campos massivos perto de buracos negros de Schwarzschild (ver a se\c{c}\~ao \ref{sec:late-time-tails}). O comportamento do primeiro sobretom \'e qualitativamente semelhante \`aquele do modo fundamental de campos massivos de buracos negros de Schwarzschild em dimens\~oes mais altas (ver a se\c{c}\~ao \ref{sec:quasi-resonance}).}

\langpar{$\bullet$ For $D=5$ ($n=1$), as $k$ grows, the imaginary part of the first overtone quickly decreases and vanishes for some threshold value of $k$, while its real part stays smaller than the threshold value. After the threshold value of $k$ is reached, the first overtone ``disappears''.}{$\bullet$ Para $D=5$ ($n=1$), conforme $k$ cresce, a parte imagin\'aria do primeiro sobretom diminui rapidamente e desaparece para algum valor de limiar de $k$, enquanto a sua parte real fica menor do que o valor de limiar. Depois que o valor de limiar de $k$ for alcan\c{c}ado, o primeiro sobretom ``desaparece''.}

\langpar{$\bullet$ For $D\geq6$ ($n\geq2$), the imaginary part of the first overtone becomes small for large $k$, while the real part asymptotically approaches $k$.}{$\bullet$ Para $D\geq6$ ($n\geq2$), a parte imagin\'aria do primeiro sobretom fica pequena para $k$ grande, enquanto a parte real se aproxima de $k$ assintoticamente.}

\begin{figure}
\includeonegraph{blackstring_n=1_instability}
\langfigurecaption{Time-domain profiles of black string perturbations for $n = 1$ $kr_+ = 0.84$ (magenta, top), $kr_+ = 0.87$ (red), $kr_+ = 0.88$ (orange), $kr_+ = 0.9$ (green), $kr_+ = 1.1$ (blue, bottom). We can see two concurrent modes: for large $k$ the oscillating one
dominates, near the critical value of $k$ the dominant mode does not oscillate, for unstable values of $k$ the dominant mode grows. The plot is logarithmic, so that straight lines correspond to an exponential growth or decay.}{Perfis temporais de perturba\c{c}\~oes de cordas negras para $n = 1$ $kr_+ = 0.84$ (magenta, topo), $kr_+ = 0.87$ (vermelha), $kr_+ = 0.88$ (laranja), $kr_+ = 0.9$ (verde), $kr_+ = 1.1$ (azul, abaixo). Pode-se ver dois modos concomitantes: para $k$ grande o modo oscilante domina, perto do valor cr\'itico de $k$ o modo dominante n\~ao oscila, para valores inst\'aveis de $k$ o modo dominante cresce. A escala \'e logar\'itmica, para que as linhas retas correspondam a um crescimento ou decrescimento exponencial.}\label{blackstring.n=1.instability}
\end{figure}

\langpar{Even though the first overtone of the spherically symmetric black strings behaves similarly to the fundamental mode of massive fields near higher-dimensional Schwarzschild black holes, the other modes have completely different behavior. The fundamental mode of black string perturbations is purely imaginary. It grows for small values of $k$, leading to instability of the black string (see fig. \ref{blackstring.n=1.instability}). At moderately large values of $k$, sufficiently far from instability, the profile
has the same form as that for massive fields, yet, when approaching the instability point, the real oscillation frequency and the decay rate decrease considerably. After crossing the instability point we observe, that starting from some tiny values, $\im(\omega)>0$ are slowly increasing (while $\re(\omega)=0$). Therefore we can conclude, that there is some static solution $\omega=0$ of the wave equation, which shows itself exactly in the threshold point of instability.}{Embora o primeiro sobretom das cordas negras esfericamente sim\'etricas se comporte de mesmo modo que o modo fundamental de campos massivos perto de buracos negros de Schwarzschild em dimens\~oes mais altas, os outros modos t\^em um comportamento completamente diferente. O modo fundamental de perturba\c{c}\~oes da corda negra \'e puramente imagin\'ario. Ele cresce para valores de $k$ pequenos, levando \`a instabilidade da corda negra (ver o fig. \ref{blackstring.n=1.instability}). Para valores de $k$ moderadamente grandes, suficientemente longe da instabilidade, o perfil tem a mesma forma que o perfil para campos massivos, mas, quando aproximando o ponto da instabilidade, a freq\"u\^encia real da oscila\c{c}\~ao e a taxa de decaimento se reduzem consideravelmente. Depois de cruzar o ponto da instabilidade observa-se que come\c{c}ando de valores muito pequenos, $\im(\omega)>0$ est\~ao aumentando lentamente (enquanto $\re(\omega)=0$). Por isso, pode-se concluir, que h\'a alguma solu\c{c}\~ao est\'atica $\omega=0$ da equa\c{c}\~ao de onda, que se mostra exatamente no ponto de limiar da instabilidade.}

\langsectionlabel{Late-time tails of massive fields}{Caudas de tempo tardio de campos massivos}{sec:late-time-tails}
\langpar{The late-time behavior of black hole perturbations was studied for the first time by R. Price, who showed that perturbations of the massless scalar and gravitational fields decay as $\propto t^{-(2l+3)}$ at asymptotically late time \cite{Price:1971fb,Price:1972pw}. In \cite{Bicak:1972} Bi\v{c}\'ak found that the scalar massless field in the Reissner-Nordstr\"om background decays as $\propto t^{-(2l+2)}$ for $|Q|<M$ and as $\propto t^{-(l+2)}$ for the extremal black hole charge $|Q|=M$. For Schwarzschild-de Sitter and Reissner-Nordstr\"om-de Sitter black holes, instead of power-law tails, the exponential tails were found \cite{Brady:1996za}.}{O comportamento de tempo tardio de perturba\c{c}\~oes de buracos negros foi estudado pela primeira vez por R. Price, quem mostrou que as perturba\c{c}\~oes gravitacionais e do campo escalar sem massa decaem como  $\propto t^{-(2l+3)}$ no tempo assintoticamente tardio \cite{Price:1971fb,Price:1972pw}. Em \cite{Bicak:1972} Bi\v{c}\'ak achou que o campo escalar sem massa no contexto de Reissner-Nordstr\"om decai como $\propto t^{-(2l+2)}$ para $|Q|<M$ e como $\propto t^{-(l+2)}$ para carga extremada do buraco negro $|Q|=M$. Para buracos negros de Schwarzschild-de Sitter e Reissner-Nordstr\"om-de Sitter, ao inv\'es das caudas da lei de pot\^encia, as caudas exponenciais foram encontradas \cite{Brady:1996za}.}

\langpar{In higher dimensional space-times the late-time behavior depends also on the number of extra dimensions. It was found that massless scalar and vector fields, and gravitational perturbations of higher-dimensional Schwarzschild black holes have the decay law $\propto t^{-(2l+D-2)}$ for odd $D$ \cite{Cardoso:2003jf} and $\propto t^{-(2l+3D-8)}$ for even $D>4$ \cite{Bizon:2007vr}. The same late-time behavior was observed for odd-dimensional Gauss-Bonnet black holes \cite{Abdalla:2005hu}. The late-time tails of the brane localised Standard Model fields were studied in \cite{Zhidenko:2008fp}. Their late-time decay law is $\propto t^{-(2l+D-1)}$.}{Em espa\c{c}os-tempo de dimens\~oes mais altas o comportamento no tempo tardio depende tamb\'em do n\'umero de dimens\~oes extra. Foi achado que campos escalar e vetorial sem massa, e perturba\c{c}\~oes gravitacionais de buracos negros de Schwarzschild em dimens\~oes mais altas t\^em a lei de decaimento $\propto t^{-(2l+D-2)}$ para $D$ \'impares \cite{Cardoso:2003jf} e $\propto t^{-(2l+3D-8)}$ para $D> 4$ pares \cite{Bizon:2007vr}. O mesmo comportamento no tempo tardio foi observado para os buracos negros de Gauss-Bonnet em dimens\~oes \'impares \cite{Abdalla:2005hu}. As caudas de tempo tardio dos campos do Modelo Padr\~ao localizados na brana foram estudadas em \cite{Zhidenko:2008fp}. A sua lei de decaimento no tempo tardio \'e $\propto t^{-(2l+D-1)}$.}

\langpar{One should note that the late-time decay of perturbations within the full non-linear gravity does not agree with the linearised theory in dimensions higher than four. If we consider a massless scalar field and take into account the back reaction of the field upon the metric, the late-time decay rate becomes smaller, coinciding with the linearised theory prediction only in four dimensions \cite{Bizon:2008iz}.}{Observemos que o decaimento no tempo tardio de perturba\c{c}\~oes na gravidade completamente n\~ao linear n\~ao concorda com a teoria linear em n\'umero de dimens\~oes maior do que quatro. Considerando um campo escalar sem massa e levando em conta a rea\c{c}\~ao de volta do campo sobre a m\'etrica, pode-se achar que a taxa de decaimento de tempo tardio \'e menor, coincidindo com a previs\~ao da teoria linear somente em quatro dimens\~oes \cite{Bizon:2008iz}.}

\langpar{The late-time behavior of massive fields is qualitatively different from massless ones: at late time the decay profile is \emph{oscillatory} inverse power law tail. Also, the field mass implies different behavior at the intermediate late time ($1\ll t/M<(\mu M)^{-3}$) and at the asymptotically late time ($t/M>(\mu M)^{-3}$).}{O comportamento de campos massivos no tempo tardio \'e qualitativamente diferente do que os sem massa: no tempo tardio o perfil de decaimento \'e a cauda da lei de pot\^encia inversa \emph{oscilat\'oria}. Tamb\'em, a massa do campo implica um comportamento diferente no tempo tardio intermedi\'ario ($1\ll t/M<(\mu M)^{-3}$) e no tempo assintoticamente tardio ($t/M>(\mu M)^{-3}$).}

\langpar{For a massive scalar field with mass $\mu$ in the background of the Schwarzschild black hole, the perturbations decay as $\propto t^{-(l+3/2)}\sin(\mu t)$ at intermediate late time \cite{Hod:1998ra} and as $\propto t^{-(5/6)}\sin(\mu t)$ at asymptotically late time \cite{Koyama:2001ee,Koyama:2001qw}. The same behavior at asymptotically late time was found also for the massive scalar field perturbations of the dilaton black hole \cite{Moderski:2001tk} and the Kerr black hole \cite{Burko:2004jn} and for the massive Dirac field in the Schwarzschild background \cite{Jing:2004zb}.}{Para um campo escalar massivo com a massa $\mu$ no contexto do buraco negro de Schwarzschild, as perturba\c{c}\~oes decaem como $\propto t^{-(l+3/2)}\sin(\mu t)$ no tempo tardio intermedi\'ario \cite{Hod:1998ra} e como $\propto t^{-(5/6)}\sin(\mu t)$ no tempo assintoticamente tardio \cite{Koyama:2001ee,Koyama:2001qw}. O mesmo comportamento no tempo assintoticamente tardio foi encontrado tamb\'em para as perturba\c{c}\~oes de campos escalares massivos do buraco negro dilat\^onico \cite{Moderski:2001tk} e do buraco negro de Kerr \cite{Burko:2004jn} e para o campo de Dirac massivo no contexto de Schwarzschild \cite {Jing:2004zb}.}

\langpar{For higher dimensional Schwarzschild black holes the intermediate late-time behavior of the massive scalar field is found to be $\propto t^{-(l+(D-1)/2)}\sin(\mu t)$ \cite{Moderski:2005hf}.}{Para buracos negros de Schwarzschild em dimens\~oes mais altas o comportamento no tempo tardio intermedi\'ario do campo escalar massivo foi achado, como sendo $\propto t^{-(l+(D-1)/2)}\sin(\mu t)$ \cite{Moderski:2005hf}.}

\langpar{The late-time behavior of the massive vector field was studied both numerically and analytically for the first time in \cite{Konoplya:2006gq}. At intermediate late time the decay law depends on the polarisation, being either $\propto t^{-(l+1/2)}\sin(\mu t)$, or $\propto t^{-(l+3/2)}\sin(\mu t)$, or $\propto t^{-(l+5/2)}\sin(\mu t)$. At the asymptotical late time the behavior is the same, as for other massive fields, $\propto t^{-(5/6)}\sin(\mu t)$. Therefore, we conclude that the asymptotically late-time decay law \emph{does not depend} on the spin of the field.}{O comportamento no tempo tardio do campo vetorial massivo foi estudado tanto numericamente como analiticamente pela primeira vez em \cite{Konoplya:2006gq}. No tempo tardio intermedi\'ario a lei de decaimento depende da polariza\c{c}\~ao, sendo ou $\propto t^{-(l+1/2)}\sin(\mu t)$, ou $\propto t^{-(l+3/2)}\sin(\mu t)$, ou $\propto t^{-(l+5/2)}\sin(\mu t)$. No tempo assintoticamente tardio o comportamento \'e o mesmo, como para outros campos massivos, $\propto t^{-(5/6)}\sin(\mu t)$. Por isso, conclui-se que a lei de decaimento no tempo assintoticamente tardio \emph{n\~ao depende} do spin do campo.}

\langpar{The late-time decay law for massless perturbations, the intermediate and asymptotically late-time behavior of massive field perturbations for various spins in asymptotically flat backgrounds are presented in the following table.}{A lei de decaimento no tempo tardio de perturba\c{c}\~oes sem massa, o comportamento no tempo tardio intermedi\'ario e assintoticamente tardio de perturba\c{c}\~oes de campos massivos para v\'arios spins em contextos assintoticamente planos \'e apresentada na tabela seguinte.}

\langlong{\begin{center}
\begin{tabular}{|l|c|c|c|}
\hline
perturbations&massless&massive (intermediate)&massive (asymptot.)\\
\hline
$s=0$ ($4D$ Schw.)&$t^{-(2l+3)}$&$t^{-(l+3/2)}\sin(\mu t)$&$t^{-(5/6)}\sin(\mu t)$\\
$s=1$ ($4D$ Schw.)&$t^{-(2l+3)}$&$\{t^{-(l+1/2)}, t^{-(l+3/2)}, t^{-(l+5/2)}\}\sin(\mu t)$&$t^{-(5/6)}\sin(\mu t)$\\
$s=2$ ($4D$ Schw.)&$t^{-(2l+3)}$&&\\
$s=0$ ($4D$ Kerr)&$t^{-(2l+3)}$&$t^{-(l+3/2)}\sin(\mu t)$&$t^{-(5/6)}\sin(\mu t)$\\
$s=0$ ($4D$ R-N)&$t^{-(2l+2)}$&&\\
\hline
$s=0$ (odd $D$)&$t^{-(2l+D-2)}$&$t^{-(l+(D-1)/2)}\sin(\mu t)$&\\
$s=0$ (even $D$)&$t^{-(2l+3D-8)}$&$t^{-(l+(D-1)/2)}\sin(\mu t)$&\\
on a $4$-brane&$t^{-(2l+D-1)}$&&\\
\hline
\end{tabular}
\end{center}
}{\begin{center}
\begin{tabular}{|l|c|c|c|}
\hline
perturba\c{c}\~oes&sem massa&massiva (intermedi\'ario)&massiva (assintot.)\\
\hline
$s=0$ ($4D$ Schw.)&$t^{-(2l+3)}$&$t^{-(l+3/2)}\sin(\mu t)$&$t^{-(5/6)}\sin(\mu t)$\\
$s=1$ ($4D$ Schw.)&$t^{-(2l+3)}$&$\{t^{-(l+1/2)}, t^{-(l+3/2)}, t^{-(l+5/2)}\}\sin(\mu t)$&$t^{-(5/6)}\sin(\mu t)$\\
$s=2$ ($4D$ Schw.)&$t^{-(2l+3)}$&&\\
$s=0$ ($4D$ Kerr)&$t^{-(2l+3)}$&$t^{-(l+3/2)}\sin(\mu t)$&$t^{-(5/6)}\sin(\mu t)$\\
$s=0$ ($4D$ R-N)&$t^{-(2l+2)}$&&\\
\hline
$s=0$ ($D$ \'impares)&$t^{-(2l+D-2)}$&$t^{-(l+(D-1)/2)}\sin(\mu t)$&\\
$s=0$ ($D$ pares)&$t^{-(2l+3D-8)}$&$t^{-(l+(D-1)/2)}\sin(\mu t)$&\\
na $4$-brana&$t^{-(2l+D-1)}$&&\\
\hline
\end{tabular}
\end{center}}

\langchapterlabel{Quasi-normal modes of black holes, whose metrics are unknown analytically}{Modos quase-normais de buracos negros cujas m\'etricas s\~ao desconhecidas analiticamente}{sec:non-analytical}
\langsectionlabel{Numerical methods}{M\'etodos num\'ericos}{sec:non-analytical-method}
\langpar{Black holes are compact objects. Therefore, it is natural to expect that their quasi-normal ringing, being the property of the black hole, at least in the dominant order, does not depend on what happens at large distance from the black hole.}{Buracos negros s\~ao objetos compactos. Por isso, \'e natural esperar que o seu toque quase-normal, sendo a propriedade do buraco negro, pelo menos na ordem dominante, n\~ao depende do que acontece a grande dist\^ancia do buraco negro.}

\langpar{It turns out that the dominating frequencies depend mostly on the black hole solution behavior in some region near the event horizon. Thus one can find them in frequency domain, even if the behavior of the solution at large distance is not known.}{Acontece que as freq\"u\^encias dominantes dependem, na maior parte, do comportamento da solu\c{c}\~ao do buraco negro em uma regi\~ao perto do horizonte de eventos. Assim pode-se ach\'a-los no dom\'inio de freq\"u\^encia, mesmo se o comportamento da solu\c{c}\~ao a grande dist\^ancia n\~ao for conhecido.}

\langpar{Two qualitatively different examples were considered in this context: the scalar hairy black hole in the anti-de Sitter background and the Einstein-Aether black hole in the asymptotically flat background. In both types of the background the behavior of the solution at large distance is not important \cite{Zhidenko:2007sj}:}{Dois exemplos qualitativamente diferentes foram considerados neste contexto: o buraco negro com cabelo escalar no contexto  anti-de Sitter e o buraco negro Einstein-Aether no contexto assintoticamente plano. Em ambos os tipos de contexto o comportamento da solu\c{c}\~ao a grandes dist\^ancias n\~ao \'e importante \cite{Zhidenko:2007sj}:}

\langpar{$\bullet$ For an asymptotically anti-de Sitter background we usually require Dirichlet boundary conditions at spatial infinity. The most significant part of the metric perturbations stays, thereby, near the black hole. That is why the solution behavior at this region causes dominant influence on the quasi-normal spectrum \cite{Zhidenko:2005mv}.}{$\bullet$ Para um contexto assintoticamente anti-de Sitter normalmente as condi\c{c}\~oes de contorno de Dirichlet s\~ao impostas no infinito espacial. A parte mais significante das perturba\c{c}\~oes m\'etricas fica, por meio disso, perto do buraco negro. Por isso, o comportamento da solu\c{c}\~ao nesta regi\~ao causa uma influ\^encia dominante no espectro quase-normal \cite{Zhidenko:2005mv}.}

\langpar{$\bullet$ For the asymptotically flat case the searching of the quasi-normal modes can be reduced to the scattering problem. Therefore, the quasi-normal frequencies are determined mainly by the form of the effective potential near its peak \cite{Konoplya:2006rv,Konoplya:2006ar}.}{$\bullet$ Para o caso assintoticamente plano a procura dos modos quase-normais pode ser reduzida ao problema de espalhamento. Por isso, as freq\"u\^encias quase-normais s\~ao determinadas principalmente pela forma do potencial efetivo perto de seu pico \cite{Konoplya:2006rv,Konoplya:2006ar}.}

\langpar{In order to calculate quasi-normal frequencies for the asymptotically anti-de Sitter background one can use the Horowitz-Hubeny method, described in the section \ref{sec:Horowitz-Hubeny}. It is clear that if $s(z)$, $t(z)$ and $u(z)$ in (\ref{HH-radial}) are series which converge at the spatial infinity ($z=1$) quickly enough, then $y_n$ is still possible to calculate because it depends insignificantly on higher terms of the series. In order to find series expansion for $s(z)$, $t(z)$ and $u(z)$, one can use equations which define the black hole solution. We can always do this because $s(z)$, $t(z)$ and $u(z)$ can be explicitly expressed in terms of the metric coefficients and their derivatives.}{Para calcular freq\"u\^encias quase-normais para o contexto assintoticamente anti-de Sitter pode-se usar o m\'etodo de Horowitz-Hubeny, descrito na se\c{c}\~ao \ref{sec:Horowitz-Hubeny}. \'E claro que se $s(z)$, $t(z)$ e $u(z)$ em (\ref{HH-radial}) forem s\'eries que convergem no infinito espacial ($z=1$) rapido o suficiente, ent\~ao ainda \'e poss\'ivel calcular $y_n$ porque ele depende insignificantemente dos termos mais altos das s\'eries. Para achar as expans\~oes para as s\'eries $s(z)$, $t(z)$ e $u(z)$, pode-se utilizar equa\c{c}\~oes que definem a solu\c{c}\~ao do buraco negro. \'E poss\'ivel fazer isto sempre porque $s(z)$, $t(z)$ e $u(z)$ podem ser expressados explicitamente em termos dos coeficientes m\'etricos e de suas derivadas.}

\langpar{Thus, the quasi-normal modes could be found without solving the equations for the black hole. It is enough to find series expansions for the metric coefficients near the horizon. It is important to note that one can control the precision of the eigenfrequencies $\omega$ by requiring the convergence of the found result with respect to increasing of the number of expansion terms of all the series \cite{Zhidenko:2005mv}.}{Assim, os modos quase-normais podem ser achados sem resolver as equa\c{c}\~oes do buraco negro. \'E suficiente encontrar expans\~oes de s\'erie para os coeficientes m\'etricos perto do horizonte. \'E importante observar que se pode controlar a precis\~ao das freq\"u\^encias pr\'oprias $\omega$ impondo converg\^encia do resultado achado com rela\c{c}\~ao ao aumento do n\'umero dos termos de expans\~ao para todas as s\'eries \cite{Zhidenko:2005mv}.}

\langpar{For asymptotically flat black hole solutions it is convenient to use the WKB method (see sec. \ref{sec:WKB}), where the asymptotic solutions of the wave equation near the event horizon and near spatial infinity are matched with Taylor expansion near the peak of the potential. We conclude therefore that the low-damping quasi-normal modes are determined mainly by the behavior of the effective potential near its peak.}{Para solu\c{c}\~oes de buracos negros assintoticamente planos \'e conveniente utilizar o m\'etodo de WKB (ver a se\c{c}\~ao \ref{sec:WKB}), onde as solu\c{c}\~oes assint\'oticas da equa\c{c}\~ao de onda perto do horizonte de eventos e perto do infinito espacial s\~ao combinadas com a expans\~ao de Taylor perto do pico do potencial. Foi concluido, por isso, que os modos quase-normais lento-decrescentes s\~ao determinados principalmente pelo comportamento do potencial efetivo perto do seu pico.}

\begin{figure}
\includeonegraph{spot}
\langfigurecaption{The effective potential for electromagnetic perturbations near the Schwarzschild black hole ($r_+ = 1$, $l = 2$) and the same potential interpolated numerically near its maximum.}{O potencial efetivo para perturba\c{c}\~oes eletromagn\'eticas perto do buraco negro de Schwarzschild ($r_+ = 1$, $l = 2$) e o mesmo potencial interpolado numericamente perto do seu m\'aximo.}\label{spot}
\end{figure}

\langpar{This statement was checked by considering the potential for the Schwarzschild black hole and also two other potentials, which lay closely to the Schwarzschild potential near its maximum, but have very different behavior far from the black hole. These two potentials are chosen in the following way. We choose some points near the maximum of the effective potential. The first potential is an interpolation of these points by cubic splines (see fig. \ref{spot}). The second potential is a fit for the points by a ratio of polynomial functions.}{Esta afirma\c{c}\~ao foi verificada considerando o potencial do buraco negro de Schwarzschild e tamb\'em dois outros potenciais, que se p\~oem pr\'oximos ao potencial de Schwarzschild perto do seu m\'aximo, mas t\^em comportamentos muito diferentes longe do buraco negro. Esses dois potenciais s\~ao escolhidos de modo seguinte. Foram escolhidos alguns pontos perto do m\'aximo do potencial efetivo. O primeiro potencial \'e uma interpola\c{c}\~ao desses pontos por splines c\'ubicos (ver a fig. \ref{spot}). O segundo potencial \'e ajustado para esses pontos por uma raz\~ao de fun\c{c}\~oes polinomiais.}

\langpar{Since the WKB formula contains the value of the effective potential and its derivatives at the potential peak, we find that the results obtained with the help of all three potentials lay very close if the interpolation and the fit were made with the appropriate precision \cite{Konoplya:2006rv}. Despite the higher derivatives of our interpolation potential are not defined, we are able to evaluate them step by step by interpolating in the same way the first and all the consequent derivatives of the potential. It turns out that the interpolation potential is very sensitive to numerical errors. Therefore, to calculate the quasi-normal modes with the appropriate precision, one must calculate the values of the potential with very high accuracy. In fact, for the practical purposes one can use fitting of the potential which does not accumulate the numerical error.}{Como a f\'ormula de WKB cont\'em o valor do potencial efetivo e as suas derivadas no pico do potencial, achamos que os resultados obtidos com a ajuda dos tr\^es potenciais constituem boa aproxima\c{c}\~ao se a interpola\c{c}\~ao e o ajuste forem feitos com a precis\~ao apropriada \cite{Konoplya:2006rv}. Apesar das derivadas mais altas do nosso potencial de interpola\c{c}\~ao n\~ao serem definidas, \'e poss\'ivel avali\'a-las passo a passo interpolando do mesmo jeito a primeira e todas as derivadas consecutivas do potencial. Resulta que o potencial de interpola\c{c}\~ao \'e muito sens\'ivel a erros num\'ericos. Por isso, para calcular os modos quase-normais com a precis\~ao apropriada, deve-se calcular os valores do potencial com uma precis\~ao muito alta. De fato, para objetivos pr\'aticos pode-se utilizar o ajuste do potencial que n\~ao acumula o erro num\'erico.}

\langpar{To test the accuracy of this approach one can use 
the fact that the sixth order WKB formula gives a smaller relative
error than the third order one (see sec. \ref{sec:WKB}). Since the
higher WKB order depends on the higher derivatives of the
effective potential, that are more sensitive to the interpolation
or fitting accuracy, the higher order WKB formula should give some
random values, if the accuracy is not enough.}{Para testar a
precis\~ao desta abordagem pode-se utilizar 
o fato que a f\'ormula de WKB da sexta ordem d\'a o erro relativo
menor do que a da terceira ordem (ver a se\c{c}\~ao
\ref{sec:WKB}). Como a ordem mais alta de WKB depende das
derivadas mais altas do potencial efetivo, que s\~ao mais
sens\'iveis \`a precis\~ao de interpola\c{c}\~ao ou ajuste, a
ordem mais alta da f\'ormula de WKB deve dar alguns valores
casuais, se a precis\~ao n\~ao for suficiente.}

\langpar{The independence of the quasi-normal modes on the behavior of the effective potential at large distance, being inspired by the WKB formula, is not related with the WKB method. However, since the WKB formula depends only on the potential near its peak, we unable to prove the statement within the described approach. That is why, it is important to check the results with a different method. In order to do this, we have found in time domain the quasi-normal modes of gravitational perturbations of black holes in the Einstein-Aether theory. The calculated frequencies show an excellent agreement with the WKB results \cite{Konoplya:2006ar}.}{A independ\^encia dos modos quase-normais do comportamento do potencial efetivo a grande dist\^ancia, que foi inspirada pela f\'ormula de WKB, n\~ao est\'a relacionada com o m\'etodo de WKB. Contudo, como a f\'ormula de WKB depende somente do potencial perto do seu pico, \'e imposs\'ivel comprovar a afirma\c{c}\~ao dentro da aproxima\c{c}\~ao descrita. Por isso, \'e importante verificar os resultados com um m\'etodo diferente. Para fazer isto, achamos no dom\'inio temporal os modos quase-normais de perturba\c{c}\~oes gravitacionais de buracos negros na teoria de Einstein-A\'eter. As freq\"u\^encias calculadas mostram o acordo excelente com os resultados de WKB \cite{Konoplya:2006ar}.}

\langsectionlabel{Quasi-normal modes of the scalar hairy black hole}{Modos quase-normais do buraco negro com cabelo escalar}{sec:hairyBH}
\langpar{Since the paper of J. Bekenstein \cite{Bekenstein:1974sf}, it is well-known that black hole can not have scalar hair within minimal coupling. However, as have been found in \cite{Winstanley:2002jt,Winstanley:2005fu}, there is a possibility of dressing a four-dimensional black hole in anti-de Sitter space-time with a non-minimally coupled classical scalar field. In \cite{Radu:2005bp} the results were extended for higher dimensional configurations. Despite we live in the de Sitter universe, black holes in the anti-de Sitter background attracted considerable interest due to the AdS/CFT correspondence \cite{Maldacena:1997re}. In this context, the quasi-normal ringing of massive non-minimally coupled scalar field to the black hole in the anti-de Sitter background was studied for the first time in \cite{Zhidenko:2005mv}.}{Desde o trabalho de J. Bekenstein \cite{Bekenstein:1974sf}, \'e bem conhecido que buracos negros n\~ao podem ter cabelo escalar dentro do acoplamento m\'inimo. Contudo, como foi encontrado em \cite{Winstanley:2002jt, Winstanley:2005fu}, h\'a uma possibilidade de vestir um buraco negro quadridimensional no espa\c{c}o-tempo anti-de Sitter com um campo escalar cl\'assico n\~ao minimamente ligado. Em \cite{Radu:2005bp} os resultados foram extendidos para configura\c{c}\~oes em dimens\~oes mais altas. Apesar de vivermos no universo de Sitter, os buracos negros no contexto anti-de Sitter atra\'iram um interesse consider\'avel por causa da correspond\^encia AdS/CFT \cite{Maldacena:1997re}. Neste contexto, o toque quase-normal do campo escalar massivo n\~ao minimamente ligado ao buraco negro no contexto anti-de Sitter foi estudado pela primeira vez em \cite{Zhidenko:2005mv}.}

\langpar{We consider the spherically symmetric solution}{Consideremos a solu\c{c}\~ao esfericamente sim\'etrica}
\begin{equation}
ds^{2} = N(r) e^{2\delta(r)} dt^2 - N(r)^{-1} dr^2 - r^2
\left( d\theta ^{2} +
\sin \theta ^{2} \, d\varphi ^{2} \right)
\end{equation}
\lang{of the action, which describes a self-interacting scalar field $\phi$ with non-minimal coupling to gravity:}{da a\c{c}\~ao, que descreve um campo escalar $\phi$ auto-interativo com o acoplamento n\~ao m\'inimo \`a gravidade,}
\begin{equation}
S=\int d^{4}x \, {\sqrt {-g}} \left[
\frac {1}{2}\left( R -2\Lambda \right)
-\frac {1}{2} \left( \nabla \phi \right) ^{2} -\frac {1}{2} \xi R
\phi ^{2} -\frac{\mu^2\phi^2}{2} \right] ,
\label{action}
\end{equation}
\lang{where $R$ is the Ricci scalar, $\Lambda $ is the cosmological constant, $\xi $ is the coupling constant, and}{onde $R$ \'e o escalar de Ricci, $\Lambda $ \'e a constante cosmol\'ogica, $\xi $ \'e a constante de acoplamento e}
$$\left(\nabla\phi\right) ^{2} = \nabla _{a }\phi\nabla ^{a } \phi.$$

\langpar{In order to simplify the equations of motion we use the following conformal transformation}{Para simplificar as equa\c{c}\~oes de movimento foi usada a transforma\c{c}\~ao conforme}
\begin{equation}
{\bar {g}}_{ab} =(1-\xi
\phi ^{2}) g_{ab}.
\label{conf_trans}
\end{equation}

\langpar{After the transformation, the action takes the form}{Depois da transforma\c{c}\~ao, a a\c{c}\~ao toma a forma}
\begin{equation}
S=\int d^{4}{\bar x} \, {\sqrt {-{\bar {g}}}} \left(
\frac{1}{2} \left( {\bar {R}}-2\Lambda  \right)
-\frac {1}{2} \left( {\bar {\nabla }} \Phi \right) ^{2} -U(\Phi )
\right),
\label{conf_trans_action}
\end{equation}
\lang{where we define}{onde foram definidos}
\begin{equation}
\Phi =\int d\phi \sqrt{
\frac {(1-\xi \phi ^{2}) + 6\xi ^{2} \phi ^{2}}{
(1-\xi \phi ^{2})^{2}} },\qquad U(\Phi ) = \frac {\frac{\mu^2\phi^2}{2}+\Lambda
\xi \phi ^{2} \left( 2- \xi
\phi ^{2} \right) }{ \left( 1- \xi \phi ^{2} \right) ^{2}}.
\label{conf_trans_field}
\end{equation}

\langpar{The metric takes the following form}{A m\'etrica toma a forma seguinte}
\begin{equation}
d{\bar {s}}^{2} = {\bar {N}}({\bar {r}}) e^{2{\bar {\delta
}}({\bar {r}})} dt ^{2} - {\bar {N}}({\bar {r}}) ^{-1} d{\bar
{r}}^{2} - {\bar {r}}^{2} \left( d\theta ^{2} +
\sin \theta ^{2} \, d\varphi ^{2} \right),
\label{conf_trans_metric}
\end{equation}
\lang{where}{onde}
\begin{eqnarray}\nonumber
{\bar {r}} &=& \left( 1 - \xi \phi ^{2} \right) ^{\frac {1}{2}}
r,\\\nonumber {\bar {N}} &=& N \left( 1- \xi \phi ^{2} -\xi r
\phi
\phi' \right) ^{2}
\left( 1- \xi \phi ^{2} \right) ^{-2},\\\nonumber
{\bar {N}} e^{\bar {2\delta }} & = & N e^{2\delta }
\left( 1- \xi \phi ^{2} \right).
\end{eqnarray}

\langpar{Varying the action, one can find the following equations of motion \cite{Winstanley:2005fu}}{Variando a a\c{c}\~ao, pode-se encontrar as equa\c{c}\~oes de movimento seguintes \cite{Winstanley:2005fu}}
\begin{subequations}
\begin{eqnarray}
\frac{d({\bar r}\bar N)}{d{\bar r}} & = & 1-\Lambda{\bar r}^2-{\bar
r}^2\left(\frac {\bar N}{2}
\left(\frac {d\Phi }{d{\bar {r}}}\right)^2
+ U(\Phi)\right),
\label{minE1}
\\
\frac {d{\bar {\delta }}}{d{\bar {r}}} & = &
\frac {{\bar {r}}}{2} \left( \frac {d\Phi }{d{\bar {r}}} \right)^{2},
\label{minE2}
\\
0 & = & {\bar {N}} \frac {d^{2}\Phi }{d{\bar {r}}^{2}} + \left(
{\bar {N}} \frac {d{\bar {\delta }}}{d{\bar {r}}} + \frac {d{\bar
{N}}}{d{\bar {r}}} + \frac {2{\bar {N}}}{{\bar {r}}} \right)
\frac {d\Phi }{d{\bar {r}}}
- \frac {dU}{d\Phi}.
\label{mincscal}
\end{eqnarray}\label{meqs}
\end{subequations}

\langpar{In order to obtain the series expansion for the metric coefficients (\ref{conf_trans_metric}), we have to impose some boundary conditions, which are associated with the black hole parameters. We are using the following parameters. The event horizon radius ${\bar r}_+$ can be chosen arbitrary in order to fix length scale. This choice defines the boundary condition for $\bar N$: ${\bar N}({\bar r}_+)=0$. In order to measure all dimensional values in units of ${\bar r}_+$, we choose ${\bar r}_+=1$. The value ${\bar\delta}({\bar r}_+)$ is chosen arbitrary in order to fix time scale. In order to introduce the boundary conditions in the same point we impose ${\bar\delta}({\bar r}_+)=0$, what differs from Winstanley's ${\bar\delta}(\infty)=0$. The last parameter $\phi({\bar r}_+)=Q$ can be associated with the scalar charge of the black hole.}{Para obter as expans\~oes de s\'erie para os coeficientes m\'etricos (\ref{conf_trans_metric}), tem-se de impor algumas condi\c{c}\~oes de contorno, que se associam com os par\^ametros do buraco negro. Foram usados os par\^ametros seguintes. O raio do horizonte de eventos ${\bar r}_+$ pode ser escolhido arbitrariamente para fixar a escala de comprimento. Esta escolha define a condi\c{c}\~ao de contorno para $\bar N$: ${\bar N}({\bar r}_+)=0$. Para medir todos os valores dimensionais em unidades de ${\bar r}_+$, escolhe-se ${\bar r}_+=1$. O valor ${\bar\delta}({\bar r}_+)$ \'e escolhido arbitrariamente para fixar a escala de tempo. Para introduzir as condi\c{c}\~oes de contorno no mesmo ponto impomos ${\bar\delta}({\bar r}_+)=0$, que \'e diferente da escolha de Winstanley ${\bar\delta}(\infty)=0$. O \'ultimo par\^ametro $\phi({\bar r}_+)=Q$ pode ser associado com a carga escalar do buraco negro.}

\langpar{Spherically symmetric perturbation equation for}{A equa\c{c}\~ao de perturba\c{c}\~ao esfericamente sim\'etrica para}
$$\delta\Phi=e^{-\imo\omega t}\frac{Q_\omega(r)}{r}$$
\lang{takes the wave-like form (\ref{eigenfunction-equation}), where the perturbation potential is given by}{toma a forma de onda (\ref{eigenfunction-equation}), onde o potencial de perturba\c{c}\~ao \'e dado por}
\cite{Winstanley:2002jt}:
\begin{equation}
V({\bar r}) =
\frac {{\bar {N}}e^{2{\bar {\delta }}}}{{\bar {r}}^{2}}
\left(
(1  - (U+\Lambda ){\bar r}^2)\left(1-{\bar r}^2\left( \frac {d\Phi
}{d{\bar {r}}} \right) ^{2}\right) - {\bar N} +2{\bar {r}}^{3}
\frac {dU}{d\Phi }
\frac {d\Phi }{d{\bar {r}}}
+ {\bar {r}}^{2} \frac {d^{2}U}{d\Phi ^{2}}\right).
\label{perturb_potential}
\end{equation}

\langpar{The tortoise coordinate is given by}{A coordenada tartaruga \'e dada por}
\begin{equation}
\frac {d{\bar r}^*}{d{\bar {r}}} =
e^{\bar\delta}{\bar N}.
\label{tortoise}
\end{equation}
\langpar{It is similar to the ordinary anti-de Sitter case:}{Esto \'e semelhante ao caso anti-de Sitter ordin\'ario:}
$$\begin{array}{rcl}
{\bar r}={\bar r}_+&\quad\Longleftrightarrow\quad&{\bar r}^*=\infty,\\
{\bar r}=\infty&\quad\Longleftrightarrow\quad&{\bar r}^*=0.
\end{array}$$

\langpar{This perturbation corresponds to infinitesimal changing of the black hole mass and is similar to the zero-multipole scalar perturbations near a hairless black hole. The crucial difference is that the perturbation of a scalar field due to the hair gives a first-order correction to the metric.}{Esta perturba\c{c}\~ao corresponde \`a modifica\c{c}\~ao infinit\'esima da massa do buraco negro e \'e semelhante \`as perturba\c{c}\~oes escalares de multipolo nulo perto de um buraco negro sem cabelo. A diferen\c{c}a crucial \'e que a perturba\c{c}\~ao de um campo escalar por causa do cabelo d\'a uma corre\c{c}\~ao de primeira ordem \`a m\'etrica.}

\begin{figure}
\begin{center}
\includetwograph[]{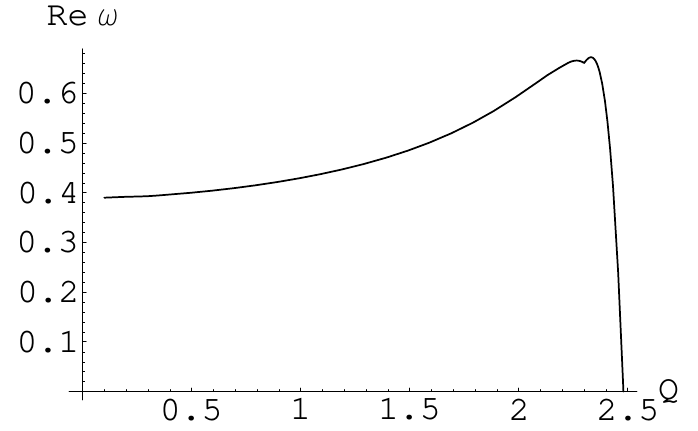}{HairyAdS_L=-0_05,xi=0_1,mu=0_Im}\\($\Lambda=-0.05$, $\xi=0.1$, $\mu=0$)\\
\includetwograph[]{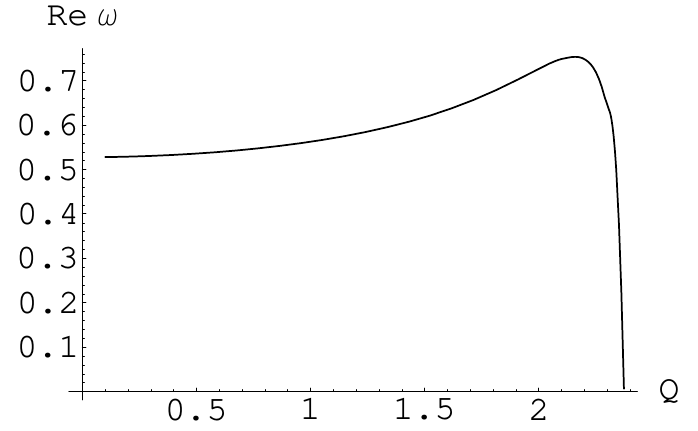}{HairyAdS_L=-0_1,xi=0_1,mu=0_Im}\\($\Lambda=-0.1$, $\xi=0.1$, $\mu=0$)
\langfigurecaption{\textbf{Dependance of the real and imaginary parts of the quasi-normal frequency on $Q$ for the scalar hairy black hole.}
The imaginary part quickly decreases as the ``scalar charge'' $Q$ increases. The real part reaches its maximum and then quickly falls
down to zero. It vanishes for some non-critical charge $Q_0<\xi^{-1/2}$. For $Q\geq Q_0$ the real part remains zero and the frequency is purely imaginary (within numerical precision).}{\textbf{Depend\^encia das partes real e imagin\'aria da freq\"u\^encia quase-normal em $Q$ para o buraco negro com cabelo escalar.} A parte imagin\'aria rapidamente diminui enquanto ``a carga escalar'' $Q$ aumenta. A parte real chega o seu m\'aximo e depois rapidamente cai at\'e o zero. Ela desaparece para algumas cargas n\~ao cr\'iticas $Q_0<\xi^{-1/2}$. Para $Q\geq Q_0$ a parte real se anula e a freq\"u\^encia \'e puramente imagin\'aria (dentro da precis\~ao num\'erica).}\label{HairyAdS.Qfig}
\end{center}
\end{figure}
\begin{figure}
\begin{center}
\includetwograph[]{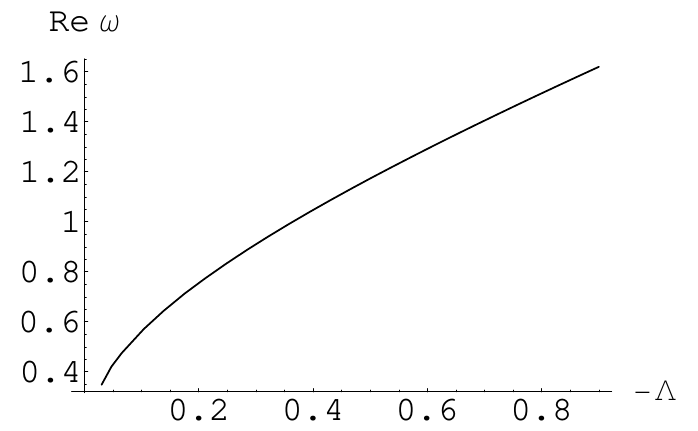}{HairyAdS_Q=1,xi=0_1,mu=0_Im}\\($Q=1$, $\xi=0.1$, $\mu=0$)\\
\includetwograph[]{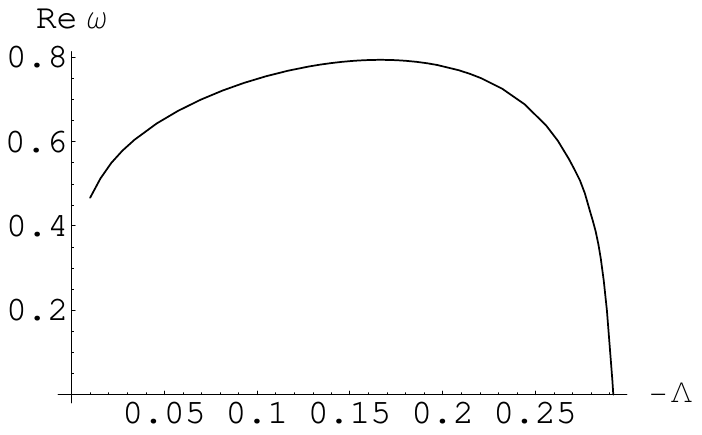}{HairyAdS_Q=2_2,xi=0_1,mu=0_Im}\\($Q=2.2$, $\xi=0.1$, $\mu=0$)
\langfigurecaption{\textbf{Dependance of the real and imaginary parts of the quasi-normal frequency on $\Lambda$ for the scalar hairy black hole.} The real part increases for relatively small ``scalar charges'' $Q$, but for large $Q=2.2$ it also reaches maximum and then falls down and vanishes for $\Lambda\approx-0.292$. For larger values of $\Lambda$ the real part remains zero and purely imaginary frequencies exist.
The imaginary part decreases almost linearly as $\Lambda$ grows.}{\textbf{Depend\^encia das partes real e imagin\'aria da freq\"u\^encia quase-normal em $\Lambda$ para o buraco negro com cabelo escalar.} A parte real cresce para ``cargas escalares'' $Q$ relativamente pequenas, mas para grande $Q=2.2$ ela tamb\'em consegue o m\'aximo e depois cai e desaparece para $\Lambda\approx-0.292$. Para valores de $\Lambda$ maiores a parte real fica nula e as freq\"u\^encias puramente imagin\'arias existem. A parte imagin\'aria diminui quase linearmente enquanto $\Lambda$ cresce.}\label{HairyAdS.Lfig}
\end{center}
\end{figure}
\begin{figure}
\begin{center}
\includetwograph[]{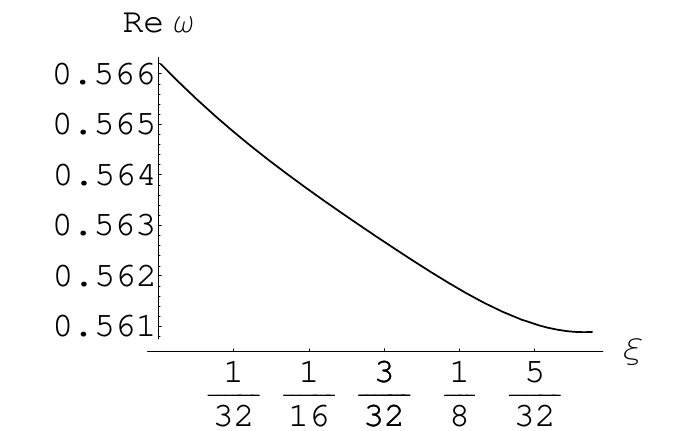}{HairyAdS_Q=1,L=-0_1,mu=0_Im}\\($Q=1$, $\Lambda=-0.1$, $\mu=0$)\\
\includetwograph[]{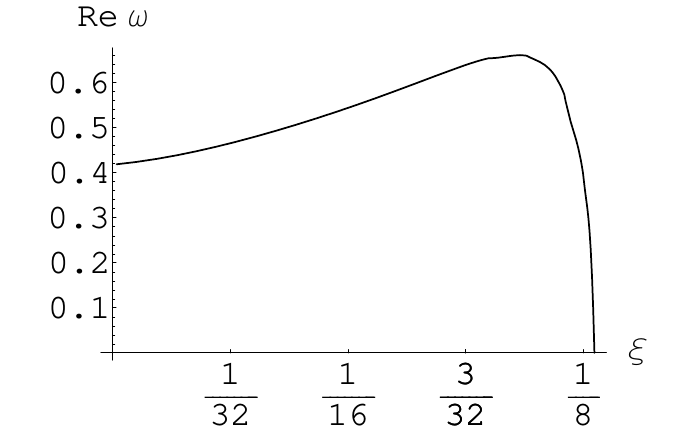}{HairyAdS_Q=2_2,L=-0_05,mu=0_Im}\\($Q=2.2$, $\Lambda=-0.05$, $\mu=0$)
\langfigurecaption{\textbf{Dependance of the real and imaginary parts of the quasi-normal frequency on $\xi<\frac{3}{16}$ for the scalar hairy black hole.} The imaginary part decreases as $\xi$ grows. The quickness of such decreasing depends on $\Lambda$ and $Q$. For some $\Lambda$ and $Q$ (as presented on the bottom graphic) the real part can fall down to zero and purely imaginary frequencies appear for larger $\xi$.}{\textbf{Depend\^encia das partes real e imagin\'aria da freq\"u\^encia quase-normal em $\xi<\frac{3}{16}$ para o buraco negro com cabelo escalar.} A parte imagin\'aria diminui enquanto $\xi$ cresce. A rapidez de tal redu\c{c}\~ao depende de $\Lambda$ e $Q$. Para alguns $\Lambda$ e $Q$ (como apresentado no gr\'afico de baixo) a parte real pode cair a zero e as freq\"u\^encias puramente imagin\'arias aparecem para maior $\xi$.}\label{HairyAdS.xifig}
\end{center}
\end{figure}
\begin{figure}
\begin{center}
\includetwograph[]{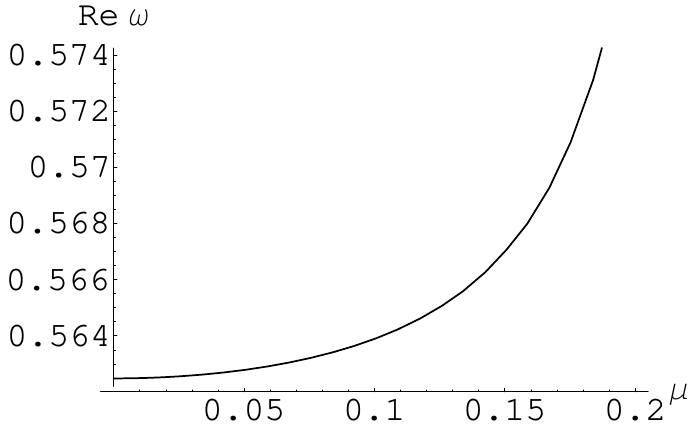}{HairyAdS_Q=1,L=-0_1,xi=0_1_Im}\\($\Lambda=-0.1$, $Q=1$, $\xi=0.1$)\\
\includetwograph[]{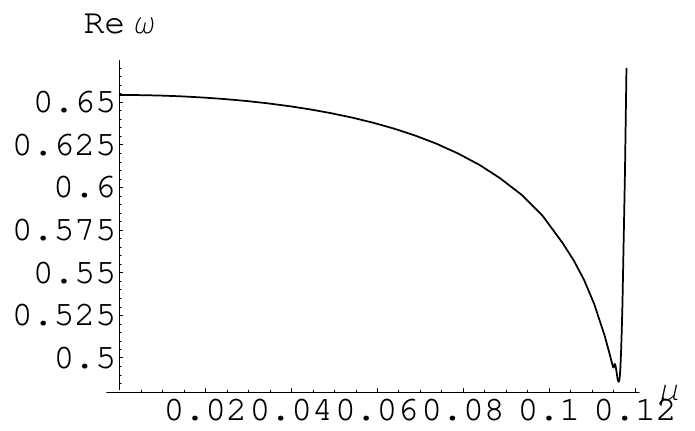}{HairyAdS_Q=2_2,L=-0_05,xi=0_1_Im}\\($\Lambda=-0.05$, $Q=2.2$, $\xi=0.1$)
\langfigurecaption{\textbf{Dependance of the real and imaginary parts of the quasi-normal frequency on $\mu$ for the scalar hairy black hole.} We consider only $0\leq\mu^2\leq-4\xi\Lambda$. The real part changes within comparatively small range. The imaginary part reaches its maximum remaining negative. For this value of $\mu$ the oscillations have the longest lifetime, but they still damp and we do not observe quasi-resonances. For larger scalar field masses the imaginary part of frequency decreases and the damping rate is higher.}{\textbf{Depend\^encia das partes real e imagin\'aria da freq\"u\^encia quase-normal em $\mu$ para o buraco negro com cabelo escalar.} Somente $0\leq\mu^2\leq-4\xi\Lambda$ foi considerado. A parte real modifica-se dentro de uma variedade comparativamente pequena. A parte imagin\'aria alcan\c{c}a o seu m\'aximo permanecendo negativa. Para este valor de $\mu$ as oscila\c{c}\~oes t\^em maior tempo de vida, mas elas ainda decaem e quase-resson\^ancias n\~ao s\~ao observadas. Para maiores massas de campo escalar a parte imagin\'aria da freq\"u\^encia diminui e a taxa de decaimento \'e mais alta.}\label{HairyAdS.mufig}
\end{center}
\end{figure}
\begin{figure}
\begin{center}
\includeonegraph[]{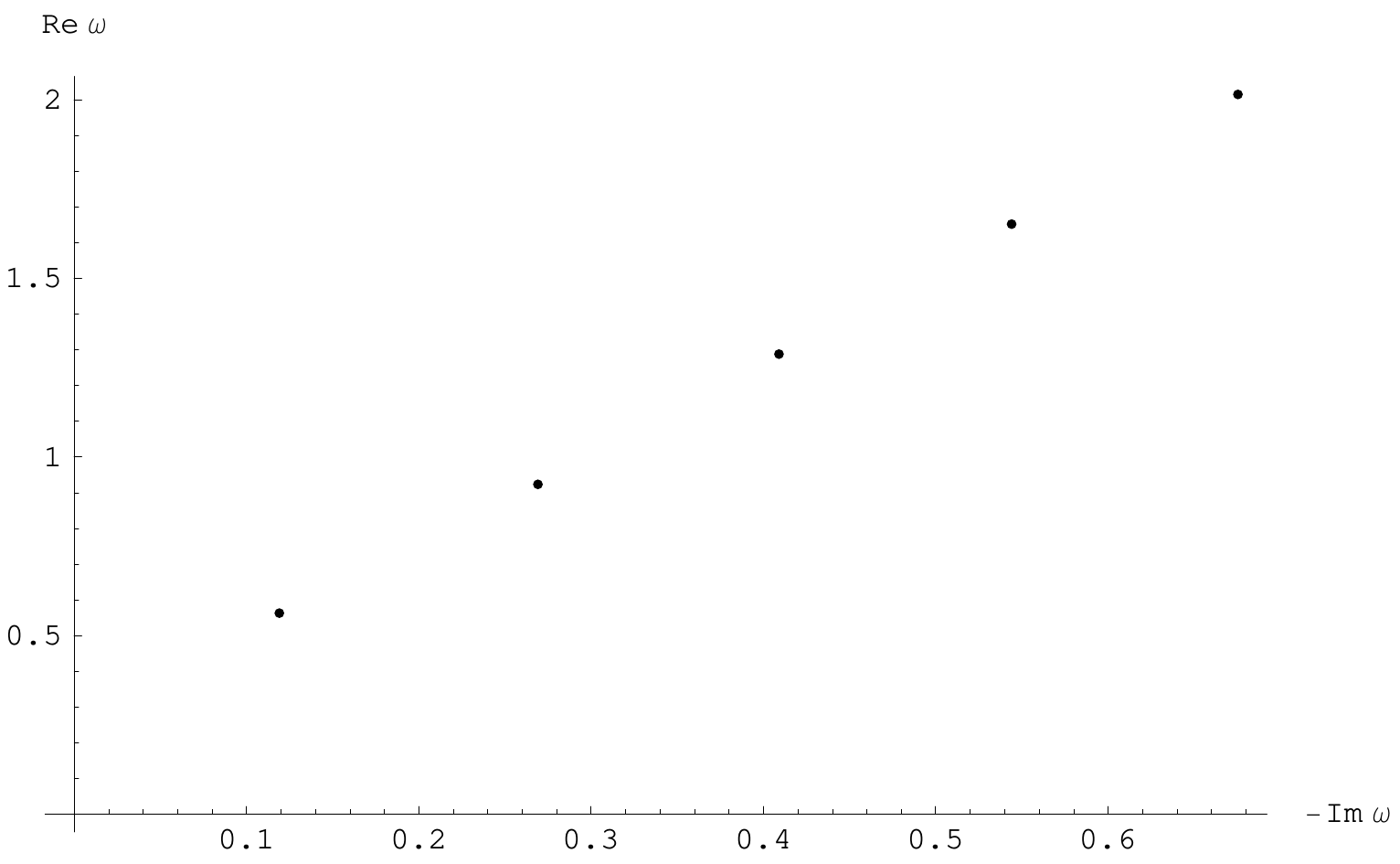}
\langfigurecaption{\textbf{First five quasi-normal frequencies for the scalar hairy black hole $\Lambda=-0.1$, $Q=1$, $\xi=0.1$, $\mu=0$.} The high overtones approach equidistant spacing.}{\textbf{As primeiras cinco freq\"u\^encias quase-normais do buraco negro com cabelo escalar $\Lambda=-0.1$, $Q=1$, $\xi=0.1$, $\mu=0$.} Os sobretons altos se aproximam de um espa\c{c}amento eq\"uidistante.}\label{HairyAdS.Nfig}
\end{center}
\end{figure}

\langpar{In order to calculate the quasi-normal modes we find the series expansions near the black hole event horizon for $\Phi$, ${\bar\delta}$ and ${\bar N}$ using the equations (\ref{meqs}). Since ${\bar N}$ grows at the spatial infinity, it is convenient to find the series for the function $g={\bar N}/{\bar r}^2$ instead of ${\bar N}$. After the series for the metric functions are known one can find the series for the effective potential (\ref{perturb_potential}) and, finally, for $s(z)$, $t(z)$ and $u(z)$ in (\ref{HH-radial}). Then we use the technique, described in the previous section.}{Para calcular os modos quase-normais s\~ao achadas as expans\~oes de s\'erie perto do horizonte de eventos do buraco negro para $\Phi$, ${\bar\delta}$ e ${\bar N}$ usando as equa\c{c}\~oes (\ref{meqs}). Como ${\bar N}$ cresce no infinito espacial, \'e conveniente achar a s\'erie para a fun\c{c}\~ao $g={\bar N}/{\bar r}^2$ em vez de ${\bar N}$. Depois que as s\'eries para as fun\c{c}\~oes m\'etricas s\~ao conhecidas pode-se achar a s\'erie para o potencial efetivo (\ref{perturb_potential}) e, finalmente, para $s(z)$, $t(z)$ e $u(z)$ em (\ref{HH-radial}). Ent\~ao utilizamos a t\'ecnica descrita na se\c{c}\~ao pr\'evia.}

\langpar{It was found \cite{Zhidenko:2005mv} that imaginary part of the quasi-normal modes changes monotonously as a function of $\Lambda$, $Q$ and $\xi$. Scalar field mass $\mu$ cause imaginary part to decrease its absolute value but not to vanish. It reaches a minimum for some particular mass and then increases very rapidly (see fig. \ref{HairyAdS.mufig}). Thus no infinitely long living oscillations appear.}{Foi achado \cite{Zhidenko:2005mv} que a parte imagin\'aria dos modos quase-normais modifica-se monotonicamente como uma fun\c{c}\~ao de $\Lambda$, $Q$ e $\xi$. A massa do campo escalar $\mu$ faz a parte imagin\'aria reduzir o seu valor absoluto mas n\~ao desaparecer. O valor alcan\c{c}a um m\'inimo para uma massa determinada e depois aumenta muito rapidamente (ver a fig. \ref{HairyAdS.mufig}). Assim, nenhuma oscila\c{c}\~ao com vida infinitamente longa aparece.}

\langpar{Real part has more complicated behavior and depends significantly on the parameters $Q$ and $\Lambda$ (see figures \ref{HairyAdS.Qfig} and \ref{HairyAdS.Lfig}). Other parameters, $\xi$ and $\mu$, change the real part within the comparable small bounds (see figures \ref{HairyAdS.xifig} and \ref{HairyAdS.mufig}). For relatively large values of $\xi$, $\Lambda$ and $Q$ purely imaginary frequencies appear.}{A parte real tem um comportamento mais complicado e depende significativamente dos par\^ametros $Q$ e $\Lambda$ (ver as figuras \ref{HairyAdS.Qfig} e \ref{HairyAdS.Lfig}). Outros par\^ametros, $\xi$ e $\mu$, modificam a parte real dentro dos limites relativamente pequenos (ver as figuras \ref{HairyAdS.xifig} e \ref{HairyAdS.mufig}). Para valores relativamente grandes de $\xi$, $\Lambda$ e $Q$ as freq\"u\^encias puramente imagin\'arias aparecem.}

\langpar{From the figure \ref{HairyAdS.Nfig} we see that higher overtones tend to equidistant spacing similar to hairless black hole spectrum behavior in the anti-de Sitter background \cite{Cardoso:2003cj}.}{A partir da figura \ref{HairyAdS.Nfig} v\^e-se que os sobretons mais altos tendem ao espa\c{c}amento eq\"uidistante semelhantemente ao comportamento do espectro de buraco negro sem cabelo no contexto anti-de Sitter \cite{Cardoso:2003cj}.}

\langsectionlabel{Perturbations and quasi-normal modes of black holes in the Einstein-Aether theory}{Perturba\c{c}\~oes e modos quase-normais de buracos negros na teoria de Einstein-A\'eter}{sec:Einstein-Aether}
\langpar{General relativity is based on the local Lorentz invariance. Yet, there appeared a lot of attempts to go beyond the local Lorentz symmetry. Aether can be considered as locally preferred state of rest at each point of space-time due-to some unknown physics. Einstein-Aether theory is general relativity coupled to a dynamical time-like vector field, which is called ``aether''. This theory is what comes instead of usual general relativity when local Lorentz symmetry is broken. Namely, the vector field breaks local boost invariance, while rotational symmetry in the preferred frame is preserved (see \cite{Eling:2004dk} for review).}{A relatividade geral \'e baseada na invari\^ancia local de Lorentz. Mas muitas tentativas de ultrapassar a simetria local de Lorentz apareceram. A\'eter pode ser considerado como um estado em repouso local preferido em cada ponto do espa\c{c}o-tempo por causa de alguma f\'isica desconhecida. A teoria de Einstein-A\'eter \'e a relatividade geral ligada a um campo vetorial din\^amico do tipo temporal, que \'e chamado ``a\'eter''. Esta teoria \'e o que vem em vez da relatividade geral usual quando a simetria local de Lorentz \'e quebrada. O campo vetorial quebra a invari\^ancia local relativ\'istica, enquanto que a simetria rotacional no sistema de refer\^encia preferencial fica conservada (ver a revis\~ao \cite{Eling:2004dk}).}

\langpar{It is important that the significant difference between Einstein and Einstein-Aether theories should be seen in the regime of strong field, for instance in observing of the characteristic quasi-normal spectrum of black holes. Thus, existence of aether could be tested in the forthcoming experiments with new generation of gravitational antennas. In this context, the quasi-normal spectrum for test scalar and electromagnetic fields \cite{Konoplya:2006rv} and for the gravitational perturbations \cite{Konoplya:2006ar} was studied.}{\'E importante que a diferen\c{c}a significante entre as teorias de Einstein e de Einstein-A\'eter deve ser vista no regime de campo forte, por exemplo, observando o espectro quase-normal caracter\'istico de buracos negros. Assim, a exist\^encia de a\'eter pode ser testada nos pr\'oximos experimentos com a nova gera\c{c}\~ao de antenas gravitacionais. Neste contexto, o espectro quase-normal dos campos de teste escalar e eletromagn\'etico \cite{Konoplya:2006rv} e para as perturba\c{c}\~oes gravitacionais \cite{Konoplya:2006ar} foi estudado.}

\langpar{The solution for a black hole metric in the Einstein-Aether theory was found numerically in \cite{Eling:2006df,Eling:2006ec}. Since the analytical solution is not known, in order to calculate the quasi-normal modes, one can apply the fitting of the effective potential as described in the section \ref{sec:non-analytical-method}.}{A solu\c{c}\~ao para uma m\'etrica de buraco negro na teoria de Einstein-A\'eter foi achada numericamente em \cite{Eling:2006df,Eling:2006ec}. Como a solu\c{c}\~ao anal\'itica n\~ao \'e conhecida, para calcular os modos quase-normais, pode-se ajustar o potencial efetivo como descrito na se\c{c}\~ao \ref{sec:non-analytical-method}.}

\langpar{The Lagrangian of the full Einstein-Aether theory that forms the most general diffeomorphism invariant action of the space-time metric $g_{ab}$ and the aether field $u^a$, involving no more than two derivatives, is given by}{O Lagrangiano da teoria de Einstein-A\'eter completa que forma a a\c{c}\~ao invariante de difeomorfismos mais geral da m\'etrica do espa\c{c}o-tempo $g_{ab}$ e do campo de a\'eter $u^a$, envolvendo n\~ao mais do que duas derivadas, \'e dado por}
\begin{equation}\label{eaft}
L=-R-K^{ab}_{~~~mn}\nabla_au^m\nabla_bu^n-\lambda(g_{ab}u^au^b-1),
\end{equation}
\lang{where $R$ is the Ricci scalar, $\lambda$ is a Lagrange multiplier, which provides the unit time-like constraint,}{onde $R$ \'e o escalar de Ricci, $\lambda$ \'e o multiplicador de Lagrange, que fornece o v\'inculo unit\'ario do tipo temporal}
$$K^{ab}_{~~~mn}=c_1g^{ab}g_{mn}+c_2\delta^a_m\delta^b_n+c_3\delta^a_n\delta^b_m+c_4u^au^bg_{mn},$$
\lang{where $c_i$ are dimensionless constants.}{onde $c_i$ s\~ao constantes adimensionais.}

\langpar{Spherical symmetry allows to fix $c_4=0$. Following \cite{Eling:2006df,Eling:2006ec}, we shall consider the so-called {\it non-reduced} Einstein-Aether theory, for which $c_3=0$, and we can use the field redefinition that fixes the coefficient $c_2$:}{A simetria esf\'erica permite fixar $c_4=0$. Seguindo \cite{Eling:2006df,Eling:2006ec}, ser\'a considerada a teoria de Einstein-A\'eter chamada {\it n\~ao-reduzida}, para a qual $c_3=0$, e pode-se usar a redefini\c{c}\~ao do campo, que fixa o coeficiente $c_2$:}
$$c_2=-\frac{c_1^3}{2-4c_1+3c_1^2},$$
\lang{so that $c_1$ is the free parameter.}{para que $c_1$ seja o par\^ametro livre.}

\langpar{The metric for a spherically symmetric static black hole is}{A m\'etrica de um buraco negro est\'atico esfericamente sim\'etrico \'e}
\begin{equation}
d s^{2} = N(r) d t^{2} -\frac{B^{2}(r)}{N(r)}d r^{2} - r^{2} d \Omega^{2}\,,
\end{equation}
\lang{where the functions $N(r)$ and $B(r)$ are given by numerical integration near the black hole event horizon.}{onde as fun\c{c}\~oes $N(r)$ e $B(r)$ s\~ao dadas pela integra\c{c}\~ao num\'erica perto do horizonte de eventos do buraco negro.}

\langpar{The perturbation equations can be reduced to the wave-like form (\ref{wave-like}) with the effective potentials}{As equa\c{c}\~oes de perturba\c{c}\~ao podem ser reduzidas \`a forma de onda (\ref{wave-like}) com os potenciais efetivos}
\begin{subequations}
\begin{eqnarray}\label{sp}
V_{s} &=& N(r)\frac{l(l + 1)}{r^{2}} + \frac{1}{r}
\frac{d} {d r_*}\frac{N(r)}{B(r)},\qquad l=0,1,2\ldots\,,\\
\label{emp}
V_{e} &=& N(r) \frac{l (l + 1)}{r^{2}},\qquad\qquad\qquad\qquad l=1,2,3\ldots\,,\\
\label{gp}
V_{g} &=& N(r) \frac{(l + 2)(l -1)}{r^{2}} + \frac{2 N^{2}(r)}{B^{2}(r) r^{2}} - \frac{1}{r} \frac{d} {d r_*}\frac{N(r)}{B(r)},\qquad l=2,3,4\ldots\,,
\end{eqnarray}
\end{subequations}
\lang{for scalar, Maxwell and axial gravitational perturbations respectively.}{para perturba\c{c}\~oes escalares, Maxwell e gravitacionais axiais, respectivamente.}

\langpar{The tortoise coordinate is defined as}{A coordenada tartaruga \'e definida como}
$$dr_\star=\frac{B(r)}{N(r)}dr.$$

\langpar{Note, that since the background value of aether coupling is small in comparison with the background characteristics of large black hole, the Schwarzschild metric is slightly corrected by the aether. That is why, when considering perturbations of the metric we neglect small perturbations of aether, keeping only linear perturbations of the Ricci tensor.}{Observe, que como o valor de fundo da liga\c{c}\~ao de a\'eter \'e pequeno em compara\c{c}\~ao com as caracter\'isticas do contexto do buraco negro grande, a m\'etrica de Schwarzschild \'e ligeiramente corrigida pelo a\'eter. Por isso, quando considerando as perturba\c{c}\~oes da m\'etrica, s\~ao desconcideradas perturba\c{c}\~oes pequenas de a\'eter, mantendo somente as perturba\c{c}\~oes lineares do tensor de Ricci.}

\begin{figure}
\includeonegraph{EA_l=2_profiles}
\langfigurecaption{Evolution of axial gravitational perturbations ($l = 2$) in time domain for non-reduced Einstein-Aether theory, $c_1 = 0.1$ (green line) and $c_1 = 0.4$ (red line), in comparison with the Schwarzschild case $c_1=0$ (blue line). The higher $c_1$ is the quicker decay of the observed perturbations.}{Evolu\c{c}\~ao temporal de perturba\c{c}\~oes gravitacionais axiais ($l = 2$) para teoria de Einstein-A\'eter n\~ao-reduzida, $c_1 = 0.1$ (linha verde) e $c_1 = 0.4$ (linha vermelha), em compara\c{c}\~ao com o caso de Schwarzschild $c_1=0$ (linha azul). Para $c_1$ mais alto, foi observado o decaimento mais r\'apido das perturba\c{c}\~oes.}\label{EA.l=2}
\end{figure}

\langpar{Since $N(r)$ and $B(r)$ are known numerically, we can use the fitting technique \cite{Konoplya:2006rv}. The fit functions for $N(r)$ and $B(r)$ were chosen as fractions of two polynomials, which are characterised by the number of terms in their numerators and denominators. There is some optimal number for which the convergence of WKB series is best. Practically, in order to find optimal number of terms for $N(r)$ we search for minimal difference between third and sixth order WKB values, first for the case $B(r) = 1$. When we have found the optimal fit for $N(r)$, in a similar fashion, i.e., by looking for best WKB convergence, we are in position to find the optimal fit for $B(r)$. Quick WKB convergence shows that higher derivatives of the metric coefficients are calculated with the best accuracy.}{Como $N(r)$ e $B(r)$ s\~ao conhecidos numericamente, pode-se utilizar a t\'ecnica de ajuste \cite{Konoplya:2006rv}. Para ajustar $N(r)$ e $B(r)$, as fun\c{c}\~oes foram escolhidas como fra\c{c}\~oes de dois polin\^omios, que s\~ao caracterizados pelo n\'umero de termos nos seus numeradores e denominadores. H\'a um n\'umero \'otimo para o qual a converg\^encia da s\'erie de WKB \'e melhor. Na pr\'atica, para achar o n\'umero \'otimo de termos para $N(r)$ procura-se a diferen\c{c}a m\'inima entre valores da terceira e da sexta ordem de WKB, primeiramente, para o caso $B(r) = 1$. Quando acharmos o ajuste \'otimo para $N(r)$ de modo semelhante, isto \'e, procurando a melhor converg\^encia de WKB, ajusta-se tamb\'em $B(r)$ de modo \'otimo. A converg\^encia r\'apida de WKB mostra que as derivadas mais altas dos coeficientes m\'etricos s\~ao calculadas com a melhor precis\~ao.}

\langpar{We find that as $c_1$ grows, the fundamental quasi-normal frequency increases the absolute value of its real and imaginary parts. This was checked also in time-domain \cite{Konoplya:2006ar}. On the figure \ref{EA.l=2} we see, that for higher $c_1$ the oscillation period and the lifetime of the perturbations decrease. We conclude that aether, if it exists, could be indirectly observed through detection of characteristic spectrum of black holes.}{Foi achado, que como $c_1$ cresce, a freq\"u\^encia quase-normal fundamental aumenta o valor absoluto das suas partes real e imagin\'aria. Isto foi verificado tamb\'em no dom\'inio de tempo \cite{Konoplya:2006ar}. Na figura \ref{EA.l=2} v\^e-se, que para $c_1$ mais alto o per\'iodo de oscila\c{c}\~ao e o tempo de vida das perturba\c{c}\~oes reduzem. Conclui-se que a\'eter, se existir, poderia ser indiretamente observado pela detec\c{c}\~ao de espectros caracter\'isticos de buracos negros.}

\langchapterlabel{Perturbations of Schwarzschild black holes in laboratories}{Perturba\c{c}\~oes de buracos negros de Schwarzschild em laborat\'orios}{sec:laval-nozzle}
\langsection{Acoustic analogue of gravity}{An\'alogo ac\'ustico de gravidade}

\langpar{In addition to the possibility to observe quasi-normal modes of black holes with the help of a new generation of gravitational antennas, there is a window for observation of the acoustic analogue of a black hole in laboratories. This is the well-known Unruh analogue of black holes \cite{Unruh:1980cg,Unruh:1994je}, which are the apparent horizons appearing in a fluid with a space-dependent velocity, in the presence of sonic points. The wave, which passed through the sonic point, cannot propagate backward, mimicking thereby, the effect of the horizon at sonic points.}{Al\'em da possibilidade de observar os modos quase-normais dos buracos negros com a ajuda de uma nova gera\c{c}\~ao de antenas gravitacionais, h\'a uma janela para observa\c{c}\~ao do an\'alogo ac\'ustico de um buraco negro em laborat\'orios. Isto \'e o bem conhecido an\'alogo de Unruh de buracos negros \cite{Unruh:1980cg,Unruh:1994je}, que s\~ao os horizontes aparentes que aparecem em um fluido com uma velocidade espa\c{c}o-dependente, na presen\c{c}a de pontos s\^onicos. A onda, que passou pelo ponto s\^onico, n\~ao pode propagar para tr\'as, imitando por meio disso, o efeito do horizonte em pontos s\^onicos.}

\langpar{The dynamic of sound waves in a fluid can be described by the propagation of a scalar field in some effective background. It turns out, that steady spherically symmetric flow cannot reproduce the Schwarzschild geometry because of the equation of continuity (see \cite{Barcelo:2005fc} for review). One could construct either an acoustic analog, which is conformal to the Schwarzschild black hole,}{A din\^amica de ondas sonoras em um fluido pode ser descrita pela propaga\c{c}\~ao de um campo escalar em algum contexto efetivo. Acontece, que o fluxo constante esfericamente sim\'etrico n\~ao pode reproduzir a geometria de Schwarzschild por causa da equa\c{c}\~ao de continuidade (ver \cite{Barcelo:2005fc} para revis\~ao). Pode-se construir ou um an\'alogo ac\'ustico, que \'e conforme ao buraco negro de Schwarzschild,}
\begin{equation}
ds^2\propto r^{-3/2}\left(\left(1-\frac{r_+}{r}\right)dt^2-\left(1-\frac{r_+}{r}\right)^{-1}dr^2-r^2(d\theta^2+\sin\theta^2 d\phi^2)\right),
\end{equation}
\lang{or an exact analog to the 7-dimensional black hole, projected on the $4$-brane (see section \ref{sec:BH-projected})}{ou um an\'alogo exato do buraco negro 7-dimensional, projetado na $4$-brana (ver a se\c{c}\~ao \ref{sec:BH-projected})}
\begin{equation}
ds^2=\left(1-\left(\frac{r_+}{r}\right)^4\right)dt^2-\left(1-\left(\frac{r_+}{r}\right)^4\right)^{-1}dr^2-r^2(d\theta^2+\sin\theta^2 d\phi^2).
\end{equation}

\langpar{If one had a complete analogy with some known solution of the Einstein equations, say, the Schwarzschild solution, he could see, in the acoustic experiments, not only qualitative, but also, up to an experimental accuracy, exact numerical coincidence with a prototype characteristics. Namely, for quasi-normal modes, which are governed by the form of the wave equation, this numerical correspondence would mean that the effective potential of the perturbations of some hydrodynamic system coincides with an effective potential of the black hole. Fortunately, the consideration of the perturbations of a gas in de Laval nozzle \cite{Okuzumi:2007hf} gives us such an opportunity of finding a system that obeys the same effective potential as a Schwarzschild black hole does \cite{Abdalla:2007dz}.}{Se tivesse uma analogia completa com alguma solu\c{c}\~ao conhecida das equa\c{c}\~oes de Einstein, por exemplo, a solu\c{c}\~ao de Schwarzschild, poder-se-ia ver, nos experimentos ac\'usticos a coincid\^encia num\'erica com umas caracter\'isticas de prototipo, n\~ao s\'o qualitativa, mas exata at\'e uma precis\~ao experimental. Para modos quase-normais, que s\~ao governados pela forma da equa\c{c}\~ao de onda, esta correspond\^encia num\'erica significaria que o potencial efetivo das perturba\c{c}\~oes de algum sistema hidrodin\^amico coincide com um potencial efetivo do buraco negro. Felizmente, a considera\c{c}\~ao das perturba\c{c}\~oes de um g\'as em bocal de Laval \cite{Okuzumi:2007hf} d\'a-nos tal oportunidade de encontrar um sistema, que obedece o mesmo potencial efetivo que um buraco negro de Schwarzschild, \cite {Abdalla:2007dz}.}

\langsection{de Laval nozzle}{Bocal de Laval}
\langpar{The canonical de Laval nozzle is a convergent-divergent tube, narrow in the middle. It allows to accelerate the gas until the sonic speed in its throat, reaching supersonic speeds after passing the throat. The perturbations of the gas in de Laval nozzle can be considered as one-dimensional if the section does not change too quickly along the length of the nozzle. Let us show that the corresponding effective potential for perturbations in a canonical de Laval nozzle can be made equal to the potential for perturbations of Schwarzschild black holes, when choosing some specific form of the nozzle.}{O bocal de Laval can\^onico \'e um tubo convergente-divergente, estreito ao meio. Ele permite acelerar o g\'as at\'e a velocidade s\^onica na sua garganta, conseguindo velocidades supers\^onicas depois de passar pela garganta. As perturba\c{c}\~oes do g\'as no bocal de Laval podem ser consideradas unidimensionais, se a sec\c{c}\~ao n\~ao se modificar r\'apido demais ao longo do comprimento do bocal. Ser\'a mostrado que o potencial efetivo correspondente para perturba\c{c}\~oes em um bocal de Laval can\^onico pode ser feito igual ao potencial para perturba\c{c}\~oes de buracos negros de Schwarzschild, escolhendo uma forma espec\'ifica de bocal.}

\langpar{We assume that a gas in the nozzle can be described by the equations of motion for the perfect fluid and that the flow is quasi-one-dimensional. The equations of continuity, the momentum and energy conservation read respectively}{Assume-se que um g\'as no bocal de Laval pode ser descrito pelas equa\c{c}\~oes de movimento para o fluido perfeito e que o fluxo \'e semi-unidimensional. As equa\c{c}\~oes de continuidade, de conserva\c{c}\~ao do momento e da energia l\^eem-se respectivamente}
\begin{subequations}
\begin{gather}
\partial_t(\rho A) + \partial_x(\rho vA) = 0 \,,
\label{eq:cont} \\
\partial_t(\rho vA) + \partial_x[(\rho v^2 + p)A] = 0 \,,
\label{eq:momentum} \\
\partial_t(\epsilon A) +  \partial_x[(\epsilon+p)vA] = 0 \,.
\label{eq:energy}
\end{gather}
\end{subequations}

\langpar{Here $\rho$ is the density, $v$ is the fluid velocity, $p$ is the pressure, $A$ is the cross section of the nozzle, and}{Aqui $\rho$ \'e a densidade, $v$ \'e a velocidade do fluido, $p$ \'e a press\~ao, $A$ \'e a sec\c{c}\~ao transversal do bocal, e}
\begin{equation}
\epsilon = \frac{1}{2}\rho v^2 + \frac{p}{\gamma-1}
\end{equation}
\lang{is the energy density. The heat capacity ratio for di-atomic molecules of air is}{\'e a densidade da energia. O coeficiente de expans\~ao adiab\'atica para mol\'eculas di-at\^omicas do ar \'e}
$$\gamma=1+2/n=7/5=1.4 \qquad (n=5).$$
\langpar{We shall assume that the flow has no entropy discontinuity. Then the fluid is isentropic}{Assume-se que o fluxo n\~ao tem qualquer descontinuidade da entropia. Ent\~ao o fluido \'e isentr\'opico}
\begin{equation}
p\propto\rho^\gamma \,.
\label{eq:isentropic}
\end{equation}

\langpar{Instead of (\ref{eq:momentum}), we can use the Euler equation}{Em vez de (\ref{eq:momentum}), pode-se usar a equa\c{c}\~ao de Euler}
\begin{equation}
\rho(\partial_t + v \partial_x)v = -\partial_x p \,.
\label{eq:Euler}
\end{equation}
\lang{For isentropic fluid, (\ref{eq:Euler}) is reduced to the Bernoulli equation}{Para o fluido isentr\'opico, (\ref{eq:Euler}) \'e reduzido \`a equa\c{c}\~ao de Bernoulli}
\begin{equation}
\partial_t\Phi + \frac{1}{2}(\partial_x\Phi)^2 + h(\rho) = 0 \,,
\label{eq:Bernoulli}
\end{equation}
\lang{where $h(\rho) \equiv \int\rho^{-1}dp$ is the specific enthalpy and $\Phi = \int v\,dx$ is the velocity potential.}{onde $h(\rho) \equiv \int\rho^{-1}dp$ \'e a entalpia espec\'ifica e $\Phi = \int v\,dx$ \'e o potencial da velocidade.}

\langpar{According to \cite{Okuzumi:2007hf}, the perturbation equations in such a nozzle can be reduced to}{Seguindo \cite{Okuzumi:2007hf}, as equa\c{c}\~oes de perturba\c{c}\~ao em tal bocal podem ser reduzidas a}
\begin{gather}
\biggl[ \frac{d^2}{dx^{*2}} + \kappa^2 - V(x^*) \biggr] H_\omega = 0, \label{eq:Sch1}\\
\kappa = \frac{\omega}{c_{s0}}, \\
V(x^*) = \frac{1}{g^2}\biggl[\; \frac{g}{2}\frac{d^2g}{dx^{*2}}
    - \frac{1}{4}\Bigl(\frac{dg}{dx^*}\Bigr)^2 \;\biggr].
\end{gather}
\lang{Here $c_{s0}=\sqrt{\gamma RT/\mu}$ is the stagnation sound speed, which is used to measure $x^*$ in meters. The variable $x^{*}$ is an acoustic analogue of the tortoise coordinate which satisfies}{Aqui $c_{s0}=\sqrt{\gamma RT/\mu}$ \'e a velocidade de estagna\c{c}\~ao do som, que foi usada para medir $x^*$ em metros. A vari\'avel $x^{*}$ \'e um an\'alogo ac\'ustico da coordenada tartaruga que satisfaz}
$$x^{*}(x=+\infty) = + \infty,\qquad x^{*}(x=0) = - \infty,$$
\lang{namely,}{cuja solu\c{c}\~ao \'e}
\begin{equation}
x^{*} = c_{s0}\int \frac{c_{s} d x}{ c_{s}^2-v^{2}}.
\end{equation}
\lang{The function $H_{\omega}$ represents small perturbations of the gas flow,}{A fun\c{c}\~ao $H_{\omega}$ representa perturba\c{c}\~oes pequenas do fluxo do g\'as,}
\begin{equation}
H_\omega(x) = g^{1/2}\int dt~e^{i\omega(t-f(x))}\phi(t,x),
\end{equation}
\begin{gather}
g = \frac{\rho A}{c_s}, \label{gdef} \\
f(x) = \int\frac{|v|\,dx}{c_s^2-v^2}.
\end{gather}
\lang{The sound speed is given by}{A velocidade de som \'e dada por}
\begin{equation}
c_s^2 = \frac{dp}{d\rho}=\frac{\gamma p}{\rho}.
\end{equation}

\langpar{Since (\ref{eq:Sch1}) is invariant with respect to the re-scaling of $g$, we can fix the coefficients in (\ref{gdef}) arbitrarily:}{Como (\ref{eq:Sch1}) \'e invariante com respeito \`a re-escalada de $g$, pode-se fixar os coeficientes em (\ref {gdef}) arbitrariamente:}
\begin{equation}\label{constfixg}
g=\frac{\rho A}{2\rho^{(\gamma-1)/2}}.
\end{equation}
\lang{Up to a coefficient, $A$ can be found in terms of $\rho$ as \cite{Abdalla:2007dz}}{At\'e um coeficiente, $A$ pode ser achado em termos de $\rho$ como \cite{Abdalla:2007dz}}
\begin{equation}\label{constfixA}
A^{-1}=\left(1-\rho^{(\gamma-1)}\right)^{1/2}\rho.
\end{equation}

\langpar{We find}{Acha-se}
\begin{equation}
g=\frac{\rho^{(1-\gamma)/2}}{2\left(1-\rho^{(\gamma-1)}\right)^{1/2}}
=\frac{\rho^{(1-\gamma)}}{2\left(\rho^{(1-\gamma)}-1\right)^{1/2}}\,.
\end{equation}
\lang{Hence it follows that}{Daqui resulta que}
\begin{equation}\label{rhodef}
\rho^{1-\gamma}=2g^2\left(1\pm
\sqrt{1-g^{-2}}\right).
\end{equation}
\lang{The sign in (\ref{rhodef}) should be chosen in order that $\rho$ be a monotonous function with respect to the transverse coordinate. As we will show later, the function $g$ for the Schwarzschild black hole can be chosen also monotonous in the $R$ region, finite at the horizon and infinite at the spatial infinity. Therefore, we choose the minus sign,}{O sinal em (\ref{rhodef}) deve ser escolhido para que $\rho$ seja uma fun\c{c}\~ao monot\^onica com respeito \`a coordenada transversal. Como ser\'a mostrado depois, a fun\c{c}\~ao $g$ para o buraco negro de Schwarzschild pode ser escolhida tamb\'em monot\^onica na regi\~ao $R$, finita no horizonte e infinita no infinito espacial. Por isso, seleciona-se o sinal negativo,}
\begin{equation}
\rho^{1-\gamma}=2g^2\left(1-\sqrt{1-g^{-2}}\right),\qquad g>1.
\end{equation}

\langpar{Substituting (\ref{rhodef}) into (\ref{constfixA}), we find the cross-section area as a function of $g$,}{Substituindo (\ref{rhodef}) em (\ref{constfixA}), encontra-se a \'area transversal como uma fun\c{c}\~ao de $g$,}
\begin{equation}\label{2}
A=\frac{\sqrt{2}\left(2g^2\left(1-\sqrt{1-g^{-2}}\right)\right)^{1/(\gamma-1)}}{\sqrt{1-\sqrt{1-g^{-2}}}}.
\end{equation}

\langpar{For the steady isentropic flow, (\ref{eq:Bernoulli}) can be rewritten as}{Para o fluxo isentr\'opico constante, (\ref{eq:Bernoulli}) pode ser reescrito como}
\begin{equation}
\frac{v^2}{c_s^2}=\frac{2}{\gamma-1}\left(\rho^{1-\gamma}-1\right)=\frac{2}{\gamma-1}\left(2g^2\left(1-\sqrt{1-g^{-2}}\right)-1\right).
\end{equation}
\lang{Since $v=c_s$ at the event horizon, $g$ must be finite there, and}{Como $v=c_s$ no horizonte de eventos, $g$ deve ser finito l\'a, e}
\begin{equation}\label{normal}
g\Biggr|_{e.h.}=\frac{\gamma+1}{2\sqrt{2}\sqrt{\gamma-1}}=\frac{3}{\sqrt{5}}>1.
\end{equation}
\lang{This requirement fixes both constants of integration.}{Esta exig\^encia fixa ambas as constantes da integra\c{c}\~ao.}

\langsection{de Laval nozzle for the Schwarzschild black hole}{Bocal de Laval para o buraco negro de Schwarzschild}
\langpar{Since $g$ allows to find the cross section $A$ and, thereby, the form of de Laval nozzle, if we find such $g(x)$ that leads to the same expression for the nozzle potential (\ref{eq:Sch1}), as the effective potential for the Schwarzschild black hole \cite{Abdalla:2007dz},}{Como $g$ permite achar a sec\c{c}\~ao transversal $A$ e, por meio disso, a forma do bocal de Laval, se acha tal $g(x)$ que leva \`a mesma express\~ao para o potencial do bocal (\ref{eq:Sch1}), como o potencial efetivo para o buraco negro de Schwarzschild \cite{Abdalla:2007dz},}
\begin{equation}\label{Schwarzschild}
V(r)=\left(1-\frac{1}{r}\right)\left(\frac{l(l+1)}{r^2}+\frac{1-s^2}{r^3}\right),\qquad (r_+=1),
\end{equation}
\lang{where $l\geq s$ is the multipole integer number. The integer $s=0,1,2$ describes the perturbations of fields of different spin: $s=0$ for the test scalar field, $s=1$ for the Maxwell field and $s=2$ for the gravitational perturbations of axial type.}{onde $l\geq s$ \'e o n\'umero inteiro de multipolo. O n\'umero inteiro $s=0,1,2$ descreve as perturba\c{c}\~oes dos campos de spin diferente: $s=0$ para o campo escalar de teste, $s=1$ para o campo de Maxwell e $s=2$ para as perturba\c{c}\~oes gravitacionais do tipo axial.}

\langpar{In order to find $g$ that produces the required potential (\ref{Schwarzschild}), we identify the ``tortoise'' coordinates of the black hole solution and of de Laval nozzle}{Para achar $g$ que produz o potencial necess\'ario (\ref{Schwarzschild}), identificam-se as coordenadas ``tartarugas'' da solu\c{c}\~ao do buraco negro e do bocal de Laval}
\begin{equation}\label{identify}
dr^*=dx^*=\frac{\rho^{(1-\gamma)/2}dx}{1-v^2/c_s^2}=\frac{\sqrt{2g^2\left(1-\sqrt{1-g^{-2}}\right)}dx}{1-\frac{2}{\gamma-1}
\left(2g^2\left(1-\sqrt{1-g^{-2}}\right)-1\right)}.
\end{equation}

\langpar{The equation (\ref{identify}) relates the real coordinate of the nozzle $x$ and the radial coordinate of the Schwarzschild solution $r$. It is convenient because we are able to find the equation for $g(r)$ explicitly,}{A equa\c{c}\~ao (\ref{identify}) relaciona a coordenada real do bocal $x$ e a coordenada radial da solu\c{c}\~ao de Schwarzschild $r$. \'E conveniente porque \'e poss\'ivel achar a equa\c{c}\~ao para $g(r)$ explicitamente,}
\begin{equation}\label{gequation}
\frac{f(r)f'(r)g'(r)+f(r)^2g''(r)}{2g(r)}-\frac{f(r)^2g'(r)^2}{4g(r)^2}=V(r).
\end{equation}
\lang{This implies that the form of de Laval nozzle is parameterised by $r$ and  $x^{*}(r)$ = $r^{*}(r)$. Note, that as we chose the radius of the event horizon to be unity, the nozzle coordinate $x$ is measured in the units of the radius of the event horizon.}{Isto implica que a forma do bocal de Laval \'e parametrada por $r$ e $x^{*}(r)$ = $r^{*}(r)$. Observe, que como o raio do horizonte de eventos foi escolhido sendo unit\'ario, a coordenada do bocal $x$ \'e medida nas unidades de raio do horizonte de eventos.}

\langpar{The general solution of the equation (\ref{gequation}) contains two arbitrary constants. They can be fixed in a unique way by the condition (\ref{normal}). Namely, the requirement that the solution must be finite at $r=1$  fixes one of the constants. Then the other constant re-scales $g(r)$, and must be fixed by its value at $r=1$. Finally, the solution of (\ref{gequation}), for arbitrary $l$ and $s$, that satisfies (\ref{normal}), is given by the following formula,}{A solu\c{c}\~ao geral da equa\c{c}\~ao (\ref{gequation}) cont\'em duas constantes arbitr\'arias. Elas podem ser fixadas de um modo \'unico pela condi\c{c}\~ao (\ref{normal}). A exig\^encia que a solu\c{c}\~ao deve ser finita em $r=1$ fixa uma das constantes. Ent\~ao a outra constante re-escala $g(r)$, e deve ser fixada pelo seu valor em $r=1$. Finalmente, a solu\c{c}\~ao de (\ref{gequation}), para $l$ e $s$ arbitr\'arios, que satisfaz (\ref{normal}), \'e dada pela f\'ormula}
\begin{eqnarray}\nonumber
g(r)&=&\frac{\gamma+1}{2\sqrt{2}\sqrt{\gamma-1}}\sum_{n=s}^{l}
\left(\frac{(-1)^{n+s} (l + n)!}{( n + s )! ( n - s )! (l - n)!}r^{n+1}\right)^2 = \\\label{galgebraic}
&=&\small\frac{\gamma+1}{2\sqrt{2}\sqrt{\gamma-1}}r^{2s+2}\left(\frac{\Gamma(1+l+s)_2F_1(s-l,s+l+1,1+2s,r)}
{\Gamma(1+l-s)\Gamma(1+2s)}\right)^2.
\end{eqnarray}
\lang{One can easily check that the above solution indeed satisfies the equation (\ref{normal}), for any fixed $l$ and $s$.}{Pode-se verificar facilmente que a solu\c{c}\~ao acima de fato satisfaz a equa\c{c}\~ao (\ref{normal}), para quaisquer $l$ e $s$ fixos.}

\langpar{The equation (\ref{identify}) allows to find the dependance of the transversal nozzle coordinate $x$ on the parameter $r$,}{A equa\c{c}\~ao (\ref{identify}) permite achar a depend\^encia da coordenada transversal do bocal $x$ no par\^ametro $r$,}
\begin{equation}
x=\intop_1^r\frac{\left(\gamma+1-4g(r)^2\left(1-\sqrt{1-g(r)^{-2}}\right)\right)dr}
{f(r)(\gamma-1)\sqrt{2g(r)^2\left(1-\sqrt{1-g(r)^{-2}}\right)}}.
\end{equation}
\lang{The integration constant is chosen so that $x$ vanishes at the sonic point.}{A constante de integra\c{c}\~ao foi escolhida para que $x$ desapare\c{c}a no ponto s\^onico.}

\begin{figure}
$s=l=0$\\
\includeonegraph{s_l_0_nozzle}\\
$s=l=1$\\
\includeonegraph{s_l_1_nozzle}\\
$s=l=2$\\
\includeonegraph{s_l_2_nozzle}\\
\langfigurecaption{The form of de Laval nozzles and the effective potential in the nozzle coordinates.}{A forma dos bocais de Laval e o potencial efetivo nas coordenadas do bocal.}\label{nozzleform}
\end{figure}

\langpar{Now we are in position to find the required form of de Laval nozzle. i.e. to find its cross-section $A(x)$. We just need to replace $g(r)$ given in (\ref{galgebraic}) and go over to the transverse nozzle coordinate $x$. The radius of the de Laval nozzle as a function of the transverse coordinate $x$ is shown in figure \ref{nozzleform}. Note that the canonical de Laval nozzle is diverging at the end of the flow trajectory, so that $A_{x=\infty}=\infty$. Indeed, the formula (\ref{galgebraic}) implies divergence at least as $\sim r^2$. Nevertheless, the diverging of the nozzle still let us keep the one-dimensional representation of the motion, because $x$ is measured in units of black hole radius, i.e. one can ``pull'' the nozzle along the transverse coordinate $x$ in order to make the area of the nozzle change as slowly as one wishes. Such a ``pulling'' simply means that we are getting the correspondence with a larger black hole.}{Agora pode-se achar a forma necess\'aria do bocal de Laval. Isto \'e, achar a sec\c{c}\~ao transversal dela $A(x)$. Somente tem-se de substituir $g(r)$, dado em (\ref{galgebraic}), e passar \`a coordenada transversal do bocal $x$. O raio do bocal de Laval como uma fun\c{c}\~ao da coordenada transversal $x$ \'e mostrado na figura \ref{nozzleform}. Observe, que o bocal can\^onico de Laval est\'a divergindo no fim da trajet\'oria de fluxo, para que $A_{x=\infty}=\infty$. De fato, a f\'ormula (\ref{galgebraic}) implica a diverg\^encia pelo menos como $\sim r^2$. Todavia, a diverg\^encia do bocal ainda deixa manter a representa\c{c}\~ao unidimensional do movimento, porque $x$ \'e medido em unidades de raio do buraco negro, isto \'e, pode-se ``puxar'' o bocal ao longo da coordenada transversal $x$ para fazer a \'area do bocal modificar-se t\~ao lentamente quanto desejar. Tal ``pux\~ao'' significa simplesmente que se obtem a correspond\^encia com um buraco negro maior.}

\langpar{In a similar fashion, one can find the de Laval nozzle form for the gravitational perturbations of polar type \cite{Abdalla:2007dz}. The form of the nozzles for modeling polar and axial gravitational perturbations are very slightly different.}{De maneira semelhante, pode-se achar a forma do bocal de Laval para as perturba\c{c}\~oes gravitacionais do tipo polar \cite{Abdalla:2007dz}. As formas dos bocais para modelar perturba\c{c}\~oes gravitacionais polares e axiais s\~ao muito ligeiramente diferentes.}

\langpar{It should be recalled also that the \emph{precise} acoustic analogy is only established for a scalar field. To be able to reproduce the potential $V(r)$ for fields of different spins, certainly, does not mean that one can reproduce all the characteristics of those fields in an acoustic model.}{Deve ser relembrado tamb\'em que a analogia ac\'ustica \emph{precisa} \'e estabelecida somente para um campo escalar. A possibilidade de se reproduzir o potencial $V(r)$ de campos de spins diferentes, certamente, n\~ao significa que se pode reproduzir todas as caracter\'isticas daqueles campos em um modelo ac\'ustico.}

\langpar{The obtained acoustic analogue for the perturbations of the Schwarzschild black holes is not limited by quasi-normal mode problems only, but allow general investigation of propagation of fields, including such processes as scattering and tunneling of waves and particles.}{O an\'alogo ac\'ustico, obtido para as perturba\c{c}\~oes dos buracos negros de Schwarzschild, n\~ao \'e limitado somente por problemas de modos quase-normais, mas permite a investiga\c{c}\~ao geral de propaga\c{c}\~ao de campos, inclusive tais processos como espalhamento e tunelamento de ondas e part\'iculas.}

\langchapter{Summary}{Sum\'ario}
\langpar{Let us summarise the results of this dissertation.}{Resumindo os resultados desta tese:}

\langpar{$\bullet$ We have performed the complete study of the influence of the cosmological constant on the quasi-normal spectrum of the Schwarzschild black hole for fields of various spins. We have found that the presence of the cosmological constant decreases the absolute value of the real and imaginary parts of quasi-normal frequencies. Also we have found analytically the large multipole limit for quasi-normal frequencies of Schwarzschild-de Sitter black holes (sec. \ref{sec:SdS}).}{$\bullet$ Foi feito o estudo completo da influ\^encia da constante cosmol\'ogica no espectro quase-normal do buraco negro de Schwarzschild para campos de v\'arios spins. Foi achado que a presen\c{c}a da constante cosmol\'ogica diminui os valores absolutos das partes real e imagin\'aria de freq\"u\^encias quase-normais. Tamb\'em foi achado analiticamente o limite de multipolo grande para freq\"u\^encias quase-normais de buracos negros Schwarzschild-de Sitter (se\c{c}\~ao \ref{sec:SdS}).}

\langpar{$\bullet$ We have studied the behavior of high overtones for the massless Dirac and massive scalar fields in the Schwarzschild background and for the test electromagnetic field and gravitational perturbations of the Schwarzschild-de Sitter black hole. We have shown numerically that the real part of the quasi-normal frequency asymptotically approaches zero for the Dirac and electromagnetic fields, but has an oscillatory behavior for the gravitational perturbations of the Schwarzschild-de Sitter black hole. These results were confirmed later analytically (sec. \ref{sec:highovertones}). The behavior of high overtones for the massive scalar field was studied both analytically and numerically. We have shown that the asymptotical behavior does not depend on the mass of the field, coinciding with the known analytical formula for the massless scalar field (sec. \ref{sec:quasi-resonance}).}{$\bullet$ Foi estudado o comportamento de sobretons altos para os campos de Dirac sem massa e escalar massivo no contexto de Schwarzschild e para o campo electromagn\'etico de teste e as perturba\c{c}\~oes gravitacionais do buraco negro Schwarzschild-de Sitter. Foi mostrado numericamente que a parte real da freq\"u\^encia quase-normal assintoticamente aproxima o zero para os campos de Dirac e eletromagn\'etico, mas tem um comportamento oscilat\'orio para as perturba\c{c}\~oes gravitacionais do buraco negro Schwarzschild-de Sitter. Esses resultados foram confirmados depois analiticamente (se\c{c}\~ao  \ref{sec:highovertones}). O comportamento de sobretons altos para o campo escalar massivo foi estudado tanto analiticamente como numericamente. Foi mostrado que o comportamento assint\'otico n\~ao depende da massa do campo e coincide com a f\'ormula anal\'itica conhecida para o campo escalar sem massa (se\c{c}\~ao \ref{sec:quasi-resonance}).}

\langpar{$\bullet$ We have investigated the quasi-normal spectrum of the electrically charged scalar and Dirac fields in the background of the Kerr-Newman-de Sitter black hole. Special attention was paid to the influence of the electromagnetic interaction between the black hole and the test field upon the quasi-normal spectrum (sec. \ref{sec:KNdS4D}).}{$\bullet$ Foi investigado o espectro quase-normal dos campos escalar e Dirac eletricamente carregados no contexto do buraco negro Kerr-Newman-de Sitter. Uma aten\c{c}\~ao especial foi prestada \`a influ\^encia da intera\c{c}\~ao eletromagn\'etica entre o buraco negro e o campo de teste no espectro quase-normal (se\c{c}\~ao \ref{sec:KNdS4D}).}

\langpar{$\bullet$ We have studied the quasi-normal spectrum of gravitational perturbations of the neutral and electrically charged black holes in higher dimensions with a positive cosmological constant. We have considered all kinds of the perturbations and prove that higher dimensional Schwarzschild-de Sitter black holes are stable. We have shown that the higher dimensional black hole is unstable if both the black hole charge and the cosmological constant are large enough. We have found the parametrical region of the instability (sec. \ref{sec:HDRNdS}).}{$\bullet$ Foi estudado o espectro quase-normal de perturba\c{c}\~oes gravitacionais dos buracos negros neutros e eletricamente carregados em dimens\~oes mais altas com uma constante cosmol\'ogica positiva. Foram consideradas todas as esp\'ecies das perturba\c{c}\~oes e comprovado que buracos negros Schwarzschild-de Sitter em dimens\~oes mais altas s\~ao est\'aveis. Foi mostrado que os buracos negros em dimens\~oes mais altas s\~ao inst\'aveis se tanto a carga do buraco negro como a constante cosmol\'ogica forem grandes o suficiente. Foi achada a regi\~ao param\'etrica da instabilidade (se\c{c}\~ao \ref{sec:HDRNdS}).}

\langpar{$\bullet$ We have considered gravitational perturbations of higher dimensional black holes in the Einstein-Gauss-Bonnet theory. We have found that the presence of the Gauss-Bonnet parameter decreases the imaginary part of quasi-normal frequencies, causing the oscillations be longer-lived. Also we have studied the developing of instability of Gauss-Bonnet black holes in five and six dimensions (sec. \ref{sec:Gauss-Bonnet}).}{$\bullet$ Foram consideradas perturba\c{c}\~oes gravitacionais de buracos negros em dimens\~oes mais altas na teoria de Einstein-Gauss-Bonnet. Foi achado que a presen\c{c}a do par\^ametro de Gauss-Bonnet diminui a parte imagin\'aria de freq\"u\^encias quase-normais, fazendo com que as oscila\c{c}\~oes sejam de vida mais longa. Tamb\'em foi estudado o desenvolvimento da instabilidade dos buracos negros de Gauss-Bonnet em cinco e seis dimens\~oes (se\c{c}\~ao \ref{sec:Gauss-Bonnet}).}

\langpar{$\bullet$ We have found quasi-normal modes and late-time tails of the rotating black holes and the non-rotating Gauss-Bonnet black holes for the Standard Model fields localised on a 4-brane (sec. \ref{sec:BH-projected}).}{$\bullet$ Foram achados modos quase-normais e caudas de tempo tardio dos buracos negros com rota\c{c}\~ao e dos buracos negros Gauss-Bonnet sem rota\c{c}\~ao para os campos do Modelo Padr\~ao localizados em uma 4-brana (se\c{c}\~ao \ref{sec:BH-projected}).}

\langpar{$\bullet$ We have studied quasi-normal modes of the Kaluza-Klein black holes with squashed horizons. We have considered the test scalar field near rotating squashed Kaluza-Klein black holes and gravitational perturbations of non-rotating squashed Kaluza-Klein black holes (sec. \ref{sec:squashedKK}).}{$\bullet$ Foram estudados modos quase-normais dos buracos negros de Kaluza-Klein com horizontes esmagados. Foram considerados o campo escalar de teste perto de buracos negros esmagados de Kaluza-Klein com rota\c{c}\~ao e as perturba\c{c}\~oes gravitacionais de buracos negros esmagados de Kaluza-Klein sem rota\c{c}\~ao (se\c{c}\~ao \ref{sec:squashedKK}).}

\langpar{$\bullet$ We have calculated the quasi-normal spectra of the massive scalar field in the backgrounds of the Tangherlini black hole (sec. \ref{sec:quasi-resonance}), the Kerr black hole (sec. \ref{sec:massive-Kerr}) and the scalar hairy asymptotically anti-de Sitter black hole (sec. \ref{sec:hairyBH}). We have found that the massive scalar field is stable in these backgrounds and provided a comprehensive discussions about properties of their quasi-normal spectrum and quasi-resonances (sec. \ref{sec:massive-properties}). Also we have studied the late-time tails for the massive vector field in the Schwarzschild background and found that the asymptotically late-time decay law does not depend on the spin of the massive field (sec. \ref{sec:late-time-tails}).}{$\bullet$ Foram calculados os espectros quase-normais do campo escalar massivo nos contextos do buraco negro de Tangherlini (se\c{c}\~ao \ref{sec:quasi-resonance}), do buraco negro de Kerr (se\c{c}\~ao \ref{sec:massive-Kerr}) e do buraco negro com cabelo escalar assintoticamente anti-de Sitter (se\c{c}\~ao \ref{sec:hairyBH}). Foi achado que o campo escalar massivo \'e est\'avel nesses contextos e foram fornecidas discuss\~oes abrangentes sobre propriedades do espectro quase-normal deles e das quase-resson\^ancias (se\c{c}\~ao \ref{sec:massive-properties}). Tamb\'em foram estudadas as caudas de tempo tardio para o campo vetorial massivo no contexto de Schwarzschild e encontrado que assintoticamente a lei de decaimento de tempo tardio n\~ao depende do spin do campo massivo (se\c{c}\~ao \ref{sec:late-time-tails}).}

\langpar{$\bullet$ We have found quasi-normal modes and tails of the gravitational perturbations of black strings. Also we have considered the developing of the long-wavelength instability of a black string in time domain (sec. \ref{sec:Gregory-Laflamme}).}{$\bullet$ Foram achados modos quase-normais e caudas das perturba\c{c}\~oes gravitacionais de cordas negras. Tamb\'em foi considerado o desenvolvimento da instabilidade de uma corda negra de ondas de comprimento longo no dom\'inio de tempo (se\c{c}\~ao \ref{sec:Gregory-Laflamme}).}

\langpar{$\bullet$ We have studied the influence of the local Lorentz symmetry breaking within the Einstein-Aether theory on the quasi-normal spectrum of the Schwarzschild black hole. We have found that the presence of the aether increases the absolute value of the real and imaginary parts of the fundamental quasi-normal frequency (sec. \ref{sec:Einstein-Aether}).}{$\bullet$ Foi estudada a influ\^encia da quebra da simetria Lorentz local dentro da teoria Einstein-A\'eter no espectro quase-normal do buraco negro de Schwarzschild. Foi achado que a presen\c{c}a do a\'eter aumenta o valor absoluto das partes real e imagin\'aria da freq\"u\^encia quase-normal fundamental (se\c{c}\~ao \ref{sec:Einstein-Aether}).}

\langpar{$\bullet$ We have proposed a possibility of observation of the acoustic analogue of the Schwarzschild black hole in a de Laval nozzle. We have found the particular forms of the de Laval nozzles for which the effective potentials will coincide with the effective potentials for test fields or gravitational perturbations of the Schwarzschild black hole (chapter \ref{sec:laval-nozzle}).}{$\bullet$ Foi proposta uma possibilidade da observa\c{c}\~ao do an\'alogo ac\'ustico do buraco negro de Schwarzschild em um bocal de Laval. Foram achadas as formas particulares dos bocais de Laval para os quais os potenciais efetivos coincidir\~ao com os potenciais efetivos para campos de teste ou perturba\c{c}\~oes gravitacionais do buraco negro de Schwarzschild (cap\'itulo \ref{sec:laval-nozzle}).}

\langpar{$\bullet$ We have developed two new numerical tools:\\
1) the generalisation of the Nollert improvement of the Frobenius method for higher dimensional problems, which provides better convergence of the numerical procedure (sec. \ref{sec:Frobenius}).\\
2) the method for the calculation of the quasi-normal frequencies of a black hole, which metric is not known analytically, but can be found as a numerical solution of a set of differential equations (sec. \ref{sec:non-analytical-method}).
}{$\bullet$ Foram desenvolvidos dois novos instrumentos num\'ericos:\\
1) a generaliza\c{c}\~ao da melhora de Nollert do m\'etodo de Frobenius para problemas em dimens\~oes mais altas, que fornece uma melhor converg\^encia do procedimento num\'erico (se\c{c}\~ao \ref{sec:Frobenius}).\\
2) o m\'etodo do c\'alculo das freq\"u\^encias quase-normais de um buraco negro, cuja m\'etrica n\~ao \'e conhecida analiticamente, mas pode ser achada como uma solu\c{c}\~ao num\'erica de um conjunto de equa\c{c}\~oes diferenciais (se\c{c}\~ao \ref{sec:non-analytical-method}).
}%

\langpar{All the results reported here were obtained by us for the first time, and are, thereby, new.}{Todos os resultados apresentados aqui foram obtidos por n\'os pela primeira vez, e s\~ao, por meio disso, novos.}


\begin{thebibliography}{100}

\bibitem{Flanagan:1997td}
E.~E. Flanagan,
\newblock (1997), [gr-qc/9804024].

\bibitem{PhysRevLett.71.2851}
P.~Anninos, D.~Hobill, E.~Seidel, L.~Smarr, and W.-M. Suen,
\newblock Phys. Rev. Lett. {\bf 71}, 2851 (1993).

\bibitem{Vishveshwara}
C.~V. Vishveshwara,
\newblock Nature {\bf 227}, 936 (1970).

\bibitem{Eling:2004dk}
C.~Eling, T.~Jacobson, and D.~Mattingly,
\newblock (2004), [gr-qc/0410001].

\bibitem{Emparan:2008eg}
R.~Emparan and H.~S. Reall,
\newblock Living Rev. Rel. {\bf 11}, 6 (2008), [0801.3471].

\bibitem{Maldacena:1997re}
J.~M. Maldacena,
\newblock Adv. Theor. Math. Phys. {\bf 2}, 231 (1998), [hep-th/9711200].

\bibitem{Horowitz:1999jd}
G.~T. Horowitz and V.~E. Hubeny,
\newblock Phys. Rev. {\bf D62}, 024027 (2000), [hep-th/9909056].

\bibitem{Policastro:2002se}
G.~Policastro, D.~T. Son, and A.~O. Starinets,
\newblock JHEP {\bf 09}, 043 (2002), [hep-th/0205052].

\bibitem{Zhidenko:2003wq}
A.~Zhidenko,
\newblock Class. Quant. Grav. {\bf 21}, 273 (2004), [gr-qc/0307012].

\bibitem{Konoplya:2004uk}
R.~A. Konoplya and A.~Zhidenko,
\newblock JHEP {\bf 06}, 037 (2004), [hep-th/0402080].

\bibitem{CastelloBranco:2004nk}
K.~H.~C. Castello-Branco, R.~A. Konoplya, and A.~Zhidenko,
\newblock Phys. Rev. {\bf D71}, 047502 (2005), [hep-th/0411055].

\bibitem{Konoplya:2004wg}
R.~A. Konoplya and A.~V. Zhidenko,
\newblock Phys. Lett. {\bf B609}, 377 (2005), [gr-qc/0411059].

\bibitem{Zhidenko:2005mv}
A.~Zhidenko,
\newblock Class. Quant. Grav. {\bf 23}, 3155 (2006), [gr-qc/0510039].

\bibitem{Konoplya:2006br}
R.~A. Konoplya and A.~Zhidenko,
\newblock Phys. Rev. {\bf D73}, 124040 (2006), [gr-qc/0605013].

\bibitem{Konoplya:2006gq}
R.~A. Konoplya, C.~Molina, and A.~Zhidenko,
\newblock Phys. Rev. {\bf D75}, 084004 (2007), [gr-qc/0602047].

\bibitem{Konoplya:2006rv}
R.~A. Konoplya and A.~Zhidenko,
\newblock Phys. Lett. {\bf B644}, 186 (2007), [gr-qc/0605082].

\bibitem{Kanti:2006ua}
P.~Kanti, R.~A. Konoplya, and A.~Zhidenko,
\newblock Phys. Rev. {\bf D74}, 064008 (2006), [gr-qc/0607048].

\bibitem{Zhidenko:2006rs}
A.~Zhidenko,
\newblock Phys. Rev. {\bf D74}, 064017 (2006), [gr-qc/0607133].

\bibitem{Konoplya:2006ar}
R.~A. Konoplya and A.~Zhidenko,
\newblock Phys. Lett. {\bf B648}, 236 (2007), [hep-th/0611226].

\bibitem{Konoplya:2007jv}
R.~A. Konoplya and A.~Zhidenko,
\newblock Nucl. Phys. {\bf B777}, 182 (2007), [hep-th/0703231].

\bibitem{Abdalla:2007dz}
E.~Abdalla, R.~A. Konoplya, and A.~Zhidenko,
\newblock Class. Quant. Grav. {\bf 24}, 5901 (2007), [0706.2489].

\bibitem{Konoplya:2007zx}
R.~A. Konoplya and A.~Zhidenko,
\newblock Phys. Rev. {\bf D76}, 084018 (2007), [0707.1890].

\bibitem{Konoplya:2008ix}
R.~A. Konoplya and A.~Zhidenko,
\newblock Phys. Rev. {\bf D77}, 104004 (2008), [0802.0267].

\bibitem{Ishihara:2008re}
H.~Ishihara, M.~Kimura, R.~A.~Konoplya, K.~Murata, J.~Soda and A.~Zhidenko,
\newblock Phys. Rev. {\bf D77}, 084019 (2008), [0802.0655].

\bibitem{Zhidenko:2008fp}
A.~Zhidenko,
\newblock Phys. Rev. {\bf D78}, 024007 (2008), [0802.2262].

\bibitem{Konoplya:2008yy}
R.~A. Konoplya, K.~Murata, J.~Soda, and A.~Zhidenko,
\newblock Phys. Rev. {\bf D78}, 084012 (2008), [0807.1897].

\bibitem{Konoplya:2008au}
R.~A. Konoplya and A.~Zhidenko,
\newblock (2008), [0809.2822].

\bibitem{Zhidenko:2007sj}
A.~Zhidenko,
\newblock (2007), [0705.2254].

\bibitem{Brill:1957fx}
D.~R. Brill and J.~A. Wheeler,
\newblock Rev. Mod. Phys. {\bf 29}, 465 (1957).

\bibitem{Mellor:1989ac}
F.~Mellor and I.~Moss,
\newblock Phys. Rev. {\bf D41}, 403 (1990).

\bibitem{PhysRev.108.1063}
T.~Regge and J.~A. Wheeler,
\newblock Phys. Rev. {\bf 108}, 1063 (1957).

\bibitem{Kodama:2003jz}
H.~Kodama and A.~Ishibashi,
\newblock Prog. Theor. Phys. {\bf 110}, 701 (2003), [hep-th/0305147].

\bibitem{Chandrasekhar}
S.~Chandrasekhar,
\newblock {\em The Mathematical Theory of Black Holes} (Oxford University
  Press, New York, 1983).

\bibitem{Nollert:1999ji}
H.-P. Nollert,
\newblock Class. Quant. Grav. {\bf 16}, R159 (1999).

\bibitem{Gundlach:1993tp}
C.~Gundlach, R.~H. Price, and J.~Pullin,
\newblock Phys. Rev. {\bf D49}, 883 (1994), [gr-qc/9307009].

\bibitem{Brady:1996za}
P.~R. Brady, C.~M. Chambers, W.~Krivan, and P.~Laguna,
\newblock Phys. Rev. {\bf D55}, 7538 (1997), [gr-qc/9611056].

\bibitem{Brady:1999wd}
P.~R. Brady, C.~M. Chambers, W.~G. Laarakkers, and E.~Poisson,
\newblock Phys. Rev. {\bf D60}, 064003 (1999), [gr-qc/9902010].

\bibitem{Kovtun:2005ev}
P.~K. Kovtun and A.~O. Starinets,
\newblock Phys. Rev. {\bf D72}, 086009 (2005), [hep-th/0506184].

\bibitem{Kodama:2003ck}
H.~Kodama and A.~Ishibashi,
\newblock (2003), [gr-qc/0312012].

\bibitem{Berti:2007dg}
E.~Berti, V.~Cardoso, J.~A. Gonzalez, and U.~Sperhake,
\newblock Phys. Rev. {\bf D75}, 124017 (2007), [gr-qc/0701086].

\bibitem{Abdalla:2006fv}
E.~Abdalla and D.~Giugno,
\newblock Braz. J. Phys. {\bf 37}, 450 (2007), [gr-qc/0611023].

\bibitem{Phys.Lett.A100.5.231}
H.-J. Blome and B.~Mashhoon,
\newblock Phys. Lett. A {\bf 100}, 231 (1984).

\bibitem{PhysRevLett.52.1361}
V.~Ferrari and B.~Mashhoon,
\newblock Phys. Rev. Lett. {\bf 52}, 1361 (1984).

\bibitem{PoschlTeller}
G.~P\"oschl and E.~Teller,
\newblock Z. Phys. {\bf 83}, 143 (1933).

\bibitem{Cardoso:2003sw}
V.~Cardoso and J.~P.~S. Lemos,
\newblock Phys. Rev. {\bf D67}, 084020 (2003), [gr-qc/0301078].

\bibitem{Molina:2003ff}
C.~Molina,
\newblock Phys. Rev. {\bf D68}, 064007 (2003), [gr-qc/0304053].

\bibitem{Astrophys.J.Lett.291.L33.1985}
B.~Schutz and C.~Will,
\newblock Astrophys. J. Lett. {\bf 291}, L33 (1985).

\bibitem{PhysRevD.35.3621}
S.~Iyer and C.~M. Will,
\newblock Phys. Rev. D {\bf 35}, 3621 (1987).

\bibitem{Konoplya:2003ii}
R.~A. Konoplya,
\newblock Phys. Rev. {\bf D68}, 024018 (2003), [gr-qc/0303052].

\bibitem{Konoplya:2003dd}
R.~A. Konoplya,
\newblock Phys. Rev. {\bf D68}, 124017 (2003), [hep-th/0309030].

\bibitem{Konoplya:2004ip}
R.~A. Konoplya,
\newblock J. Phys. Stud. {\bf 8}, 93 (2004).

\bibitem{Froeman:1992gp}
N.~Froeman, P.~O. Froeman, N.~Andersson, and A.~Hoekback,
\newblock Phys. Rev. {\bf D45}, 2609 (1992).

\bibitem{Leaver:1985ax}
E.~W. Leaver,
\newblock Proc. Roy. Soc. Lond. {\bf A402}, 285 (1985).

\bibitem{PhysRevD.47.5253}
H.-P. Nollert,
\newblock Phys. Rev. D {\bf 47}, 5253 (1993).

\bibitem{Rostworowski:2006bp}
A.~Rostworowski,
\newblock Acta Phys. Polon. {\bf B38}, 81 (2007), [gr-qc/0606110].

\bibitem{Suzuki:1998vy}
H.~Suzuki, E.~Takasugi, and H.~Umetsu,
\newblock Prog. Theor. Phys. {\bf 100}, 491 (1998), [gr-qc/9805064].

\bibitem{Ohashi:2004wr}
A.~Ohashi and M.-a. Sakagami,
\newblock Class. Quant. Grav. {\bf 21}, 3973 (2004), [gr-qc/0407009].

\bibitem{Kodama:2003kk}
H.~Kodama and A.~Ishibashi,
\newblock Prog. Theor. Phys. {\bf 111}, 29 (2004), [hep-th/0308128].

\bibitem{Ishibashi:2003ap}
A.~Ishibashi and H.~Kodama,
\newblock Prog. Theor. Phys. {\bf 110}, 901 (2003), [hep-th/0305185].

\bibitem{Kokkotas:1999bd}
K.~D. Kokkotas and B.~G. Schmidt,
\newblock Living Rev. Rel. {\bf 2}, 2 (1999), [gr-qc/9909058].

\bibitem{PhysRevLett.81.4293}
S.~Hod,
\newblock Phys. Rev. Lett. {\bf 81}, 4293 (1998).

\bibitem{Barbero:1994ap}
J.~F. Barbero~G.,
\newblock Phys. Rev. {\bf D51}, 5507 (1995), [gr-qc/9410014].

\bibitem{Immirzi:1996dr}
G.~Immirzi,
\newblock Nucl. Phys. Proc. Suppl. {\bf 57}, 65 (1997), [gr-qc/9701052].

\bibitem{Motl:2002hd}
L.~Motl,
\newblock Adv. Theor. Math. Phys. {\bf 6}, 1135 (2003), [gr-qc/0212096].

\bibitem{Jing:2005dt}
J.-l. Jing,
\newblock Phys. Rev. {\bf D71}, 124006 (2005), [gr-qc/0502023].

\bibitem{Cardoso:2004up}
V.~Cardoso, J.~Natario, and R.~Schiappa,
\newblock J. Math. Phys. {\bf 45}, 4698 (2004), [hep-th/0403132].

\bibitem{Hod:1997mt}
S.~Hod and T.~Piran,
\newblock Phys. Rev. {\bf D58}, 024017 (1998), [gr-qc/9712041].

\bibitem{Hod:1997fy}
S.~Hod and T.~Piran,
\newblock Phys. Rev. {\bf D58}, 024018 (1998), [gr-qc/9801001].

\bibitem{Konoplya:2002ky}
R.~A. Konoplya,
\newblock Phys. Rev. {\bf D66}, 084007 (2002), [gr-qc/0207028].

\bibitem{Konoplya:2002wt}
R.~A. Konoplya,
\newblock Phys. Lett. {\bf B550}, 117 (2002), [gr-qc/0210105].

\bibitem{Zhou:2003ts}
W.~Zhou and J.-Y. Zhu,
\newblock Int. J. Mod. Phys. {\bf D13}, 1105 (2004), [gr-qc/0309071].

\bibitem{Jing:2004zb}
J.~Jing,
\newblock Phys. Rev. {\bf D72}, 027501 (2005), [gr-qc/0408090].

\bibitem{He:2006jv}
X.~He and J.~Jing,
\newblock Nucl. Phys. {\bf B755}, 313 (2006), [gr-qc/0611003].

\bibitem{ArkaniHamed:1998rs}
N.~Arkani-Hamed, S.~Dimopoulos, and G.~R. Dvali,
\newblock Phys. Lett. {\bf B429}, 263 (1998), [hep-ph/9803315].

\bibitem{ArkaniHamed:1998nn}
N.~Arkani-Hamed, S.~Dimopoulos, and G.~R. Dvali,
\newblock Phys. Rev. {\bf D59}, 086004 (1999), [hep-ph/9807344].

\bibitem{Antoniadis:1998ig}
I.~Antoniadis, N.~Arkani-Hamed, S.~Dimopoulos, and G.~R. Dvali,
\newblock Phys. Lett. {\bf B436}, 257 (1998), [hep-ph/9804398].

\bibitem{Randall:1999ee}
L.~Randall and R.~Sundrum,
\newblock Phys. Rev. Lett. {\bf 83}, 3370 (1999), [hep-ph/9905221].

\bibitem{Randall:1999vf}
L.~Randall and R.~Sundrum,
\newblock Phys. Rev. Lett. {\bf 83}, 4690 (1999), [hep-th/9906064].

\bibitem{Abdalla:2006qj}
E.~Abdalla, B.~Cuadros-Melgar, A.~B. Pavan, and C.~Molina,
\newblock Nucl. Phys. {\bf B752}, 40 (2006), [gr-qc/0604033].

\bibitem{Cremades:2002dh}
D.~Cremades, L.~E. Ibanez, and F.~Marchesano,
\newblock Nucl. Phys. {\bf B643}, 93 (2002), [hep-th/0205074].

\bibitem{Kokorelis:2002qi}
C.~Kokorelis,
\newblock Nucl. Phys. {\bf B677}, 115 (2004), [hep-th/0207234].

\bibitem{Argyres:1998qn}
P.~C. Argyres, S.~Dimopoulos, and J.~March-Russell,
\newblock Phys. Lett. {\bf B441}, 96 (1998), [hep-th/9808138].

\bibitem{Banks:1999gd}
T.~Banks and W.~Fischler,
\newblock (1999), [hep-th/9906038].

\bibitem{Giddings:2001bu}
S.~B. Giddings and S.~D. Thomas,
\newblock Phys. Rev. {\bf D65}, 056010 (2002), [hep-ph/0106219].

\bibitem{Dimopoulos:2001hw}
S.~Dimopoulos and G.~L. Landsberg,
\newblock Phys. Rev. Lett. {\bf 87}, 161602 (2001), [hep-ph/0106295].

\bibitem{Tangherlini:1963bw}
F.~R. Tangherlini,
\newblock Nuovo Cim. {\bf 27}, 636 (1963).

\bibitem{Boulware:1985wk}
D.~G. Boulware and S.~Deser,
\newblock Phys. Rev. Lett. {\bf 55}, 2656 (1985).

\bibitem{Konoplya:2004xx}
R.~Konoplya,
\newblock Phys. Rev. {\bf D71}, 024038 (2005), [hep-th/0410057].

\bibitem{Abdalla:2005hu}
E.~Abdalla, R.~A. Konoplya, and C.~Molina,
\newblock Phys. Rev. {\bf D72}, 084006 (2005), [hep-th/0507100].

\bibitem{Dotti:2005sq}
G.~Dotti and R.~J. Gleiser,
\newblock Phys. Rev. {\bf D72}, 044018 (2005), [gr-qc/0503117].

\bibitem{Gleiser:2005ra}
R.~J. Gleiser and G.~Dotti,
\newblock Phys. Rev. {\bf D72}, 124002 (2005), [gr-qc/0510069].

\bibitem{Beroiz:2007gp}
M.~Beroiz, G.~Dotti, and R.~J. Gleiser,
\newblock Phys. Rev. {\bf D76}, 024012 (2007), [hep-th/0703074].

\bibitem{Kanti:2002nr}
P.~Kanti and J.~March-Russell,
\newblock Phys. Rev. {\bf D66}, 024023 (2002), [hep-ph/0203223].

\bibitem{Kanti:2002ge}
P.~Kanti and J.~March-Russell,
\newblock Phys. Rev. {\bf D67}, 104019 (2003), [hep-ph/0212199].

\bibitem{Kanti:2005xa}
P.~Kanti and R.~A. Konoplya,
\newblock Phys. Rev. {\bf D73}, 044002 (2006), [hep-th/0512257].

\bibitem{Kanti:2004nr}
P.~Kanti,
\newblock Int. J. Mod. Phys. {\bf A19}, 4899 (2004), [hep-ph/0402168].

\bibitem{Myers:1986un}
R.~C. Myers and M.~J. Perry,
\newblock Ann. Phys. {\bf 172}, 304 (1986).

\bibitem{Rychkov:2004sf}
V.~S. Rychkov,
\newblock Phys. Rev. {\bf D70}, 044003 (2004), [hep-ph/0401116].

\bibitem{Ishihara:2007ni}
H.~Ishihara and J.~Soda,
\newblock Phys. Rev. {\bf D76}, 064022 (2007), [hep-th/0702180].

\bibitem{Kimura:2007cr}
M.~Kimura, K.~Murata, H.~Ishihara, and J.~Soda,
\newblock Phys. Rev. {\bf D77}, 064015 (2008), [arXiv:0712.4202 [hep-th]].

\bibitem{Hod:1998ra}
S.~Hod and T.~Piran,
\newblock Phys. Rev. {\bf D58}, 044018 (1998), [gr-qc/9801059].

\bibitem{Konoplya:2007yy}
R.~A. Konoplya and R.~D.~B. Fontana,
\newblock Phys. Lett. {\bf B659}, 375 (2008), [0707.1156].

\bibitem{Simone:1991wn}
L.~E. Simone and C.~M. Will,
\newblock Class. Quant. Grav. {\bf 9}, 963 (1992).

\bibitem{Cho:2003qe}
H.~T. Cho,
\newblock Phys. Rev. {\bf D68}, 024003 (2003), [gr-qc/0303078].

\bibitem{Konoplya:2005hr}
R.~A. Konoplya,
\newblock Phys. Rev. {\bf D73}, 024009 (2006), [gr-qc/0509026].

\bibitem{Beyer:2000fz}
H.~R. Beyer,
\newblock Commun. Math. Phys. {\bf 221}, 659 (2001), [astro-ph/0008236].

\bibitem{Gregory:1993vy}
R.~Gregory and R.~Laflamme,
\newblock Phys. Rev. Lett. {\bf 70}, 2837 (1993), [hep-th/9301052].

\bibitem{Gregory:1994bj}
R.~Gregory and R.~Laflamme,
\newblock Nucl. Phys. {\bf B428}, 399 (1994), [hep-th/9404071].

\bibitem{Hovdebo:2006jy}
J.~L. Hovdebo and R.~C. Myers,
\newblock Phys. Rev. {\bf D73}, 084013 (2006), [hep-th/0601079].

\bibitem{Price:1971fb}
R.~H. Price,
\newblock Phys. Rev. {\bf D5}, 2419 (1972).

\bibitem{Price:1972pw}
R.~H. Price,
\newblock Phys. Rev. {\bf D5}, 2439 (1972).

\bibitem{Bicak:1972}
J.~Bi\v{c}\'ak,
\newblock Gen. Relativ. Gravit. {\bf 3}, 331 (1972).

\bibitem{Cardoso:2003jf}
V.~Cardoso, S.~Yoshida, O.~J.~C. Dias, and J.~P.~S. Lemos,
\newblock Phys. Rev. {\bf D68}, 061503 (2003), [hep-th/0307122].

\bibitem{Bizon:2007vr}
P.~Bizon, T.~Chmaj, and A.~Rostworowski,
\newblock Phys. Rev. {\bf D76}, 124035 (2007), [0708.1769].

\bibitem{Bizon:2008iz}
P.~Bizon, T.~Chmaj, and A.~Rostworowski,
\newblock (2008), [0812.4333].

\bibitem{Koyama:2001ee}
H.~Koyama and A.~Tomimatsu,
\newblock Phys. Rev. {\bf D64}, 044014 (2001), [gr-qc/0103086].

\bibitem{Koyama:2001qw}
H.~Koyama and A.~Tomimatsu,
\newblock Phys. Rev. {\bf D65}, 084031 (2002), [gr-qc/0112075].

\bibitem{Moderski:2001tk}
R.~Moderski and M.~Rogatko,
\newblock Phys. Rev. {\bf D64}, 044024 (2001), [gr-qc/0105056].

\bibitem{Burko:2004jn}
L.~M. Burko and G.~Khanna,
\newblock Phys. Rev. {\bf D70}, 044018 (2004), [gr-qc/0403018].

\bibitem{Moderski:2005hf}
R.~Moderski and M.~Rogatko,
\newblock Phys. Rev. {\bf D72}, 044027 (2005), [hep-th/0508175].

\bibitem{Bekenstein:1974sf}
J.~D. Bekenstein,
\newblock Ann. Phys. {\bf 82}, 535 (1974).

\bibitem{Winstanley:2002jt}
E.~Winstanley,
\newblock Found. Phys. {\bf 33}, 111 (2003), [gr-qc/0205092].

\bibitem{Winstanley:2005fu}
E.~Winstanley,
\newblock Class. Quant. Grav. {\bf 22}, 2233 (2005), [gr-qc/0501096].

\bibitem{Radu:2005bp}
E.~Radu and E.~Winstanley,
\newblock Phys. Rev. {\bf D72}, 024017 (2005), [gr-qc/0503095].

\bibitem{Cardoso:2003cj}
V.~Cardoso, R.~Konoplya, and J.~P.~S. Lemos,
\newblock Phys. Rev. {\bf D68}, 044024 (2003), [gr-qc/0305037].

\bibitem{Eling:2006df}
C.~Eling and T.~Jacobson,
\newblock Class. Quant. Grav. {\bf 23}, 5625 (2006), [gr-qc/0603058].

\bibitem{Eling:2006ec}
C.~Eling and T.~Jacobson,
\newblock Class. Quant. Grav. {\bf 23}, 5643 (2006), [gr-qc/0604088].

\bibitem{Unruh:1980cg}
W.~G. Unruh,
\newblock Phys. Rev. Lett. {\bf 46}, 1351 (1981).

\bibitem{Unruh:1994je}
W.~G. Unruh,
\newblock Phys. Rev. {\bf D51}, 2827 (1995).

\bibitem{Barcelo:2005fc}
C.~Barcelo, S.~Liberati, and M.~Visser,
\newblock Living Rev. Rel. {\bf 8}, 12 (2005), [gr-qc/0505065].

\bibitem{Okuzumi:2007hf}
S.~Okuzumi and M.-a. Sakagami,
\newblock Phys. Rev. {\bf D76}, 084027 (2007), [gr-qc/0703070].

\end{thebibliography}
\end{document}